%% file: 41114_fjgg_arxiv_corr.tex
\documentclass{aa}  
\usepackage{rotating}
\usepackage{graphicx, multirow, tablefootnote, longtable}
\usepackage{times}
\usepackage{txfonts}
\usepackage{xspace}
\usepackage{epsfig}
\usepackage{natbib}
\usepackage{dcolumn}
\usepackage[caption=false]{subfig}
\usepackage{xcolor, tikz}
\usepackage{hyperref}
\usepackage{pdflscape}                 
\usepackage{tabularx}
\usepackage[percent]{overpic}
\usepackage{ulem}

\begin{document}

   \title{Lithium depletion boundary,  
          stellar associations, and 
          \textit{Gaia}\thanks{Table E.4. is only available in electronic form at the CDS via anonymous ftp to cdsarc.u-strasbg.fr (130.79.128.5) or via http://cdsweb.u-strasbg.fr/cgi-bin/qcat?J/A+A/} 
         }
                                                          
\author{F. J. Galindo-Guil      \inst{ 1, 2}
\and    D. Barrado              \inst{ 3}
\and    H. Bouy                 \inst{ 4}
\and    J. Olivares             \inst{ 4, 5}
\and    A. Bayo                 \inst{ 6, 7, 8}
\and    M. Morales-Calder\'on   \inst{ 3}
\and    N. Hu\'elamo            \inst{ 3}
\and    L. M. Sarro             \inst{ 9}
\and    P. Rivi\`ere-Marichalar \inst{10}
\and    H. Stoev                \inst{11}
\and    B. Montesinos           \inst{ 3}
\and    J. R. Stauffer          \inst{12, $\textdagger$} \thanks{\textdagger\ Deceased\ : https://baas.aas.org/pub/2021i0306/release/1}
}

   \authorrunning{F. J. Galindo-Guil et al.}
   \offprints{pgalindo@hotmail.cl; pgalindo@cefca.es}
\institute{
Nordic Optical Telescope, Rambla Jos\'e Ana Fern\'andez P\'erez 7, E-38711 San Antonio, Bre\~na Baja, Santa Cruz de Tenerife, Spain.\relax                                  \and
Centro de Estudios de F\'isica del Cosmos de Arag\'on (CEFCA), Plaza San Juan 1, 44001 Teruel, Spain.\relax                                                                 \and
Departamento de Astrof\'isica, Centro de Astrobiolog\'ia (CAB, CSIC-INTA), ESAC Campus, Camino Bajo del Castillo s/n, 28692 Villanueva de la Ca\~nada, Madrid, Spain.\relax \and
Laboratoire d’astrophysique de Bordeaux, Univ. Bordeaux, CNRS, B18N, All\'ee Geoffroy Saint-Hilaire, 33615 Pessac, France.\relax                                            \and
Instituto de Astrof\'isica de Canarias, E-38205 La Laguna, Tenerife, Spain; Universidad de La Laguna, Dpto. Astrof\'isica, E-38206 La Laguna, Tenerife, Spain.\relax        \and
Inst. F\'isica y Astronom\'ia, Fac. Ciencias, Universidad de Valpara\'iso, Gran Breta\~na 1111, Valpara\'iso, Chile.\relax                                                  \and
N\'ucleo Milenio Formaci\'on Planetaria - NPF, Valpara\'iso, Chile\relax                                                                                                    \and  
European Organisation for Astronomical Research in the Southern Hemisphere (ESO), Karl-Schwarzschild-Str. 2, 85748, Garching bei M\"unchen, Germany.\relax                   \and
Departamento de Inteligencia Artificial, UNED, Juan del Rosal 16, 28040 Madrid, Spain.\relax                                                                                \and  
Observatorio Astron\'omico Nacional (OAN-IGN) -- Observatorio de Madrid, Alfonso XII, 3, 28014 Madrid, Spain.\relax                                                         \and
Fundaci\'on Galileo Galilei -- INAF, Rambla Jos\'e Ana Fern\'andez P\'erez 7, 38712 Bre\~na Baja, Santa Cruz de Tenerife, Spain.\relax                                      \and
Spitzer Science Center, California Institute of Technology, Pasadena, CA 91125, USA.\relax                                                                                  
}

\titlerunning{Lithium depletion boundary and \textit{Gaia}}


  \abstract
   { 
     Stellar ages are key to improving our understanding of different astrophysical phenomena. 
     However, many techniques to estimate stellar ages are highly model-dependent. 
    The lithium depletion boundary (LDB), based on the presence or absence of lithium in low-mass stars, 
    can be used to derive ages in stellar associations of between 20 and 500~Ma. 
   }
   {
     The purpose of this work is to revise former LDB ages in stellar associations  
    in a consistent way, taking advantage of the homogeneous \textit{Gaia} parallaxes as well as 
    bolometric luminosity estimations that do not rely on monochromatic bolometric corrections.        
   }
   {
      We studied nine open clusters and three moving groups characterised by a previous determination of the LDB age.    
    We gathered all the available information from our data and the literature: 
    membership, distances, photometric data, reddening, metallicity, and surface gravity. 
      We re-assigned membership and calculated bolometric luminosities and effective temperatures 
    using distances derived from \textit{Gaia} DR2 and multi-wavelength photometry for individual objects around the former LDB. We located the LDB using a homogeneous method for all the stellar associations.   
      Finally, we estimated the age by comparing it with different evolutionary models.  
   }
   {
      We located the LDB for the twelve stellar associations and derived their ages using  
    several theoretical evolutionary models. 
      We compared the LDB ages among them, along with data obtained with other techniques, 
    such as isochrone fitting, ultimately finding some discrepancies among the various approaches. 
     Finally, we remark that the 32 Ori MG is likely to be composed 
    of at least two populations of different ages.
   }
   {
   }

   \maketitle

\section{Introduction}\label{intro}
One of the most fundamental parameters necessary for the characterisation of any astrophysical object, 
from planets, stars, and galaxies up through the whole universe, is age.
  There are only two true pillars holding up the scale of astrophysical age. 
 The first is the age of the Sun, which comes from radiometric measurements of 
the oldest Solar System material, which yields $4.55\pm0.07$~Ga 
(see \citealt{c_patterson1956} or the recent value $4.5672\pm0.6$~Ga from \citealt{y_amelin2012d}).
 The second is the age of the universe, with an estimate of 13.8 Ga, 
based on the Lambda cold dark matter cosmological model 
($\Lambda$CDM, \citealt{planck2018_part06}).

    All other ages are subject to significant uncertainties, 
and their determinations face multiple observational and theoretical challenges. 
  Ages are generally better determined in stellar associations\footnote{We use 
the term stellar association to refer both to open clusters and moving groups.
 A moving group is a stream of stars with common age and motion through the Milky Way
and with no overdensity of stars discernible in any region \citep{zuckerman2004c}. 
 An open cluster is a group of gravitationally bound stars that shows a clear concentration 
above the surrounding stellar background \citep{e_moraux2016c}.}
than in isolated stars and, more precisely, open clusters are very convenient for this purpose. 
They usually provide a significant sample of objects when testing the hypothesis that 
they were born simultaneously from the same molecular cloud and with an identical composition. 
  Two excellent and complete summaries of techniques for estimating ages in  
ensembles of stars can be found in \cite{mermilliod2000a} and 
\cite{soderblom2010c}.

    There are several methods and techniques aimed at assigning ages at present, however,
each of them is valid within a specific age interval. 
  Ages can vary as much as $50\%$ depending on the technique, 
as in the case of stellar associations with ages below 150~Ma 
(see \citealt{stauffer1995a} or \citealt{barrado1999b}).
  Even when using the same technique, there can be broad variations depending solely on 
the different grids of theoretical models.

    Several features and characteristics are taken into account when deriving ages in stellar associations.
The isochrone fitting technique is the most widely used 
since it covers all ages and masses, making use of colour-magnitude or Hertzsprung-Russell diagrams (HRDs).  
The turn-off isochrone fitting consists of reproducing the morphology of the cluster sequence close to the turn-off region,
  which is where stars are approaching the end of hydrogen burning at their core.
 The pre-main-sequence (PMS) isochrone fitting consists of reproducing 
the position of PMS stars in the cluster.
  However, there are significant deviations 
between ages derived using 
PMS stars still contracting onto the main sequence, and 
high-mass ones that have already evolved away from it 
(\citealt{w_lyra2006}, \citealt{t_naylor2009}). 
  These differences can be due to the effects of binarity and rotation of the upper main-sequence stars  
\citep{mermilliod2000a} or due to the fact that 
the effective temperatures predicted by current PMS evolutionary models are overestimated 
due to the magnetic fields and star-spots 
(\citealt{f_dantona2000e}, \citealt{stauffer2003}, \citealt{rj_jackson2014a}, and \citealt{e_franciosini2022}).
  The rotation of massive stars increases the estimated age of a stellar association 
about $25\%$ \citep{g_meynet2000e}, with this effect being more pronounced in younger associations, 
such as Alpha Persei and the Pleiades, than in older ones such as Coma Berenices, Praesepe, or the Hyades 
\citep{a_maeder1971b}.
  The isochrone fitting ages estimated from models that take into account 
the magnetic field are systematically higher 
than those estimated from models that ignore this effect \citep{l_malo2014b}.

    There are other methods used to derive stellar ages. 
  The kinematic age technique is based 
on the motions of a group of stars due to its expansion 
from their birthplace without experiencing any forces 
(see the seminal works from \citealt{va_ambartsumian1946} and \citeyear{va_ambartsumian1949}, 
\citealt{a_blaauw1946}, and \citeyear{a_blaauw1964}, and 
the subsequent works with a refined methodology: \citealt{aga_brown1997b} or \citealt{n_miretroig2018}).
This technique is applicable to stellar associations younger than 20~Ma \citep{soderblom2010c}.

    The Str\"omgen photometric system $uvby\beta$ allows us to derive 
effective temperatures and surface gravities for BAF-type stars and 
they are used to estimate the age 
comparing with theoretical evolutionary tracks 
(see \citealt{i_song2001b} and \citealt{tj_david2015a}). 
  Also, the strength of gravity-sensitive spectrophotometric features 
is an age proxy in stellar associations within the 5 to 15~Ma range, 
with an uncertainty of 2~Ma \citep{wa_lawson2009b}.
  Another technique, namely, gyrochronology, derives ages for main-sequence and low-mass stars 
using their rotation periods and colours, for ages older than 600~Ma \citep{sa_barnes2007}.
  Finally, the white dwarf cooling age exploits the slope and position
of its thermal evolution relative to the main-sequence with an age range of validity 
from 150~Ma to 4 Ga \citep{s_degennaro2009}.

    The lithium depletion boundary (LDB) 
is a technique that allows for an independent age determination 
for a stellar association \citep{basri1996}, valid for values between 20 to few hundred Ma. 
  It is based on the presence or absence of lithium in low-mass stars and brown dwarfs \citep{rebolo1992}.  
Lithium burns inside the stellar nucleus (\citealt{f_dantona1984})
and stars with masses below about $0.3 M_{\odot}$ are fully convective.
The convection makes that the lithium surface abundance to decrease with age, 
and for low-mass objects, the decay is very rapid. 
For masses below \mbox{$\sim 0.07 M_{\odot}$} (depending on the evolutionary theoretical model), 
the core temperature is never hot enough to destroy it, and its abundance
(both internal or in the stellar or sub-stellar atmosphere) remains the same. 
The burning process is so fast and depends so strongly on mass that there is a sharp boundary between 
lithium-rich and lithium-poor members (\citealt{f_dantona1994}, \citealt{l_bildsten1997d}, and \citealt{g_ushomirsky1998b}).  
  The LDB technique was successfully applied to 
the Pleiades \citep{stauffer1998b} and Alpha Persei \citep{stauffer1999} clusters,  
and since then, this method has been used in a dozen of stellar associations.

    Similarly to other model-dependant techniques, 
the LDB age is also affected by possibly assumptions made in the input physics 
of the stellar models it relies on
(\citealt{jeffries2001a}, \citealt{cj_burke2004}, and \citealt{e_tognelli2015b}).
The expected uncertainties on LDB age depends on the luminosity of the star at the LBD,
ranging from about  $3\%$ percent for faint old stars to about $15\%$ for brighter young stars
(\citealt{cj_burke2004} and \citealt{e_tognelli2015b}).
  In addition, the LDB technique is indirectly affected by other uncertainties: 
(a) stellar distances, 
(b) the contamination by field stars, 
(c) the transformation from observational quantities, magnitude, and colours, as well as theoretical ones such as bolometric luminosities and effective temperatures \citep{jeffries2001a}, 
(d) the presence of binaries or multiple systems,     
(e) the magnetic activity and rotation of the objects 
(\citealt{soderblom1993h_SJB93h}, \citealt{g_chabrier2007c}, \citealt{l_malo2014b}), 
(f) the identification of a significant number of low-mass cluster members within a small magnitude range, and 
(g) the low quality and/or low resolution of the used stellar spectra. 

    The availability of photometric all-sky or wide sky surveys, such as \textit{Gaia} DR2 \citep{aga_brown2018}, 
SDSS \citep{s_alam2015a}, 2MASS \citep{skrutskie2006}, WISE \citep{cutri2013}, or Pan-STARRS \citep{panstarrs1} 
has signalled the dawning of a new era and allowed us to address several 
of the factors mentioned above from a new perspective.
  The high-quality astrometric data of \textit{Gaia} DR2 represent a significant advance for at least three reasons: 
they allow us 
(a) to calculate, for the first time, individual distances for each low-mass cluster member, 
(b) to refine the membership criteria, and in some cases 
(c) to determine whether an object is a multiple system.
  Furthermore, the use of wide-area multi-wavelength imaging surveys with ranges from ultraviolet to infrared 
allows us to calculate the total flux of each object 
without assuming any monochromatic bolometric correction, 
which is a noteworthy source of systematic uncertainties in the LDB age.
Empirical bolometric corrections are derived from main-sequence
stars and our objects have different $\log {\rm g}$ values; in addition, bolometric
corrections vary as a function of metallicity, especially for temperatures less than $3\,000$~K 
(\citealt{jeffries2001a}, \citealt{cj_burke2004}, and \citealt{e_tognelli2015b}).

    In this work, we present, for the first time, a homogeneous analysis for all the stellar associations 
with former LDB ages aimed at creating a complete set of an LDB age scale.
  This analysis includes new photometric data in a broad range of wavelengths, 
individual accurate parallaxes from \textit{Gaia} DR2, and the effects of metallicity and gravity. 
    The work is structured as follows.
  In Section \ref{sec:stellarassociations_sample}, 
we introduce the stellar associations under consideration. 
  We provide information of the data and evolutionary models 
used in Section \ref{sec:data_and_models}.
  Section \ref{sec:method_ldb} shows the method scheme in detail. 
  We analyse each stellar association, 
explain additional information from literature, identify outliers, 
and locate the LDB in Section \ref{sec:analysis_ldb}.
  Section \ref{sec:ldb_error_budget} covers 
the effects of systematic uncertainties on the derived LDB 
including a comparison between the LDB ages and those derived from other techniques.
  Section \ref{sec:conclusions_summary} presents our conclusions.
    Additional material is presented in the Appendices. 
In Appendix \ref{app:plots_stellarassociations}, we made  all the plots for each stellar association available, along with the vector point diagrams (VPDs), parallax distributions, and HRDs. 
In Appendix \ref{app:ldb_av}, we show the effect on the LDB of the adopted reddening. 
A short description for some objects belonging to the Beta Pictoris moving group are given in Appendix \ref{app:bpmg_objects},  
and a brief study of the 32 Ori moving group in Appendix \ref{app:32orimg_one_or_two}. 
Partial tables with data (names, astrometry, lithium equivalent widths, photometry, and quantities calculated) are given in Appendix \ref{app:additional_tables}.

\section{Stellar associations sample}\label{sec:stellarassociations_sample}

For our study, we considered twelve stellar associations with a previous LDB age estimation.
Nine of them are young open clusters and three are moving groups. 
Table \ref{tab:sample_clusters} includes the main properties of the studied targets that are briefly discussed below.

\subsection{Alpha Persei}\label{sub:alphaper}

Alpha Persei is an open cluster located at a distance of \mbox{$175$~pc} 
with an age of $71$~Ma \citep{c_babusiaux2018}
and with chemical composition values ranging from $\rm{[Fe/H]} =-0.054\pm0.046$~dex \citep{am_boesgaard1990}
to $\rm{[Fe/H]}=+0.18$~dex \citep{h_pohnl2010}.
\cite{mermilliod2008c} noticed the existence of a extended co-moving stream of stars, 
recently confirmed and characterised by \cite{vv_nikiforova2020}.
The age ranges between 50 Ma to 80 Ma: 
 $51.3$~Ma \citep{mermilliod1981a},     
 52~Ma assuming $E(B-V)=0.055$~mag \citep{vv_makarov2006a}, 
 $80\pm10$~Ma \citep{prosser1992}, 
 and $56^{+2}_{-2}$~Ma, assuming $(m-M)_{0}=6.21$~mag and $A_{V}=0.28$~mag \citep{d_bossini2019}.
  
    Regarding the LDB technique, 
  \citet{BM99} computed a lower age limit of $66$~Ma based on one object, AP270.
  \citet{stauffer1999} using a large sample, calculated $90\pm10$~Ma 
assuming $176$~pc and $A_{I}=0.17$~mag, $E(R-I)_{C}=0.07$~mag. 
  Later, \citet{barrado2004d} recomputed it and yielded $85\pm10$~Ma 
assuming $(m-M)_{0}=6.23$~mag and $E(B-V)=0.096$~mag. 
  Finally, \cite{cj_burke2004} derived $87\pm9$~Ma, 
using a $(m-M)=6.23\pm0.10$~mag and $A_{I}=0.17\pm0.04$~mag.

\subsection{NGC 1960 (M 36)}\label{sub:ngc1960}

NGC 1960 is a northern hemisphere open cluster located at 1\,157~pc \citep{cantatgaudin2018b} 
with an estimated age of 25~Ma (\citealt{d_bossini2019}, 
assuming $(m-M)_{0}=10.256^{+0.020}_{-0.019}$~mag and $A_{V}=0.657$~mag).
Several works have determined the distance, age, and reddening of NGC 1960: 
 $\leq 22.4$~Ma \citep{mermilliod1981a}, 
 $16^{+10}_{-5}$~Ma, assuming $(m-M)_{0}=10.6\pm0.2$~mag ($1\,318\pm120$~pc) and $E(B-V)=0.25\pm0.02$~mag \citep{j_sanner2000}, 
 $25$~Ma assuming $1\,330$~pc and $E(B-V)=0.22$~mag \citep{s_sharma2006}, 
 20~Ma using $(m-M)=10.33^{+0.02}_{-0.05}$ and $E(B-V)=0.20$~mag \citep{nj_mayne2009}, 
 and finally $26.3^{+3.2}_{-5.2}$~Ma 
and $20^{+1}_{-1}$~Ma 
\citep{cpmbell2013}. \cite{jeffries2013} calculated \textbf{a} LDB age of $22\pm4$~Ma, 
assuming a distance modulus of 10.33~mag ($1\,164^{+11}_{-26}$~pc) 
and $E(B-V)=0.20$~mag.

\subsection{IC 4665}\label{sub:ic4665}

IC 4665 is located at 346~pc \citep{c_babusiaux2018}, 
with an age of $38^{+5}_{-2}$~Ma \citep{d_bossini2019} and 
a metallicity of $\rm{[Fe/H]}=-0.03\pm0.04$ \citep{ws_dias2002e}. The age ranges between 25 to 50~Ma:
 $36.3$~Ma \citep{mermilliod1981a}, 
 $25\pm5$~Ma \citep{manzi2008} 
 $36\pm9$~Ma, assuming a distance of $360\pm12$~pc 
and $42\pm12$~Ma, with a distance of $357\pm12$~pc 
\citep{pa_cargile2010b}, 
 and between 31 to 34~Ma \citep{randich2018}.
\cite{manzi2008} computed an LDB age of $27.7^{+4.2}_{-3.5}\pm1.1\pm2$~Ma, 
assuming $370\pm50$~pc and $E(B-V)=0.18$~mag ($A_{I_{c}}=0.333$~mag). 
 \cite{randich2018} re-calculated the LDB age, $23.2^{+3.5}_{-3.1}$~Ma, 
using $(m-M)_{0}=7.82\pm0.25$~mag and $E(B-V)=0.226\pm0.080$~mag.

\subsection{NGC 2547}\label{sub:ngc2547}

NGC 2547 is located in the Vela complex \citep{oj_eggen1986f} 
at 394~pc with $E(B-V)=0.040$~mag, 
an age of $40^{+11}_{-10}$~Ma \citep{c_babusiaux2018}, 
and a metallicity of 
$\rm{[Fe/H]}=-0.16\pm0.09$ \citep{ws_dias2002e}. The NGC 2547 age ranges between 14~Ma to 57~Ma:   
  $14\pm4$~Ma 
and $55\pm25$~Ma 
(\citealt{jeffries1998c}, using a distance modulus of $8.1$~mag and $E(B-V)=0.06$~mag), 
  an age range between 20 to 35~Ma 
(\citealt{t_naylor2002}, assuming a distance modulus range of $8.00-8.15$~mag), 
  two age ranges, from 32 to 50~Ma and from $63\pm30$ to $75\pm25$~Ma, depending on the technique 
(\citealt{w_lyra2006}, assuming $E(B-V)=0.03\pm0.02$~mag and $390\pm25$~pc),
  $38$~Ma 
and $48^{+14}_{-21}$~Ma 
(\citealt{t_naylor2009}, assuming $(m-M)_{0}=8.03$~mag and $E(B-V)=0.04$~mag),
  36-40~Ma \citep{randich2018}, and   
  $27^{+1}_{-1}$~Ma (\citealt{d_bossini2019} assuming $(m-M)_{0}=7.98$~mag, and $A_{V}=0.124$~mag. 
  
    \cite{jm_oliveira2003b} made the first attempt to locate the LDB     
resulting in an age range between $35-54$~Ma.
 Later, \cite{jeffries2005} calculated an LDB age of $34-36$~Ma, 
assuming $(m-M)_{0}=8.10\pm0.10$~mag and $E(B-V)=0.06\pm0.02$~mag.   
 \cite{cj_burke2004} calculated an age of $38\pm7$~Ma, 
using $(m-M)=8.15\pm0.15$~mag and $A_{I}=0.05\pm0.03$~mag. 
 Finally \cite{randich2018} re-calculated the LDB age 
using $(m-M)_{0}=7.81\pm0.25$~mag and $E(B-V)=0.080\pm0.024$~mag: 
$37.7^{+5.7}_{-4.8}$~Ma.

\subsection{IC 2602}\label{sub:ic2602}

IC 2602, the brightest Southern Hemisphere open cluster, is located 
at a distance 152~pc ($6.571\pm0.007$~mas), 
$E(B-V)=0.031$~mag, 
with an age of $40^{+11}_{-10}$~Ma \citep{c_babusiaux2018} and 
$\rm{[Fe/H]}=-0.05\pm0.05$ \citep{randich2001a}. 

    The age ranges between 25 to 63.2~Ma. 
Other values are: $36.3$~Ma \citep{mermilliod1981a}, 
 25~Ma 
and 35~Ma 
(\citealt{stauffer1997b} depending on the technique),
 $44^{+18}_{-16}$~Ma (\citealt{t_naylor2009}, assuming a distance modulus of $5.88$~mag, $150$~pc and $E(B-V)=0.02$~mag), 
 29-32~Ma \citep{randich2018}, and 
 $35.3^{+1.3}_{-1.1}$~Ma (\citealt{d_bossini2019}, assuming $(m-M)_{0}=5.91$~mag and $A_{V}=0.10$~mag). 

   \cite{dobbie2010} computed an LDB age of $46^{+6}_{-5}$~Ma 
assuming $(m-M)_{0}=5.86\pm0.1$~mag (149~pc), and $E(B-V)=0.035$~mag. 
 \cite{randich2018} recalculated the LDB age using a distance modulus of 
$5.85\pm0.10$~mag and \mbox{$E(B-V)=0.068\pm0.025$~mag:} 
$43.7^{+4.3}_{-3.9}$~Ma.

\subsection{IC 2391}\label{sub:ic2391}

This cluster is located at a distance of 152~pc ($6.597\pm0.007$~mas), 
$E(B-V)=0.030$~mag, with an age of $50^{+14}_{-13}$~Ma \citep{c_babusiaux2018} 
and $\rm{[Fe/H]}=-0.03\pm0.07$ \citep{randich2001a}. It is an open cluster with similar age and distance to IC 2602
\citep{stauffer1997b}, 
but they are located in different sky regions, see Fig.\ref{fig:pos_ic2602_ic2391}.
  Several works have studied both clusters in a parallel way: 
\cite{stauffer1997b}, 
\cite{randich2001a}, 
and \cite{dorazzi2009} 

    The IC 2391 age ranges between 25 to 50~Ma:
$36.3$~Ma \citep{mermilliod1981a}, 
$25$ to $35$~Ma isochrone fitting age \citep{stauffer1997b}, 
40-45~Ma \citep{randich2018}, and 
$36.4^{+2.0}_{-2.0}$~Ma, assuming $(m-M)_{0}=5.91$~mag and $A_{V}=0.09$~mag \citep{d_bossini2019}.
    
    The LDB age was initially estimated by \cite{barrado1999b}: 
$53\pm5$~Ma, assuming $A_{I}=0.02$~mag and $(m-M)_{0}=5.95\pm0.1$~mag. 
A further refinement in \cite{barrado2004d} produced a value of $50\pm5$~Ma, 
assuming the same distance and $E(B-V)=0.06$~mag. 
  \cite{cj_burke2004} calculated an age of $48\pm5$~Ma, 
using a $(m-M)=5.95\pm0.10$~mag and $A_{I}=0.02\pm0.02$~mag. 
  \cite{randich2018} re-computed the LDB age 
using $(m-M)_{0}=5.82\pm0.10$~mag and $E(B-V)=0.088\pm0.027$~mag: 
$51.3^{+5.0}_{-4.5}$~Ma.

\subsection{The Pleiades}\label{sub:thepleiades}

The Pleiades cluster is located at $136$~pc ($7.364\pm0.005$~mas), $E(B-V)=0.045$~mag,  
and an age of $110^{+22}_{-14}$~Ma \citep{c_babusiaux2018}. 
Metallicity values, $\rm{[Fe/H]}$, 
range from \mbox{$-0.034\pm0.024$} \citep{am_boesgaard1990} 
to \mbox{$0.06\pm0.06$} \citep{m_gebran2008b}. 
  Excellent summaries of previous works can be found in 
\cite{stauffer2007}, \cite{h_bouy2015b}, and \cite{n_lodieu2019c}.

    Although it has been widely studied, its distance and age have been a matter of controversy, 
ranging the distance between 115 to 140~pc, depending on the work and the technique used:  
$115.9\pm2.7$~pc \citep{vanleeuwen1997f}, 
$131.8\pm2.5$~pc \citep{mh_pinsonneault1998}, 
$139.1\pm3.5$~pc \citep{j_southworth2005a}, 
$120.2\pm1.9$~pc assuming $E(B-V)=0.04$~mag \citep{vanleeuwen2009}, 
$136.2\pm1.2$~pc \citep{c_melis2014a}, 
and $134.4^{+2.9}_{-2.8}$~pc \citep{pab_galli2017}. 
    The age ranges between 70 to 150 Ma:
$77.6$~Ma \citep{mermilliod1981a}, $150$~Ma \citep{mazzei1989} 
$100$~Ma, assuming $(m-M)_{0}=5.60$~mag and $E(B-V)=0.04$~mag \citep{g_meynet1993}, 
$70\pm10$~Ma \citep{stauffer1995b}\footnote{The uncertainty is an estimate from their Figure 5c and 5d.}, 
$115^{+3}_{-11}$~Ma, assuming $(m-M)_{0}=5.35$~mag and $E(B-V)=0.02$~mag \citep{t_naylor2009}, 
$132\pm2$~Ma \citep{cpmbell2014}, 
$134^{+9}_{-10}$~Ma \citep{pa_cargile2014}, 
$86^{+6}_{-3}$~Ma, assuming $(m-M)_{0}=5.67$~mag and $A_{V}=0.14$~mag \citep{d_bossini2019}, 
and $132^{+26}_{-27}$~Ma, assuming $E(B-V)=0.045$~mag \citep{n_lodieu2019c}.

    \cite{basri1996} attempted to determine the LDB age and reported a lower limit of $\sim115$~Ma 
based on the detection of lithium in the binary system PPL-15 \citep{basri1999c}.
 Later, the exhaustive study of \cite{stauffer1998a}, yielded a value of $125\pm8$~Ma
assuming $(m-M)_{0}=5.60$~mag and $A_{I}=0.06$~mag.  
  In addition, \cite{barrado2004d} re-calculated the value at $130\pm20$~Ma, 
 whereas \cite{cj_burke2004} re-calculated an age of $126\pm11$~Ma, 
both using a $(m-M)=5.60\pm0.10$~mag and $A_{I}=0.06\pm0.03$~mag. 
  Finally, \cite{se_dahm2015} derived an age of $112\pm5$~Ma, 
assuming $136.2\pm1.2$~pc \citep{c_melis2014a} and 
$A_{V}=0.12$~mag, $A_{I}=0.06$~mag, $A_{J}=0.03$~mag, and $A_{K}=0.01$~mag.

\subsection{Blanco 1}\label{sub:blanco1}

Blanco 1 is located at $237$~pc with $E(B-V)=0.001$~mag and 
an age of $115^{+23}_{-15}$~Ma \citep{c_babusiaux2018}\footnote{
  They also report an age of $200^{+ 40}_{- 26}$~Ma, 
  but with this value the lower main sequence is poorly fitted.
}. 
It has an estimated metallicity of $\rm{[Fe/H]}=+0.03\pm0.07$\footnote{This 
metallicity value has been a matter of controversy,
see: \citealt{be_westerlund1988c}, \cite{b_edvardsson1995}, and \cite{a_ford2005}.}
\citep{m_netopil2016}.
The age ranges between 90 to 200~Ma: 
 $90\pm25$~Ma \citep{pm_panagi1997}, 
 $80\pm20$~Ma (\citealt{pa_cargile2009}, assuming $240\pm10$~pc and $E(B-V)=0.016$~mag), 
 $94^{+5}_{-7}$~Ma (\citealt{d_bossini2019}, assuming $(m-M)_{0}=6.876$~mag and $A_{V}=0.031$~mag), 
 and $146^{+13}_{-14}$~Ma \citep{pa_cargile2014}. 
 
    \cite{pa_cargile2010c} identified the LDB at 
\mbox{$\log L_{\rm bol}=-2.94\pm0.12$ $L_{\sun}$} or 
\mbox{$\log L_{\rm bol}=-2.90\pm0.09$ $L_{\sun}$} (depending on monochromatic bolometric corrections) 
and determined an age of $132\pm24$~Ma, 
assuming a distance modulus of $6.58\pm0.12$~mag and 
$E(I-K_{s})=0.02$~mag ($A_{I}=0.03$~mag). 
  Later, \cite{aj_juarez2014} used empirical activity corrections 
to investigate the effects of magnetic activity on the LDB age. 
They updated the LDB age to $126^{+13}_{-14}$~Ma, and 
taking into account the activity, they gave a value of $114^{+9}_{-10}$~Ma.

\subsection{The Hyades}\label{sub:thehyades}

The Hyades cluster is the nearest open cluster to the Sun, 45~pc \citep{jhj_de_bruijne2001}. 
It has an estimated age of $625\pm50$~Ma \citep{mac_perryman1998}
and $\rm{[Fe/H]}=+0.127\pm0.022$ \citep{am_boesgaard1990}. 
Metallicity ranges from $\rm{[Fe/H]}=+0.05\pm0.05$ \citep{m_gebran2010} to 
$\rm{[Fe/H]}=+0.13\pm0.05$, derived using 61 sources \citep{m_netopil2016}.
  Additional information can be found in \cite{mac_perryman1998} and \cite{n_lodieu2019b}.
Recently, \cite{s_meingast2020a} and \cite{s_roser2020a} 
reported the detection of the tidal tails.
The age ranges between  500 to 1\,200~Ma: 
$661$~Ma \citep{mermilliod1981a}, 
an age range between 500 to 1\,000~Ma \citep{oj_eggen1998d}, 
$648\pm+45$~Ma \citep{s_degennaro2009}, 
$800$~Ma \citep{td_brandt2015a}, 
$794^{+161}_{-102}$~Ma \citep{c_babusiaux2018}, 
and $640^{+67}_{-49}$~Ma \citep{n_lodieu2019b} \cite{el_martin2018} identified the LDB in the Hyades and derived an age of $650\pm70$~Ma.     
  More recently, \cite{n_lodieu2018c} re-analysed the age adding to the study the source 2M0418
and inferred an age range of 
$580$-$775$~Ma (mean of 700 Ma) from the bolometric luminosity, and 
$580$-$950$~Ma (mean of 760 Ma) from the effective temperature.

\subsection{\textbf{Beta Pictoris moving group}}\label{sub:bpmg}

The Beta Pictoris moving group, hereafter BPMG, named after the star $\beta$ Pictoris,   
is a $20\pm10$~Ma co-moving group of stars \citep{barrado1999a}
with a metallicity of $\rm{[Fe/H]}=-0.01\pm0.08$ \citep{vianaalmeida2009}.
Several studies have addressed the BPMG age with different results: 
$20\pm10$~Ma \citep{barrado1999a}, 
$12^{+8}_{-4}$~Ma \citep{zuckerman2001d}, 
$11.5$~Ma \citep{vg_ortega2002}, 
$\sim12$~Ma \citep{i_song2003}, 
15~Ma \citep{cao_torres2006}, 
$22\pm12$~Ma \citep{vv_makarov2007c}, 
an age range between $13\pm5$~Ma 
 to $21\pm9$~Ma 
 \citep{e_mentuch2008b}, 
$22\pm3$~Ma \citep{mamajek2014} 
$24\pm3$~Ma \citep{cpmbell2015} 
40~Ma \citep{j_macdonald2010}, 
an age range between 15 and 28~Ma \citep{l_malo2014b}, 
$13^{+7}_{-0}$~Ma \citep{n_miretroig2018} 
which was updated to $18.5^{+2.0}_{-2.4}$~Ma \citep{n_miretroig2020b}.
    
    \cite{i_song2002c} estimated the LDB in $<20$~Ma. 
  Then, \cite{as_binks2014} added to the study eight low-mass members and   
re-calculated the age: $21\pm4$~Ma with an additional model-dependent uncertainty of $\pm1$~Ma. 
  \cite{s_messina2016d}, after de-correlating the lithium equivalent width from the rotation period, 
calculated an age of $25\pm3$~Ma using the hot side of the LDB.
  Finally, other LDB ages have also been calculated: 
$\geq30$~Ma \citep{jc_yee2010},   
$26\pm3$~Ma \citep{l_malo2014b} and 
$22\pm6$~Ma \citep{e_shkolnik2017}. 

\subsection{\textbf{Tucanae-Horologium moving group}}\label{sub:thmg}

At first, the Tucanae-Horologium moving group (THMG) was identified as two separate associations: 
the Tucanae association \citep{zuckerman2000a} 
with ten co-moving stars placed at a distance $\sim45$~pc, 
and the Horologium association \citep{cao_torres2000} 
located at $\sim60$~pc. 
 \cite{vianaalmeida2009} reported a metallicity of $\rm{[Fe/H]}=-0.03\pm0.05$ using nine members.
The age ranges between 5-45~Ma: 
$40$~Ma \citep{zuckerman2000a}, 
$\sim30$~Ma \citep{cao_torres2000}, 
an age range between 10-30~Ma \citep{b_stelzer2000b}, 
$\sim20$~Ma \citep{al_kraus2014c}, 
$45\pm4$~Ma \citep{cpmbell2015}, 
and $5^{+23}_{-0}$~Ma \citep{n_miretroig2018}. \cite{al_kraus2014c} confirmed 129 new late-type members with spectral types between K3 to M6 and these authors calculated an LDB age of $40$~Ma.

\subsection{\textbf{32 Ori moving group}}\label{sub:32orimg}

The 32 Ori moving group, (32 Ori MG, also \textbf{known} as Mamajek 3), 
was noticed as a group of X-ray-bright late-type stars 
from the ROSAT All-Sky Survey \citep{jm_alcala2000a} with similar
proper motions and moving with the massive binary 32 Ori (B5V+B7V). 
 \cite{mamajek2007a} defined the moving group with about ten sources and 
estimated an age of 25 Ma (with no technique specifically given). 

    \cite{cpmbell2015} calculated an isochronal fitting PMS age of $22^{+4}_{-3}$~Ma 
adopting a distance of $91.86\pm2.42$~pc for all the stars. 
With the new census, \cite{cpmbell2017} calculated 
a PMS isochronal fitting age of $25\pm5$~Ma 
using $E(B-V)=0.03\pm0.01$~mag ($A_{V}=0.10\pm0.03$~mag) and 
the kinematic or trigonometric distances derived for each source. 

    \cite{cpmbell2017} calculated an LDB age of $23\pm4$~Ma
using $E(B-V)=0.03\pm0.01$~mag ($A_{V}=0.10\pm0.03$~mag) and 
the kinematic or trigonometric distances derived for each source. 
They combined both results, PMS age and LDB age, and released an age of 
$24\pm4$~Ma ($\pm4$~Ma statistical and $\pm2$~Ma systematic).

\setcounter{table}{0}
\begin{table*}
    \caption[Data from the literature for the twelve stellar associations with former LDB ages]
            {Data from the literature for the twelve stellar associations with LDB ages.}
    \label{tab:sample_clusters}
    \small{
    \center{
    \begin{tabular}{l cr  crcr rcc c cc}
\hline \hline    
Stellar     &    LDB         &Ref.&\multicolumn{7}{c}{\citet{c_babusiaux2018}}                                         &    [Fe/H]      & Other              &Ref. \\  
                                   \cline{4-10}                          
association &    age         &    &    Age                   &Ref.&      Parallaxes       &   DM &   $d$ &$E(B-V)$&Ref.&                &  age               &     \\       
            &   [Ma]         &    &    [Ma]                  &    &        [mas]          & [mag]&  [pc] &  [mag] &    &    [dex]       & [Ma]               &     \\  
\hline \hline
Alpha Persei& $90\pm10$      & (8)&\tiny{ $71^{+21}_{-18}$  }&(23)&\tiny{$5.718\pm0.005$} & 6.214 & 175  & 0.090  &(23)&\tiny{$+0.14\pm0.11$}&$ 51.3^{+12}_{- 8}$&(29)\\
\smallskip 
            & $85\pm10$      & (7)&                          &    &                       &       &      &        &    &                &                    &     \\
\hline
NGC 1960    & $22\pm4$       & (1)&\tiny{ $25$              }&(22)&\tiny{$0.835\pm0.064$} &10.391 &1\,197& 0.20   &(26)&             -  &$           <22.4 $ &(29) \\
\hline
IC 4665     & $28\pm4$       & (2)&\tiny{ $38^{+5}_{-2}$    }&(22)&\tiny{$2.891\pm0.003$} & 7.694 & 346  & 0.17   &(27)&\tiny{$-0.03\pm0.08$}&$ 36.3^{+12}_{- 8}$&(29)\\
\smallskip
        &$23.2^{+3.5}_{-3.1}$& (3)&                          &    &                       &       &      &        &    &                &                    &     \\ 
\hline
NGC 2547    & $35\pm1$       & (4)&\tiny{ $40^{+11}_{-10}$  }&(23)&\tiny{$2.543\pm0.002$} & 7.980 & 394  & 0.040  &(23)&\tiny{$-0.01\pm0.01$}&$ 57\pm5.7   $ &(30) \\
\smallskip
        &$37.7^{+5.7}_{-4.8}$& (3)&                          &    &                       &       &      &        &    &                &                    &     \\ 
\hline
IC 2602       &$46^{+6}_{-5}$& (5)&\tiny{ $40^{+11}_{-10}$  }&(23)&\tiny{$6.571\pm0.007$} & 5.914 & 152  & 0.031  &(23)&\tiny{$-0.02\pm0.02$}&$ 36.3^{+12}_{- 8}$&(29)\\
\smallskip 
        &$43.7^{+4.3}_{-3.9}$& (3)&                          &    &                       &       &      &        &    &                &                    &     \\
\hline
              &$53\pm5$      & (6)&                          &    &                       &       &      &        &    &                &                    &     \\ 
IC 2391       &$50\pm5$      & (7)&\tiny{ $50^{+14}_{-13}$  }&(23)&\tiny{$6.597\pm0.007$} & 5.908 & 152  & 0.030  &(23)&\tiny{$-0.01\pm0.03$}&$ 36.3^{+12}_{- 8}$&(29)\\
\smallskip 
        &$51.3^{+5.0}_{-4.5}$& (3)&                          &    &                       &       &      &        &    &                &                    &     \\
\hline  
              &$125\pm8$     & (9)&                          &    &                       &       &      &        &    &                &                    &     \\
The Pleiades  &$130\pm20$    & (7)&\tiny{$110^{+22}_{-15}$  }&(23)&\tiny{$7.364\pm0.005$} & 5.667 & 136  & 0.045  &(23)&\tiny{$-0.01\pm0.05$}&$ 77.6^{+13}_{-15}$&(29)\\
\smallskip 
              &$112\pm5$     &(10)&                          &    &                       &       &      &        &    &                &                    &     \\
\hline        
Blanco 1      &$132\pm24$    &(11)&\tiny{{$115^{+23}_{-15}$}}&(23)&\tiny{$4.216\pm0.003$} & 6.876 & 237  & 0.010  &(23)&\tiny{$+0.03\pm0.07$}&$ 90\pm25    $ &(31) \\
\smallskip 
           &$126^{+13}_{-14}$&(12)&\tiny{{$200^{+40}_{-26}$}}&(23)&                       &       &      &        &    &                &                    &     \\
\hline     
The Hyades    &$650\pm70$    &(13)&\tiny{$794^{+161}_{-102}$}&(23)&\tiny{$21.052\pm0.065$}& 3.389 &  48  & 0.001  &(23)&\tiny{$+0.13\pm0.05$}&$661_{-215}  $ &(29) \\ 
\smallskip
              &$678\pm98$    &(14)&                          &    &                       &       &      &        &    &                &                    &     \\
\hline
              & $<20$        &(15)&                          &    &                       &       &      &        &    &                &                    &     \\ 
              & $21\pm4$     &(16)&                          &    &                       &       &      &        &    &                &                    &     \\ 
BPMG          & $26\pm3$     &(17)&\tiny{12-22              }&(24)&\tiny{              - }&      -&  9-73&       -&    &\tiny{$-0.01\pm0.08$}&$ 20\pm10    $ &(32) \\ 
              & $25\pm3$     &(18)&                          &    &                       &       &      &        &    &                &                    &     \\ 
\smallskip
              & $22\pm6$     &(19)&                          &    &                       &       &      &        &    &                &                    &     \\
\hline  
THMG          & $40\pm3$     &(20)&\tiny{10-40              }&(24)&\tiny{              - }&      -& 36-71&       -&    &\tiny{$-0.03\pm0.05$}&$ 30         $ &(33) \\
\hline
32 Ori MG     & $23\pm4$     &(21)&\tiny{$22^{+4}_{-3}$     }&(25)&\tiny{              - }&      -&70-110&$0.03\pm0.01$&(28)&         - &$ 25.0\pm2.5      $ &(34) \\
\hline \hline
$\,$
    \end{tabular}
    }
    }\\
\begin{flushleft}
 $\,$
  {\bf Notes: } 
    The ``Parallaxes'' and ``DM'' values comes from \cite{c_babusiaux2018} with the exception of NGC 1960 that comes from \cite{cantatgaudin2018b}.
    The ``$d$'' values come from \cite{c_babusiaux2018} with the exception of BPMG and THMG \citep{l_malo2013}; and 32 Ori MG \citep{cpmbell2017}.   
    We took the ``[Fe/H]`` values from \cite{m_netopil2016}, with the exception of NGC 2547 \citep{randich2018}, and BPMG and THMG \citep{vianaalmeida2009}. 
    We assumed solar abundances for NGC 1960 and the 32 Ori MG because lack metallicity values in the literature.
    \cite{mermilliod1981a} does not provide age errors, so we estimated them from the $\log t$.\\ 
 $\,$    
  {\bf References: }
      (1) \cite{jeffries2013};
      (2) \cite{manzi2008};
      (3) \cite{randich2018};
      (4) \cite{jeffries2005}; 
      (5) \cite{dobbie2010}; 
      (6) \cite{barrado1999b}; 
      (7) \cite{barrado2004d}; 
      (8) \cite{stauffer1999}; 
      (9) \cite{stauffer1998a}; 
     (10) \cite{se_dahm2015}; 
     (11) \cite{pa_cargile2010c}; 
     (12) \cite{aj_juarez2014};
     (13) \cite{el_martin2018}; 
     (14) \cite{n_lodieu2018c};
     (15) \cite{i_song2002c}; 
     (16) \cite{as_binks2014}; 
     (17) \cite{l_malo2014b}; 
     (18) \cite{s_messina2016d};
     (19) \cite{e_shkolnik2017}; 
     (20) \cite{al_kraus2014c}; 
     (21) \cite{cpmbell2017};  
     (22) \cite{d_bossini2019};
     (23) \cite{c_babusiaux2018}; 
     (24) \cite{l_malo2013}; 
     (25) \cite{cpmbell2015}; 
     (26) \cite{cantatgaudin2018b};
     (27) \cite{pa_cargile2010b};
     (28) \cite{cpmbell2017}; 
     (29) \cite{mermilliod1981a}; 
     (30) \cite{jj_claria1982}; 
     (31) \cite{pm_panagi1997}; 
     (32) \cite{barrado1999a};
     (33) \cite{cao_torres2008};
     (34) \cite{mamajek2007a}.
\end{flushleft}     
\end{table*}

\section{Data and evolutionary models}\label{sec:data_and_models}

\subsection{Initial census for the stellar associations}\label{sub:ldb_sample}

Most of the open clusters we analysed here are included in the \cite{c_babusiaux2018}, 
so we have built the initial sample of cluster members with data from that work. 
There are two exceptions: NGC 1960 and 32 Ori MG.
For these two, we selected the members from \cite{cantatgaudin2018b} for NGC 1960
and \cite{cpmbell2017} for the 32 Ori MG. 

    We created a new list restricting to only members with measured lithium equivalent widths and close to the expected LDB
-- hereafter referred to as the LDB sample (see Table E.4.). 
The selected members must satisfy the following requirements: 
they should have available spectral data with enough resolving power ($\mathcal{R}>1\,000$) and 
a moderate signal-to-noise ratio ($S/R>20$) 
around the lithium feature located at 6\,707.8~\AA, (see Figure A.2. from \citealt{bayo2011}).
  Unfortunately, the lithium data for the same cluster are not homogeneous, 
as they have been obtained by several authors using different analyses and 
with spectra with different resolving power or signal-to-noise ratio. We compiled the dataset using the astronomical databases 
\textit{Astrophysics Data System} (ADS, \citealt{ads}), SIMBAD \citep{egret1991}, and VizieR \citep{ochsenbein2000}.

\subsection{\textit{Gaia} DR2: parallaxes, proper motions, and photometry}\label{sub:gaiadr2_data}

The second \textit{Gaia} \citep{timoprusti2016} data release, hereafter \textit{Gaia} DR2, 
is an all-sky survey catalogue with celestial positions and $G$ photometry for approximately $1.7\cdot10^9$ sources;  
for $1.3\cdot10^9$ of those sources, parallaxes and proper motions are also available \citep{aga_brown2018}.
The quality and quantity of data released by \textit{Gaia} DR2 are unprecedented. 
  However, we should take into account some caveats related to the astrometry:
a) all the sources have been considered as single stars;
so, if an object is a multiple system 
it likely lacks an astrometric solution or its value is corrupted;
b) some sources have negative parallaxes (see \citealt{x_luri2018}); and 
c) objects with separations $\sim0.2-0.3^{\prime\prime}$ and not resolved in \textit{Gaia} DR2  
can contain spurious parallax values with small uncertainties. 

    The \textit{Gaia} DR2 catalogue includes photometry in three optical bands $G,G_{BP},G_{RP}$, 
with a bright limit of $G\approx3$~mag and a limiting magnitude of $G=21$~mag \citep{dw_evans2018}. 
The analysis from \cite{maizapellaniz2018e} reports a systematic linear trend in the photomotric system, 
for the range $G\in[6, 16]$~mag, estimated to be $3.5\pm0.3$~mmag mag$^{-1}$
  These values can be greater in crowded regions, 
in the vicinity of bright sources, and at values of $G\geq19$~mag. 
  This mainly affects $G_{BP}$ and to a lesser extent $G_{RP}$. 
  \cite{rl_smart2019} indicated that for their sample of cool dwarfs 
(with spectral types ranging from M8 to T6), $G_{BP}$ is heavily affected. 
So, our faintest and furthest sources might suffer from overestimation in this band.
  The flux-excess factor (named \verb+phot_bp_rp_excess_factor+ and included in the catalogue) 
gives a quantitative indication of this effect and 
can be used to filter out unreliable values (see recommendations from \citealt{dw_evans2018}).

    We queried the \textit{Gaia} DR2\footnote{https://gea.esac.esa.int/archive/} archive 
without imposing any restriction, cross-matching the positions on the sky 
(right ascension and declination) from our sources using a $2.5^{\prime \prime}$ radius.   
  When finding two or more sources for the same input, 
we made certain which counterpart was the correct one and 
we carefully investigated the images and the entries 
using ESASky \citep{d_baines2017} and Aladin \citep{bonnarel2000}.
  We followed a different approach for the Hyades and the MGs because 
we lost a lot of sources in the cross-matching. 
This was due to two factors: 
a) the proximity and high proper motion of the sources, and 
b) our input positions on the sky are in epoch 2000, while \textit{Gaia} DR2 uses as 
reference epoch 2015.5. 
In such cases we identified each source 
using the 2MASS source name and the SIMBAD database, 
then we checked with ESASky and Aladin whether there was a \textit{Gaia} DR2 counterpart or not. 
Once we knew the \textit{Gaia} DR2 name, we obtained the rest of the quantities, 
astrometric and photometric data from the archive.
Throughout this work, we performed the analysis with the \textit{Gaia} DR2 data.
However, we carried out an additional check with the \textit{Gaia} EDR3 data \citep{aga_brown2021a}.

\subsection{Photometry}\label{sub:phot_data}

We gathered photometric data from the literature. 
For each association, these came from studies carried out with 
different telescopes and instruments and, therefore, 
with heterogeneous uncertainties and different completeness and limiting magnitudes
    Additionally, we gathered data from all-sky or wide-area multi-wavelength imaging surveys,
with ranges from the ultraviolet to the infrared, 
using, whenever possible, the Virtual Observatory SED Analyzer (VOSA, \citealt{bayo2008}), 
together with the SIMBAD and VizieR databases. 
  We cross-matched the celestial position of each source with the catalogues provided  
using a matching radius of $3.0^{\prime \prime}$. 
  The catalogues adopted for the analysis are the following:
Tycho-2  \citep{hog2000} $B_{T}V_{T}$ bands; 
2MASS \citep{cutri2003} $JHKs$ bands; 
CMC 14 \citep{dw_evans2002} $r$ band; 
DENIS \citep{denisconsort2005} $iJK_{s}$ bands; 
UKIDSS \citep{lawrence2007} bands $ZYJHK_{s}$; 
APASS DR9 \citep{henden2009h} $BV\ gri$ bands; 
GALEX \citep{l_bianchi2000b} $FUV$ $NUV$ bands; 
WISE \citep{cutri2013} $W1\ W2\ W3\ W4$ bands; 
IPHAS DR2 \citep{barentsen2014a} $ri\rm{H}\alpha$ bands; 
SDSS DR12 \citep{s_alam2015a} $ugriz$ bands; 
VPHAS DR2 \citep{je_drew2016} $ugri$ bands; and 
Pan-STARRS1 \citep{panstarrs1} $g_{P1}r_{P1}i_{P1}z_{P1}y_{P1}$ bands.
  As described in Section \ref{sub:gaiadr2_data}, 
here we also checked all the images to avoid any misidentification.
  We show the collected photometric bands in Appendix \ref{app:additional_tables}.

\subsection{Evolutionary theoretical models and LDB}\label{sub:ev_models_ldb}

To estimate the LDB age, we used the luminosity-age relationships from:  
 (a) \cite{cj_burke2004}, calculated by assuming a solar metallicity of $Z=0.0188$ and $Y=0.27$; and 
 (b) \cite{e_tognelli2015b} with Z$=0.013$ and Z$=0.016$.\footnote{In the \cite{e_tognelli2015b} models, 
   [Fe/H]$=0.0$ correspond to Z$=0.013$, and [Fe/H]$=+0.1$ to Z$=0.016$. 
   We estimated the ages for Alpha Persei and the Hyades using Z$=0.016$ and for the rest Z$=0.013$, 
   due to their metallicities, see Table \ref{tab:sample_clusters}.}
In both the cases, the LDB has been defined as the model 
in which lithium is depleted by a factor of 100 from its initial value.
Moreover, we took the mass-tracks or the isochrones provided from other works 
and calculated the LDB bolometric luminosity-age relationships as well as 
the point where lithium has been depleted by a factor of 100: 
 (c) \citeauthor{a_burrows1993a} (\citeyear{a_burrows1993a} and \citeyear{a_burrows1997b}) 
calculated it on the basis of solar metallicities \citep{e_anders1989a}; 
 (d) \cite{f_dantona1994} 
calculated it based on solar metallicities \citep{e_anders1989a}, 
using the \cite{vm_canuto1991b} convection model and opacities from \cite{dr_alexander1983};
 (e) BT-Settl models \citep{f_allard2012} using $[\mathrm{M/H}]=0.0$ (following \citealt{m_asplund2009}); and 
 (f) \cite{siess2000} using solar composition Z$=0.02$, 
These relationships also provide other stellar parameters: mass, radius, 
surface gravity and effective temperature.

  Also, we used isochrone models from \cite{e_tognelli2011}, 
and a Zero Age Main Sequence from \cite{sw_stahler_f_palla2005}, 
to identify background, foreground sources or multiple systems in the HRDs.

\section{Method}\label{sec:method_ldb}

    We built an HRD to locate the LDB in each stellar association, 
including only those sources with a moderate signal-to-noise ratio ($S/N>20$),
in the region of the lithium absorption line.
The way we derived the bolometric luminosities will be discussed in Section \ref{sub:lbol_teff}. 
According to \cite{al_kraus2014c}, the LDB determined using 
bolometric luminosities or bolometric magnitudes 
is more unambiguous due to their large dynamical range 
over which it varies across the relevant age scales.
Thus, we demarcated the LDB in terms of bolometric luminosity.

\begin{figure}[t]
   \includegraphics[width=9cm]{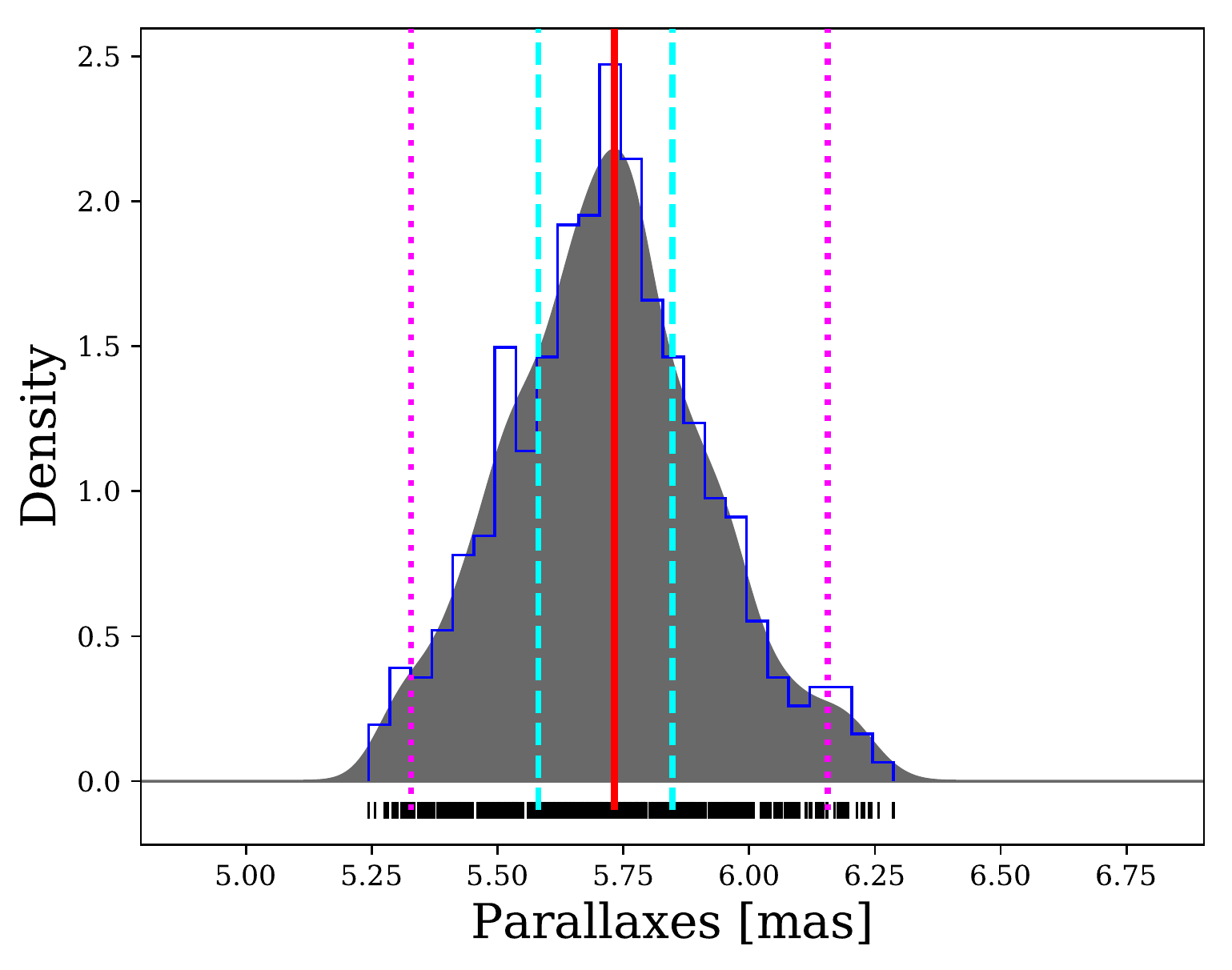}
     \caption[Parallax PDF of the Alpha Persei members.]
             {Parallax PDF for the Alpha Persei members (taken from \citealt{c_babusiaux2018}).
              The modelled PDF is shown in gray and a superimposed histogram in blue. 
              The two magenta vertical dotted lines point the 2.5th and 97.5th percentiles; 
              the two cyan vertical dashed lines point the quartiles and 
              the the red vertical line shows the mode of the PDF. 
              The small vertical black lines below the horizontal axis (below $y=0.0$) 
              show the parallaxes of each source.            
             }
     \label{fig:pdf_plxs_alphapersei_prev} 
\end{figure}

    To estimate subsequent quantities, we first need to estimate the distance of the cluster. 
To do this, using the member list 
we evaluated the median distance and size of each open cluster constructing a 
kernel density estimation\footnote{We used the library \texttt{scikit-learn} \citep{scikitlearn} to calculate it.} 
 (KDE) function, as suggested in \cite{x_luri2018}.
  Following this method, we obtained 
the mode parallax ($\varpi_{\rm\ mode}$), 
the median, the weighted median, 
the weighted variance, the quartiles, 
the 2.5th percentile, and the 97.5th percentile.
  Finally, we calculated distances at different percentiles with $d=1/\varpi$. 
  In Table \ref{tab:oc_plx_dist_prev_mem_kde} we provide the estimated parallaxes  
together with the calculated distances for each stellar association.
As an example, Figure \ref{fig:pdf_plxs_alphapersei_prev} shows the parallax probability density function (PDF) for Alpha Persei.
  Unlike open clusters, the sources belonging to BPMG, THMG, and 
32 Ori MG are positioned across a wide range of distances (see Table \ref{tab:sample_clusters} and \citealt{l_malo2013}).
For this reason, we avoided calculating the average value and we used the distance (parallax) available for each source.

\setcounter{table}{1}
\begin{table*}[ht] 
  \caption[Parallax distributions from the initial sample of cluster members using a KDE based on \textit{Gaia} DR2 data.]
          {Parallax distributions from the initial sample of cluster members using a KDE based on \textit{Gaia} DR2 data.} 
  \label{tab:oc_plx_dist_prev_mem_kde}
  \tiny
  \center{
  \begin{tabular}{lrcrrrrcrcrrrr}
\hline \hline
Stellar     &$\varpi_{\rm\ mode}$&&&\multicolumn{2}{c}{$\varpi_{\rm\ KDE}$ [mas]}& & &$d_{\rm\ mode}$& & &\multicolumn{2}{c}{$d_{\rm\ KDE}$ [pc]}&  \\
association &   [mas]          & &  2.5th  &  25th   &  75th   &  97.5th & &   [pc]  & &  2.5th  &  25th   &  75th   &  97.5th   \\
\hline        
Alpha Persei&$ 5.73192$        & & 5.32825 & 5.58102 & 5.84774 & 6.15589 & & 174.461 & & 187.678 & 179.178 & 171.006 & 162.446 \\
NGC 1960    &$ 0.82873$        & & 0.65783 & 0.77649 & 0.87735 & 0.99241 & &1\ 206.665& &1\ 520.149&1\ 287.846&1\ 139.795&1\ 007.648 \\
IC 4665     &$ 2.88239$        & & 2.34501 & 2.72233 & 2.99779 & 3.28228 & & 346.934 & & 426.437 & 367.332 & 333.579 & 304.666 \\
NGC 2547    &$ 2.52757$        & & 1.98269 & 2.40987 & 2.61619 & 2.89936 & & 397.379 & & 504.365 & 414.960 & 382.235 & 344.903 \\
IC 2602     &$ 6.57185$        & & 6.11623 & 6.42974 & 6.70321 & 7.08069 & & 152.164 & & 163.499 & 155.527 & 149.182 & 141.229 \\
IC 2391     &$ 6.59159$        & & 6.14352 & 6.44991 & 6.71406 & 7.11775 & & 151.708 & & 162.773 & 155.040 & 148.941 & 140.494 \\
The Pleiades&$ 7.35915$        & & 6.77853 & 7.18755 & 7.51509 & 8.02017 & & 135.885 & & 147.524 & 139.129 & 133.065 & 124.685 \\      
Blanco 1    &$ 4.17949$        & & 3.98254 & 4.13539 & 4.30490 & 4.47278 & & 239.264 & & 251.096 & 241.815 & 232.293 & 223.574 \\
The Hyades  &$21.08226$        & &16.44997 &19.67124 &22.48718 &27.34499 & &  47.433 & &  60.790 &  50.835 &  44.469 &  36.569 \\
32 Ori MG   &$ 9.62252^{a}$    & & 8.30029 & 8.97758 &10.19470 &10.76314 & & 103.922$^{b}$&&120.477&111.388&  98.090 &  92.909 \\
\hline
$\,$
    \end{tabular}
    }\\ 
\begin{flushleft}
  $^{a}$ This value is not the $\varpi_{\rm mode}$ it is the 50th percentile of the parallax PDF, $\varpi_{\rm KDE}$.\\
  $^{b}$ The value is calculated using the value marked as $^{a}$.
\end{flushleft} 
\end{table*}

\subsection{Outliers based on proper motions and parallaxes}\label{sub:outliers}

\begin{figure}[t]
   \includegraphics[width=9cm]{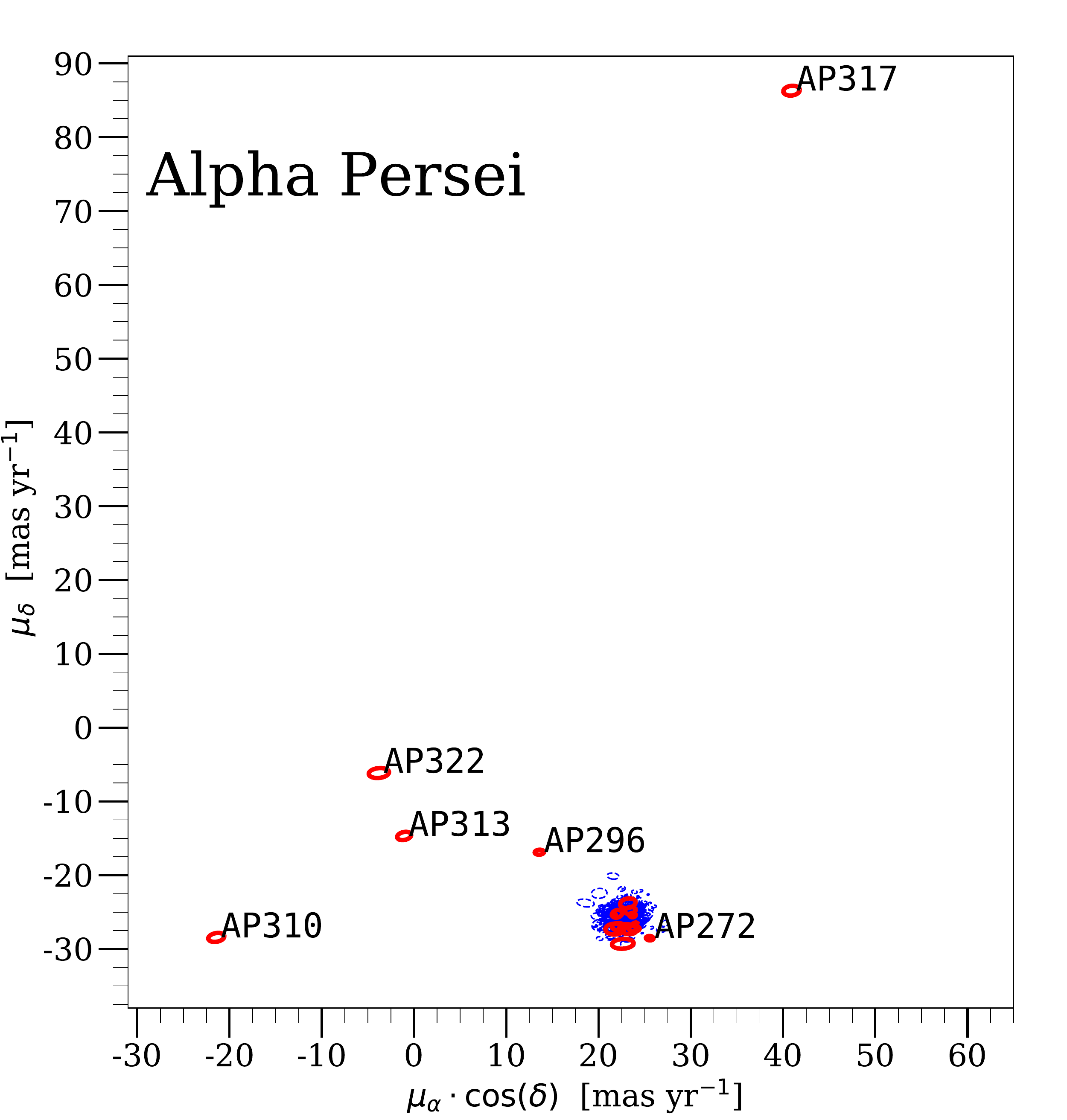}
   \includegraphics[width=9cm]{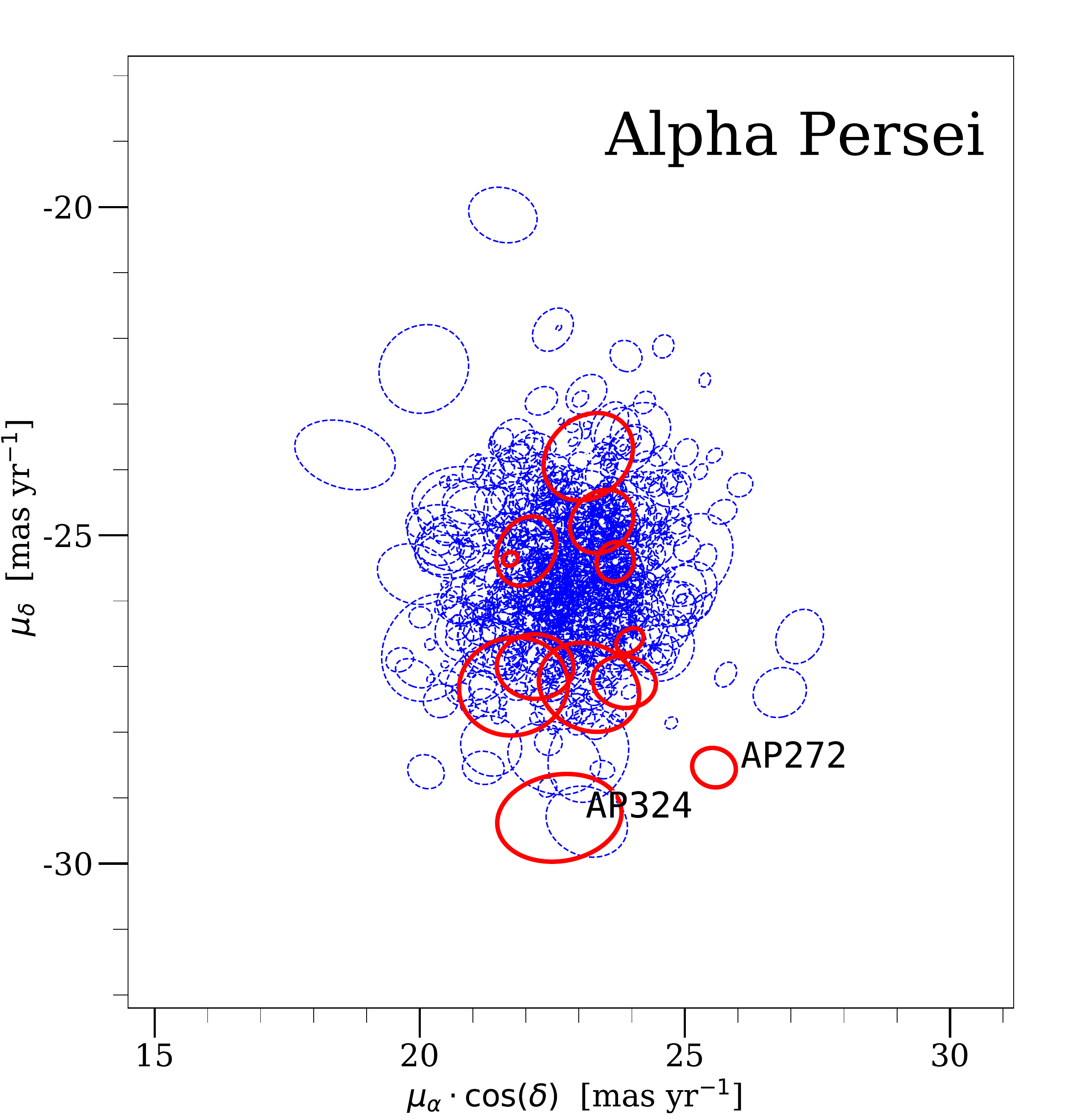}
     \caption[VPD for Alpha Persei.]
             {\textbf{VPD for Alpha Persei.}
              {\bf Top: }Blue dashed ellipses are the members taken from \citealt{c_babusiaux2018}; 
                red ones are the LDB sample objects.
                Five sources (AP296, AP310, AP313, AP317, and AP322) possess a proper motion different from that of the rest of the members.
                AP272 is also shown (see the text). 
                Uncertainties in the proper motions and its correlations are taken into a
                account and shown as ellipses.
                All data have been taken from the \textit{Gaia} DR2 catalogue.
              {\bf Bottom: }Zoom on the previous plot.
             }
         \label{fig:vpd_alphapersei}
\end{figure}

\begin{figure}
\centering
   \includegraphics[width=9cm]{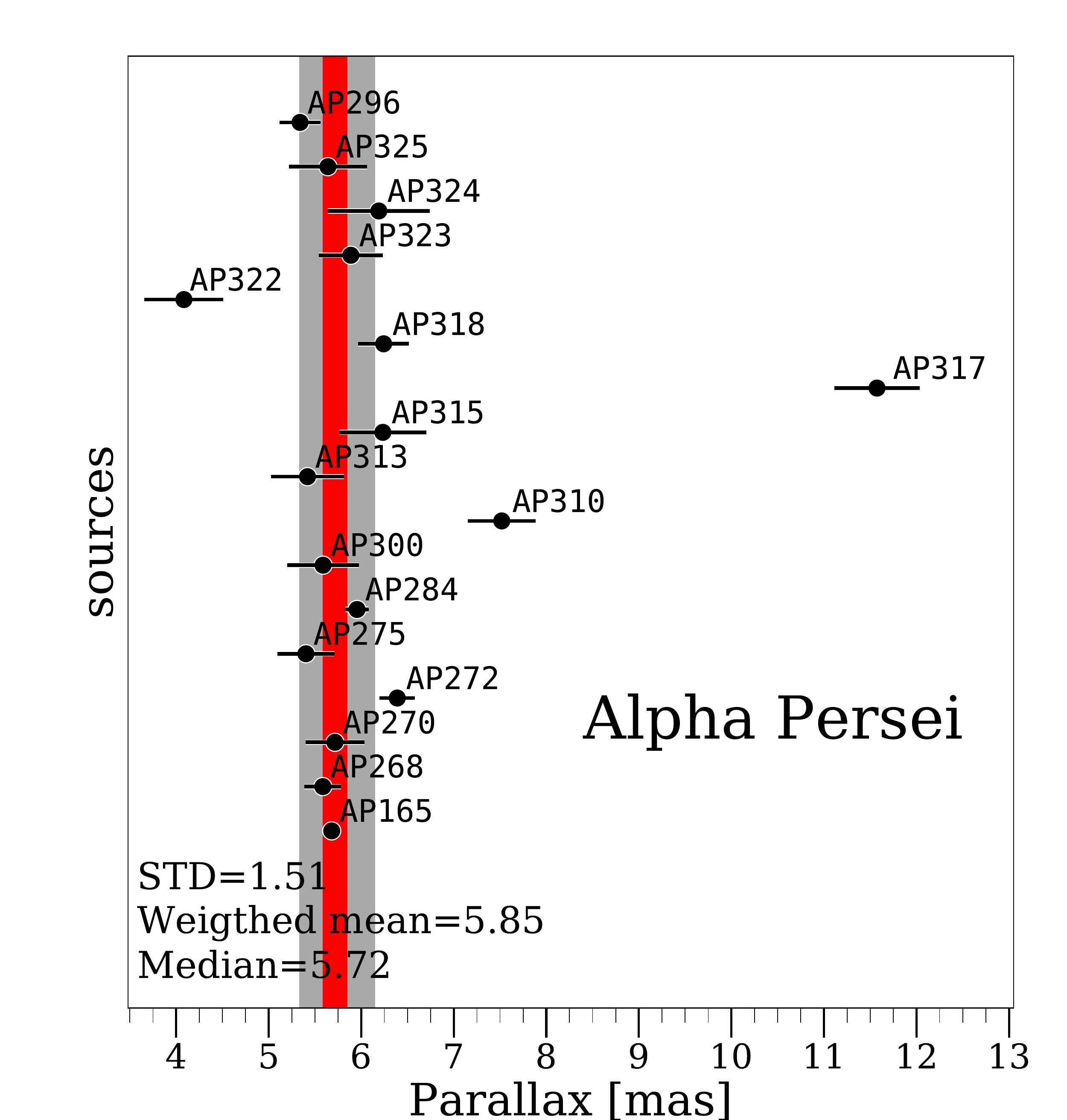}
   \caption[Parallaxes Alpha Persei members.]
             {Parallaxes for 17 selected Alpha Persei sources close to the LDB.
              The grey filled area shows the parallax distribution from \cite{c_babusiaux2018} 
              between the percentiles 2.5th and 97.5th,  
              while the distribution quartiles are delimited by a red filled area superimposed.         
              All the parallaxes have been taken from the \textit{Gaia} DR2 catalogue.
              The standard deviation (STD), the weighted mean, and the median 
              of the LDB sample are also displayed.
             }
     \label{fig:plxs_alphapersei}
\end{figure}

The accurate proper motions and parallaxes obtained from \textit{Gaia} DR2 
allowed us remove possible outliers in the different LDB samples.
  As an example, Figure \ref{fig:vpd_alphapersei} shows the proper motion vector point diagram (VPD)
for the Alpha Persei with sources from the LDB sample and the comparison sample. 
Several sparse sources do not share proper motions with the rest, 
so we considered them, based on visual inspection, as outliers and were discarded as members. 
  We plotted the proper motions as ellipses because we took into account 
the uncertainties and the correlation coefficients between them, 
as recommended in \cite{aga_brown1997c} 
(see \citeauthor{l_lindegren2016} \citeyear{l_lindegren2016} \& \citeyear{l_lindegren2018},
\citealt{x_luri2018} and examples in \citealt{cantatgaudin2018a}). 
    Similarly, we compared the parallaxes, see Figure \ref{fig:plxs_alphapersei} 
for Alpha Persei. 
  We discarded in the LDB analysis the sources with parallaxes outside 
the range delimited by the values at the  2.5th and 97.5th percentiles, 
taking into account the uncertainties of the individual measurements.
  This values are shown with a grey filled area and 
the quartiles with a red superimposed area.
  The lack of radial velocities with enough resolving power for all objects in all associations 
prevents us from performing a further complete analysis using positions, parallaxes, proper motions, and radial velocities.
We proceeded in the same way in all the open clusters.   
The figures can be found in Appendix \ref{app:plots_stellarassociations}.

\subsection{Distances from parallaxes}\label{sub:distances}

One of the innovations of this work is the inclusion of individual distances for each object. 
The conversion of $d=1/\varpi$ might not be used to assess the distance and to deal with this issue.
Previous works (see \citealt{maizapellaniz2005b}, \citealt{cal_bailerjones2015b} and \citealt{x_luri2018})
suggested that we derive distances as an inference problem using a Bayesian approach. 
  Thus, we inferred the distance of each source individually,
drawn from the parallax using the \textit{Kalkayotl} code \citep{j_olivares2020} after $10\,000$ iterations. 
We tested all prior families (see \citealt{j_olivares2020}) 
and found that the Gaussian prior proves to be the one most suited to describe the parallax distribution. 
The Gaussian prior is defined by two parameters: 
the \verb+prior_loc+ and the \verb+prior_scale+. 
The \verb+prior_loc+ is the open cluster distance value taken 
from \cite{c_babusiaux2018} or \cite{cantatgaudin2018b} and 
the \verb+prior_scale+ is five times the maximum radius of the open cluster \citep{j_olivares2020}.

  We assumed the same prior for all objects belonging to the same stellar association. 
For each open cluster, we calculated the maximum radius using the comparison lists, 
comparing their sky-positions with the sky-position of the centre derived from \cite{c_babusiaux2018}
(private communication F. van Leeuwen). 
  On the other hand, in the MGs, 
the members are distributed in a wide range of distances without being grouped or `clustered' in a particular area of the sky. 
For instance, BPMG members are located between $9$-$73$~pc, 
THMG members between $36$-$71$~pc (see \citealt{l_malo2013}) and
32 Ori MG members between $71$-$131$~pc (see \citealt{cpmbell2017}). 
We chose the median value of the distance for the \verb+prior_loc+, 
and a \verb+prior_scale+ of 200 (roughly two times the larger distance).
In Table \ref{tab:kalkayotl_priors} we show the priors used in the calculations. 

    We obtained a distance distribution for each source  
where the maximum-a-posteriori describes the distance of the object ($d_{MAP}$), 
and the two values at the 2.5th and 97.5th percentiles 
determine its minimum and maximum distances ($d_{min}$ and $d_{max}$). 

    For those objects without parallaxes or with negative values, 
we assumed the distribution corresponding to the association they belong. 
  The parallax is the mode of the PDF, $\varpi_{\rm\ mode}$, and 
the uncertainties are the values at the 2.5th and the 97.5th percentiles.
We calculated the distances and the uncertainties by taking the reciprocal of the parallax values.
  Regarding 32 Ori MG, we proceeded differently. 
We took all the members from \cite{cpmbell2017} and calculated the parallax PDF. 
 In this way, we realized that several sources are outliers, 
and we considered a sub-sample of objects with parallaxes in the range of $\varpi \in(8.0, 11.0)$~mas. 
We used this sub-sample to derive the 32 Ori MG distance.
  We did not apply this procedure to BPMG and THMG due to the wide range of parallaxes of their members.

    Other works, such as that of \cite{n_lodieu2019b}, determine parallaxes for the faintest Hyades objects. 
We computed distances from these parallaxes in the same way as before, 
by using the same prior (see Table \ref{tab:kalkayotl_priors}).
Table \ref{tab:oc_plx_dist_prev_mem_kde} lists the values of the parallaxes (mode, KDE), 
the percentiles, and the derived distance with its uncertainties for each analysed stellar association.

\subsection{Bolometric luminosities and effective temperatures}\label{sub:lbol_teff}

The bolometric luminosity and effective temperatures  were evaluated 
reconstructing the spectral energy distribution (SED) of each analysed source. 
To do this, we relied on the several photometric bands in order to have a complete coverage 
of the stellar spectra in a broad range of wavelengths. 
We used the data of the photometric surveys listed in Section \ref{sub:phot_data}.
     
    The SED was built using the VOSA tool \citep{bayo2008}.
  In a first step, VOSA brings together all the photometric values of each source and 
transforms the magnitudes into fluxes. 
The total observed flux is computed adding all the photometric bands. 
When measurements of overlapping bands are taken, 
VOSA estimates the overlapping wavelength regions and weights 
the fluxes accordingly to avoid overestimating the total flux.
  In addition, each photometric value is de-reddened 
taking into account the values of $A_V$\footnote{The $A_V$ were calculated from $E(B-V)$ with $R_{V}=3.1$.}
and the extinction law from \citet{indebetouw2005a}.
  Secondly, VOSA compares the flux values with synthetic photometry from several grids of theoretical spectra,
using a $\chi^2$ test and determines which model reproduces the observed data.  
  Thirdly, the best-fitting model infers the flux in the wavelength range where there are no data, 
calculates the final total flux for the source (\texttt{Ftot}), 
its uncertainty (\texttt{Ferr}), 
and its stellar parameters with their respective uncertainties.
  Blue and infrared excesses  are not taken into account, 
and the analysis does not include UV or $u$ bands.

    We used two sets of theoretical models to fit the SEDs: Kurucz \citep{castelli1997} and BT-Settl \citep{f_allard2013} grids. 
Both grids and the range of validity of each one have been studied before
(see \citealt{h_bouy2015b}, \citealt{barrado2016b}, 
\citealt{bayo2017a}, and \citealt{e_solano2019} and \citeyear{e_solano2021}). 
  We restricted the parameters to:
$T_{\rm eff}\geq 4\,500$~K for Kurucz models and $T_{\rm eff}\leq 4\,500$~K for BT-Settl, 
where they work best.
We fixed the metallicity to the value of each association, see Table \ref{tab:sample_clusters}, 
and the surface gravity to the fair value of $\log {\rm g}=4.5$~dex, 
except for the Hyades members where we assumed $\log {\rm g}=5.0$~dex. 
  This assumption is in line with the BT-Settl evolutionary models \citep{f_allard2012}: 
very low-mass stars with masses from 0.006 to 1.4~$M_{\odot}$ and ages between 10~Ma to 1~Ga, 
have $\log {\rm g}\in[4.4, 5.3]$~dex. 
From an empirical point of view, 
\cite{in_reid_s_hawley2006} pointed out that the values for 
M0-M5 spectral types are ranged $\log g \in [4.5, 5.0]$~dex.
  Based on our previous experience with VOSA, 
we fixed these values to explore the effective temperature range 
more effectively\footnote{Information related to VOSA quality tests 
and caveats are in \cite{c_cifuentes2020}, 
and in the VOSA Help and Documentation:\\
\url{http://svo2.cab.inta-csic.es/svo/theory/vosa50/helpw4.}
\url{php?otype=star&action=help&what=fit\#fit:bollum}\\
\url{http://svo2.cab.inta-csic.es/svo/theory/vosa50/helpw4.} 
\url{php?otype=star&action=help&what=qua_libraries}.}.
  A large number of photometric bands, as well as the wide spectral range they cover, 
allows us to say that the vast majority of the {\tt Ftot} from the object comes 
from observations ($\geq 70\%$) and not from any model. 
  We checked all the SEDs to detect and avoid spurious photometric bands, 
false detections, or contamination due to background sources. 
  We fitted a cubic spline over the five best models, minimizing the $\chi^{2}$
in order to avoid the discrete values in $T_{\rm eff}$ returned by VOSA \citep{barrado2016b}.
  The analysis of the uncertainties on our calculated bolometric luminosities and effective temperatures, 
 as well as the assumed values for metallicity, surface gravity, and reddening,  
are discussed in Section \ref{sec:ldb_error_budget}.

    We calculated bolometric luminosities using \mbox{$L_{\rm bol}=4\pi d^{2}{\tt Ftot}$}, 
where {\tt Ftot} is the total observed flux obtained from VOSA, and 
$d$ is the distance of the object. 
    In the case of sources with parallaxes, we calculated bolometric luminosities
generating a Gaussian random sample of {\tt Ftot} considering its uncertainty {\tt Ferr},  
and using its distance distribution 
as we mentioned in Section \ref{sub:distances}.
The maximum-a-posteriori of this distribution is the $L_{\rm bol}$ 
and the 2.5th and 97.5th percentiles determine its uncertainties.

\subsection{LDB loci determination}\label{sub:ldb_loci_determination}

Previous works (\citealt{jeffries2005}, \citealt{aj_juarez2014} or \citealt{as_binks2014}) 
have defined the LDB loci in the HRD, ($T_{\mathrm{eff\ LDB}}$, $L_{\rm bol\ LDB}$)
as the mid-point between 
the coldest lithium-poor source and the hottest lithium-rich one 
(in terms of effective temperature) and 
the faintest lithium-poor source and the brightest lithium-rich one 
(in terms of bolometric luminosity). 

    Using the LDB samples defined previously, 
we discarded all the non-members (NM) using the \textit{Gaia} data
and confirmed multiple systems (hereafter CMS, such as spectroscopic binaries) 
that contribute to blur the LDB for this analysis.
  We took the faintest lithium-poor object and considered all the sources 
whose bolometric luminosities were in the interval between $L_{\rm bol}{\rm\ 97.5th}$ and $L_{\rm bol}{\rm\  2.5th}$.  
Likewise, we did the same for the brightest lithium-rich object. 
Some objects can be considered as multiple systems because of their location in the HRD or analogous diagrams. 
However, there is no explicit confirmation of whether they are\footnote{
 Throughout this work, we call these objects suspected multiple systems.}.
In this case, we have performed the LDB analysis considering both cases:
(a) assuming it is a single star,
(b) or it is a multiple system and was excluded in the determination of the LDB.

    Applying this procedure, we obtain (for any stellar association) a sample of n-sources close to the LDB locus.
  In some cases (IC 2602, Blanco 1), 
we can select only two sources, a lithium-poor and a lithium-rich one, and this is insufficient for the analysis.  
We would require at least two lithium-poor sources and two lithium-rich ones. 
Thus, the resulting sub-samples should have at least four sources.

    To estimate the LDB and quantify the uncertainty associated, 
we used a jackknife method with a bootstrap re-sampling. 
We split the sub-sample of $n$ sources into $n$ sets containing all but one source: 
the first set contains all but source 1, the second set contains all but source 2, and so forth.
  For each set, we generated 10\,000 simulated bootstrap samples.
A bootstrap sample contains $n-1$ bolometric luminosities, one from each source.
They were generated using a Gaussian random value of the {\tt Ftot} considering its uncertainty $\Delta{\tt Ftot}$, 
and a random value of the distance derived from its distribution (Section \ref{sub:distances}). 
  
\begin{figure}
\centering
   \includegraphics[width=9cm]{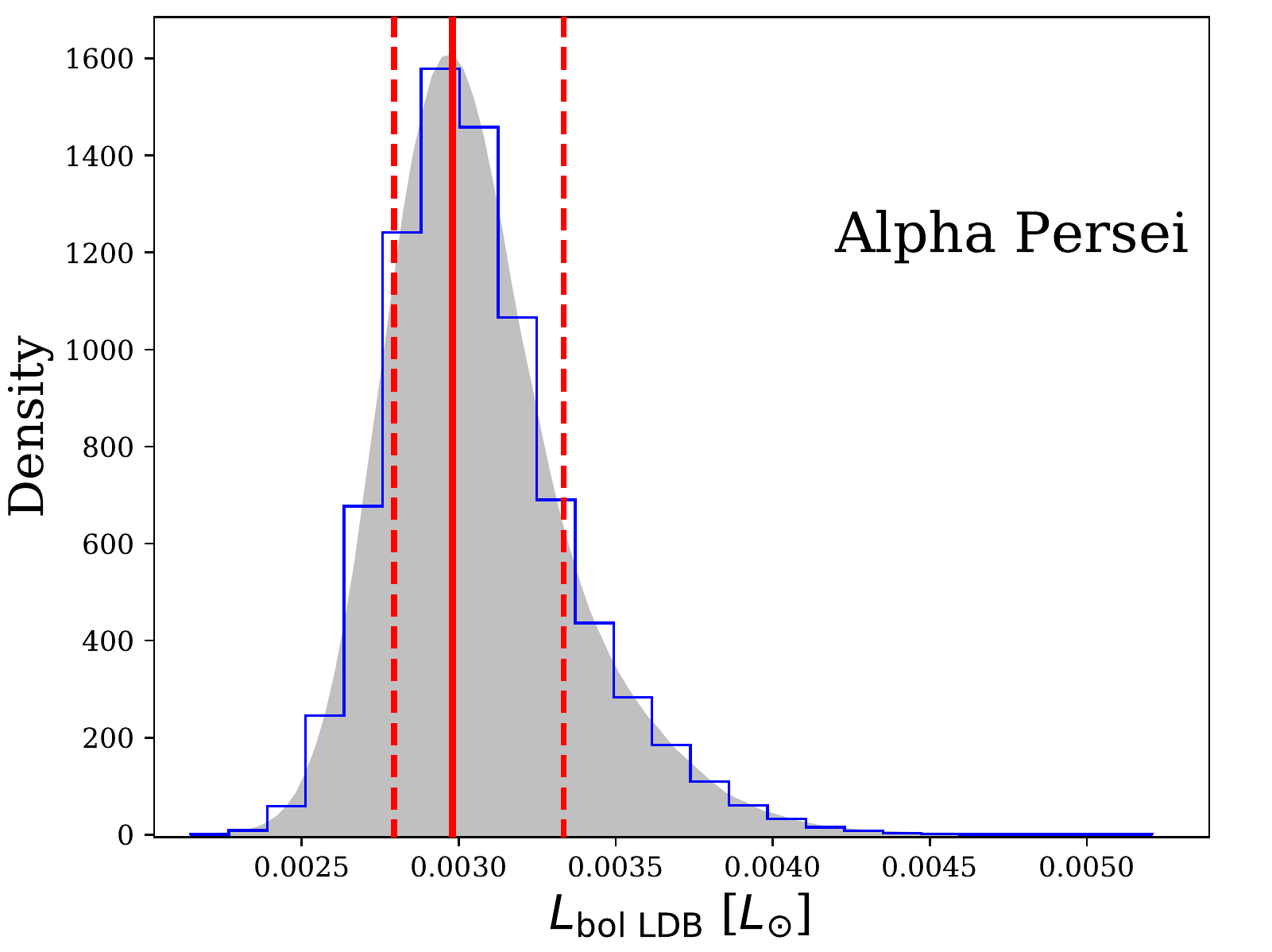}
     \caption[Bolometric luminosities PDF for the Alpha Persei LDB.]
             {Bolometric luminosity PDF for the Alpha Persei LDB,
              after run 60\,000 bootstrap samples, see Section \ref{sub:ldb_loci_determination}.              
              We show all the sample distribution as a blue histogram, and 
              we model a PDF with all the values in grey.
              We highlight the mode of the PDF with a red line and               
              the percentiles at 16th and 84th with two red dashed ones.
             }
     \label{fig:histogram_kde_lbol_ldb_alphapersei}
\end{figure}

     We calculated the LDB luminosity as the equal middle point between 
the bolometric luminosity of the faintest lithium-poor object and
the brightest lithium-rich object. 
  The result is a bolometric luminosity distribution of the LDB loci obtained from all the simulations, 
where the LDB is located at the mode, 
and the uncertainties are given at 16th and 84th percentiles, 
as we show in Figure \ref{fig:histogram_kde_lbol_ldb_alphapersei}.
Finally, we derived ages using the evolutionary models presented in Section \ref{sub:ev_models_ldb}.
  Also, we determined the $T_{\rm eff\ LDB}$ as the mid-point in terms of effective temperature 
between the coldest lithium-poor source and the hottest lithium-rich one.  
We estimated an uncertainty due to the models of $\pm71$~K 
based on the mean squared error of the two sources.
  As an example, we show in Fig. \ref{fig:hrd_alphapersei_discussion} the HRD for Alpha Persei 
with the LDB sample, the comparison sample, and our estimated $L_{\rm bol\ LDB}$. 
Similar figures for the remaining stellar associations are shown in Appendix \ref{app:plots_stellarassociations}.

\section{Analysis of the LDB}\label{sec:analysis_ldb}

In the following subsections, we analyse in detail each stellar association and locate its LDB. 
We apply the method presented in Section \ref{sec:method_ldb}.
The results are summarised in Table \ref{tab:ldb_ages_lbol} and 
Figure \ref{fig:mbol_ldb_vs_ldb_age}.

\subsection{Alpha Persei}\label{sub:alphaper_ldb}

We collected sources from former works around the LDB (with \mbox{$K_{s}>12.0$~mag}). 
In total, our compilation includes 17 sources: 
nine without lithium and six with it, from \cite{stauffer1999}
(first work with a complete census of members around the LDB); 
one with lithium, AP270, from \cite{BM99} and 
one without lithium, AP272, from \cite{zapateroosorio1996}.
    
    Photometric data from previous works were used: 
$R_{c}I_{c}K_{s}$ (\citealt{stauffer1999} and \citealt{barrado2002b}), and 
$F814W$ and $F785L$ from the \textit{Hubble Space Telescope} \citep{el_martin2003c}.
In addition, we recovered $VI_{c}$ from \citeauthor{prosser1992} 
(\citeyear{prosser1992} and \citeyear{prosser1994b}), and 
$R_{c}I_{c}$ from \citet{deacon2004}.
  All the sources have parallaxes and proper motions in \textit{Gaia} DR2.

    \cite{stauffer1999} included remarks about some sources:  
 \mbox{AP317} could be a background star or a possible member; 
 \mbox{AP322} (\mbox{2MASSJ 03330848+4937550})
is likely to be a non-member; and
 \mbox{AP323} and \mbox{AP325} could be binary systems.

\begin{figure*}
      {\includegraphics[width=0.5\textwidth,scale=0.50]{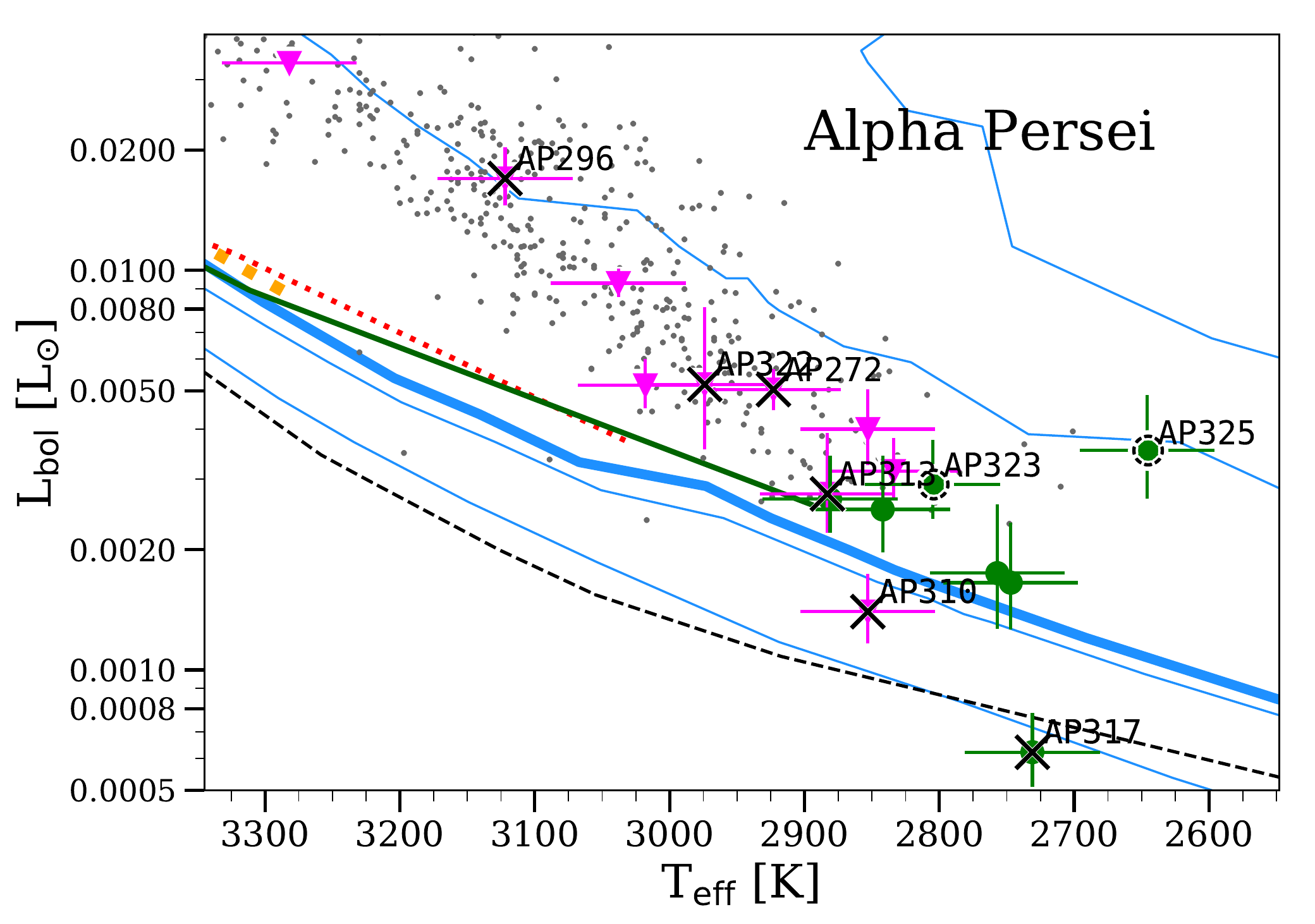}} \quad     
      {\includegraphics[width=0.5\textwidth,scale=0.50]{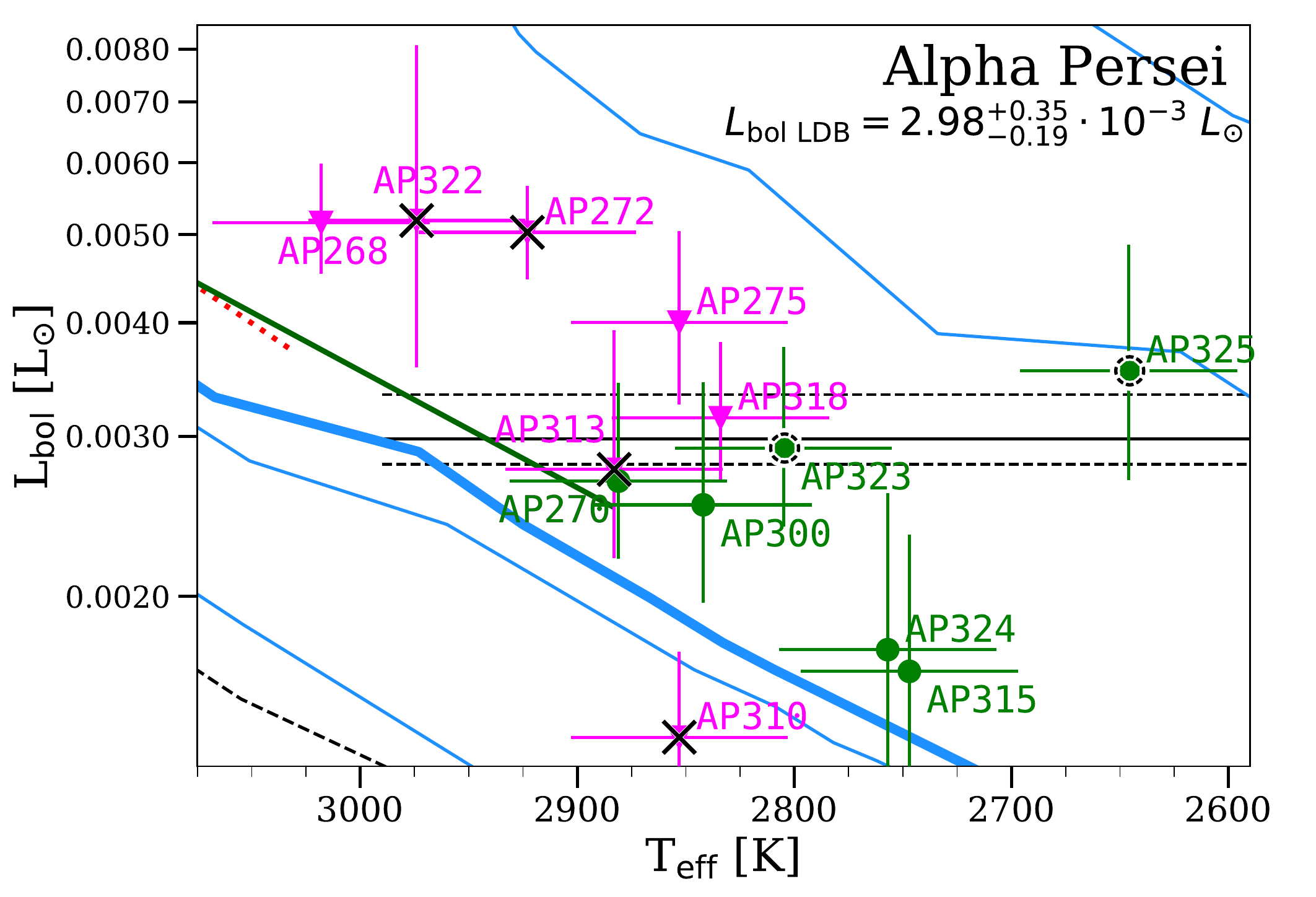}} \\
    \caption[HRDs and the LDB for Alpha Persei.]
          {HRD and the LDB for Alpha Persei.
          {\bf Left: }Symbols are as follows: 
            green solid circles are lithium-rich sources and 
            magenta triangles lithium-poor ones, both from our LDB sample;
            empty large broken black circles are suspected multiple systems; 
            over-imposed big black crosses are sources discarded as members after this work;
            and grey points are known members from \cite{c_babusiaux2018}.  
             Thin blue lines correspond to isochrones of 1, 10, 100 Ma, and 1 Ga 
            from the BT-Settl models \citep{f_allard2013}, while the thick blue one to 80 Ma.            
             The figure also includes \mbox{70~Ma} isochrones from:  
            \citealt{f_dantona1994} (black dashed line), 
            \citealt{e_tognelli2011} (orange dashed dot line), 
            \citealt{siess2000} (red dotted line), and 
            a Zero Age Main Sequence from \citealt{sw_stahler_f_palla2005} (the green solid line).
           {\bf Right: }Zoom on the left plot.  
            A black horizontal line marks the $L_{\rm bol\ LDB}$,  
            together with the 16th and 84th percentiles (dotted lines). 
            Alpha Persei is $79.0^{+1.5}_{-2.3}$~Ma old using the BT-Settl bolometric luminosity-age relationship \citep{f_allard2012}.
            For clarity, members from \cite{c_babusiaux2018} are not shown. 
          }
  \label{fig:hrd_alphapersei_discussion}
\end{figure*}

\begin{figure}
\centering
   \includegraphics[width=9cm]{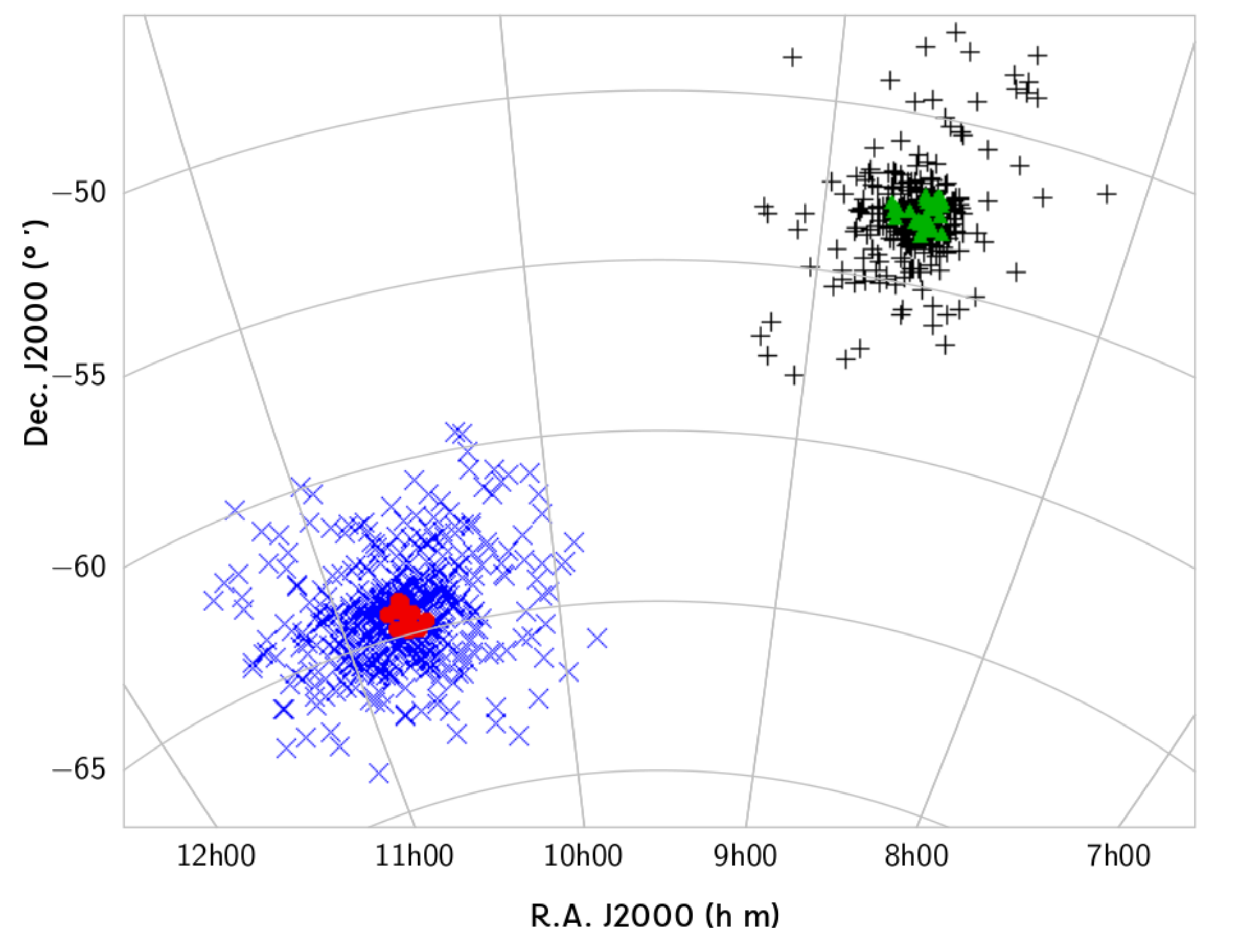}
     \caption[Projected position for IC 2602 and IC 2391 members.]
             {Projected positions for IC 2391 
              (black pluses are members from \citealt{c_babusiaux2018} and 
              green triangles objects from the LDB sample) 
              and IC 2602  
              (blue crosses are members from \citealt{c_babusiaux2018} and 
              red circles sources from the LDB sample). 
             }
   \label{fig:pos_ic2602_ic2391}             
\end{figure}

    The proper motions (Fig. \ref{fig:vpd_alphapersei}) and parallaxes (Fig. \ref{fig:plxs_alphapersei}) 
suggest that AP310 at 7.518~mas and AP317 at 11.573~mas, 
are foreground objects that do not belong to Alpha Persei, 
whereas we confirmed that AP322 is a background source with $\varpi=4.083\pm0.426$~mas. 
 AP272 looks like a foreground object, 6.389~mas, 
but its proper motion could agree with the rest of the members. 
Its position on the HRD does not affect the LDB locus.
 AP313 shows a discrepant proper motion but a parallax (5.419~mas) consistent with 
the members from the comparison sample, so we considered it as a non-member. 
  Lastly, \cite{stauffer1999} surmised that AP325 was a binary system, and 
we considered it as a suspected multiple system due to its HRD location 
(Fig. \ref{fig:hrd_alphapersei_discussion}).  
 
    \citet{stauffer1999} located the LDB with the sources 
AP323 (lithium-rich but probably a binary system), 
AP310 (lithium-poor) and AP300 (lithium-rich), 
and they estimated a \mbox{$M_{\rm bol\ LDB}=11.40$~mag} 
(\mbox{$L_{\rm bol\ LDB}=2.168\cdot 10^{-3}$~$L_{\sun}$}).
  However, \cite{barrado2004d} assume that the LDB is defined by AP310, AP322, and AP300 
and located the LDB at \mbox{$M_{\rm bol\ LDB}=11.31$~mag} 
\mbox{($L_{\rm bol\ LDB}=2.355\cdot 10^{-3}$~$L_{\sun}$}).
  With the new analysis, we located the LDB at $L_{\rm bol\ LDB}=2.98^{+0.35}_{-0.19}\cdot 10^{-3}$ $L_{\sun}$, 
and $T_{\mathrm{eff\ LDB}}=2\,857\pm71$~K (see the HRD in Fig. \ref{fig:hrd_alphapersei_discussion}).

    Although we considered AP323 and AP325 as suspected binaries, 
they may play a vital role in the LDB loci, 
therefore it would be necessary to study their multiplicity more reliably.

\subsection{NGC 1960 (M 36)}\label{sub:ngc1960_ldb}

We created the LDB sample collecting 32 members from \citet{jeffries2013}.  
  We used $gri$ photometric data from \citet{jeffries2013}, and 
from our database (Galindo-Guil et al. in prep.). 
  The source 4165 is the only one without a counterpart in \textit{Gaia} DR2.
Candidate members identified as 3028, 3590, and 3596 have counterparts, but lack parallaxes and proper motions, and 
the source 3081 possesses negative parallax. 
  \cite{jeffries2013} point out that 
3080 shows evidence of being a binary system, 
and 3081 (the brightest object with lithium) could have depleted the $\sim99\%$ of 
its initial lithium content. 

    From the VPD (Fig. \ref{fig:vpd_plxs_ngc1960} left),  
we considered as non-members  2188, 2249, 2663, 3080, and 3150. 
Other sources, namely, 1859, 1871, 2696, and 3073, are 
inside the bulk of the confirmed  members by the narrowest of margins.
  We discarded 1269, 1859 and 2703 in the location of the LDB, 
because lie outside the percentiles of the parallax PDF 
(see Fig. \ref{fig:vpd_plxs_ngc1960} right).
  We note that 2171 is almost outside the percentiles of the parallax PDF, 
and 3612 possesses a large parallax uncertainty;    
 3081 has negative parallax, and the bolometric luminosity was calculated based on 
the distance derived from the comparison sample.

    We note that the abrupt transition
between lithium-poor and lithium-rich objects is not visible in NGC 1960, 
(see the HRD in Fig. \ref{fig:hrd_ngc1960_general}),  
and compared with other stellar associations  
(see Sects. \ref{sub:ic4665}, \ref{sub:ngc2547} and subsequent sections).
  \citet{jeffries2013} locate the LDB with three lithium-rich objects: 3080, 3081, and 3596 
at \mbox{$L_{\rm bol\ LDB}=31.6^{+6.4}_{-5.3}\cdot 10^{-3}$~$L_{\sun}$}. 
However the combination of undetected binary systems, 
photometric uncertainties and contamination, 
prevented them to detect a reliable LDB loci.
  Two out of the three lithium-rich sources, 
3596 and 3081, do not have individual parallaxes in \textit{Gaia} DR2, 
so we assumed the mean distance to the cluster (see Table \ref{tab:oc_plx_dist_prev_mem_kde}), 
and 
 we considered 3080 as a non-member because of its proper motion. 
 
    The scarcity of lithium-rich sources and the poor astrometric quality of the data 
(due to the faintness of the sources) did not allow us to derive a bona fide LDB age.
 In any case, we proceeded in the same way as we did with Alpha Persei. 
We located the LDB at 
$L_{\rm bol\ LDB}=23.6^{+12}_{-4.9}\cdot 10^{-3}$~$L_{\sun}$, and  
$T_{\rm{eff\ LDB}}=3\,123\pm71$~K, see Fig. \ref{fig:hrd_ngc1960_general}.
  However, if we consider the three lithium-rich objects (3080, 3081 and 3596) 
as multiple systems and discard them, 
then the faintest lithium-poor source, 4165, gives an upper limit in the LDB location at 
$L_{\rm bol\ LDB}<16.27\cdot 10^{-3}$~$L_{\sun}$ and at $T_{\rm{eff\ LDB}}<3\,016$~K 
(see Fig. \ref{fig:hrd_ngc1960_zoom_ldb}).
This object would push the minimum age to $37.4^{+8.3}_{-8.1}$~Ma.
(see Fig. \ref{fig:hrd_ngc1960_zoom_ldb}).

    NGC 1960 needs a deeper photometric survey followed by a spectroscopic study 
to determine lithium abundances in fainter objects than our LDB sample. 
This is because the abrupt transition between lithium-poor and lithium-rich objects 
is not visible in contrast to other associations.
  Our LDB loci might be a tentative one, and we do not discard the possibility that 
it could be a possible upper limit.

\subsection{IC 4665}\label{sub:ic4665_ldb}

We recovered data from different works: 
39 members from \citet{manzi2008}, 13 members from \citet{prosser1993} and \citet{jeffries2009d} 
and 15 members only from \citet{jeffries2009d}. 
All sources from \citet{manzi2008} and some from \citet{jeffries2009d} have been detected 
previously in \citet{prosser1993}, \citet{m_giampapa1998} and \citet{wj_dewit2006a}.
  \cite{manzi2008} did not provide equivalent widths of lithium for their faintest members,  
they highlighted the presence or absence of lithium and classified them
in three categories: 
`Y' for a lithium-rich source, 
`N' for a lithium-poor one, and 
`?' when the signal-to-noise ratio did not allow to confirm or reject the presence of lithium. 
   Our LDB sample comprises a total of 67 sources and all of them
have counterparts in \textit{Gaia} DR2.  
\mbox{P333} (with lithium) and \mbox{P373} (without lithium) lack 
parallax and proper motion measurements. 
  In addition, we collected photometric data in filters 
$I_{\rm CFHT12}$ and $z_{\rm CFHT12}$ \citep{wj_dewit2006a} and 
$VI_{c}$ \citep{jeffries2009d}. 

    The proper motions are illustrated in Figure \ref{fig:vpds_ic4665} 
and parallaxes in Figure \ref{fig:plxs_ic4665}.
  In terms of proper motions, we rejected: 
\mbox{JCO1-530}, \mbox{JCO3-285},  
\mbox{JCO3-357}, \mbox{JCO4-337}, \mbox{JCO9-120}, 
\mbox{P002}, 
\mbox{P030}, \mbox{P059}, \mbox{P065}, \mbox{P139} \mbox{P215}, 
\mbox{P290}, \mbox{P331}, \mbox{P377}, \mbox{P398,} and \mbox{P411}.
  Some objects lie outside the percentiles of the parallax PDF. 
\mbox{Then, A.09.30.47}, \mbox{JCO4-337}, \mbox{P396}, and \mbox{P398}
were considered foreground objects, 
while
\mbox{JCO1-530}, \mbox{JCO2-213}, \mbox{JCO3-285}, 
\mbox{P002}, \mbox{P059}, \mbox{P065}, \mbox{P139}, \mbox{P215}, 
\mbox{P290}, \mbox{P331}, and \mbox{P377} were taken as background sources;                                   
  \mbox{P344} is on the borderline of the bulk of the proper motions 
but we still considered it as a member.

    \mbox{P065}, \mbox{P290}, \mbox{P313,} and \mbox{P315} 
show variable radial velocities or evidence for a double line system, or both \citep{manzi2008}; consequently, we considered them as multiple systems.   
 We note that \mbox{P344}, as in the case of \mbox{P313}, could be a multiple system because 
both share close positions in the HRD (see Figure \ref{fig:hrds_ic4665}) 
and have similar parallax values, 
so it is considered a suspected multiple system. 
 \mbox{A.09.30.14} and \mbox{P238} are marked as `?' 
in \cite{manzi2008} and their lithium detection is uncertain. Then,
 \mbox{P372} and \mbox{P344} are marked as `Y?' \citep{manzi2008} 
and we treated them as lithium-rich objects. 

    \citet{manzi2008} locate the LDB between the sources 
\mbox{P350} (lithium-poor) and \mbox{P338} (lithium-rich) at
\mbox{$L_{\rm bol\ LDB}=24.9^{+11.1}_{-8.1}\cdot 10^{-3}$~$L_{\sun}$}. 
  We located the LDB by considering two different scenarios. 
In the first one, we considered \mbox{P344} as a single source (Fig.\ref{fig:hrd_ic4665_zoom_ldb_v1}), 
while the second scenario, we considered \mbox{P344} as a suspected multiple system (Fig. \ref{fig:hrd_ic4665_zoom_ldb_v2}).  
  In the first scenario, we located the LDB at 
\mbox{$L_{\rm bol\ LDB}=25.8^{+3.3}_{-1.9}\cdot 10^{-3}$~$L_{\sun}$} and 
\mbox{$T_{\mathrm{eff\ LDB}}=3\,110\pm71$~K}, while 
  the second scenario resulted in 
\mbox{$L_{\rm bol\ LDB}=23.6^{+2.1}_{-1.6}\cdot 10^{-3}$~$L_{\sun}$} 
and with equal $T_{\mathrm{eff\ LDB}}$. 
For this open cluster, we decided to apply this second scenario.

    We remark that the four objects, 
\mbox{A.09.30.14}, \mbox{P238} \mbox{P372,} and \mbox{P344,} 
located around the LDB, must be studied in detail
with high-quality spectra to determine their lithium abundances. 
  Also, it would be interesting to have parallaxes for 
\mbox{P333} (with lithium) and \mbox{P373} (without lithium).

\subsection{NGC 2547}\label{sub:ngc2547_ldb}

We took the sample of 52 radial velocity members from \citet{jeffries2005} 
as our LDB sample. 
  We collected photometric data in 
$BVI_{c}$ filters from \cite{t_naylor2002}, 
$R_{c}I_{c}Z$ from \cite{jeffries2004} and  
the \textit{Spitzer} photometric bands 3.6$\mu$m, 4.5$\mu$m, 5.8$\mu$m, 8.0$\mu$m, and 24$\mu$m from \cite{n_gorlova2007}.
All the sources have counterparts in \textit{Gaia} DR2 with parallax and proper motion measurements. 

    \cite{jeffries2005} identified six possible multiple systems: 
\mbox{JO-21}, \mbox{JO-27}, \mbox{JO-30}, \mbox{JO-31}, \mbox{JO-39,} and \mbox{JO-56}. 
  In addition, three of them, (\mbox{JO-27}, \mbox{JO-30,} and \mbox{JO-31}) 
together with \mbox{JO-26}, 
are lithium-rich objects located at brighter luminosities than the LDB 
(see Figures 5, 6, and 7 from \citealt{jeffries2005}). 
  So, \mbox{JO-26} could be an unresolved multiple system, and it was excluded from our calculation.

    The analysis of the proper motions and the parallax distribution (see Figure \ref{fig:vpd_plxs_ngc2547}) allowed us to affirm that 
\mbox{JO-26} and \mbox{JO-30} are non-members;  
\mbox{JO-27} and \mbox{JO-31} could be multiple systems or fast rotators.
  We note that \mbox{JO-62} has a very high uncertainty in proper motions;
and \mbox{JO-02}, \mbox{JO-07}, \mbox{JO-37}, \mbox{JO-45,} and \mbox{JO-57} are near  
the borders of the bulk of previous members.
  Overall, the objects that we considered as non-members 
based on proper motions are
\mbox{JO-26} and \mbox{JO-30};
those based on parallaxes-:
\mbox{JO-45}, \mbox{JO-57}, and again \mbox{JO-30}.

    \cite{jeffries2005} located the LDB at
\mbox{$L_{\rm bol\ LDB}=11.59^{+1.59}_{-1.84}\cdot 10^{-3}$~$L_{\sun}$}.
  We located the LDB at 
\mbox{$L_{\rm bol\ LDB}=11.13^{+2.4}_{-0.89} \cdot 10^{-3}$~$L_{\sun}$} and 
\mbox{$T_{\mathrm{eff\ LDB}}=3\,004\pm71$~K} (see Figure \ref{fig:hrds_ngc2547}).

\subsection{IC 2602}\label{sub:ic2602_ldb}

We selected 14 sources from \cite{dobbie2010}.
Six of them are lithium-rich objects and six are lithium-poor. 
\cite{dobbie2010} considered two objects: \mbox{10420419-6404236} and \mbox{10443357-6415455}
as possible members, and the rest as bona fide members.
  From \cite{dobbie2010}, we have recovered photometric data in the band $I_{c}/Iwp_{ESO845}$. 
However, we did not use them in our SED fitting,  
given the large uncertainties in their photometric values, as the authors pointed out.   
All the sources have a counterpart in \textit{Gaia} DR2, but
\mbox{10452984-6444543} lacks parallax and proper motions.

    \mbox{10443357-6415455} is considered as possible member by \citet{dobbie2010} 
and we confirmed that it is a background object due to its parallax (see Fig. \ref{fig:vpd_plxs_ic2602} right). Then,
  \mbox{10370251-6444416} can also be considered as a background object.
  Besides, \cite{dobbie2010}, suggest that two sources could be 
binary systems or young interlopers 
due to their positions above the cluster sequence in the colour-magnitude diagram, 
and with clear lithium detections:
\mbox{10452984-6444543} and \mbox{10391188-6456046}. 
  The position of \mbox{10391188-6456046} in the HRD (Fig. \ref{fig:hrds_ic2602})
indicates that it may be an equal-mass binary system. 
  \mbox{10452984-6444543} is detected in \textit{Gaia} DR2 but without astrometric solution,
for this reason and taking into account its HRD position,
we suggest that it is a suspected multiple system.
 We note that \textit{Gaia} DR2 has revealed that this source has a fainter background counterpart, 
\mbox{5239800972309874176}, located at $1.8^{\prime\prime}$, with a parallax of $0.6386\pm0.5577$~mas. 
The fact that it is only identified in \textit{Gaia} DR2 suggests that 
the photometric data released in previous surveys are a combination of two sources. 
    Finally, 10452984 and 10391188 are two suspected binaries 
that need solid confirmation about their multiplicity.
\cite{dobbie2010} located the LDB between 
\mbox{10430890-6356228} and \mbox{10430236-6402132}, 
at \mbox{$M_{J}=8.26\pm0.23$~mag} (\mbox{$L_{\rm bol\ LDB}=6.2^{+1.5}_{-1.2}$~$L_{\sun}$}). 
  We located the LDB at 
\mbox{$L_{\rm bol\ LDB}=7.14^{+1.2}_{-0.57}\cdot 10^{-3}$~$L_{\sun}$} and 
\mbox{$T_{\mathrm{eff\ LDB}}=2\,926\pm71$~K.}

\subsection{IC 2391}\label{sub:ic2391_ldb}

We collected a total of 28 sources from our previous works 
(\citeauthor{barrado1999b} \citeyear{barrado1999b}, \citeyear{barrado2001d}, and \citeyear{barrado2004d} 
and \citet{s_boudreault2009}). 
  We added some objects to the LDB sample: 
\mbox{CTIO-106}, \mbox{CTIO-202,} and \mbox{CTIO-205}, 
which are members but lack lithium feature measurements  
because the low-quality spectra do not allow us 
to discern the presence or absence of it \citep{barrado2004d}. 
  We used our photometric data $VR_{c}I_{c}Z$ from \cite{barrado2001d} 
and we collected $VR_{c}I_{c}$ from \cite{patten1999}. 
  All the sources have a counterpart with proper motions and parallaxes in \textit{Gaia} DR2.  
  Our membership criterion is in contrast to that of \cite{s_boudreault2009}, 
who considered as members objects without the H$\alpha$ line in emission:
\mbox{20-001}, \mbox{20-009}, \mbox{20-024,} and \mbox{20-029}.
Below, we will see that these sources can be considered outliers 
due to their parallax, proper motions, or both.
  Furthermore, they suggest that values of EW Li$<0.1$~\AA, 
as it happens with \mbox{20-009} and \mbox{20-014} (see their Table 6), 
indicate the presence of the lithium feature at \mbox{6\,708~\AA}. 
However they do not provide a reliable measurement of the equivalent width.
In the same way, we go on to see that these sources are outliers.

    From \citeauthor{barrado1999b}, \citeyear{barrado1999b}, and \citeyear{barrado2004d},  
we collected some remarks about some objects: 
 \mbox{PP-07} and \mbox{CTIO-206} could be possible binaries or outliers; 
 \mbox{CTIO-062} is a lithium-rich source  
(also studied in \citealt{s_boudreault2009} with name \mbox{20-022}); 
 the lithium detection of \mbox{CTIO-077} is quite uncertain \citep{barrado2004d}, 
but we considered it as a lithium-rich object.

    We confirmed two foreground objects: \mbox{PP-07} and \mbox{20-024}, 
along with eight background sources: 
\mbox{CTIO-017}, \mbox{CTIO-059}, \mbox{20-001}, \mbox{20-009}, 
\mbox{20-012}, \mbox{20-014}, \mbox{20-018,} and \mbox{20-029} 
(see Figure \ref{fig:vpd_ic2602_ic2391_plxs_ic2391} right). 
  We show the proper motions from IC 2602 and IC 2391 LDB samples in the same VPD, 
Figure \ref{fig:vpd_ic2602_ic2391_plxs_ic2391} left. 
 There are ten sources that we considered 
as non-members because they are far away from the bulk of the members: 
\mbox{20-001}, \mbox{20-009}, \mbox{20-012}, \mbox{20-014}, 
\mbox{20-018}, \mbox{20-024}, \mbox{20-029}, 
\mbox{CTIO-017}, \mbox{CTIO-0 59,} and \mbox{PP-07}; these are the same objects that are not members in terms of parallaxes. 
  We note that there is a crowded region in the VPD around 
$\mu_{\alpha}\cdot \cos(\delta)=-12.9$ and $\mu_{\delta}=11.5$~mas yr$^{-1}$, 
with six sources: 
\mbox{PP-07}, \mbox{CTIO-017}, \mbox{20-001}, \mbox{20-017}, \mbox{20-018,} 
and \mbox{20-029}. 

    In addition, CTIO-062 (20-022), a lithium-rich member 
(\citealt{barrado2004d} and \citealt{s_boudreault2009}), 
has proper motions and parallax similar to other members, as well as H$\alpha$ in emission, 
but it is located below the cluster sequence in the HRD. 
It shows a fainter bolometric luminosity than other members with the same 
effective temperature, so it could be an outlier. 
  The objects without measurements of the lithium feature 
(\mbox{CTIO-106}, \mbox{CTIO-202}, \mbox{CTIO-205,} and \mbox{15-005}) 
are not relevant in the LDB determination, except for \mbox{CTIO-205} 
that could be a multiple system due to its position in the HRD.
\mbox{CTIO-081} could also be a multiple system (see the HRD, Figure \ref{fig:hrds_ic2391}).
  According to \cite{s_boudreault2009}, 
the lithium-poor M7.5 dwarf \mbox{15-005} fulfils our membership criteria, 
but due to its position in the HRD and its spectral type, it would be expected to have lithium. 
The reason for the non-detection is the low signal-to-noise ratio of the spectrum ($S/N=8.3$).
We considered this object a member with doubtful detection, although it does not affect the LDB location.
  Lastly, \mbox{CTIO-077}, a member with an uncertain lithium detection 
\citep{barrado2004d}, considered as a lithium-rich object, 
is not key in the determination of LDB loci. 

    \cite{barrado1999b} located the LDB between the sources 
PP-14, CTIO-096, and CTIO-038 at \mbox{$M_{\rm bol\ LDB}=10.29$}~mag  
(\mbox{$L_{\rm bol\ LDB}=6.026\cdot 10^{-3}$~$L_{\sun}$}).
  Subsequently, \cite{barrado2004d} located the LDB at \mbox{$M_{\rm bol\ LDB}=10.24$}~mag  
(\mbox{$L_{\rm bol\ LDB}=6.31\cdot 10^{-3}$~$L_{\sun}$}).   
  We located the LDB at 
\mbox{$L_{\rm bol\ LDB}=5.93^{+0.23}_{-0.13}\cdot 10^{-3}$~$L_{\sun}$} and 
\mbox{$T_{\mathrm{eff\ LDB}}=2\,956\pm71$~K.}

\subsection{The Pleiades}\label{sub:thepleiades_ldb}

We selected 35 sources 
from different works: 
\cite{gw_marcy1994}, \cite{rebolo1996}, 
\cite{oppenheimer1997}, \citeauthor{stauffer1998a} (\citeyear{stauffer1998a} and \citeyear{stauffer1998b}), 
and \cite{se_dahm2015}, 
together with photometry and lithium quantities from the DANCe database 
(\citealt{h_bouy2015b} and \citealt{barrado2016b} and other works: 
\cite{ia_steele1993}, \cite{j_bouvier1998}, 
\cite{zapateroosorio1999a}, \cite{el_martin2000c}, and \cite{stauffer2007}). 
  With the exception of Roque 5 and Roque 25, 
the rest of the sources have a counterpart in the \textit{Gaia} DR2 catalogue;
and with the exception of \mbox{CFHT-Pl-18} and \mbox{MHObd 3,} 
all the sources have parallaxes and proper motions\footnote{\mbox{CFHT-Pl-9} and \mbox{MHObd 6} 
are the same source 
(\citealt{stauffer1998a} and \citealt{j_bouvier1998})
even though the SIMBAD database treats them as two independent sources. 
We carried out a visual check of the charts from \cite{stauffer1998a} 
and \cite{j_bouvier1998} to confirm it.
Also the counterpart of \mbox{CFHT-Pl-10} in 2MASS catalogue is \mbox{2MASS J03443231+2525181}.}. 

    \cite{j_olivares2018b} studied the probability of a source being an equal-mass binary
and eight sources fulfil their criterion:
\mbox{HCG 509}, \mbox{HHJ 6}, \mbox{HHJ 3},  
\mbox{PPL-15}, \mbox{PPL-15}, \mbox{Roque 13},  
\mbox{MHObd 3}, \mbox{CFHT-Pl-12,} and \mbox{CFHT-Pl-16}. 
  Some of these sources are confirmed binary systems: 
 \mbox{CFHT-Pl-12} \citep{h_bouy2006a}, 
 \mbox{CFHT-Pl-18} \citep{el_martin1998e},  
 \mbox{PPL-15} \citep{basri1999c}, and 
 \mbox{HCG 509} \citep{stauffer2020}. 
Others are possible suspected binaries:  
 \mbox{PPL-1} with a strong absorption lithium feature, 
 \mbox{HHJ-6} and 
 \mbox{BPL-169} \citep{stauffer1998a}, 
 and \mbox{Roque 12} 
\cite[which might be an eclipsing binary brown dwarf as suggested by][based on a single eclipse]{a_scholz2020}.
 
  \citet{el_martin2000c} claimed that the lithium-rich source \mbox{CFHT-Pl-16} 
could be an unresolved binary (with identical components) or it could be a non-member. 
 \mbox{PPL-14} was observed once during a flare event 
(see Table 1, \citealt{stauffer1995a}). \cite{se_dahm2015} studied its radial velocity, 
compatible with the cluster, and its membership.
  We suggest that \mbox{MHObd 3} is a suspected multiple system due to 
the absence of astrometric solution in \textit{Gaia} DR2 and 
its position in the HRD (Fig. \ref{fig:hrd_thepleiades_general}), 
as previously suggested \cite{stauffer1998a}.
  Two sources, \mbox{MHObd 1} and \mbox{MHObd 3,} present low signal-to-noise ratio lithium detections
(\citeauthor{stauffer1998a} \citeyear{stauffer1998a} and \citeyear{stauffer1998b}),  
which would indicate that they are brown dwarfs or, at least, transition objects that have burned most of its initial lithium. 
\mbox{IPL 43} has a low signal-to-noise ratio spectrum near 6\,700~\AA\ that precludes an assessment 
of the lithium equivalent width \citep{se_dahm2015}.
  Another significant aspect is the three lithium-poor objects studied by \cite{el_martin2000c}:    
Roque 12 (spectral type M7.5), Roque 5 (M9) and Roque 25 (L0), 
all of them fainter than the LDB assumed by us (see the HRDs, Figure \ref{fig:hrds_thepleiades}). 
This fact might be due to the low quality of the spectra.
 
    From the parallax distribution, Figure \ref{fig:plxs_thepleiades}, 
we discarded as members 
\mbox{HCG 332}, \mbox{HCG 509}, 
\mbox{Calar 3}, \mbox{MHObd 1,} and \mbox{CFHT-Pl-12}. 
About \mbox{HCG 332} (HHJ 339) and \mbox{HCG 509} (HHJ 430), 
\cite{oppenheimer1997} suggested that 
they could simply be foreground weak-line T Tauri stars 
with the same proper motion and only a parallax measurement could make the membership 
assertion more certain. 
Thus, they are not members of the Pleiades as \cite{stauffer2020} recently confirmed.
  \mbox{CFHT-Pl-15}, \mbox{HCG 332,} and \mbox{HCG 509} have proper motions that are very different 
from the rest of the objects, as we show in Figure \ref{fig:vpds_thepleiades}.
  \mbox{Furthermore, HCG 509}, \mbox{HCG 332}, \mbox{CFHT-Pl-15,} and \mbox{CFHT-Pl-18} 
are non-members\footnote{\cite{j_olivares2018b} use a membership probability threshold \mbox{$p=0.84$} 
(the one with higher accuracy and only for classification purposes).  
On the other hand, \cite{h_bouy2015b} use a threshold of \mbox{$p=0.75$}.} 
due to the low membership probabilities in the DANCe catalogue 
(\citealt{h_bouy2015b} and \citealt{j_olivares2018b}) and in the work of \cite{stauffer2020}.
  In short, we show all the remarks of each object in Table E.4. 
 
    \citealt{stauffer1998a} located the LDB with the sources \mbox{CFHT-Pl-10} and \mbox{CFHT-Pl-11}, 
at $M_{\rm bol\ LDB}=11.99$~mag (\mbox{$L_{\rm bol\ LDB}=1.259\cdot 10^{-3}$~$L_{\sun}$}).
 On the other hand, \citet{barrado2004d} located the LDB with the sources 
\mbox{CFHT-Pl-09}, \mbox{CFHT-Pl-10}, \mbox{Roque 16} (\mbox{CFHT-Pl-11}), 
and \mbox{Teide 2} (\mbox{CFHT-Pl-13}), at 
\mbox{$M_{\rm bol\ LDB}=12.14$~mag} (\mbox{$L_{\rm bol\ LDB}=1.096\cdot 10^{-3}$~$L_{\sun}$}).
 \cite{se_dahm2015} located the LDB at $M_{J}=9.73\pm0.05$~mag 
(\mbox{$L_{\rm bol\ LDB}=1.464^{+0.069}_{-0.066}\cdot 10^{-3}$~$L_{\sun}$}), 
assuming the distance from \cite{c_melis2014a}. 
  We located the LDB at 
\mbox{$L_{\rm bol\ LDB}=1.321^{+0.19}_{-0.089}\cdot 10^{-3}$~$L_{\sun}$} and 
\mbox{$T_{\mathrm{eff\ LDB}}=2\,728\pm71$~K.}

    The lithium feature in Roque 5, Roque 12, Roque 25, and IPL 43 
needs to be studied in more detail. 
 \mbox{MHObd 3} lacks astrometric solution in \textit{Gaia} DR2 and requires a further study 
in terms of membership due to its position in the HRD 
(Figure \ref{fig:hrds_thepleiades}) and its lithium content. 
 \mbox{MHObd 1} was considered a non-member based on its parallax, 
even though this one is not very different from the previous members. 
It likely belongs to the tidal tail of the Pleiades.

\subsection{Blanco 1}\label{sub:blanco1_ldb}

We selected 15 sources: 
9 members studied both in \citet{pa_cargile2010c} and \citet{aj_juarez2014}; 
1 member studied only in \citet{pa_cargile2010c}
and 
5 members only studied in \citet{aj_juarez2014}.
We did not find any counterpart to \mbox{B1opt-6335} in any catalogue; 
consequently, our sample ultimately contains 14 objects.  
Five targets have lithium detection and the rest lack any. 
We took the coordinates for each source from \citet{e_moraux2007}
since
it was not possible to find the counterparts using the coordinates from \citet{aj_juarez2014}.
  We gathered photometric bands: 
$I_{c}z$ from \citet{e_moraux2007}, and 
$BV$ from \citet{pa_cargile2009} for the object \mbox{JC0-F18-88}.

  All the objects have a counterpart in \textit{Gaia} DR2, 
except the faintest object \mbox{CFHT-BL-49}.
 \mbox{Then, CFHT-BL-36} has  a negative parallax, $\varpi=-0.28\pm1.35$, and 
 \mbox{CFHT-BL-45} is detected in \textit{Gaia} DR2 but lacks astrometric solution.

    From the VPD, Fig. \ref{fig:vpd_plxs_blanco1} left,  
we discarded as members \mbox{CFHT-BL-29} and \mbox{CFHT-BL-36}. 
We also note that sources \mbox{CFHT-BL-38} and \mbox{CFHT-BL-43} have large uncertainties;
 \mbox{CFHT-BL-16}, \mbox{CFHT-BL-36,} and \mbox{CFHT-BL-38} have  parallaxes outside the 
percentiles of the parallax PDF (see Fig. \ref{fig:vpd_plxs_blanco1}) 
and we considered them as non-members. 
We note that although \mbox{CFHT-BL-43} is within the parallax distribution, 
it has huge uncertainties.
  All the objects from \cite{aj_juarez2014} have a radial velocity 
within the range of 2 to +10~km s$^{-1}$ in $3\sigma$
(see Figure 4 from \citealt{aj_juarez2014}), 
however, some objects are not within this range in $1\sigma$.
  Under this criterion, 
\mbox{CFHT-BL-22}, \mbox{CFHT-BL-25}, \mbox{CFHT-BL-29}, 
\mbox{CFHT-BL-36,} and \mbox{CFHT-BL-45} 
fall into the category of non-members.
  Lastly, we note that \mbox{CFHT-BL-22} shows lithium in its spectrum 
even though it is hotter and brighter than other lithium-poor objects 
(beyond effective temperature and bolometric luminosity uncertainties).
It fulfils the proper motion, distance membership criterion and, 
according to \cite{aj_juarez2014}, it is currently depleting lithium.

    \cite{pa_cargile2010c} delimited the LDB between \mbox{CFHT-BL-24} 
and \mbox{CFHT-BL-38}, at \mbox{$L_{\rm bol\ LDB}=1.26^{+0.40}_{-0.30}\cdot 10^{-3}L_{\sun}$}. 
  \cite{aj_juarez2014} used the sources \mbox{CFHT-BL-22} and \mbox{CFHT-BL-38} 
to establish the LDB boundaries, so the updated Blanco 1 LDB 
is located at 
$\log (L_{\rm bol\ LDB}/L_{\sun})=-2.910^{+0.062}_{-0.105}$ 
(\mbox{$L_{\rm bol\ LDB}=1.230^{+0.189}_{-0.264}\cdot 10^{-3}$ $L_{\sun}$}).
  We located the LDB at 
$L_{\rm bol\ LDB}=1.68^{+0.62}_{-0.26}\cdot 10^{-3}L_{\sun}$ and 
$T_{\mathrm{eff\ LDB}}=2\,746\pm71$~K (see the HRD, Figure \ref{fig:hrd_blanco1_zoom_ldb_v1}).  
  If we discarded those sources whose radial velocities are not within 
the range proposed by \cite{aj_juarez2014} in $1\sigma$ 
(see text above and Figure \ref{fig:hrd_blanco1_zoom_ldb_v2}), 
we located the LDB at 
$L_{\rm bol\ LDB}=1.182^{+0.69}_{-0.099}\cdot 10^{-3}L_{\sun}$ and 
$T_{\mathrm{eff\ LDB}}=2\,746\pm71$~K. 
We consider the last criterion because it contains the cleanest sample. There are few objects around the LDB and 
it would be interesting to perform a monitoring process with a wealth of objects.

\subsection{The Hyades}\label{sub:thehyades_ldb}

We gathered 34 sources from different works: 
 12 sources from \cite{e_hogan2008}, from which 
6 have a spectrum to determine if they are lithium-rich objects \citep{el_martin2018};  
 1 source from \cite{a_perezgarrido2017}, 
the L5 lithium-rich brown dwarf \mbox{2MASS J04183483+2131275} 
(2M0418, \citealt{n_lodieu2018c}) and 
21  sources from \cite{j_bouvier2008a}, 
of which only  
CFHT-Hy-10, CFHT-Hy-11, CFHT-Hy-12, CFHT-Hy-13, 
CFHT-Hy-19, CFHT-Hy-20, and CFHT-Hy-21           
have spectra, but do not cover the lithium feature.
  It is important to note that Hya10 is not considered 
a definitive lithium-rich object in \cite{el_martin2018} due to its poor detection. 
Nevertheless, we studied both situations, lithium-rich and lithium-poor object, 
intending to examine the location of the LDB and possible variations.
  In short, there are only lithium measurements for 7 objects. 
  From our selection,   
Hya09, Hya10, Hya12, 2M0418, CFHT-Hy-20, and CFHT-Hy-21    
do not have a counterpart in \textit{Gaia} DR2, 
 whilst, 
Hya02, Hya04, Hya05, Hya07, Hya11, and CFHT-Hy-16
do have a counterpart but lack proper motions and parallaxes due to their faintness.

    We collected additional data from different works:
 proper motions and distances from \citet{e_hogan2008}; 
 photometric bands $IzJHK$ and proper motions from \cite{j_bouvier2008a};
 photometric data in the $J$ band, proper motions and distances from \cite{n_lodieu2014c};
proper motions from \cite{a_perezgarrido2017}; 
 spectrophotometric distances calculated using the photometry from \cite{el_martin2018};
 and ground based parallaxes and proper motions from \cite{n_lodieu2019b}. 
  
    We considered in our study 
 \mbox{CFHT-Hy-19} (19 in Fig. \ref{fig:vpd_plxs_thehyades}) and 
 \mbox{CFHT-Hy-20} (20 in Fig. \ref{fig:vpd_plxs_thehyades}) as foreground sources. 
In addition, we noted that CFHT-Hy-21 and Hya11 have large uncertainties.  
We did not rule out any source from the VPD because all sources fall within the bulk of previous members.
 
  On the question of distances and parallaxes,
different works carried out calculations with different approaches: 
 spectroscopic distances based on an absolute magnitude versus spectral type relationship 
(\citealt{n_lodieu2014c}, \citealt{a_perezgarrido2017} and \citealt{el_martin2018}), or
 from the proper motion of the object and the cluster velocity 
(see Equation 3 from \citealt{e_hogan2008}). We followed the same method that we used before for the open clusters 
with the \textit{Gaia} DR2 data (see Section \ref{sub:lbol_teff}).
We show all the information in Appendix \ref{app:additional_tables}. 

    \cite{el_martin2018} located the LDB between the lithium-poor  
Hya11 and the lithium-rich sources Hya09 and Hya12 at
\mbox{$L_{\rm bol\ LDB}=0.079^{+0.028}_{-0.021}\cdot 10^{-3}$~$L_{\sun}$} 
  They calculated bolometric luminosities of the sources assuming calibrations 
for field L dwarfs to avoid any dependence on the distance. 
  Later, \cite{n_lodieu2018c} added the lithium-rich source 2M0418 and locate the LDB at 
\mbox{$L_{\rm bol\ LDB}=0.0501^{+0.0088}_{-0.0074}\cdot 10^{-3}$~$L_{\sun}$}.
  We located the LDB at 
\mbox{$L_{\rm bol\ LDB}=0.125^{+0.019}_{-0.013}\cdot 10^{-3}$~$L_{\sun}$}, and    
\mbox{$T_{\mathrm{eff\ LDB}}=2\,021\pm71$~K} (see Figure \ref{fig:hrd_thehyades_zoom_ldb_v1}).
  If we considered that Hya10 is not a lithium-rich object (see text above), 
the LDB location does not change, only the uncertainties change slightly:  
\mbox{$L_{\rm bol\ LDB}=0.124^{+0.019}_{-0.016}\cdot 10^{-3}$~$L_{\sun}$} 
(Fig. \ref{fig:hrd_thehyades_zoom_ldb_v2}). 

    The Hyades is the only stellar association whose LDB is placed at L spectral types.
 Another fact is that all sources except 2M0418 does not have the H$\alpha$ feature in emission. 
It would be interesting to study whether this effect takes place in L spectral type objects 
in younger associations and establish whether it can be an  indicator of age.

\subsection{\textbf{Beta Pictoris moving group}}\label{sub:bpmg_ldb}

 \begin{figure}
  \includegraphics[width=9cm, height=6.0cm]{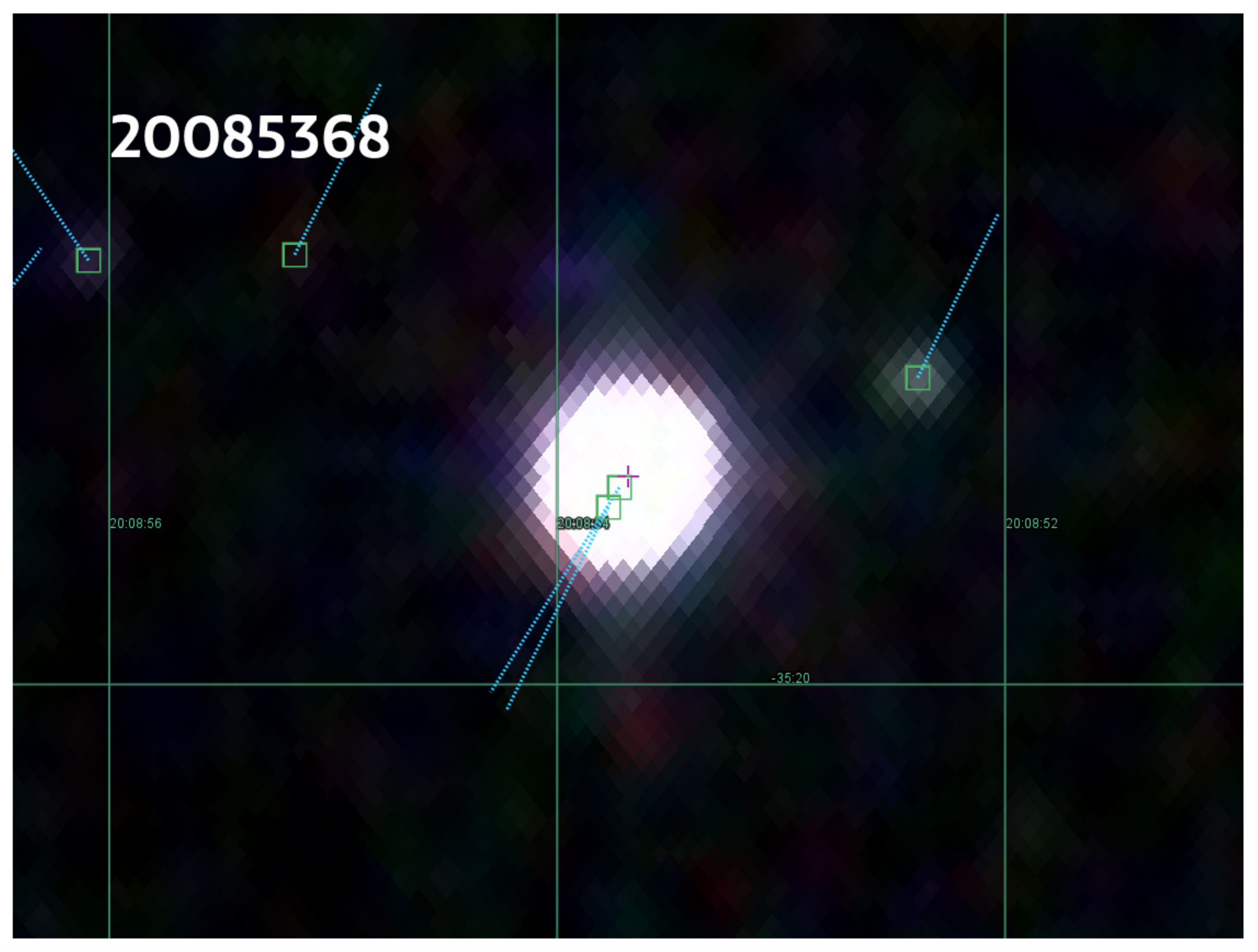}
  \includegraphics[width=9cm, height=6.0cm]{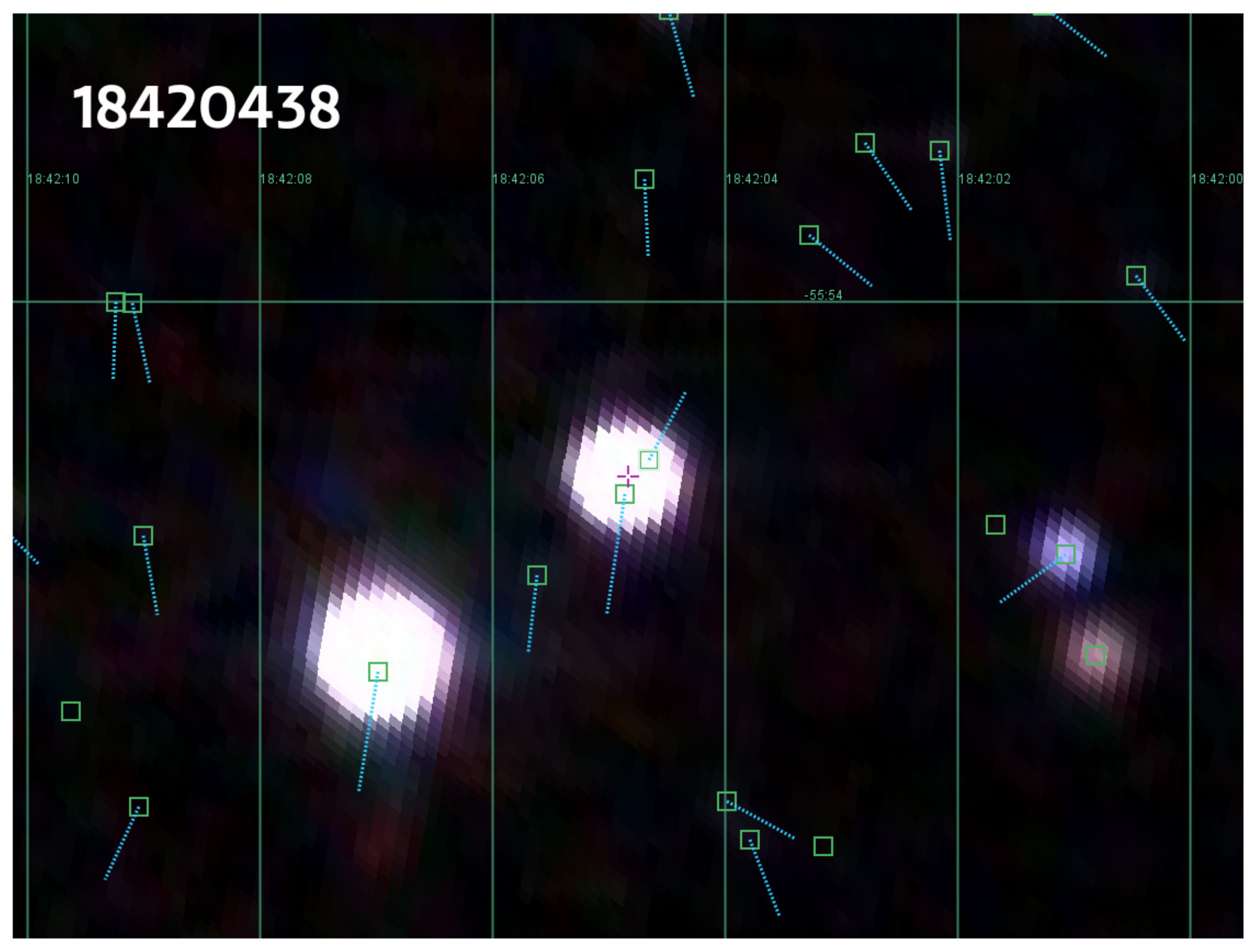}
  \includegraphics[width=9cm, height=6.0cm]{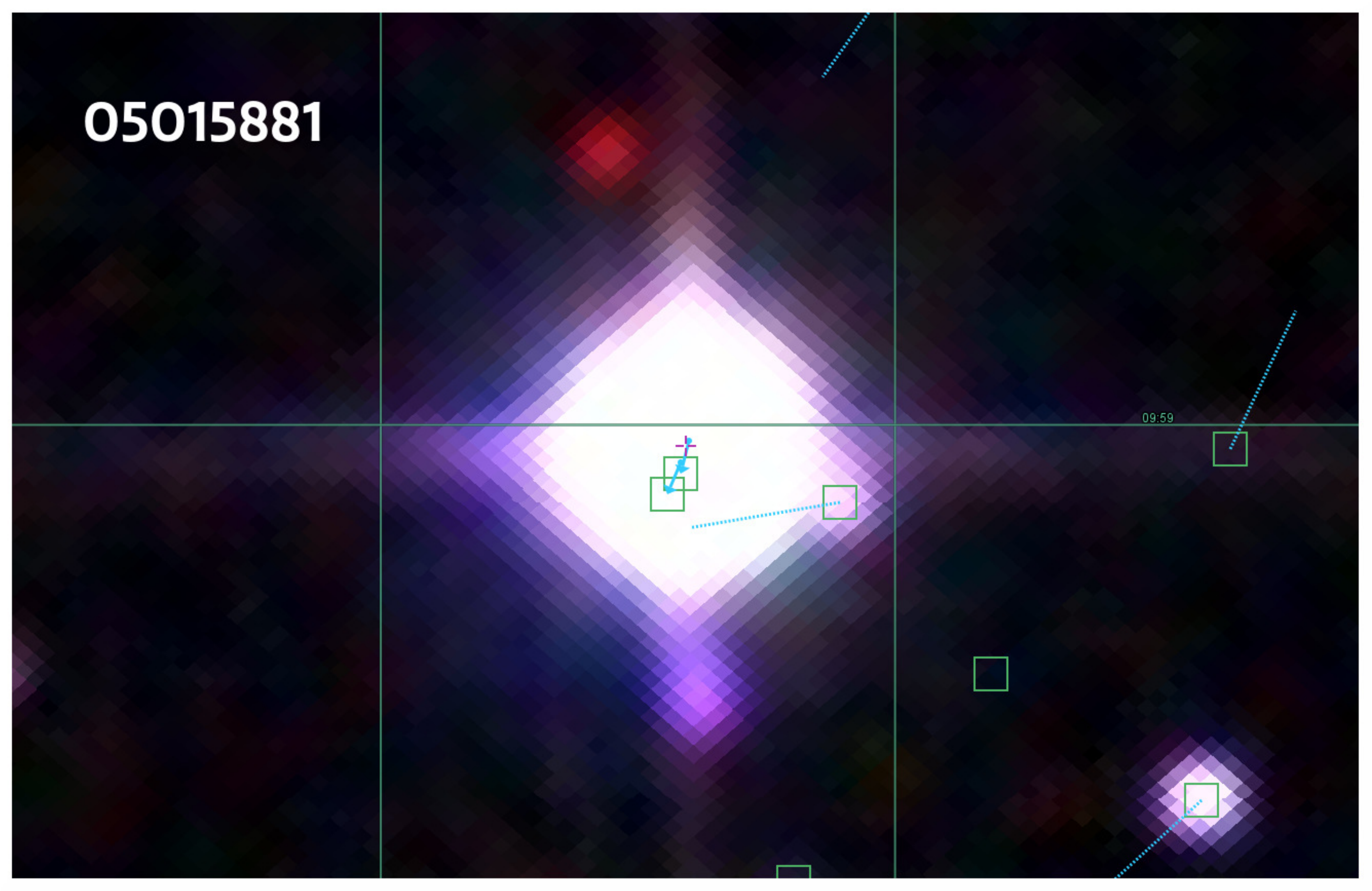}
      \caption[2MASS RGB images of BPMG sources.]
              {2MASS RGB images of three remarkable sources belonging to BPMG. 
                The main source, 2MASS catalogue, is shown with a purple cross, 
                the \textit{Gaia} DR2 counterparts are shown with green squares
                and proper motions- with cyan dotted lines. 
               {\bf Top: } 20085368 is a binary system resolved with \textit{Gaia} DR2.
               {\bf Medium: } 18420438 is a source whose photometry may be contaminated by a background source. 
               {\bf Bottom: } 05015881 is a binary system whose photometry is resolved in \textit{Gaia} DR2 
                 and there is a very close background source.
              }
         \label{fig:gaiadr2_xmatch_config}
\end{figure}
    
    We took a sample of 127 sources, all M dwarfs, 
collected from \cite{e_shkolnik2017}. 
This work is our starting point because it gathers information about multiple systems, 
possible outliers, and equivalent widths of the lithium feature. 
  We cross-matched the 127 sources with \textit{Gaia} DR2, 
by visual inspection using ESASky.

   Our final LDB sample contains 138 sources. 
Some of them have several counterparts associated in the \textit{Gaia} DR2 catalogue, and they may be multiple systems.
  In some cases, the multiple system is only well resolved in \textit{Gaia} DR2 
(with individual parallaxes, proper motions and photometric bands), for instance, 20085368 (Fig. \ref{fig:gaiadr2_xmatch_config}).  
Consequently, the object is a merge of several sources and its stellar parameters are affected. 
These objects are considered multiple systems in our study 
(marked as `CMS' in Table E.4.). 
  In others, the initial source has several associated background objects, and they might contaminate the photometry of the main object, for instance, 18420438 (Fig. \ref{fig:gaiadr2_xmatch_config}).
  We can also have a combination of both cases: 
a multiple system only well resolved in \textit{Gaia} with several associated background objects, 
for instance, 05015881 (Fig. \ref{fig:gaiadr2_xmatch_config}) or 03393700.
The photometric data are probably blended and could have erroneous stellar parameters.  
We marked them as contaminated photometry (CP) in Table E.4. 
  Finally, we searched other works to identify additional multiple systems in our LDB sample: 
\cite{bd_mason2001}, \cite{jl_beuzit2004}, \cite{bergfors2010a}, and successive references.

  Only three sources, (01112542, 03323578, and 05241914\footnote{
 These objects are not extremely faint and cool, have a counterpart in other surveys like 2MASS, and the SEDs do not have strange behaviours. 
 The most plausible reason that they do not appear in \textit{Gaia} DR2 is 
 that the astrometric solution does not converge, because they are multiple systems.})
do not have a counterpart in \textit{Gaia} DR2 
and nine lack astrometric solutions. 
  All the information related to binary, multiple systems, and background sources 
associated with any object that belongs to our LDB sample 
is shown in the Appendix \ref{app:bpmg_objects}. 

    The members are sparse and spread over the sky, located between $9$-$73$~pc \citep{l_malo2013}, 
so there is not any over-density in the sky and 
we cannot discard any member based on their VPD.
  Initially, we did not discard any source based on their parallaxes. 
Nevertheless, some of them might be outliers because they have significantly smaller parallaxes 
than other members: 03255277, 03370343, 11493184, 15063505, and 18435838. 
    
    We show in the HRD, Figure \ref{fig:hrds_bpmg},
that most of the sources follow the main sequence and 
recognise the region delimited by the lithium-poor sources. 
  The faintest ones give an idea of the suspected LDB, 
which is illustrated in Figure \ref{fig:hrd_bpmg_zoom}. 
There may be several explanations of why a lithium-rich source lies 
above the single star locus in a HRD,  
(e.g. multiplicity, high rotational velocities, or activity)  
and there is an explanation of why a lithium-poor source lies below the single star locus:
it is probably a field star, older than the association (a non-member).
  There are six lithium-poor objects below the main sequence;
three of them are: 
01365516, 05363846, and 23512227, which 
are located at fainter bolometric luminosities than the suspected LDB 
and they might be non-members.
  In contrast, the distribution of lithium-rich sources along the main sequence is blurred. 
After discarding multiple systems and sources with possible contaminated photometry (see Figure \ref{fig:hrd_bpmg_zoom}), 
we identified eight lithium-rich sources brighter than the suspected LDB: 
05082729, 20333759, 
00501752, 17173128, 18030409, 
02450826, 18011345, and 05061292.
  Two of them have undergone high activity periods: 
 05082729 with $L_{\rm x}/L_{\rm bol}=-3.19$ and 
 20333759 \footnote{Also known as SCR J2033-2556 AB}, with $L_{\rm x}/L_{\rm bol}=-3.15$  
exhibited a strong X-ray to bolometric luminosity \citep{s_messina2017c} 
and the multi-epoch radial velocity measurements \citep[see][]{l_malo2014b} have 
ruled out its binary nature.
This object suffered a flare with a magnitude increase of $\Delta V=3.29$~mag. 
  The rest of sources are above the suspected LDB, 
three with \mbox{$L_{\rm bol}\sim50$-$60\cdot 10^{-3}$~$L_{\sun}$} 
(00501752, 17173128, and 18030409) and 
the other three with \mbox{$L_{\rm bol}\sim30\cdot 10^{-3}$~$L_{\sun}$}  
(02450826, 18011345, and 05061292). 
We may need further analysis to find out if they are outliers or 
if their positions are related to other effects, for instance, multiplicity.

    \cite{i_song2002c} located the LDB between 
the lithium-poor \mbox{WW PsA} (22445794) and 
the lithium-rich \mbox{Tx PsA} (22450004), 
with both sources constituting the wide binary system \mbox{HIP 112312}, at 
\mbox{$L_{\rm bol\ LDB}=40.2^{+20.6}_{-13.6}\cdot 10^{-3}L_{\sun}$}. 
  \cite{as_binks2014} located the LDB as a rectangular region separating 
lithium-poor from lithium-rich sources 
in several colour-magnitude diagrams and calculated from the position of the central points of the boxes
at \mbox{$L_{\rm bol\ LDB}=37.6^{+12.7}_{-9.5}\cdot 10^{-3}L_{\sun}$}.
  \cite{l_malo2014b}, using a maximum likelihood method for determining the LDB 
in terms of bolometric luminosity, located it at  
$\log (L_{\rm bol\ LDB}/L_{\sun})=-1.49\pm0.08$ (\mbox{$L_{\rm bol\ LDB}=32.36^{+6.54}_{-5.44}\cdot 10^{-3}L_{\sun}$}). 
  Finally, \cite{e_shkolnik2017} added new members and refined the LDB age 
(following the same procedure as \citealt{al_kraus2014c}  in THMG; see Section \ref{sub:thmg}), 
but these authors did not find a clear gap in luminosities where there are lithium-poor and lithium-rich objects, 
They identified the LDB at \mbox{$M_{K\rm{\ LDB}}=5.6\pm0.4$~mag}, 
at \mbox{$L_{\rm bol\ LDB}=28.576^{+12.729}_{-8.806}\cdot 10^{-3}L_{\sun}$},
where there are an equal number of lithium-poor and lithium-rich sources, 
with eight stars in each group. 

    We located the LDB at 
$L_{\rm bol\ LDB}=43.61^{+0.90}_{-0.81}\cdot 10^{-3}L_{\sun}$ and
$T_{\mathrm{eff\ LDB}}=3\,095\pm71$~K, 
using the following sources: 
20083784, 00482667, 00193931, 18092970, 22450004,
22085034, 04480085, 04480258, 00194303, 23314492,
01303534, 05061292, 18011345, 21200779, 02450826,
21384755, 20333759, 05015665, 19300396, 05082729,
01351393, 22445794, 17173128, 02175601, 13215631,
18030409, 00501752, 05335981, 18055491, 21100535,
06131330, and 18151564, 
(see Figure \ref{fig:hrd_bpmg_zoom_ldb_v5}).
  However, if we take into account the faintest lithium-poor sources, 
we can locate the LDB at 
$L_{\rm bol\ LDB}=21.26^{+0.42}_{-0.21}\cdot 10^{-3}L_{\sun}$ and 
$T_{\mathrm{eff\ LDB}}=2\,986\pm 81$~K, see Figure \ref{fig:hrd_bpmg_zoom_ldb_v3}.  
In this scenario, 
we discarded all the lithium-rich sources above the suspected LDB. 
All of them are bona-fide members, 
but two sources undergone high activity periods 
and the rest could be considered as multiple systems (see previous paragraphs).
  In this case, the LDB locus is located at cooler effective temperatures and 
fainter bolometric luminosities $\Delta L_{\rm bol}=25.35\cdot 10^{-3}$ $L_{\sun}$, 
which leads to a 7 to 10$\%$~Ma age variation. 
  A comparison of both criteria reveals a difference in the LDB locus 
of a $119\%$ in terms of bolometric luminosity and 
a $36\%$ in terms of effective temperature. 
  We decided to use the first scenario because there is no reason
to rule out the lithium-rich sources. 

    In addition, several blended objects require actual bolometric luminosity 
and effective temperature estimations. 
  In addition, there are nine lithium-rich objects in the HRD region where 
lithium content should have been depleted:  
02272804, 02450826, 00501752, 18011345, 
17173128, 18030409, 05061292, 05082729, and 20333759, so 
further data are required to determine exactly how the multiplicity or activity in the X-Rays domain (or, perhaps, erroneous membership classification) might be associated with this effect.
Due to the proximity of these objects, the quality of the astrometric parallax is quite good. 
A new membership assignment using excellent astrometric and spectroscopic data as 
\cite{n_miretroig2020b} for less massive objects might resolve this jigsaw puzzle.

\subsection{\textbf{Tucanae-Horologium moving group}}\label{sub:thmg_ldb}

We collected a total of 110 sources with parallax measurements from \textit{Gaia} DR2, 
all of them confirmed as members, with spectral type M and  
lithium information taken from \cite{al_kraus2014c}.
 We have followed the same criteria as them to determine 
whether the object is lithium-rich or not:
one lithium-rich object has $\rm{EW_{Li}}>0.100$~\AA\ and 
one lithium-poor object has $\rm{EW_{Li}}<0.100$~\AA.
 We gathered all the photometric data through VOSA. 
  
    There are two binary systems confirmed through spectroscopic data \citep{al_kraus2014c}: 
\mbox{02070176\-4406380} and \mbox{04515303\-4647309}. 
We deemed these objects as multiple systems, marked as ``CMS'' in Table E.4. 
  Two sources deviate in the HRD (Fig. \ref{fig:hrds_thmg}):
\mbox{03050976-3725058} (5.465~mas, 178~pc) and 
\mbox{21354554-4218343} (0.735~mas, 563~pc),
they are located quite above the rest of sources, so we considered them as non-members.
  The source \mbox{22444835-6650032} might be a multiple system due to its position in the HRD. 
  Other sources with a similar spectral type (our source spectral type is M4.8)
    have bolometric luminosities half of that of our source.
However, \cite{j_gagne2015a} and \cite{ac_schneider2018a} do not suggest 
anything related to possible multiple systems. 

    \citet{al_kraus2014c} quantified the location of the LDB at 
\mbox{$M_{\rm bol}=9.89\pm0.10$~mag},  
\mbox{$L_{\rm bol\ LDB}=8.710^{+0.840}_{-0.767} \cdot 10^{-3}$~$L_{\sun}$}. 
They defined the LDB 
identifying the limit where equal numbers of lithium-depleted
and lithium-bearing stars encroach onto the opposite side of the boundary. 
  We located the LDB in the same way as we did in the rest of associations at  
$L_{\rm bol\ LDB}=8.572^{+0.028}_{-0.11}\cdot 10^{-3}$ $L_{\sun}$ and 
$T_{\mathrm{eff\ LDB}}= 2\,935\pm71$~K (see Fig. \ref{fig:hrd_thmg_zoom_ldb_v1}).  
  If we assumed that \mbox{22444835-6650032} is a multiple system, 
the LDB location changes in terms of bolometric luminosity:  
$L_{\rm bol\ LDB}=7.486^{+0.054}_{-0.11}\cdot 10^{-3}$ $L_{\sun}$, 
a change of $15\%$ in terms of bolometric luminosities that produces an 
age increase of 2 and 6 Ma (6 and $10\%$)
depending on the evolutionary model used (see Fig. \ref{fig:hrd_thmg_zoom_ldb_v2}).
  This last case is the one we selected.

    The membership selection from \cite{al_kraus2014c} based on lithium measurements 
might introduce a bias in the location of the LDB (see Section 5 from that work).
A possible solution would be to carry out a new membership assignment, 
as we propose for the BPMG low-mass objects.

\subsection{\textbf{32 Ori moving group}}\label{sub:32orimg_ldb}

To assess the LDB, we collected 33 M dwarfs from \cite{cpmbell2017}.
However, this number rises to 36 thanks to the multiplicity of three sources
discovered after we cross-matched our input catalogue with \textit{Gaia} DR2.
Then, we have   
THOR-14Aa (3341625121978020736) and 
THOR-14Ab (3341625126275164032) instead of THOR 14A; 
THOR-17Aa (3389908598860532096) and 
THOR-17Ab (3389908598858685696) instead of THOR-17A; 
and
THOR-40a (3341538436655619840) and 
THOR-40b (3341538436653898240) instead of THOR-40.  
  THOR-14Aa and THOR-14Ab are two sources in \textit{Gaia} DR2 
with different astrometric parameters,  
unlike the rest of the photometric surveys, where they are blended.
The bolometric luminosity is obtained with the total observed blended flux
and calculated at both distances. 
  All the sources have a counterpart in the \textit{Gaia} DR2 catalogue, 
but 
THOR-17Aa, THOR-17Ab, THOR-18, THOR-35, THOR-36, THOR-40a and THOR-40b 
lack of parallaxes and proper motions.
 
    Previous works (\citealt{cpmbell2017} and \citealt{sj_murphy2020a})
reported additional information that we took into account.
  THOR-17A (so THOR-17Aa, and THOR17Ab), THOR-18, THOR-21, THOR-26, and THOR-37 
showed strong \ion{He}{i} $\lambda$~5876\AA\ and 6876~\AA\ emission lines.
  There are several confirmed spectroscopic binaries: 
 THOR-18, 
 THOR-31 (a binary system), 
 THOR-33 (a $5^{\prime\prime}$ visual binary), and recently, 
 \cite{sj_murphy2020a} reported that THOR-42
(2MASS J05525572-0044266) 
is a young eclipsing binary comprising two pre-main sequence M dwarfs 
with combined spectral type M3.5. 
   Some objects are suspected spectroscopic binary systems: 
 THOR-15 
and 
 THOR-37; and  
others are considered fast rotators:
 THOR-05, THOR-07, THOR-23, and THOR-34. 

\begin{figure}
  \includegraphics[width=0.46\textwidth,scale=0.50]{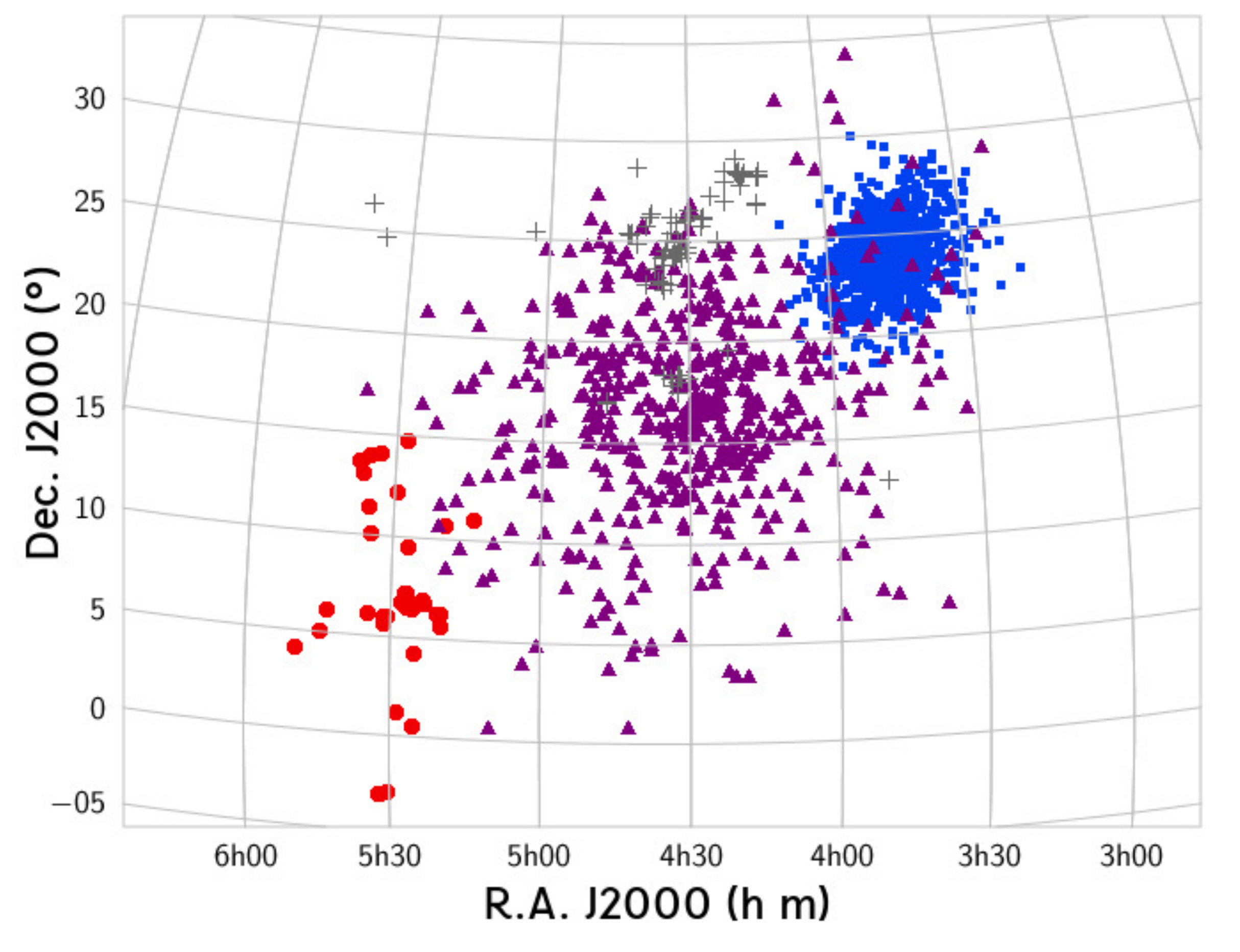}  
  \caption[Distance 32 Ori MG members.]
           {Location of all 32 Ori MG sources (red circles from \citealt{cpmbell2017}), 
            together with the Hyades (purple triangles), 
            the Pleiades (blue squares), and 
            Taurus members (grey pluses from \citealt{c_ducourant2005}). 
            The Hyades and the Pleiades members are taken from \cite{c_babusiaux2018}. 
           }
         \label{fig:sky_32orimg}
   \end{figure}

\begin{figure*}
   \centering
   \includegraphics[width=0.95\textwidth,scale=0.75]{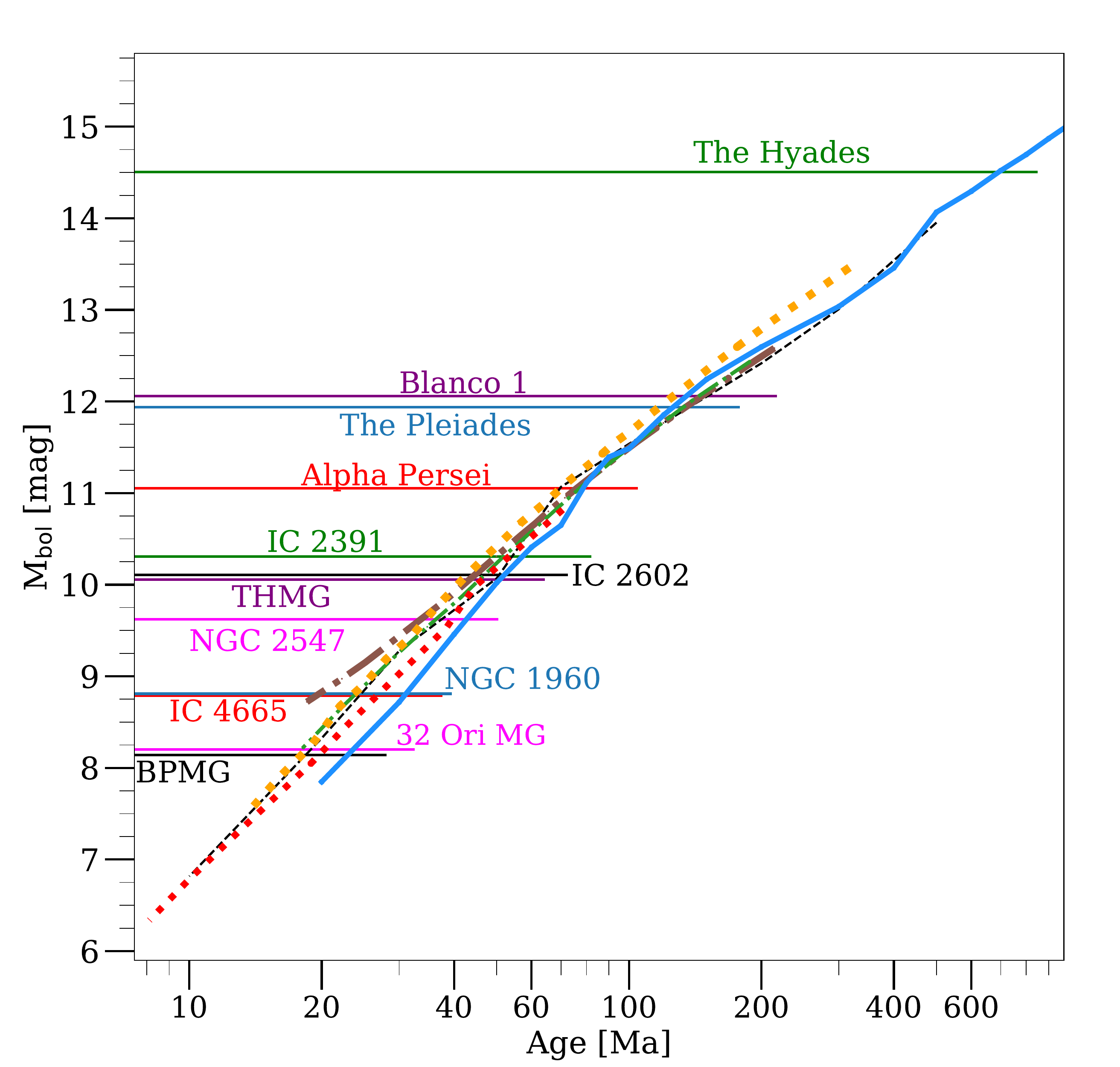}   
      \caption[LBD loci vs. age]
              {Location of the LDB for all studied stellar associations and its relation with the age. 
               We show bolometric magnitudes of the LDB for each stellar association as horizontal lines (several colours).
               Uncertainties are not shown for clarity.
               The lines that cross diagonally from the lower left corner 
               to the upper right one correspond to different LDB models: 
               the dashed black line is the \cite{f_dantona1994} LDB, 
               the dashed dotted brown thicker line is the \cite{a_burrows1997b} LDB, 
               the dashed dotted green thinner line is the \cite{cj_burke2004} LDB, 
               the dotted red line thicker corresponds to the Pisa LDB \citep{e_tognelli2015b},  
               the dotted red line thiner correspond to the \cite{siess2000} LDB and 
               the blue line is the BT-Settl LDB \citep{f_allard2013}.
              }
         \label{fig:mbol_ldb_vs_ldb_age}
\end{figure*}

    32 Ori MG members are located in a bounded sky area  
(as we show in Figure \ref{fig:sky_32orimg}), unlike other MGs 
(see SACY by \citealt{cao_torres2006} or \citealt{l_malo2013}).
We performed a similar analysis to the one done with the open clusters, 
  We discarded those sources that have 
parallaxes outside the interval \mbox{$\varpi \in(8.0, 11.0)$~mas}: 
THOR-27 (a background source),
along with three foreground sources, 
2MASS J05053333+0044034, THOR-33, and THOR-34. 
  We did not discard any source in the VPD because none of these sources are placed
outside the area described by other members (as can be checked in Figure \ref{fig:vpds_32orimg}. 
With regard to 2MASS J05053333+0044034 (plotted as 05053333), THOR-33, and THOR-34, 
they show discrepant proper motions. 

    \cite{cpmbell2017} located the LDB with the sources 
\mbox{THOR 30} (lithium-poor object) and  \mbox{THOR 32} (lithium-rich object)
at \mbox{$L_{\rm bol\ LDB}=31.7^{+1.6}_{-1.5}\cdot 10^{-3}L_{\sun}$}. 
  We located the LDB at 
\mbox{$L_{\rm bol\ LDB}=41.3^{+2.4}_{-2.1}\cdot 10^{-3}L_{\sun}$} and 
\mbox{$T_{\mathrm{eff\ LDB}}=3\ 135\pm71$~K}, see Figure \ref{fig:hrd_32orimg_zoom_ldb_v1}.

    \cite{stauffer2020} found that \mbox{HCG 332} (HHJ 339) and \mbox{HCG 509} (HHJ 430), 
as described in Section \ref{sub:thepleiades}, share space motions that match those of 32 Ori MG. 
The inclusion of these objects in our analysis does not affect to the LDB loci, 
because \mbox{HCG 509} is a binary system \citep{stauffer2020} and \mbox{HCG 332} is fainter than the LDB. 
We also show these objects in the HRD, Figure \ref{fig:hrd_32orimg_zoom_ldb_v1}.
  In addition, \mbox{THOR-35} and \mbox{THOR-36} lack parallaxes and are key objects around the LDB.
So, it would be interesting to obtain parallaxes for them. 

    Finally, we studied the possibility that 32 Ori MG is made up of two populations, 
see the details in the Appendix \ref{app:32orimg_one_or_two}.
Briefly, from our analysis we distinguished two populations:  
  32 Ori MG-Pop 1 with a LDB located at
\mbox{$L_{\rm bol\ LDB}=41.2^{+2.2}_{-14}\cdot 10^{-3}L_{\sun}$} and 
\mbox{$T_{\mathrm{eff\ LDB}}=3\ 157\pm71$~K} (Fig. \ref{fig:hrd_32orimg_zoom_ldb_pop1}); and 
  32 Ori MG-Pop 2 with a LDB located at 
\mbox{$L_{\rm bol\ LDB}=30.06^{+6.1}_{-0.32}\cdot 10^{-3}L_{\sun}$} and 
\mbox{$T_{\mathrm{eff\ LDB}}=3\ 098\pm71$~K} (Fig. \ref{fig:hrd_32orimg_zoom_ldb_pop2});
    In addition we did an extra analysis. 
The 32 Ori MG sources belong to two different stellar groups from \cite{m_kounkel2019b}:
  Theia 133 and Theia 370.
  We located the LDB at 
\mbox{$L_{\rm bol\ LDB}=39.4^{+3.9}_{-1.4}\cdot 10^{-3}L_{\sun}$} and 
\mbox{$T_{\mathrm{eff\ LDB}}=3\ 139\pm71$~K}, for Theia 133 (Fig. \ref{fig:hrd_theia133_zoom_ldb_v1}); and 
\mbox{$L_{\rm bol\ LDB}=29.8^{+2.4}_{-2.7}\cdot 10^{-3}L_{\sun}$} and 
\mbox{$T_{\mathrm{eff\ LDB}}=3\ 087\pm71$~K}, for Theia 370 (Fig. \ref{fig:hrd_theia370_zoom_ldb_v1}).

\setcounter{table}{2}
\begin{table*} \begin{center}
\tiny
\caption[The LDB loci in terms of $L_{\rm bol}$, $T_{\rm {eff}}$ and the ages derived using several evolutionary models.]
        { LDB loci in terms of $L_{\rm bol}$, $T_{\rm {eff}}$ and the ages derived using several evolutionary models.} 
\label{tab:ldb_ages_lbol}
\begin{tabular}{lcc cccccc}
\hline \hline
Stellar     &                              &                       &\multicolumn{6}{c}{LDB ages based on $L_{\rm{bol}}$}                                                                                           \\
                                                                    \cline{4-9}                                                                                                                                   
Association &$      L_{\rm{bol\ LDB} }    $&$  T_{\rm{eff\ LDB}}  $& D\&M                  & Burrows               & Siess              & Burke                 &   BT-Settl             &   Pisa                   \\
            &$10^{-3}$ [$L_{\sun}$]        &         [K]           & [Ma]                  & [Ma]                  & [Ma]               & [Ma]                  &   [Ma]                 &   [Ma]                  \\
\hline\hline
            &                              &                       &                        &                  &                      &                         &                        &                                 \\    
Alpha Persei&$ 2.98^{+0.35}_{-0.19}       $&$      2\,857         $&$ 69.8^{+4.5}_{-1.6} $&$  76.1^{+3.3}_{-5.2} $&$  >70             $&$   77.8^{+2.5}_{-4.9}$&$    79.0^{+1.5}_{-2.3}$&$   70.1^{+2.8}_{-4.2}   $\\
\medskip
            &                              &                       &   (3\,222)             &   (3\,113)        &   ($<$3\,230)        &    (3\,109)             &     (2\,983)             &    (3\,001)                \\
NGC 1960    &$23.6^{+12}_{-4.9}           $&$      3\,123         $&$  26.2^{+2.3}_{-5.4}$&$  19.8^{+4.0}       $&$27.7^{+3.5}_{-5.6}$&$   24.4^{+3.4}_{-4.6}$&$    31.6^{+3.9}_{-4.7}$&$   23.4^{+3.3}_{-3.8}   $       \\
\medskip
            &                              &                       &   (3\,461)             &   (3\,462)       &   (3\,546)           &    (3\,368)             &     (3\,370)             &     (3\,344)                \\
IC 4665     &$23.6^{+2.1}_{-1.6}          $&$      3\,110         $&$  26.1^{+0.8}_{-1.0}$&$   19.7^{+1.2}       $&$27.7^{+0.9}_{-1.1}$&$   24.4^{+1.1}_{-1.5}$&$    31.6^{+1.3}_{-1.6}$ &$   23.4^{+0.6}_{-0.8}   $    \\
\medskip
            &                              &                       &   (3\,461)             &   (3\,462)       &   (3\,546)           &    (3\,369)             &     (3\,370)           &       (3\,345)             \\
NGC 2547    &$11.13^{+2.4}_{-0.89}        $&$      3\,004         $&$  40.6^{+2.3}_{-6.1}$&$   35.1^{+1.5}_{-4.0}$&$40.1^{+1.3}_{-3.1}$&$   36.7^{+1.9}_{-4.0}$&$    43.5^{+1.7}_{-4.0}$&$   34.1^{+1.2}_{-3.0}   $     \\
\medskip    
            &                              &                       &   (3\,377)              &   (3\,329)        &   (3\,440)            &    (3\,236)            &     (3\,267)             &   (3\,229)                \\
IC 2602     &$ 7.14^{+1.2}_{-0.52}        $&$      2\,926         $&$ 51.1^{+2.3}_{-3.5}$&$   45.6^{+1.8}_{-4.3}$&$50.2^{+3.1}_{-6.0}$&$   47.6^{+1.8}_{-4.1}  $&$    52.5^{+2.2}_{-3.7}$&$   42.5^{+1.7}_{-3.1}   $        \\
\medskip
            &                              &                      &   (3\,342)              &   (3\,265)        &   (3\,351)           &    (3\,191)             &     (3\,203)             &    (3\,160)                 \\
IC 2391     &$ 5.93^{+0.23}_{-0.13}       $&$      2\,956         $&$  56.5^{+0.6}_{-1.0}$&$   49.9^{+0.7}_{-0.8}$&$57.4^{+0.8}_{-1.4}$&$   52.0^{+0.9}_{-1.1}$&$    57.7^{+0.5}_{-1.0}$&$   47.0^{+0.6}_{-1.0}   $        \\
\medskip    
            &                              &                      &   (3\,307)              &   (3\,240)        &   (3\,308)           &    (3\,162)             &     (3\,170)             &    (3\,134)               \\
The Pleiades&$ 1.321^{+0.19}_{-0.089}     $&$      2\,728         $&$155.5^{+8.4}_{-17}  $&$ 134.5^{+7.4}_{-13}  $&$  >70             $&$  134.9^{+4.7}_{-9.7}$&$   127.4^{+6.3}_{-10} $&$  117.6^{+5.6}_{-9.7}   $         \\
\medskip
            &                              &                       &   (3\,057)             &   (2\,922)        &   ($<$3\,230)        &    (2\,959)             &     (2\,793)             &    (2\,834)                \\
Blanco 1    &$ 1.182^{+0.69}_{-0.099}     $&$      2\,746         $&$168.6^{+9.3}_{-65}  $&$ 146^{+10}_{-42}     $&$  >70             $&$  146^{+12}_{-40}      $&$   137.1^{+7.0}_{-33} $&$  126.6^{+7.8}_{-33}    $         \\
\medskip
            &                              &                       &   (3\,031)             &   (2\,891)        &   ($<$3\,230)        &    (2\,935)             &     (2\,758)             &    (2\,800)                \\ 
The Hyades  &$ 0.124^{+0.019}_{-0.016}    $&$      2\,021         $&$>500             $&$>210               $&$  >70             $&$ >188                $&$   695^{+85}_{-67}    $&$ >317             $\\
\medskip
            &                              &                       &   ($<$2\,265)          &   ($<$2\,741)     &   ($<$3\,230)        &    ($<$2\,818)          &     (1\,969)             &    ($<$2\,331)             \\
BPMG        &$43.61^{+0.90}_{-0.81}       $&$      3\,095         $&$  19.4^{+0.1}_{-0.1}  $&$ <18.5              $&$20.0^{+0.2}_{-0.2}$&$  <18.1               $&$   24.3^{+0.3}_{-0.3}$ &$17.4^{+0.2}_{-0.2}$        \\     
\medskip
            &                              &                       &   (3\,559)             &($>$3\,482)           &   (3\,641)         & ($>$3\,467)            &     (3\,432)          &       (3\,440)            \\
THMG        &$ 7.486^{+0.054}_{-0.11}     $&$      2\,935         $&$  49.7^{+0.3}_{-0.1}$&$   44.3^{+0.4}_{-0.2}$&$48.1^{+0.7}_{-0.3}$&    $46.4^{+0.4}_{-0.2}$&$   51.0^{+0.5}_{-0.2}$&$   41.3^{+0.4}_{-0.2}   $  \\
\medskip
            &                              &                      &   (3\,350)              &   (3\,273)        &   (3\,363)           &    (3\,200)             &     (3\,212)             &    (3\,168)       \\
32 Ori MG   &$41.3^{+2.4}_{-2.1}          $&$      3\,135         $&$  19.6^{+0.2}_{-0.2}$&$  <18.5              $&$20.6^{+0.5}_{-0.6}$&$   18.1^{+0.5}       $&$    25.0^{+0.7}_{-0.8} $&$17.9^{+0.6} _{-0.5}$      \\
\medskip
            &                              &                       &   (3\,553)             &  ($>$3\,482)     &   (3\,631)           &    (3\,467)             &     (3\,426)           &       (3\,431)             \\
\hline
32 Ori MG-P1&$41.8^{+2.2}_{-14}           $&$      3\,157         $&$  19.6^{+4.6}_{-0.2}  $&$  <18.5             $&$20.5^{+5.0}_{-0.6}$&$   18.1^{+4.0}        $&$    24.9^{+4.5}_{-0.7} $&$17.8^{+4.}_{-0.4}$         \\
\medskip
            &                              &                       &   (3\,555)             &  ($>$3\,482)     &   (3\,634)           &    (3\,467)             &     (3\,427)           &       (3\,433)              \\
32 Ori MG-P2&$30.06^{+6.1}_{-0.33}        $&$      3\,098         $&$  23.1^{+0.2}_{-2.9}$&$  <18.5              $&$24.3^{+0.2}_{-2.4}$&$   21.4^{+0.1}_{-1.9}$&$    28.6^{+0.1}_{-2.0}$ &$   21.2^{+0.1}_{-1.9}$      \\
\medskip
            &                              &                       &   (3\,501)             &  ($>$3\,482)     &   (3\,582)           &    (3\,409)             &     (3\,396)           &       (3\,380)              \\
Theia 133   &$39^{+18}_{-17}              $&$      3\,139         $&$  19.8^{+7.0}_{-1.6}$&$  <18.5^{+2.2}       $&$21.1^{+7.3}_{-3.3}$&$   18.5^{+6.7}       $&$    25.6^{+7.0}_{-5.5}$&$   18.4^{+5.5}_{-3.4}   $      \\
\medskip
            &                              &                       &   (3\,549)             &  ($>$3\,482)     &   (3\,624)           &    (3\,460)             &     (3\,421)             &     (3\,423)                \\
Theia 370   &$29.8^{+8.3}_{-8.4}          $&$      3\,087         $&$  23.3^{+3.9}_{-3.4}$&$ <18.5^{+2.9}        $&$24.4^{+4.4}_{-3.0}$&$   21.5^{+4.4}_{-2.6}$&$    28.7^{+4.7}_{-2.6}$&$   21.3^{+3.2}_{-2.5}   $         \\
\medskip
            &                              &                       &   (3\,499)             &  ($>$3\,482)     &   (3\,580)           &    (3\,407)             &     (3\,396)             &     (3\,378)                \\    
\hline
\end{tabular}
\\ 
 \begin{flushleft}
 $\,$
          $L_{\rm{bol\ LDB}}$ and $T_{\rm{eff\ LDB}}$ are the locations of the LDB in terms of bolometric luminosity and effective temperature.  
            The rest of the columns are the ages estimated using $L_{\rm{bol\ LDB}}$ and the next evolutionary models: 
             \citealt{f_dantona1997a} (``D\&M''), 
             \citeauthor{a_burrows1993a} \citeyear{a_burrows1993a} and \citeyear{a_burrows1997b} (`Burrows'),  
             \citealt{siess2000} (`Siess'), \citealt{cj_burke2004} (`Burke'),  
             \citealt{f_allard2012} (`BT-Settl'), and \citealt{e_tognelli2015b} (`Pisa'). We add in parenthesis, below each estimated age, the $T_{\rm{eff\ LDB}}$ at which the LDB would be expected to be found, 
             see Section \ref{sub:age_comparison}.\\ 
  \textbf{Notes: }              
          We added the LDB ages of the 32 Ori MG subsamples and sub-populations for completeness 
             (see Apendix \ref{app:32orimg_one_or_two}).
 \end{flushleft}  
\end{center} 
\end{table*} 
 
\clearpage
\onecolumn
\clearpage
\twocolumn

\section{LDB error budget}\label{sec:ldb_error_budget}

    In this section, we discuss 
the sources of error in our LDB loci and their derived ages. 
We study the accuracy of our derived stellar parameters, 
 the importance of reddening in the LDB location,    
 and we compare our LDB ages with those derived using other techniques. 
We also discuss some caveats in obtaining our estimations.

\subsection{Accuracy of the stellar parameters}\label{sub:accuracy_stellar_params}

    In order to estimate the accuracy of our stellar parameters, both assumed and calculated 
(metallicity, surface gravity, reddening, effective temperature, and distance), 
we made a comparison with the values from \cite{fede_anders2019}\footnote{We accessed the data 
via the \textit{Gaia} mirror archive at the Leibniz-Institut f{\"u}r Astrophysik Potsdam: 
{\tt https://gaia.aip.de/}, using the \textit{Gaia} DR2 {\tt source\_id}.} 
and from \cite{randich2018}. 
  In addition, we employed two grids of model atmospheres to study
their effect on the derivation of the total flux of each object.

\subsubsection{Comparison with \cite{fede_anders2019}}\label{subsub:comparison_starhorse}

\cite{fede_anders2019} derived stellar parameters, distances, and reddenings for $285\cdot 10^{6}$ stars down to $G=18$~mag, 
using the \texttt{StarHorse} code \citep{aba_queiroz2018} and 
combining photometric catalogues and parallaxes from \textit{Gaia} DR2. We calculated the difference between 
the {\tt StarHorse} stellar parameters and 
the ones we either assumed or derived. 
We applied four filters to the {\tt StarHorse} dataset:
the first one with all the sources in common,  
the second one with a clean sub-sample that fulfils   
{\tt SH\_GAIAFLAG=``000''} and {\tt SH\_OUTFLAG=``00000''} (see \citealt{fede_anders2019} for an explanation),
the third one only shows the members after our work, and 
the fourth one is the clean sub-sample of the previous one. 
We provide the results grouped into stellar associations in Table \ref{tab:vosa_vs_starhorse}.
  No object was found in the case of the NGC 2547 LDB sample, 
only two objects from Alpha Persei and the Pleiades, and 
only the hotter and brighter objects with $T_{\rm eff}>3\,100$~K, from IC 4665 and Blanco 1. 
 On the other hand, the least massive objects with later spectral types were found 
due to their proximity in the youngest (32 Ori MG and BPMG) and oldest (the Hyades) associations. 

    We assumed a homogeneous metallicity for all the objects in a stellar association given by the mean value (the values are shown in Table \ref{tab:sample_clusters}). 
  Some of them are not tabulated in the models, so we have taken the closest one. 
Our fixed values and those calculated with {\tt StarHorse} agree. 

    A comparison of the two approaches reveals that the fixed value of 
\mbox{$\log g=4.5$} for all the objects is consistent.  
However, we discern that the {\tt StarHorse} values in the Hyades, the oldest association,
are close to \mbox{$\log g=5.0$~dex} instead of \mbox{$\log g=4.5$~dex}.
   We calculated the stellar parameters of the Hyades using \mbox{$\log g=4.5$} and \mbox{$\log g=5.0$~dex}. 
The total flux, {\tt Ftot}, is not affected by the change.  
The differences in terms of percentages are $0.7\%$, with a standard deviation of $3\%$. 
This fact is not reflected in the $T_{\rm eff}$, 
as the calculated values using \mbox{$\log g=5.0$~dex} are hotter by 26~K on average 
(with standard deviation 115~K). 
So, we fixed \mbox{$\log g=5.0$~dex} for the Hyades and \mbox{$\log g=4.5$~dex} for the rest of the associations.

    The comparison between our calculated distances and that in {\tt StarHorse} must be done carefully. 
  While \cite{fede_anders2019} re-calibrated the parallaxes and their uncertainties
(see further explanation in Table 1 and Section 2), we decided not to apply any corrections. 
  The \textit{Gaia} DR2 parallaxes show several systematic uncertainties  
on a small scale (depending on magnitude, colour, and position) and 
on a large scale 
(as the global parallax zero-point offset of $-0.029$~mas, 
see further explanations from \citealt{f_arenou2018} and \citealt{l_lindegren2018}).
  These last two papers suggest treating the parallax zero-point as an adjustable parameter, 
adding that this is not always possible for very small samples or when the distance is nearly constant in the sample, 
as in a stellar cluster. 
So, we decided not to apply this global offset in our analysis\footnote{Other works, as \cite{s_meingast2021},  
do not apply the parallax-offset value, 
because it is a global average and can be significantly different on local scales.}.
Even so, we tested the impact of this offset in Sect. \ref{sub:effect_d_ldb}.

\begin{figure}[t]
   \includegraphics[width=9cm]{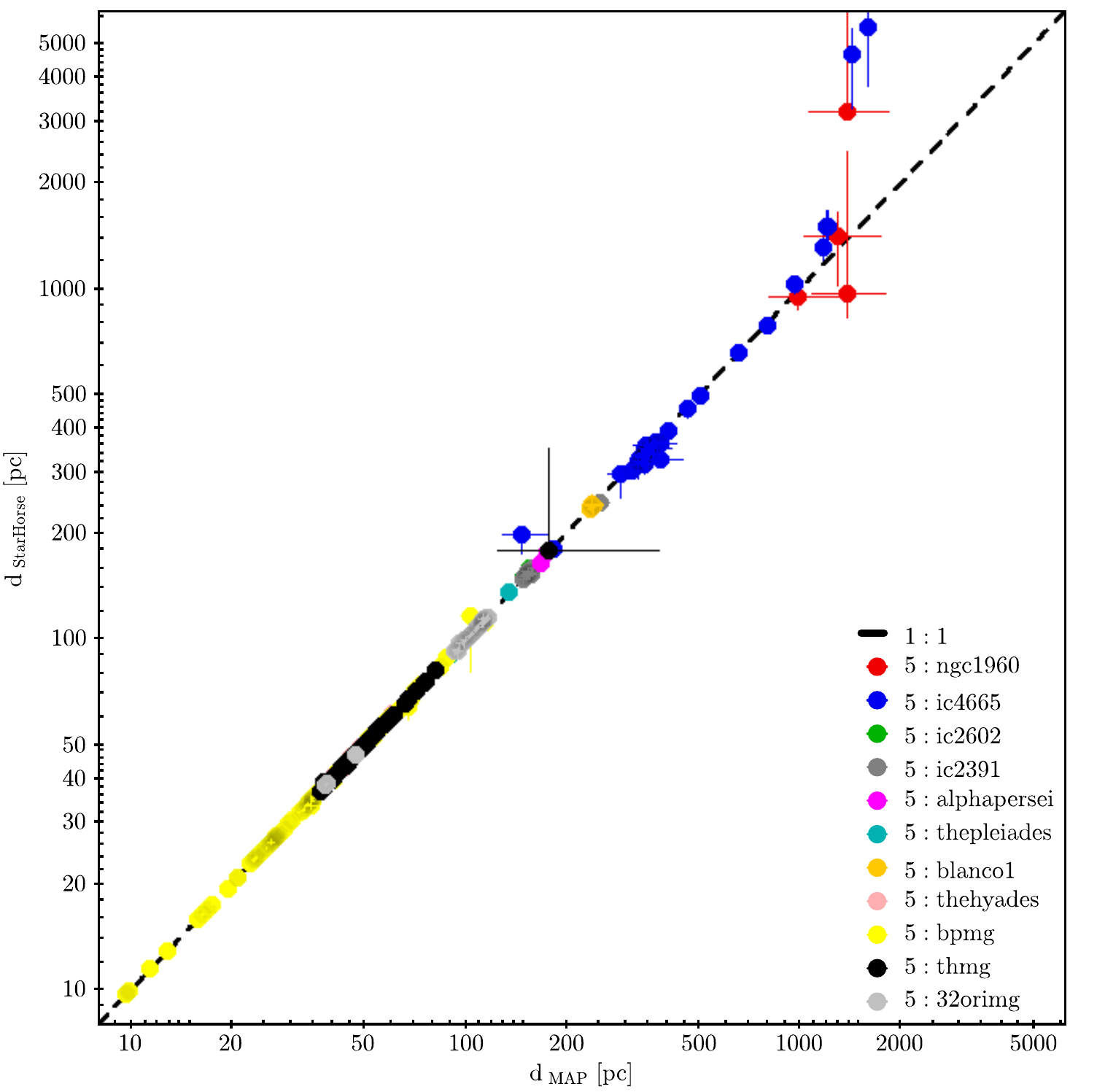}
     \caption[Comparison of our calculated distances and those derived in \cite{fede_anders2019}.]
             {Comparison of our calculated distances, $d_{\rm MAP}$ and 
              those derived with {\tt StarHorse} \citep{fede_anders2019}.
             }
     \label{fig:dist_starhorse_vs_kalkayotl} 
\end{figure}

    Concerning the distance calculations, 
\cite{fede_anders2019} adopted a Bayesian approach using elaborate priors with the aim of 
estimating distances for a complex variety of objects in all the Milky Way 
(priors include information about stellar evolutionary models and 
density laws for the main components of the Milky Way as the thin and thick disc, bulge, and halo, 
as well as the broad metallicity and age ones for those components).
  By contrast, our goal is to calculate the distance of an object (as well as possible) belonging to a stellar association, thus, we used a suitable Gaussian prior 
(see Section \ref{sub:distances} for the explanation).
 Here,    {\tt StarHorse} gives the distance values in the 5th, 16th, 50th and 95th percentiles, 
where {\tt dist50} is considered the main estimator of the distance, and the uncertainties are given by {\tt dist16} and {\tt dist84}. 
In our case, the maximum-a-posteriori is the main estimator of the distance, 
and the 2.5th and 97.5th percentiles give the uncertainties.
However, in Figure \ref{fig:dist_starhorse_vs_kalkayotl}, all the uncertainties were determined at 16th and 84th percentiles, showing a good agreement between both methods.
The most significant discrepancies are found at distances \mbox{$d_{\rm MAP}>950$~pc}.
  The poorest agreement is found in objects from NGC 1960 and IC 4665, 
which are the furthest associations.
  The NGC 1960 sample only contains four objects, two of them with low quality 
as it is shown by the {\tt StarHorse} flags.
  The IC 4665 sample contains a a large number of sources considered as non-members after this work. 
These objects have systematically higher {\tt StarHorse} distances than our calculated values.
 As a result of the comparison, 
we can conclude that this effect affects non-member background objects. 

    Therefore, the mere fact of using the Bayesian approach is not enough to calculate reliable distance measurements. 
We assumed that all the objects belong to the open cluster and chose a prior accordingly.
 While bona fide members have accurate distances, the background objects and outliers 
with parallaxes higher than $20\%$ are likely to be biased, 
as we see when making comparisons with {\tt StarHorse} results. 
  This suggests that it is necessary to discard outliers using the parallaxes 
instead of the calculated distances.

\begin{figure}[t]
   \includegraphics[height=7.2cm,width=9cm]{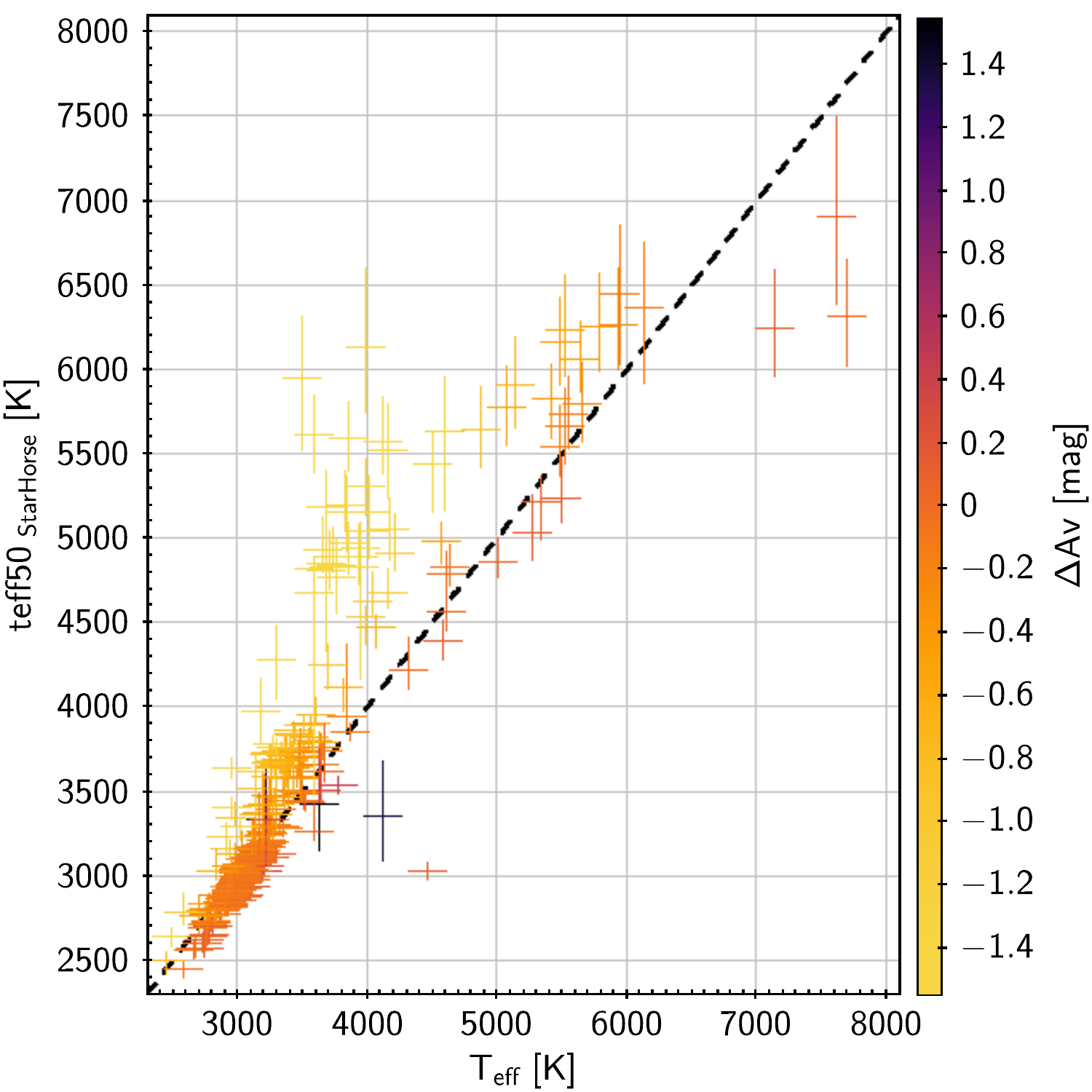}
   \includegraphics[height=7.2cm,width=9cm]{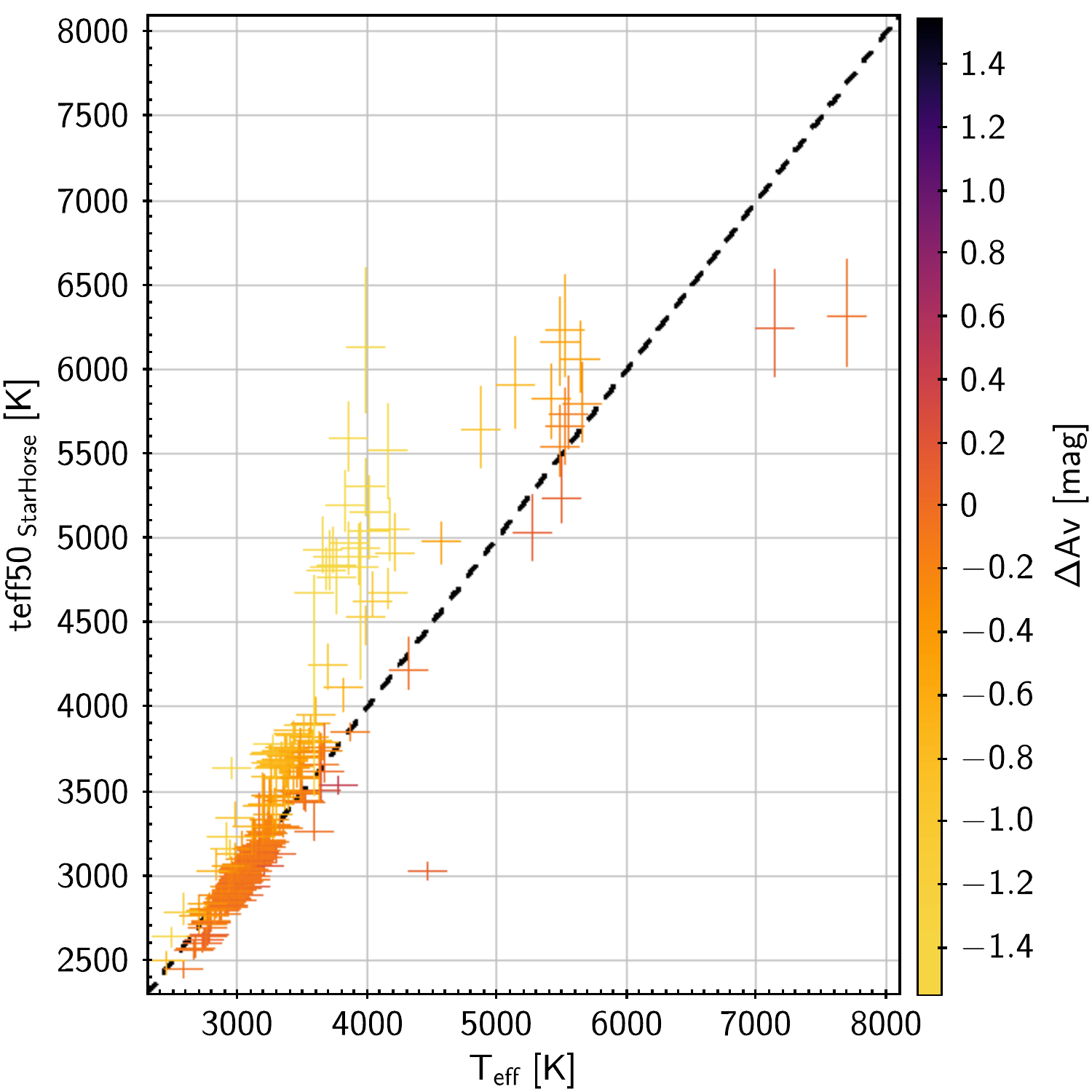}
     \caption[Comparison of effective temperatures derived with other methods.]
             {Comparison of effective temperatures derived in \cite{fede_anders2019} and from VOSA.
              {\bf Top: }Sources from the LDB samples. 
                 Note we did not find any NGC 2547 counterpart in StarHorse. 
              {\bf Bottom: }Only those sources considered as members after this work.
             }
     \label{fig:teff_starhorse_vs_vosa} 
\end{figure}

    Another quantity, namely, interstellar reddening, affects the stellar parameters and 
in particular, the effective temperature.  
  We show our results for the effective temperatures with those from {\tt StarHorse} 
together with the reddening, $A_{V}$ in Figure \ref{fig:teff_starhorse_vs_vosa}. 
We note that $T_{\rm eff}$ is our calculated effective temperature, and 
{\tt teff50} is the ${\tt StarHorse}$ effective temperature at 50th percentile.
  As a matter of fact, we can distinguish three different zones depending on the effective temperature: 
 (1) zone A with \mbox{$T_{\rm eff}<3\,200$~K}, populated by the objects that are considered in the LDB location,  
 (2) zone B with \mbox{$T_{\rm eff}\in (3\,200, 3\,700)$~K}, and 
 (3) zone C with \mbox{$T_{\rm eff}>3\,700$~K}.
  In zone A, there is no significant difference between the two effective temperatures 
$\Delta T_{\rm eff}=-9$~K (standard deviation 127~K) and a   
$\Delta A_{V}=-0.20$~mag (standard deviation 0.34~mag).  
There are objects from all the stellar associations included, but there is a greater presence of objects from the Hyades, BPMG, and THMG. 
One implication of this is the possibility that the best match corresponds to those 
objects with very low or null values of $A_{V}$, for instance, the Hyades, THMG, and BPMG.
  In zone B, we measured an offset of 
$\Delta T_{\rm eff}=-330$~K (standard deviation 468~K) and  
$\Delta A_{V}=-0.63$~mag (standard deviation 0.76~mag), and we systematically found lower values than those from {\tt StarHorse}.
  In zone C, in overall terms, there is an offset of 
$\Delta T_{\rm eff}=-440,$ on average (standard deviation 710~K), and 
$\Delta A_{V}=-0.68$~mag on average (standard deviation 0.79~mag).
  We observed two trends: 
 objects very close to the 1:1 line with small differences 
between $T_{\rm eff}$ and $A_{V}$, and 
 a second group with $\Delta T_{\rm eff}>500$~K. 
In the latter case, most of the objects are from IC 4665 and THMG.

     The four NGC 1960 sources with counterparts in {\rm StarHorse} are located 
inside the range of \mbox{$T_{\rm eff}\in (3\,600, 4\,500)$~K} and ${\tt teff50}>4\,500$~K.
  In the range of \mbox{$T_{\rm eff}\in (3\,250, 3\,600)$~K} and ${\tt teff50}>4\,500$~K, 
there are only sources from IC 4665, BPMG, THMG, and 32 Ori MG.
  In brief, sources with $T_{\rm eff} < {\tt teff50}$ have
values of the extinction {\tt AV50}\footnote{{\tt AV50} is the {\tt StarHorse} line-of-sight extinction at 
$\lambda=5\,420$~\AA, $A_{V}$, 50th percentile} greater than the ones we have assumed.  
  We distinguish three zones depending on their tendency even though 
there is no definitive explanation.
  The $T_{\rm eff}-A_{V}$ relation might be affected by the use of 
synthetic photometry from the Kurucz grid together with the extinction law used in {\tt StarHorse}, 
the differential reddening in some stellar associations
(e.g. the Pleiades: see \citealt{m_breger1986}, \citealt{stauffer1987c}, 
\citealt{sj_gibson2003a} and \citeyear{sj_gibson2003b}) or both.

\subsubsection{Comparison with \cite{randich2018}}\label{subsub:comparison_randich2018}

    We compared our results with those obtained in \cite{randich2018}. 
In this work they employ the stellar parameters obtained from the \textit{Gaia}-ESO Survey, 
GES \citep{g_gilmore2012a}, and derive reddening values 
comparing several grids of models with various colour-magnitude diagrams. 
  We calculated the difference between their values and ours, 
for the four associations, we have in common: IC 4665, NGC 2547, IC 2602, and IC 2391 (see Table \ref{tab:randich2018_vs_vosa}). As with the {\tt StarHorse} values, 
the assumed $\log g=4.5$~dex is consistent,  
although we seem to have an offset of $-0.36\pm0.61$~dex.
 
\begin{figure}[t]
   \includegraphics[width=9cm]{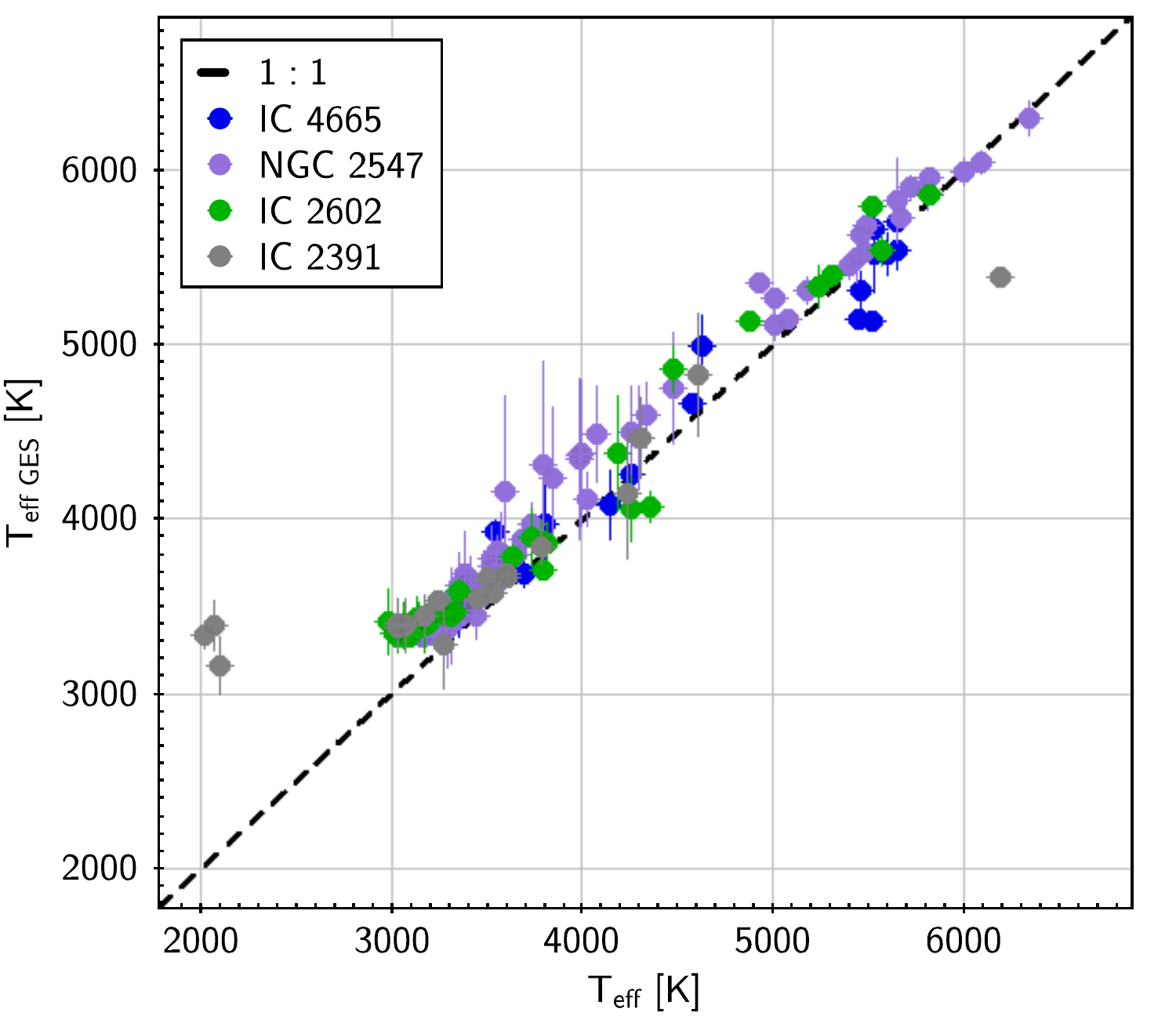}
   \includegraphics[width=9cm]{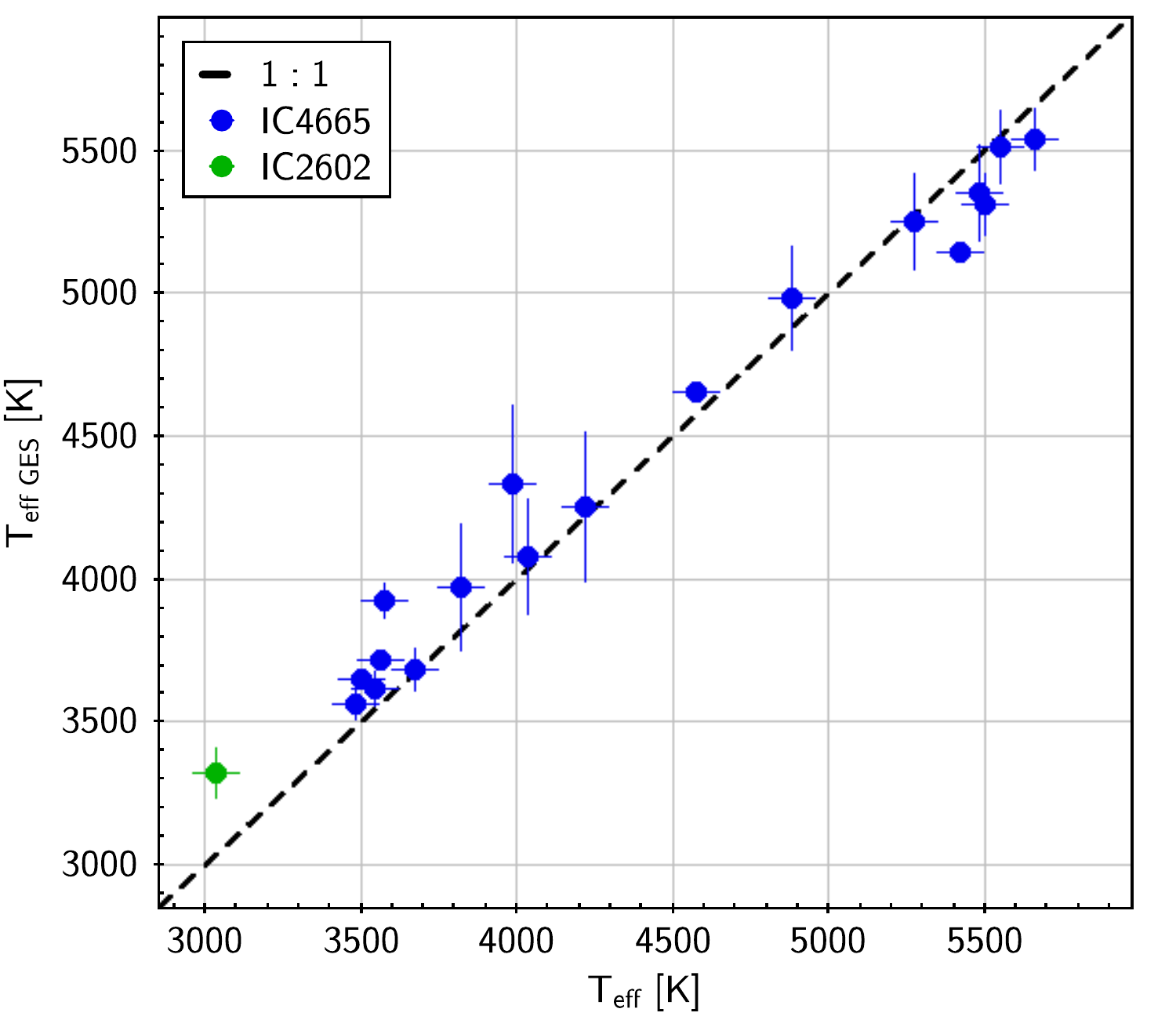}
     \caption[Comparison of effective temperatures derived in \cite{randich2018} and VOSA.]
             {Comparison of effective temperatures derived in \cite{randich2018} and from VOSA.
              {\bf Top: }All the sources are members from \cite{c_babusiaux2018} 
                with counterpart in \cite{randich2018}.      
              {\bf Bottom: }Only sources considered as members after this work from the LDB samples 
                with counterpart in \cite{randich2018}.  
             }
     \label{fig:teff_randich2018_vs_vosa} 
\end{figure}

    These authors $E(B-V)$ values are systematically higher than ours, 
of about 0.0477~mag, with a standard deviation of 0.0093~mag.
This value is within the \cite{c_babusiaux2018} uncertainties for the excess. 
  As we will see in Section \ref{sub:effect_av_ldb} and Appendix \ref{app:ldb_av}, 
higher reddening values produced hotter effective temperatures.
  Figure \ref{fig:teff_randich2018_vs_vosa} shows a temperature offset 
of approximately 200 K, probably caused by the values $E(B-V)$. 
  Our effective temperature values were obtained from the SED after 
applying the assumed $E(B-V)$ value, while values from \cite{randich2018} are obtained 
from particular lines in the spectrum \citep{f_damiani2014}.

\subsubsection{Impact of different model atmosphere grids on the total flux.}\label{subsub:comparison_grid_model_atm}

The amount of flux inferred from just one colour using grids of models is huge in typical bolometric corrections.
  In contrast, VOSA offers a panchromatic bolometric correction 
based on the numerous photometric bands collected and their uncertainties 
(included in the quantity {\tt Ferr}, see Sect. \ref{sub:phot_data}).
  VOSA releases the {\tt Fobs/Ftot} for all adjustments. 
This value is higher than 0.7 in our objects, 
so less than $30~\%$ of the total flux comes from model atmospheres. 
As such, we do not expect a large effect due to the adopted synthetic spectra.
  Besides the effective temperature, synthetic spectra depend also on surface gravity and metallicity. 
For these parameters, we assumed constant values, 
namely, [Fe/H]$=0.0$ and $\log {\rm g}=4.5$~dex ($\log {\rm g}=5.0$~dex for the Hyades), 
consistent with what has been done in previous works 
(see Sects. \ref{subsub:comparison_starhorse} and \ref{subsub:comparison_randich2018}). 
It is likely that it does not increase the uncertainty in the {\tt Ftot} estimation.
  However, to check the impact of the adopted spectra in the determination of bolometric luminosity, 
we compared the results obtained using two different grids of model atmospheres.

    Our reference grid is the BT-Settl (\citealt{f_allard2012},  
see Section \ref{sub:lbol_teff}). 
We compared it with the MARCS grid \citep{b_gustafsson2008}, 
using the standard metallicity class ([Fe/H]$=0.0$) and 
$\log {\rm g}=4.5$~dex\footnote{BT-Settl CIFIST:\\ {\tt http://svo2.cab.inta-csic.es/svo/theory/newov2/index.php}\\
MARCS: {\tt https://marcs.astro.uu.se/index.php}}. 
  We calculated the {\tt Ftot} in the wavelength range common to both models: from 1\,300~\AA\ to 20~$\mu{\rm m}$.
The results for models with $T_{\rm eff}\in[2\,500, 3\,500]$~K are given in Table \ref{tab:marcs_btsettl_models},
where {\tt Ftot} is the integrated flux, $\Delta${\tt Ftot} is the difference of fluxes of the MARCS and BT-Settl models, 
and $\%\Delta$ is the difference in percentage terms.
  The maximum difference is $5\%$, being less than $1\%$ in most cases. 
  Moreover, these effects are even smaller because in most of the targets 
the flux between $u$ and $K_{s}$ (and up to the $W2$ photometric band) 
is determined by the observations only -- and not by any model. 
Thus, we did not find any significant difference between the two grids 
in the effective temperature range of the LDB.

\begin{table}[ht] 
  \caption[Differences in fluxes between MARCS and BT-Settl model atmosphere.]
          {Differences in fluxes between MARCS and BT-Settl model atmosphere.}
  \label{tab:marcs_btsettl_models}
  \small
  \center
  \begin{tabular}{lccrc}
\hline \hline
{$T_{\mathrm{eff}}$}&\multicolumn{2}{c}{{\tt Ftot}} &$\Delta${\tt Ftot}& $\%\Delta$ \\
                    &   MARCS   & BT-Settl&         &      \\                  
                     \cline{2-4}
{[K]}               &\multicolumn{3}{c}{$10^{7}$[erg/cm$^2$/s/\AA]}& \\
\hline
  2500              &$ 221.53 $&$ 219.26 $&$  2.26 $& 1.03 \\
  2600              &$ 259.12 $&$ 264.74 $&$ -5.61 $& 2.12 \\
  2700              &$ 301.45 $&$ 317.76 $&$-16.31 $& 5.13 \\
  2800              &$ 348.60 $&$ 355.87 $&$ -7.27 $& 2.04 \\
  2900              &$ 401.11 $&$ 401.46 $&$ -0.35 $& 0.08 \\
  3000              &$ 459.35 $&$ 457.79 $&$  1.55 $& 0.33 \\
  3100              &$ 523.72 $&$ 527.43 $&$ -3.70 $& 0.70 \\
  3200              &$ 594.63 $&$ 595.41 $&$ -0.77 $& 0.13 \\
  3300              &$ 672.53 $&$ 670.34 $&$  2.18 $& 0.32 \\
  3400              &$ 757.86 $&$ 757.62 $&$  0.23 $& 0.03 \\
  3500              &$ 851.03 $&$ 844.93 $&$  6.09 $& 0.72 \\
\hline
    \end{tabular}
\end{table}

\subsection{Effects of reddening in the LDB}\label{sub:effect_av_ldb}

The reddening values vary greatly from one work to another for the same stellar association. 
Even if the same technique is used to calculate it, the isochrone fitting, 
the result is quite sensitive to the set of stellar models used to derive it, 
(see e.g. \citealt{c_babusiaux2018}, \citealt{randich2018}, \citealt{d_bossini2019}.)
  In our analysis, 
we used the values of the reddening computed by \cite{c_babusiaux2018}, 
whenever possible, neglecting the uncertainties\footnote{Extinction values from 
\cite{c_babusiaux2018} have an uncertainty of $\Delta E(B-R)=0.04$~mag.}.
  We used these values in order to compare age values determined 
with the same reddening and different technique (see Section \ref{sub:age_comparison}).

    Different methodologies reveal different reddening values, and 
a kaleidoscope study from diverse perspectives would be interesting. 
   Another effect that may be related is differential reddening,
as occurs in the Pleiades (see Section \ref{subsub:comparison_starhorse}), 
but that is beyond the scope of this work.

    In order to quantify and make a rough estimate
of how different values affect the location of the LDB (see Appendix \ref{app:ldb_av}), 
we studied Alpha Persei assuming two very different single reddening values: 
$A_{V}=0.279$~mag \citep{c_babusiaux2018} and 
$A_{V}=0.055$~mag \citep{d_bossini2019}.
We chose this open cluster because the LDB is very clearly located. 
All the calculations are presented in Appendix \ref{app:ldb_av}. 
If we assume $A_{V}=0.055$~mag, the LDB locus is at: 
$L_{\rm bol\ LDB}=2.74^{+0.40}_{-0.15}\cdot 10^{-3}$ $L_{\sun}$ and 
$T_{\mathrm{eff\ LDB}}=2\,806\pm71$~K, 
which it means a variation of $L_{\rm bol\ LDB}$ of about $-0.24\cdot 10^{-3}$ $L_{\sun}$.   
So, the LDB is located at fainter luminosities, cooler effective temperatures, and 
the resulting age is older: $81.3^{+2.4}_{-3.4}$~Ma (BT-Settl model from \cite{f_allard2012}.
Table \ref{tab:ldb_ages_alphapersei_av} shows the age differences between models.
If we make a comparison with the ages determined using $A_{V}=0.279$~mag, 
we find age differences ranging from 3 to 6~Ma (about $3$ to $9~\%$).
 However a more in-depth and systematic study of this variable is needed for each stellar association.

\subsection{Effects of the parallax zero-point offset in the LDB}\label{sub:effect_d_ldb}

We studied how the inclusion of the \textit{Gaia} DR2 global parallax zero-point offset 
(see \citealt{f_arenou2018} and \citealt{l_lindegren2018}) affects the the average stellar association distance and its LDB.
In addition, we studied possible differences between \textit{Gaia} DR2 and EDR3. 

    We tested the impact of the global parallax zero-point offset on the average stellar association distance    
by adding $+0.029$~mas to all parallaxes before performing our distance estimation.
    For the Hyades, we obtained 
$d_{\rm\ mode}=47.368$~pc and the values at the 
2.5th, 25th, 75th, and 97.5th percentiles are 
60.684, 50.761, 44.412, and 36.530~pc, respectively (see Table \ref{tab:oc_plx_dist_prev_mem_kde}).
In the case of more distant stellar associations such as 
  Alpha Persei, we calculated 
$d_{\rm\ mode}=173.571$~pc with  
2.5th, 25th, 75th, and 97.5th percentile values of  
186.653, 178.254, 170.162, and 161.682~pc; 
  and for the Pleiades we obtained  
$d_{\rm\ mode}=135.352$~pc with percentile values   
of 146.899, 138.571, 132.554, and 124.234~pc.
In these cases, the zero-point offset carries a change of up to a maximum of 1~pc.
  However the effect is larger in the most distant NGC 1960, 
with a $d_{\rm\ mode}=1\,165.908$~pc and percentile values of 
1\,455.688, 1\,241.557, 1\,103.338, 978.990~pc, 
which makes it a 40~pc closer.

    In addition, we tested their impact on the derived $L_{\rm bol\ LDB}$.  
  We found that the effect is negligible for the Hyades   
$L_{\rm bol\ LDB}=0.125^{+0.018}_{-0.017}\cdot 10^{-3}$ $L_{\sun}$, and the Pleiades: 
$L_{\rm bol\ LDB}=1.32^{+0.17}_{-0.10}\cdot 10^{-3}$ $L_{\sun}$, 
where the LDB loci changes $\Delta L_{\rm bol\ LDB}=0.001\cdot 10^{-3}$~$L_{\sun}$, 
less than $0.1\%$.
  After we applied the offset, 
we located the LDB for Alpha Persei at 
$L_{\rm bol\ LDB}=2.94^{+0.36}_{-0.17}\cdot 10^{-3}$ $L_{\sun}$, 
which implies $\Delta L_{\rm bol\ LDB}=0.04\cdot 10^{-3}$~$L_{\sun}$ fainter, 
and a change of $1.4\%$ in $L_{\rm bol\ LDB}$.
  The most pronounced case is that of NGC 1960 
because it is the cluster with the smaller parallax and the most distant. 
After adding the offset, NGC 1960 it is placed 40~pc nearest and 
we located the LDB at 
$L_{\rm bol\ LDB}=23.4^{+12}_{-4.8}\cdot 10^{-3}$ $L_{\sun}$ 
that implies an LDB loci $\Delta L_{\rm bol\ LDB}=0.199\cdot 10^{-3}$~$L_{\sun}$
smaller, and account for a $0.84\%$ change in $L_{\rm bol\ LDB}$.
  We conclude that the effect produced by zero-point offset parallax on the LDB locus
is more pronounced for more distant stellar associations. 
However, it is not the predominant  source of the uncertainty.

    To conclude this analysis, we studied the effect of \textit{Gaia} EDR3 parallaxes on 
the Alpha Persei distance and its LDB location. 
We obtained a $d_{\rm\ mode}=173.840$~pc, 
with the following values at the 2.5th, 25th, 75th, 97.5th percentiles: 
190.660, 179.165, 169.287, 160.242~pc.
In this case, the distance is 0.62~pc closer ($\Delta d_{\rm\ mode}$). 
We do not see a substantial change in the location and size of Alpha Persei 
using the \textit{Gaia} EDR3 parallaxes, although the parallax uncertainties are smaller in EDR3. 
    We also checked the effect on the LDB location. We obtained: 
$L_{\rm bol\ LDB}=3.11^{+0.25}_{-0.15}\cdot 10^{-3}$ $L_{\sun}$, 
which implies an LDB loci $\Delta L_{\rm bol\ LDB}=0.125\cdot 10^{-3}$~$L_{\sun}$ 
brighter, and account for a $4.1\%$ change in $L_{\rm bol\ LDB}$. 
Furthermore, the age decreases by 0.9~Ma: 
$78.1^{+1.0}_{-1.6}$~Ma, using the BT-Settl models.

\subsection{Comparison between ages: several models and techniques}\label{sub:age_comparison}

We locate the LDB in terms of the bolometric luminosity. 
The choice is due to the large dynamic range over which it varies in relevant age scales.
    We have found that the $L_{\rm bol\ LDB}$ and $T_{\rm eff\ LDB}$ produce 
different age values. 
  In Table \ref{tab:ldb_ages_lbol}, we present our calculated $L_{\rm bol\ LDB}$ and $T_{\rm eff\ LDB}$,   
and the age estimated using $L_{\rm bol\ LDB}$ and several grids of evolutionary models. 
In addition, as a test, 
we have calculated the $T_{\rm eff\ LDB}$ from the previously estimated age 
(between parentheses in Table \ref{tab:ldb_ages_lbol}). 
The purpose of that is to highlight the 
differences between the ages estimated from the $L_{\rm bol\ LDB}$ and the $T_{\rm eff\ LDB}$ 
(see \citealt{i_song2016}). 
  This effect is significantly larger at brighter $L_{\rm bol\ LDB}$ (younger stellar associations) and 
we also show an evolutionary model dependence.
  A possible explanation is that the radii of these objects are larger than predicted by the models 
(\citealt{e_berger2006l}, \citealt{m_lopezmorales2007a}, \citealt{i_ribas2008a}), 
of up to $14\pm2\%$ for a given bolometric luminosity for a sample of fast-rotating pleiades \citep{rj_jackson2018}. 
  The rotation, accretion, and magnetic fields are effects 
that may be involved in this fact.
  In summary, it is necessary to determine precisely how radii affect effective temperatures and bolometric luminosities 
and, hence, the LDB. 

\begin{figure}[t]
   \includegraphics[width=9cm]{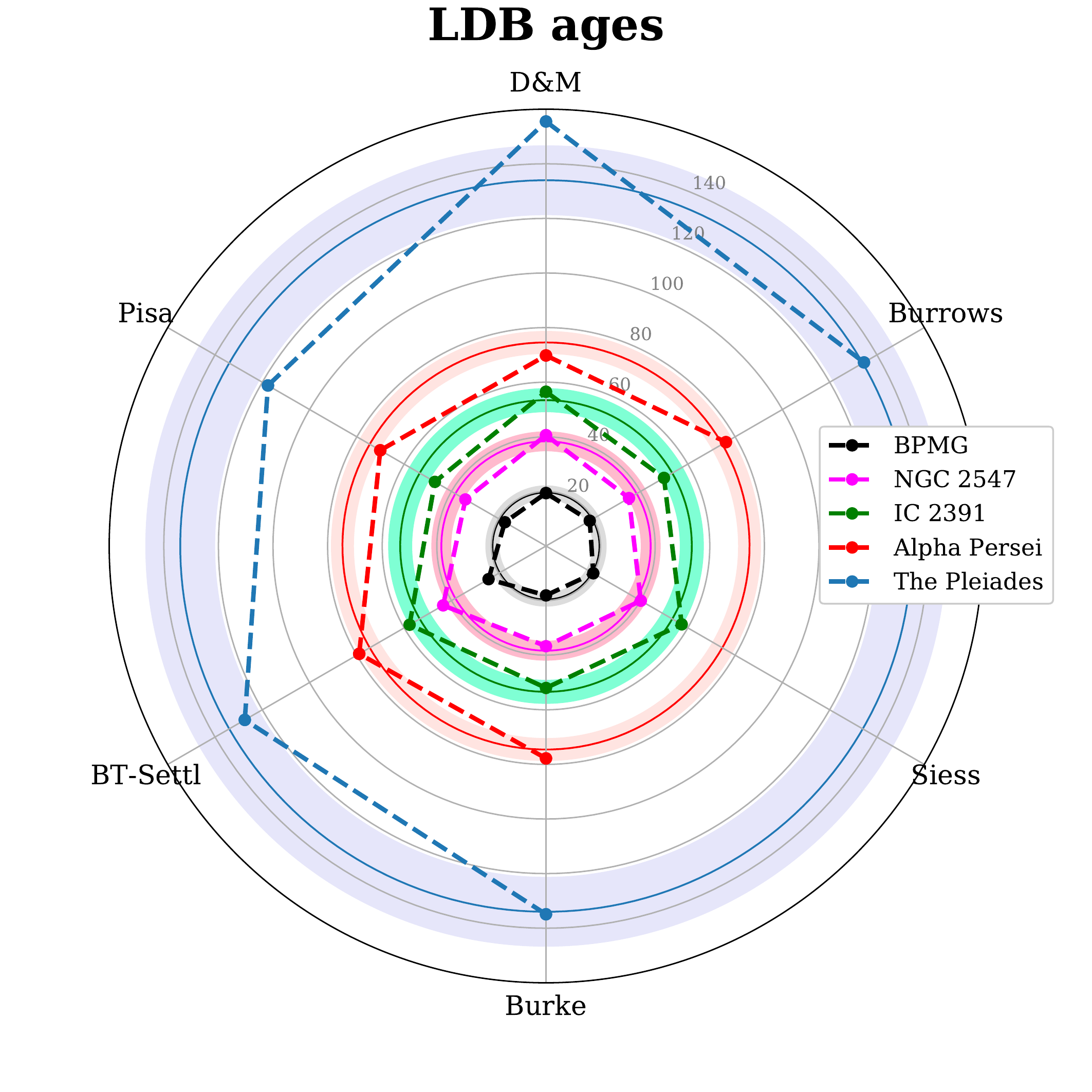}
     \caption[Radar chart with LDB ages.]
             {Radar chart with LDB ages estimated using several evolutionary grid models. 
              The age values showed as filled circles and connected with dashed lines are the same as Table \ref{tab:ldb_ages_lbol}. 
              We also show the mean age value (thin solid line) and its standard deviation (the shadow region)
              for each association.       
             }
     \label{fig:radarchart_ldb_ages} 
\end{figure}

    What can be clearly seen in Table \ref{tab:ldb_ages_lbol} and Figure \ref{fig:radarchart_ldb_ages}  
are the differences between ages calculated with different grids of evolutionary models, 
so there is a clear dependence. 
Any grid of models starts with a different 
equation-of-state, opacities, nuclear reaction rates and hypothesis,  
such as the convection model or the model atmosphere and leads to these age differences.
  We find several general remarks: 
 (a) the Pisa and Siess LDB give ages younger than BT-Settl LDB;
 (b) the largest differences, 12~Ma, are found at $M_{\rm bol}\in[8.0, 9.5]$~mag, 
around 20~Ma, and account for $50\%$ of the age, 
as we show for the BPMG with the Burrows and BT-Settl models.

\begin{figure*}\centering
   \includegraphics[scale=0.49]{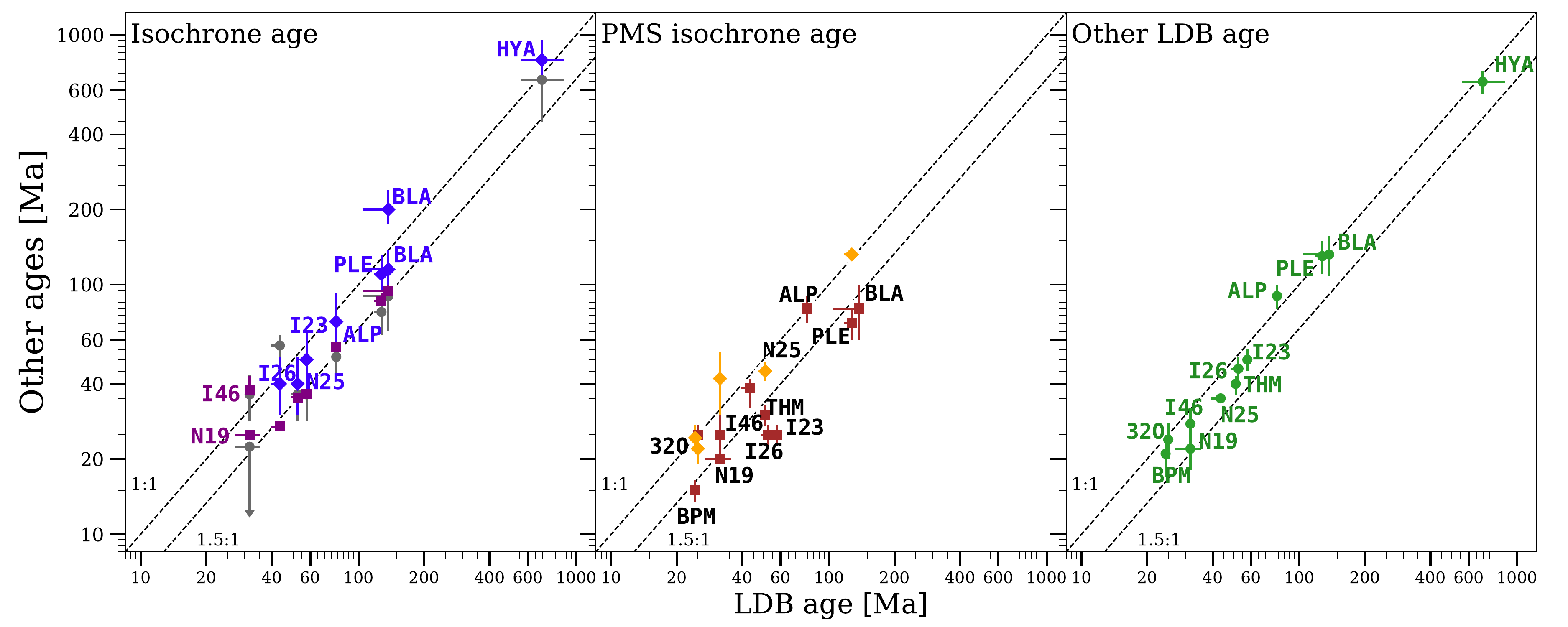}
      \caption[Comparison between our LDB age and the age estimated using other techniques.] 
              {Comparison between our LDB age and the age estimated using other techniques.
              \textbf{Left: }Ages derived from the isochrone fitting technique. 
               The ages come from: 
                \citealt{c_babusiaux2018} (blue diamonds),              
                \citealt{d_bossini2019} (purple squares), and 
                \citealt{mermilliod1981a} (gray circles).    
              \textbf{Middle: }Ages derived from the Pre Main-Sequence isochrone fitting technique.                
                Orange diamonds represent PMS isochrone fitting ages: 
                 the Pleiades, BPMG, THMG and 32 Ori MG.- \citeauthor{cpmbell2013} (\citeyear{cpmbell2013}, 
                  \citeyear{cpmbell2014}, and \citeyear{cpmbell2015}); and 
                 IC4665.- \cite{pa_cargile2010b}.
                Brown squares are the rest of PMS isochrone fitting ages: 
                 NGC 1960.- \cite{cpmbell2013}; 
                 IC 4665.- \cite{manzi2008}; 
                 NGC 2547.- \cite{t_naylor2006b}; 
                 IC 2602, IC 2391.- \cite{stauffer1997b}; 
                 Alpha Persei.- \cite{prosser1992}; 
                 the Pleiades.- \cite{stauffer1995b}; 
                 Blanco 1.- \cite{pa_cargile2009}; 
                 BPMG and THMG.- \cite{cao_torres2006}; and 
                 32 Ori MG.- \cite{mamajek2007a}. 
              \textbf{Right: } LDB ages derived from previous works. 
                See the references in Table \ref{tab:sample_clusters}, its legend, and 
                Section \ref{sec:stellarassociations_sample}. 
              }
         \label{fig:comparison_otherages_vs_ldbages}
\end{figure*}

We compared the LDB age scale derived using the BT-Settl luminosity-age relation 
with other age values (see Table \ref{tab:sample_clusters}) in 
Fig. \ref{fig:comparison_otherages_vs_ldbages}. 
  In general, the LDB ages are older than those estimated using a different method.
When we compare our LDB ages with previous works using the same methodology, 
(as first pointed out \citealt{basri1996}, \citealt{stauffer1998a})
there is still some discrepancy, although not so pronounced. 
  This result is probably related to the use of 
(a) different grid of models to derive ages, 
(b) individual parallaxes and to the fact that we discarded multiple systems 
together with foreground and background objects, and
(c) in some cases, different reddening values.

    On the other hand, we noted that the Pleiades PMS isochrone age from \cite{cpmbell2014} fits well 
with our LDB age because these models were re-calculated using this fact as input; 
  \cite{c_babusiaux2018} provided two age values for Blanco 1. 
From the main-sequence turn-off, they obtained an age of $200^{+ 40}_{- 26}$~Ma. 
However, if they used the LDB age of $115^{+23}_{-15}$~Ma \citep{aj_juarez2014}, 
they would reproduce the lower main sequence, with a marginal fit to the upper main sequence.
We show both values in Figure \ref{fig:comparison_otherages_vs_ldbages}.

    A summary is presented in Table \ref{tab:summary_ldb_ages}.
  Also, concerning studies on active stars, e.g. \cite{a_reiners2012b},
works like \cite{l_malo2014b} partly resolved the age inconsistency between LDB and isochrone-fitting age 
for the BPMG. 
It would be interesting to extend this study to the rest of the associations.

\begin{figure*}
      {\includegraphics[width=0.5\textwidth,scale=0.50]{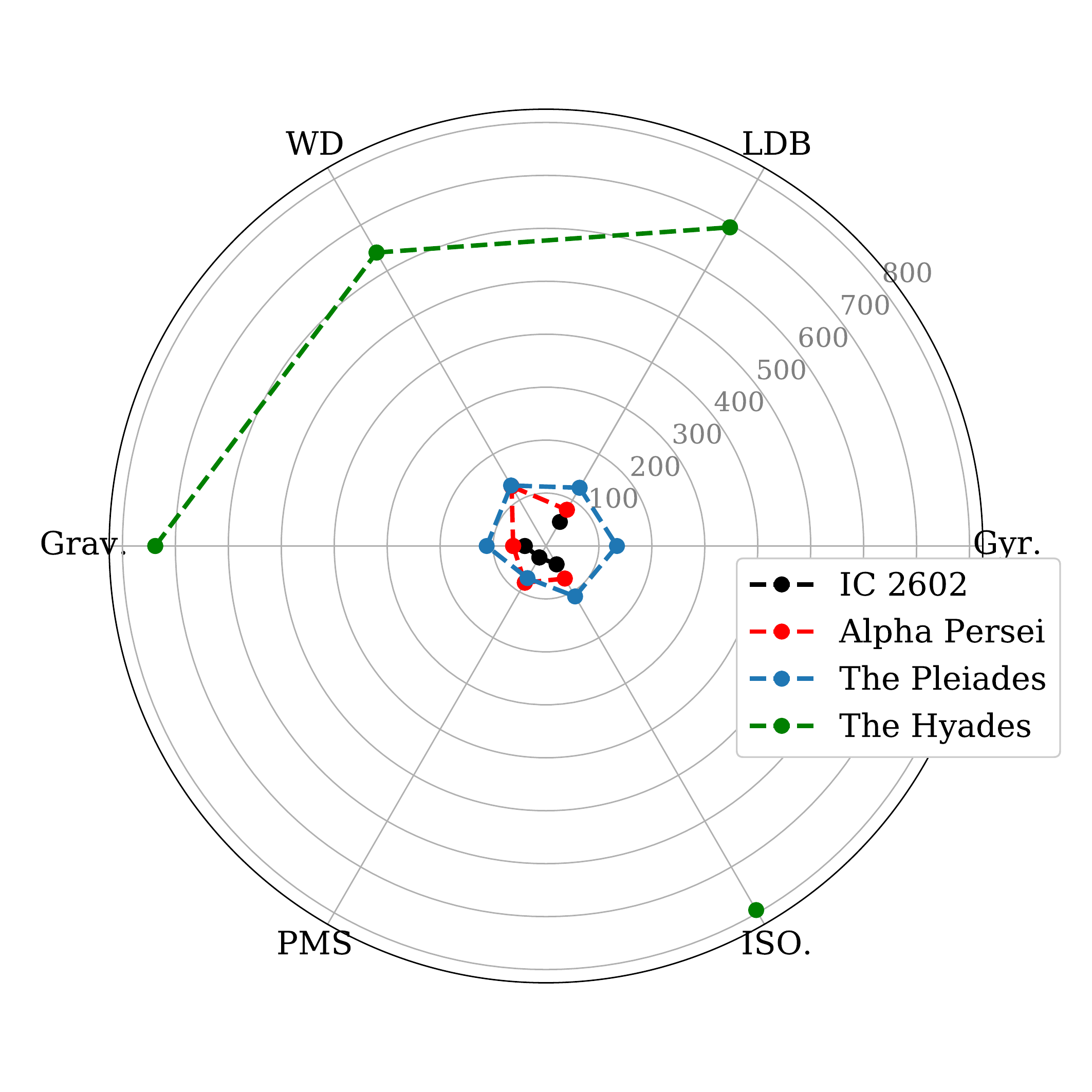}} \quad
      {\includegraphics[width=0.5\textwidth,scale=0.50]{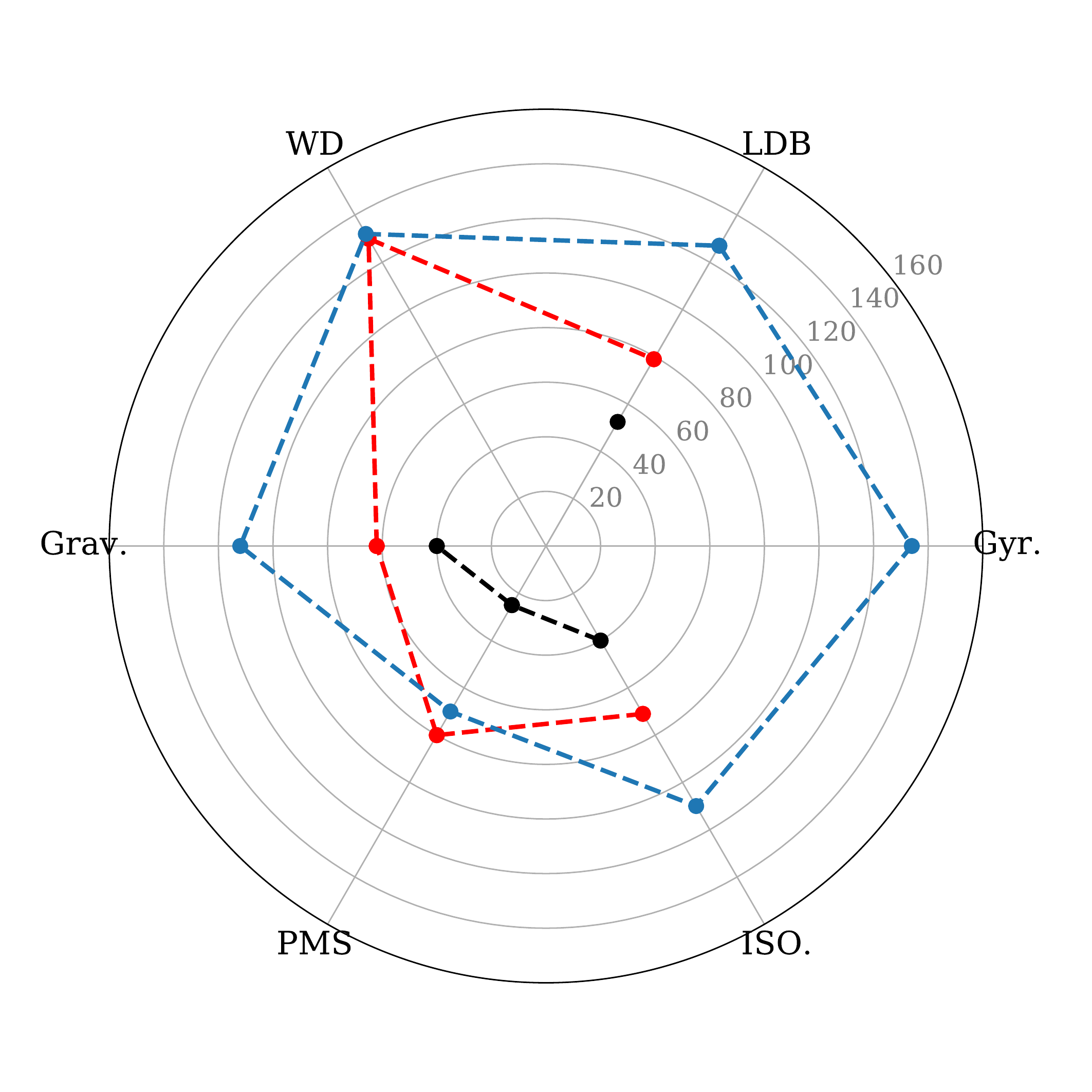}} \\
    \caption[Radar charts with ages calculated using different techniques.]
          {Radar charts with ages in [Ma] calculated using different techniques: `LDB' refers to our calculated ages using the BTSettl (see Table \ref{tab:ldb_ages_lbol}); `Gyr' is the gyrochronology ages \citep{pa_cargile2014}; 
           `ISO' is isochrone fitting ages \citep{c_babusiaux2018}; `PMS' is PMS isochrone fitting ages from several works: IC 2602.- \citealt{stauffer1997b}; 
                   Alpha Persei.- \citealt{prosser1992}; the Pleiades.- \citealt{stauffer1995b}. 
           Then, `Grav'' is the gravity ages \citep{tj_david2015a};
           `WD' is the ages derived using the white dwarf cooling sequence age Alpha Persei and the Pleiades.- \citealt{n_lodieu2019c}; 
                   the Hyades.- \cite{n_lodieu2019b}.
           {\bf Right: }Zoom onto the age range $\in[0, 160]$ Ma. 
          }
  \label{fig:radarchart_several_age_techniques}
\end{figure*}

    In Figure \ref{fig:radarchart_several_age_techniques} 
we plot our LDB ages together with other ages calculated with other techniques for a selection of stellar associations. 
  Here, we also see the differences in age between the different techniques. 
It is worth noting that in some cases, parameters like reddening and distance 
are completely different from those used in our work.
This fact may be the reason for the age discrepancies.

\setcounter{table}{5}
\begin{table*}[ht]
    \caption[Properties of the Lithium Depletion Boundary for the twelve stellar associations.] 
    {Properties of the Lithium Depletion Boundary for the 12 stellar associations.}
    \label{tab:summary_ldb_ages}
    \center{
    \begin{tabular}{lc rrrrc} 
\hline     
             & & \multicolumn{5}{c}{Summary of the LDB data}                                                 \\
                \cline{3-7}                                                                                  \\
Stellar      & & $L_{\mathrm{bol\ LDB}}$&$T_{\mathrm{eff\ LDB}}$&$d_{\rm\ mode}$ &$E(B-V)$&          LDB age \\       
association  & &$10^{-3}$  [$L_{\sun}$] &              [K]      & [pc]           &  [mag] &             [Ma] \\  
\hline \hline \\[0.7mm]
Alpha Persei & &$   2.98^{+0.35}_{-0.19}   $&$  2\,857        $&    174 &  0.090 &$  79.0^{+1.5}_{-2.3}     $\\[1.5mm]
NGC 1960     & &$  23.6^{+12}_{-4.9}       $&$  3\,123        $& 1\,207 &  0.200 &$  31.6^{+3.9}_{-4.7}     $\\[1.5mm]
IC 4665      & &$  23.6^{+2.1}_{-1.6}      $&$  3\,110        $&    347 &  0.170 &$  31.6^{+1.3}_{-1.6}     $\\[1.5mm]
NGC 2547     & &$  11.13^{+2.4}_{-0.89}    $&$  3\,004        $&    397 &  0.040 &$  43.5^{+1.7}_{-4.0}     $\\[1.5mm]
IC 2602      & &$   7.14^{+1.2}_{-0.52}    $&$  2\,926        $&    152 &  0.031 &$  52.5^{+2.2}_{-3.7}     $\\[1.5mm]
IC 2391      & &$   5.93^{+0.35}_{-0.19}   $&$  2\,956        $&    152 &  0.030 &$  57.7^{+0.5}_{-1.0}     $\\[1.5mm]
The Pleiades & &$   1.321^{+0.19}_{-0.089} $&$  2\,728        $&    136 &  0.045 &$ 127.4^{+6.3}_{-10}      $\\[1.5mm]
Blanco 1     & &$   1.182^{+0.069}_{-0.099}$&$  2\,746        $&    239 &  0.010 &$ 137.1^{+7.0}_{-33}      $\\[1.5mm]
The Hyades   & &$   0.124^{+0.019}_{-0.016}$&$  2\,021        $&     47 &  0.001 &$ 695  ^{+85}_{-67}       $\\[1.5mm]
BPMG         & &$  43.61^{+0.90}_{-0.81}   $&$  3\,095        $& 9 - 73 &  0.000 &$  24.3^{+0.3}_{-0.3}     $\\[1.5mm]
THMG         & &$   7.486^{+0.054}_{-0.11} $&$  2\,935        $& 36 - 71&  0.000 &$  51.0^{+0.5}_{-0.2}     $\\[1.5mm]
32 Ori MG    & &$  41.3^{+2.4}_{-2.1}      $&$  3\,135        $&70 - 100&  0.030 &$  25.0^{+0.7}_{-0.8}     $\\[1.5mm]
\hline
32 Ori MG-P1 & &$  41.8^{+4.8}_{-15}       $&$  3\ 106        $&        &  0.030 &$  24.9^{+4.8}_{-1.5}     $\\[1.5mm]       
32 Ori MG-P2 & &$  30^{+9}_{-0.8}          $&$  3\ 106        $&        &  0.030 &$  28.6^{+0.3}_{-2.9}     $\\[1.5mm]        
Theia 133    & &$  39^{+18}_{-17}          $&$  3\,139        $&   112  &  0.030 &$  25.6^{+7.0}_{-5.5}     $\\[1.5mm]       
Theia 370    & &$  29.8^{+8.3}_{-8.4}      $&$  3\,087        $&    97  &  0.030 &$  28.7^{+4.7}_{-2.6}     $\\[1.5mm]      
\hline
$\,$
    \end{tabular}
    }\\ 
$\,$
\begin{flushleft}
 $\,$
  \textbf{Note: }  
    The `LDB age' was derived using the BT-Settl models, see Table \ref{tab:ldb_ages_lbol}.\\
\end{flushleft}     
\end{table*}

\section{Conclusions and summary}\label{sec:conclusions_summary}

    The origin of the data and membership selection is very different for each stellar association and 
this fact can result in bias effects on the location of the LDB. 
In the case of the BPMG (\textbf{such us} \citealt{cao_torres2006} or \citealt{e_shkolnik2017}) and the THMG \citep{al_kraus2014c}, 
the membership selection was based on lithium measurements, 
so it is very likely that it might be a bias.
  We have tried to mitigate this possible effect by studying the membership of these objects based on the \textit{Gaia} DR2 data: 
positions, parallaxes, and proper motions. 

    In line with our aim of creating a complete set of LDB age scale and following 
the example of \cite{mermilliod1981a}, 
we present a systematic and homogeneous study of twelve stellar associations. 
Our main results are as follows:\begin{enumerate}
    \item Twelve stellar associations with former LDB ages have been thoroughly studied
conducting a careful and extensive bibliographic review to recover additional data 
to determine possible outliers, as well as multiple systems 
and sources with unique features that can affect the LDB location.\\

    \item We calculated individual distances for each source 
using a Bayesian statistic approach (\textit{Kalkayotl}), and 
derived individual bolometric luminosities and 
effective temperatures for 500 sources, from their SEDs, using Gaia DR2 data 
including photometric data in a broad range of wavelengths, 
as well as the effects of metallicity and surface gravity.
All the SEDs and images in several photometric bands 
were checked individually to determine special features.\\

    \item We fine-tuned previous memberships  
using individual parallaxes and proper motions from \textit{Gaia} DR2, 
in addition to an HRD. 
  The LDB has been determined using a jackknife method with a bootstrap re-sampling 
that takes into account the bolometric luminosities distributions of objects. \\ 
  
    \item We derived LDB ages using different theoretical evolutionary models. 
The discrepancies between them are due to the input physics. 
  Also, we found a discrepancy between the ages derived using bolometric luminosities 
and those derived using effective temperatures.
Since the ages from the bolometric luminosities are more robust to variations of $A_{v}$ than the rest, 
we have selected them as the reference set.\\

    \item Our work reveals $L_{\rm bol\ LDB}$ values that are more precise than former ones in general terms,  
because our method is quite robust since we statistically determine them. 
  Nevertheless, it is affected when: 
 (a) the sequence of M-dwarfs (or L-dwarfs) objects is scarce, 
for instance, there are few objects with spectroscopic measurements that allow us to discern their lithium abundance; and 
 (b) the sources that determine the LDB have larger bolometric luminosities uncertainties 
due to larger parallax uncertainties. 
  In specific terms, our calculated $L_{\rm bol\ LDB}$ are in agreement with former ones for 
IC 4665, NGC 2547, IC 2602, IC 2391, the Pleiades, and Blanco 1. 
  In contrast to earlier calculations, for the BPMG, the 32 Ori MG, Alpha Persei, and the Hyades, 
our $L_{\rm bol\ LDB}$ calculation is brighter. 
On the other hand, we found fainter $L_{\rm bol\ LDB}$ for NGC 1960 and THMG. 
  All these discrepancies are mainly due to a new membership assignment, 
the update distance determination from individual parallaxes, or both 
-- as in the case of the 32 Ori MG, NGC 1960, THMG, Alpha Persei, or the Hyades.\\

    \item We identified two peaks in the parallax distribution of the 32 Ori MG members.
Besides, we checked that the 32 Ori MG members belong to two different stellar groups: 
Theia 133 and Theia 370, after \cite{m_kounkel2019b}.
Finally, we located a tentative LDB and age for Theia 133 and Theia 370.\\ 

    \item The final results are summarized in Table \ref{tab:summary_ldb_ages}, 
where the best age determinations are listed. 
Those were derived homogeneously and provide a well established relative age scale. 
\end{enumerate}

    Based on the theoretical calculations (\citealt{l_bildsten1997d} and \citealt{g_ushomirsky1998b}),   
we would expect an abrupt transition between lithium-poor and lithium-rich objects. 
Nevertheless, this is not always the case, as we have found in our study of several stellar associations.
    One possible explanation may be the presence of unresolved multiple systems  
that could affect the interpretation and location of the LDB loci. 
    Another explanation is that other phenomena or mechanisms come into play. 
The accretion, the rotation, and the magnetic fields (all  correlated with each other) 
prevent lithium depletion or generate different abundances 
for stars with the same mass (see 
\citealt{soderblom1993h_SJB93h}, \citealt{g_chabrier2007c}, \citealt{g_somers2014}, 
\citealt{rj_jackson2014b}, \citealt{barrado2016b}, and \citealt{j_bouvier2018}). 
  Although only a few works have studied the LDB with some of these effects in mind 
(\cite{l_malo2014b}, \cite{aj_juarez2014}, \cite{s_messina2016d}), 
it would be interesting to extend it to the other associations, 
and to carry out a study focussed on the multiplicity around the LDB.  
  The stellar collision or the engulfment of a planetary object by a main star or brown dwarf 
(\citealt{c_abia2020b} and \citealt{g_israelian2001b}),   
 or the accretion of planetesimals leaving lithium abundance abnormalities 
in the atmosphere of the main object \citep{bc_kaiser2021} 
might be another possible scenario.  
  However, it remains to be seen how  the change in the lithium abundance, 
the stellar parameters may be quantified, and how common these phenomena may be overall.

    In a future work, we will analyse the LDB loci in terms of 
effective temperature and bolometric luminosity together, and    
  we will incorporate data to help explain and quantify 
the effects of rotation and magnetic fields in the LDB age.  
  It would be interesting to apply the \cite{l_malo2014b} procedure to the rest of the associations, 
and check if the inconsistency is resolved between LDB and isochrone-fitting ages. 
  It is necessary to populate some areas of the $L_{\mathrm{bol\ LDB}}$ range and study stellar associations, 
with ages ranging between that of the Pleiades and the Hyades, 
such as Praesepe or Coma Berenices (see \citealt{el_martin2020}). 
  There are no significant variations in our results based on the \textit{Gaia} EDR3 data. 
However, we plan to update these results following the launch of \textit{Gaia} DR3 in 2022.

\begin{acknowledgements}
A special thanks to Alcione Mora for his help with the \textit{Gaia} data, 
Floor van Leeuwen for his help with the size of the clusters,
Ana Mar\'ia \'Alvarez Garc\'ia, Mar\'ia Teresa Galindo Guil and Irene Pintos-Castro for their comments and suggestions, 
Agnes Monod-Gayraud for her careful reading of the manuscript and English corrections, 
and the anonymous referee for the comments that helped to improve the quality of this manuscript. 
%
%
FJGG acknowledges support from Johannes Andersen Student Programme at the Nordic Optical Telescope. 
This research has been funded by the Spanish State Research Agency (AEI) Project No.PID2019-107061GB-C61 and No. MDM-2017-0737 Unidad de Excelencia “Mar\'ia de Maeztu”- Centro de Astrobiolog\'ia (INTA-CSIC);
AB acknowledges support by ANID, -- Millennium Science Initiative Program -- NCN19\_171 and BASAL project FB210003, and from FONDECYT Regular 1190748;  
VOSA, developed under the Spanish Virtual Observatory project supported from the Spanish MINECO through grant AyA2017-84089;
the  astronomical java software TOPCAT \citep{marktaylor2005} and STILTS \citep{marktaylor2006};  
NASA's Astrophysics Data System Bibliographic Services; 
ESASky, developed by the ESAC Science Data Centre (ESDC) team and maintained alongside other ESA science mission's archives at ESA's European Space Astronomy Centre (ESAC, Madrid, Spain); 
the SVO Filter Profile Service (http://svo2.cab.inta-csic.es/theory/fps/) supported from the Spanish MINECO through grant AyA2014-55216; 
data from the European Space Agency (ESA) mission {\it Gaia} (\url{https://www.cosmos.esa.int/gaia}), processed by the {\it Gaia} Data Processing and Analysis Consortium (DPAC,\url{https://www.cosmos.esa.int/web/gaia/dpac/consortium}). Funding for the DPAC has been provided by national institutions, in particular the institutions participating in the {\it Gaia} Multilateral Agreement; 
The Pan-STARRS1 Surveys (PS1) and the PS1 public science archive have been made possible through contributions by the Institute for Astronomy, the University of Hawaii, the Pan-STARRS Project Office, the Max-Planck Society and its participating institutes, the Max Planck Institute for Astronomy, Heidelberg and the Max Planck Institute for Extraterrestrial Physics, Garching, The Johns Hopkins University, Durham University, the University of Edinburgh, the Queen's University Belfast, the Harvard-Smithsonian Center for Astrophysics, the Las Cumbres Observatory Global Telescope Network Incorporated, the National Central University of Taiwan, the Space Telescope Science Institute, the National Aeronautics and Space Administration under Grant No. NNX08AR22G issued through the Planetary Science Division of the NASA Science Mission Directorate, the National Science Foundation Grant No. AST-1238877, the University of Maryland, Eotvos Lorand University (ELTE), the Los Alamos National Laboratory, and the Gordon and Betty Moore Foundation;  
the SIMBAD database, operated at CDS, Strasbourg, France; 
Funding for the Sloan Digital Sky Survey IV has been provided by the Alfred P. Sloan Foundation, the U.S. Department of Energy Office of Science, and the Participating Institutions. SDSS acknowledges support and resources from the Center for High-Performance Computing at the University of Utah. The SDSS web site is www.sdss.org. 
SDSS is managed by the Astrophysical Research Consortium for the Participating Institutions of the SDSS Collaboration including the Brazilian Participation Group, the Carnegie Institution for Science, Carnegie Mellon University, the Chilean Participation Group, the French Participation Group, Harvard-Smithsonian Center for Astrophysics, Instituto de Astrofísica de Canarias, The Johns Hopkins University, Kavli Institute for the Physics and Mathematics of the Universe (IPMU) / University of Tokyo, the Korean Participation Group, Lawrence Berkeley National Laboratory, Leibniz Institut für Astrophysik Potsdam (AIP), Max-Planck-Institut für Astronomie (MPIA Heidelberg), Max-Planck-Institut für Astrophysik (MPA Garching), Max-Planck-Institut für Extraterrestrische Physik (MPE), National Astronomical Observatories of China, New Mexico State University, New York University, University of Notre Dame, Observatório Nacional / MCTI, The Ohio State University, Pennsylvania State University, Shanghai Astronomical Observatory, United Kingdom Participation Group, Universidad Nacional Autónoma de México, University of Arizona, University of Colorado Boulder, University of Oxford, University of Portsmouth, University of Utah, University of Virginia, University of Washington, University of Wisconsin, Vanderbilt University, and Yale University; 
data products from the Wide-field Infrared Survey Explorer, which is a joint project of the University of California, Los Angeles, and the Jet Propulsion Laboratory/California Institute of Technology,  funded by the National Aeronautics and Space Administration; 
data products from the Two Micron All Sky Survey, which is a joint project of the University of Massachusetts and the Infrared Processing and Analysis Center/California Institute of Technology, funded by the National Aeronautics and Space Administration and the National Science Foundation 
\end{acknowledgements}

\bibliographystyle{aa} 
\bibliography{bibliografia}

\clearpage
\begin{appendix}
\section{Parallaxes distributions, VPDs, and HRDs for all the stellar associations}\label{app:plots_stellarassociations}

In this appendix, we present the parallax distributions the VPDs and the 
HRDs for all the stellar associations. 
   In some cases, we studied different possibilities of locating the LDB and for these, we present different HRDs, that is, a different one for each location.

\onecolumn
\begin{figure}
   \includegraphics[width=0.45\textwidth,scale=0.50]{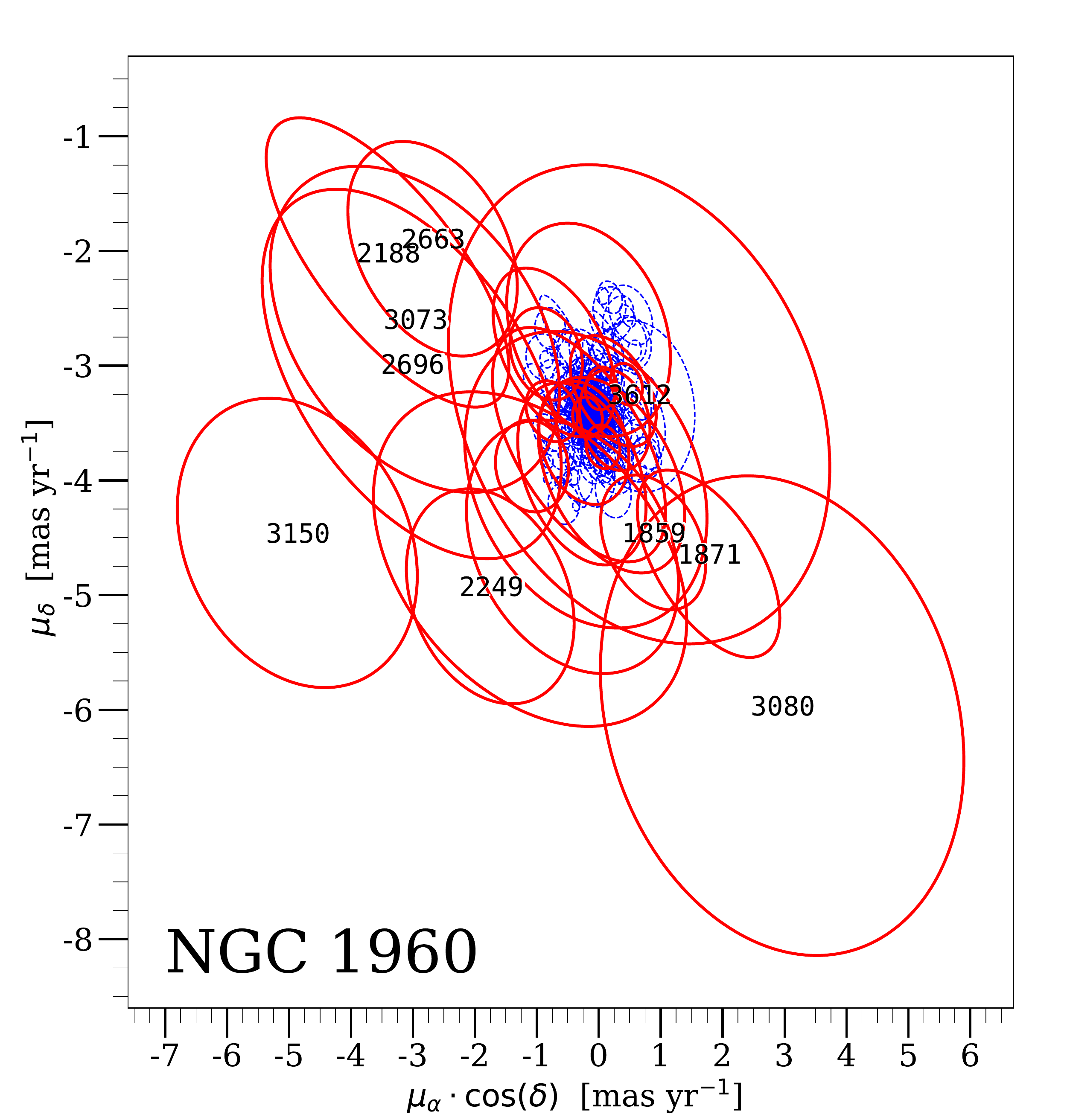}
   \includegraphics[width=0.45\textwidth,scale=0.50]{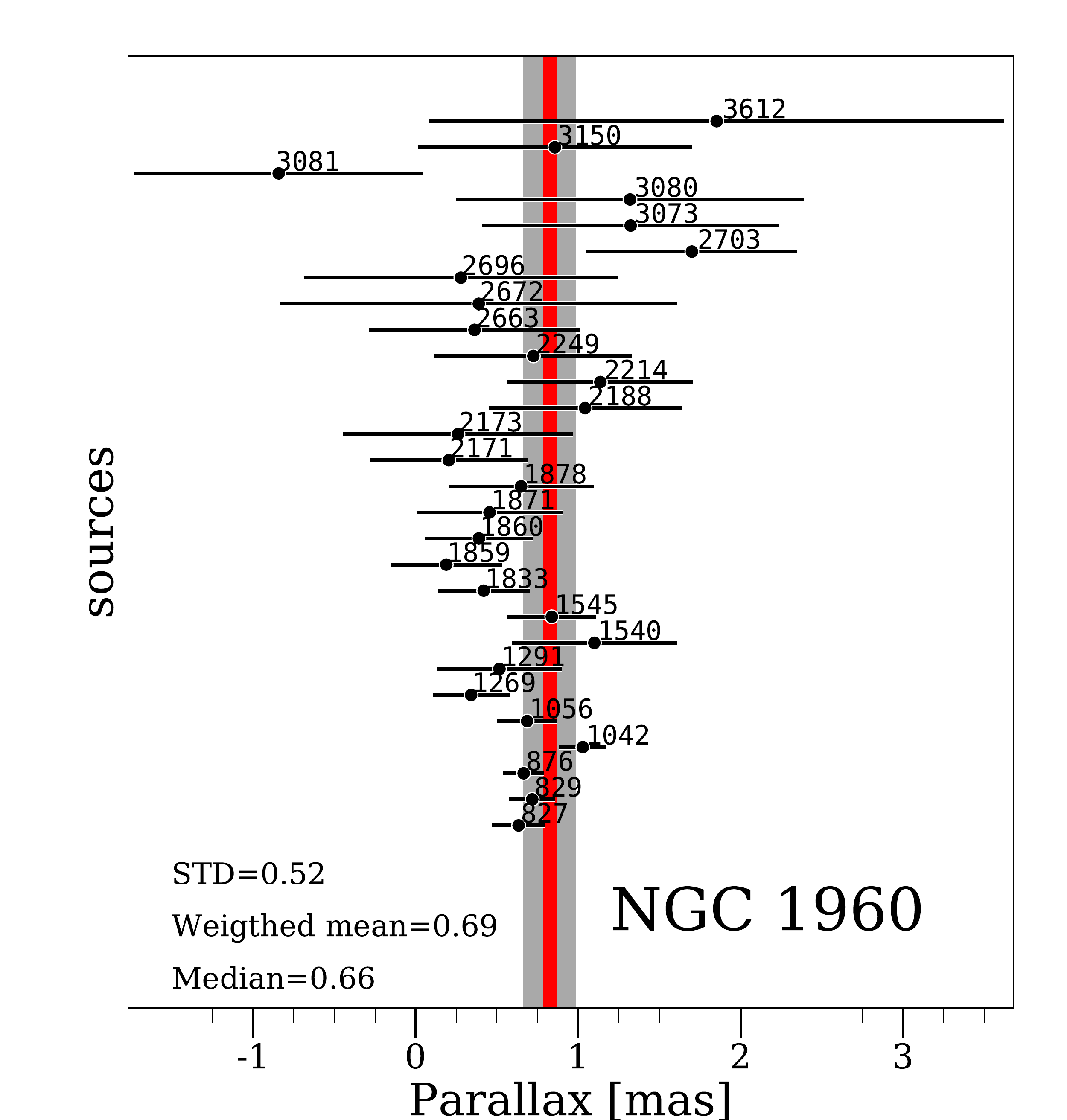}
      \caption[VPD and parallax distribution for NGC 1960.]     
              {VPD and parallaxes for NGC 1960.              
               {\bf Left: }VPD, for NGC 1960.  
                Blue dashed ellipses are the members taken from \citealt{cantatgaudin2018b}; 
                red ellipses are our LDB sample. 
                The rest is the same as in Fig. \ref{fig:vpd_alphapersei} but for NGC 1960.                
               {\bf Right: }Same as in Fig. \ref{fig:plxs_alphapersei} but for NGC 1960.
              }
         \label{fig:vpd_plxs_ngc1960}
\end{figure}

\begin{figure}
    \subfloat[\label{fig:hrd_ngc1960_general}] {\includegraphics[width=0.5\textwidth,scale=0.50]{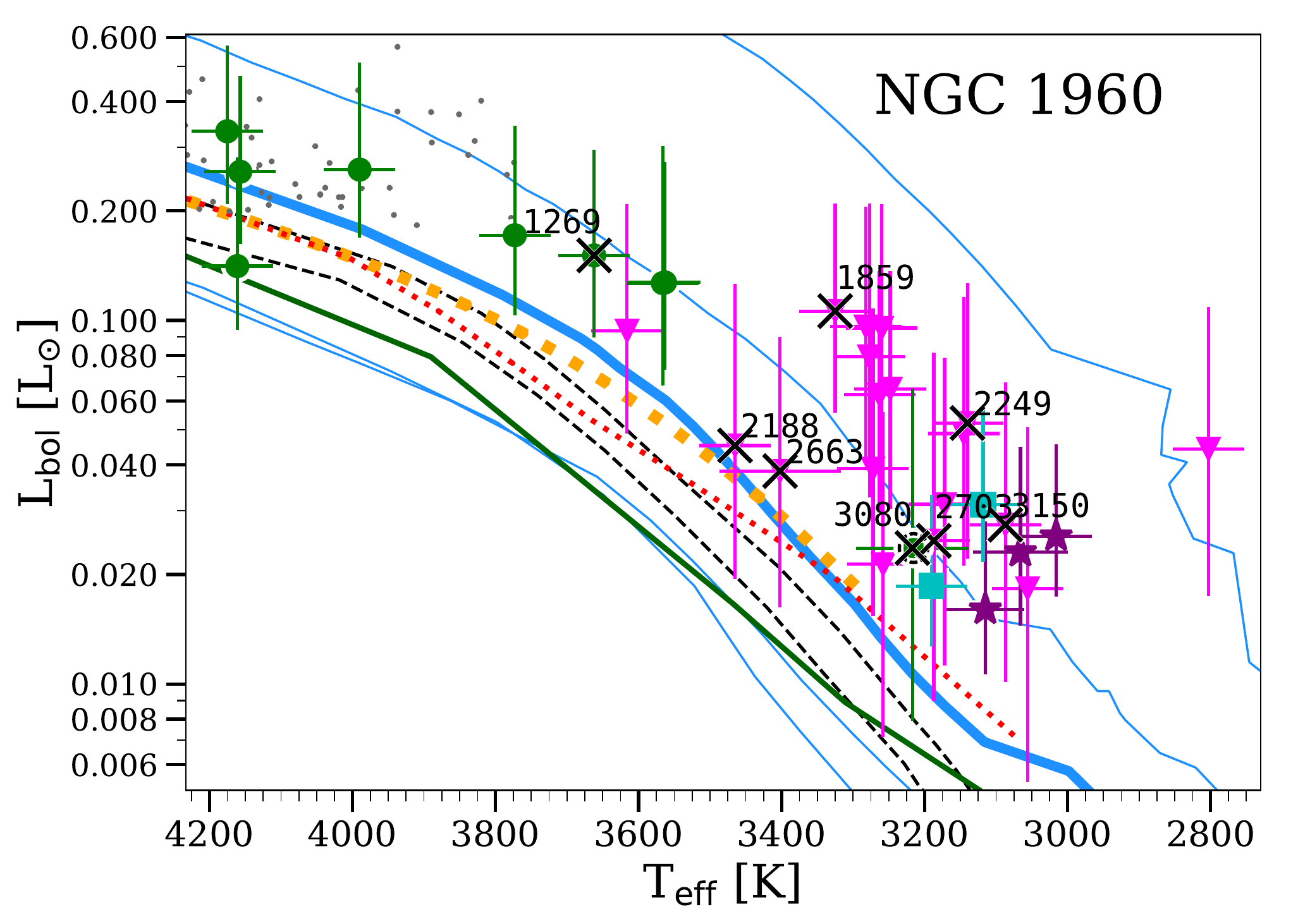}} \quad     
    \subfloat[\label{fig:hrd_ngc1960_zoom_ldb}]{\includegraphics[width=0.5\textwidth,scale=0.50]{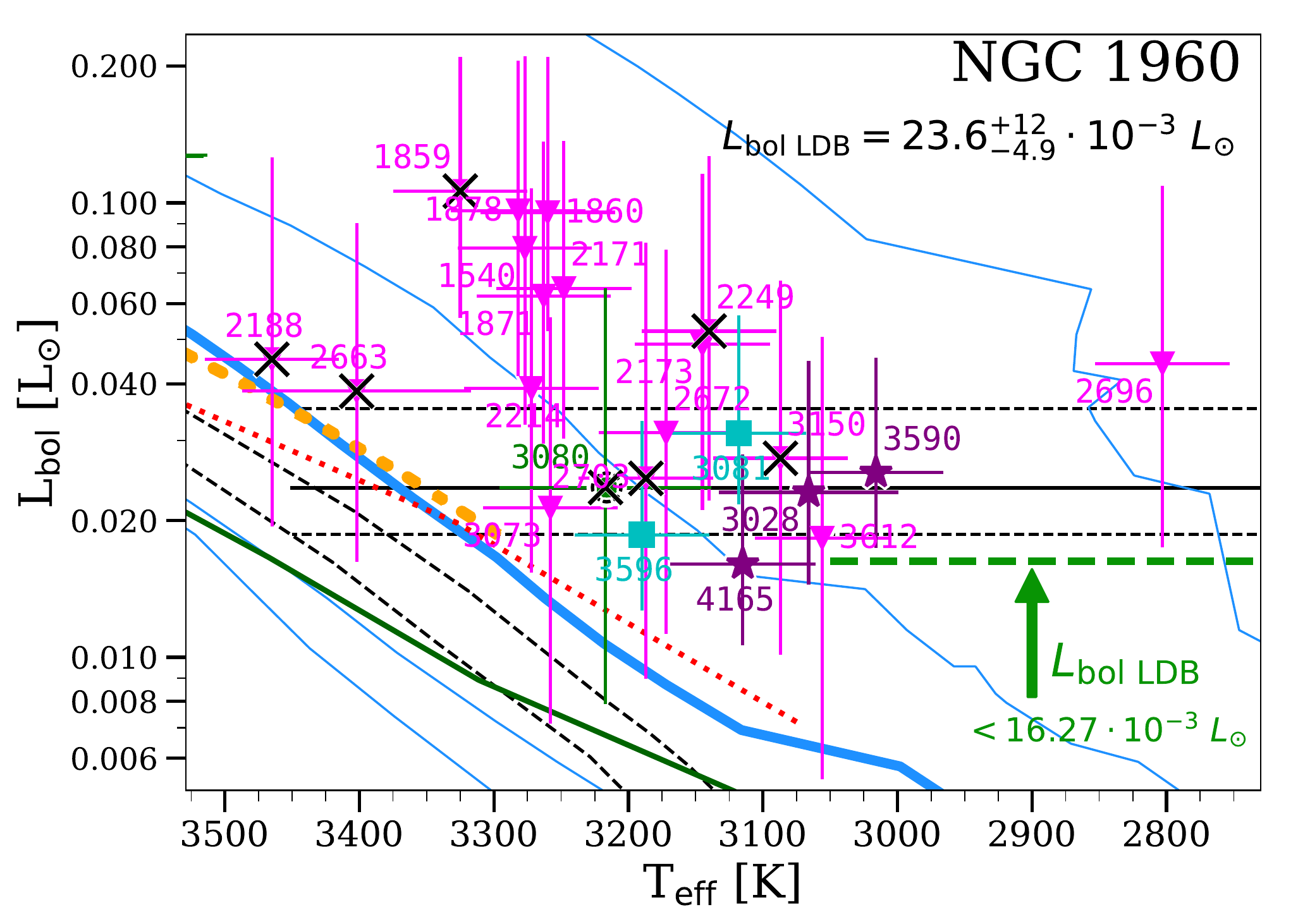}} \\            
    \caption[HRDs and the LDB for NGC 1960.]
          {HRDs and the LDB for NGC 1960.                     
           {\bf Left: }Same as Fig. \ref{fig:hrd_alphapersei_discussion}. 
            Cyan squares are lithium-rich sources without \textit{Gaia} DR2 parallaxes and  
            purple stars are lithium upper limits or lithium-poor sources without \textit{Gaia} DR2 parallaxes,  
            both from our LDB sample (see Section \ref{sub:distances}); 
            empty large black circles are confirmed multiple systems and  
            broken lines are suspected multiple systems;
            over-imposed big black crosses are sources discarded as members after this work.
            Moreover, grey points are known members from \cite{cantatgaudin2018b}. 
             Thin blue lines correspond to isochrones of 1, 10, 100 Ma, and 1 Ga 
            from the BT-Settl models \citep{f_allard2013}, while the thick blue one corresponds to 30 Ma. 
             The figure includes:  
            a 20~Ma and a 30~Ma isochrones from \citealt{f_dantona1994} (black dashed line), 
            a 25~Ma isochrone from \citealt{e_tognelli2011} (orange dashed dot line), 
            a 30~Ma isochrone from \citealt{siess2000} (red dotted line) and 
            a Zero Age Main Sequence from \citealt{sw_stahler_f_palla2005} (green solid line).           
           {\bf Right: }Zoom on the left plot.  
            A black horizontal line marks the $L_{\rm bol\ LDB}$, 
            together with the 16th and 84th percentiles (dotted lines).
            NGC 1960 is $31.6^{+3.9}_{-4.7}$~Ma old using the BT-Settl bolometric luminosity-age relationship \citep{f_allard2012}.
            An extensive dashed horizontal green line indicates a possible LDB upper limit (see Section \ref{sub:ngc1960_ldb}).            
          }
  \label{fig:hrds_ngc1960}
\end{figure}

\clearpage
\begin{figure}
   \centering
   \includegraphics[width=0.45\textwidth,scale=0.50]{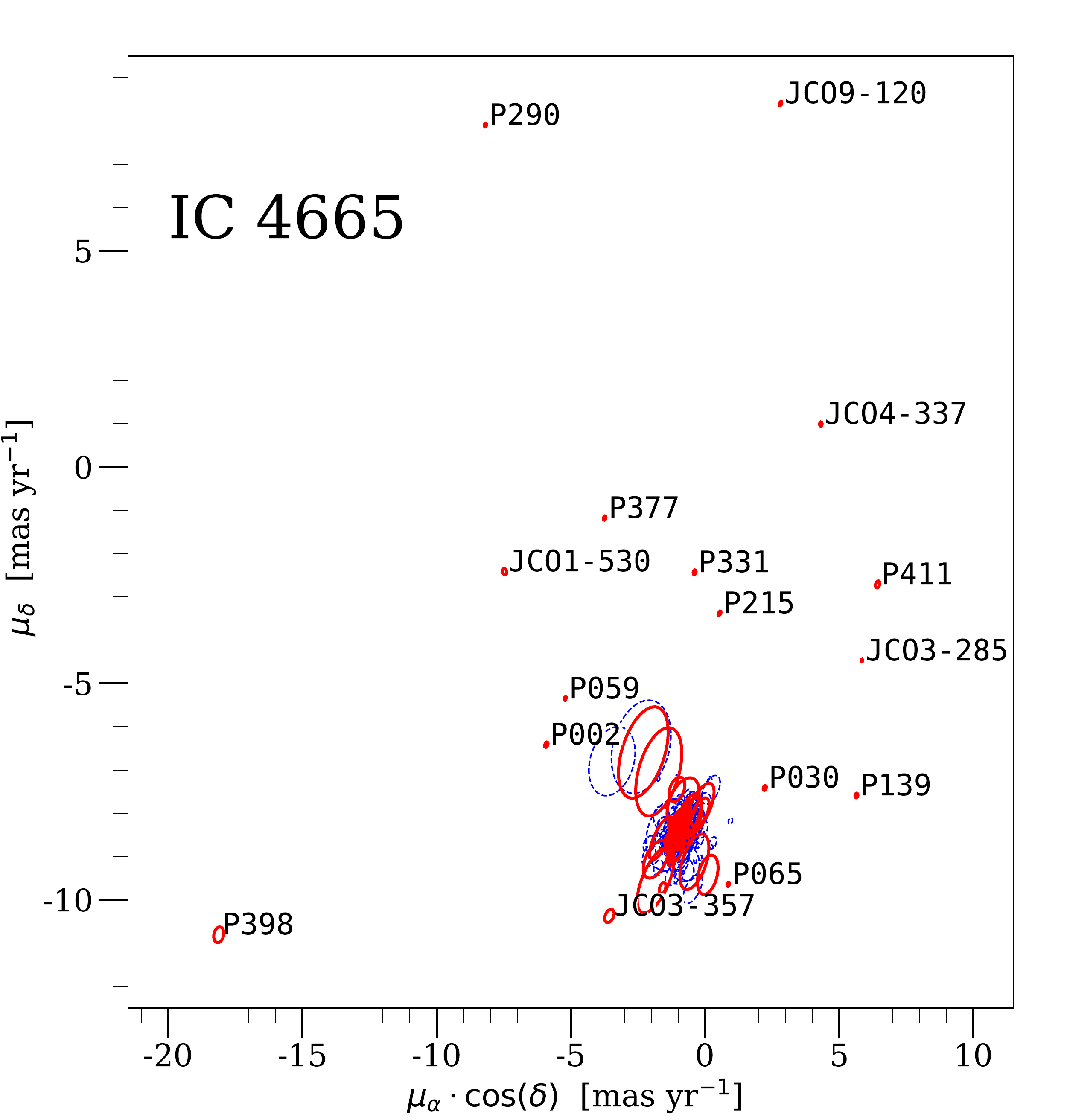}
   \includegraphics[width=0.45\textwidth,scale=0.50]{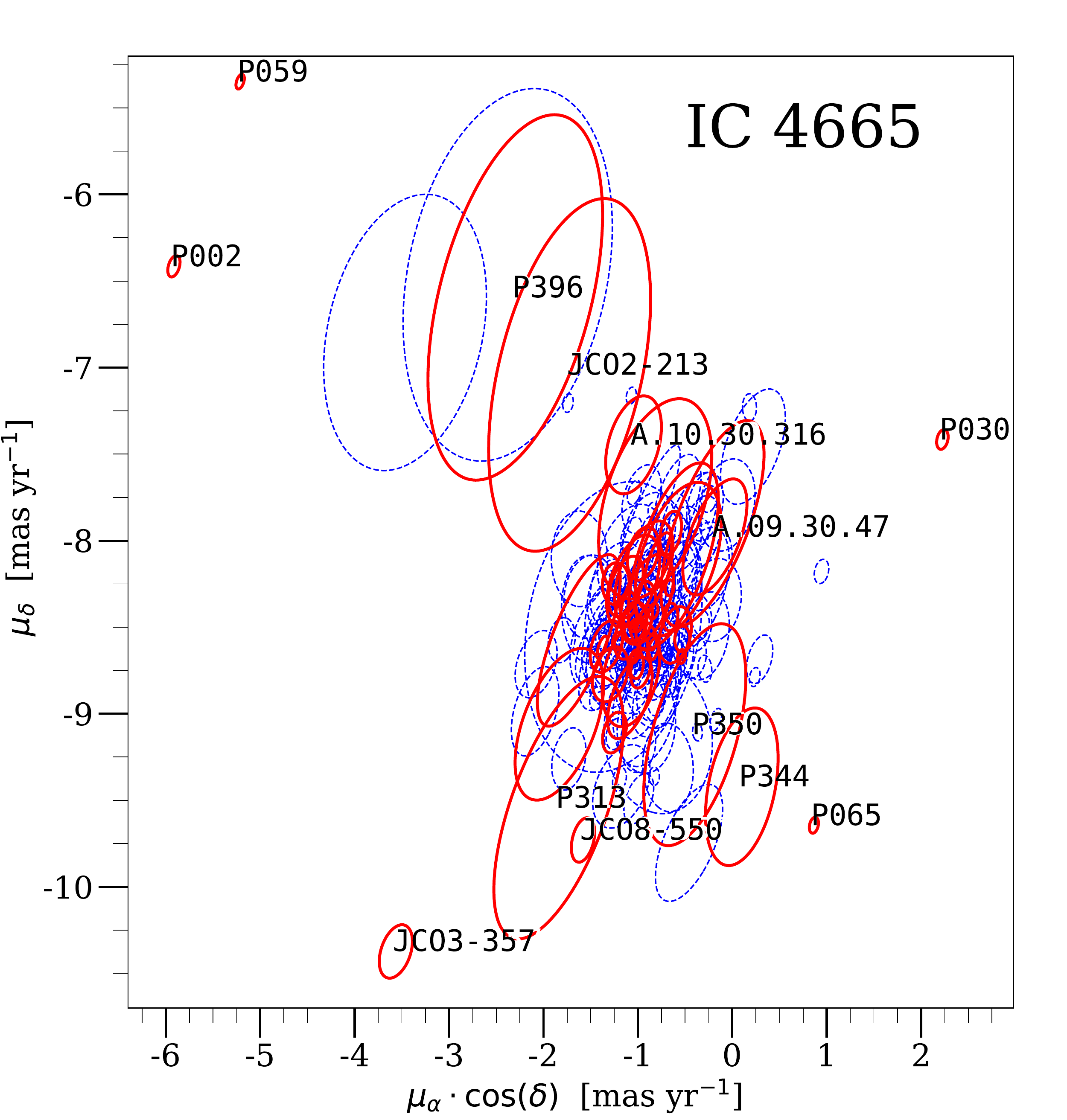}
      \caption[VPD for IC 4665.]
              {VPD for IC 4665. 
               {\bf Left: }Details are the same as in Fig. \ref{fig:vpd_alphapersei} but for IC 4665.
               {\bf Right: }Zoom on the left figure. 
              }
         \label{fig:vpds_ic4665}
\end{figure}

\begin{figure}
   \centering
   \includegraphics[width=0.45\textwidth,scale=0.50]{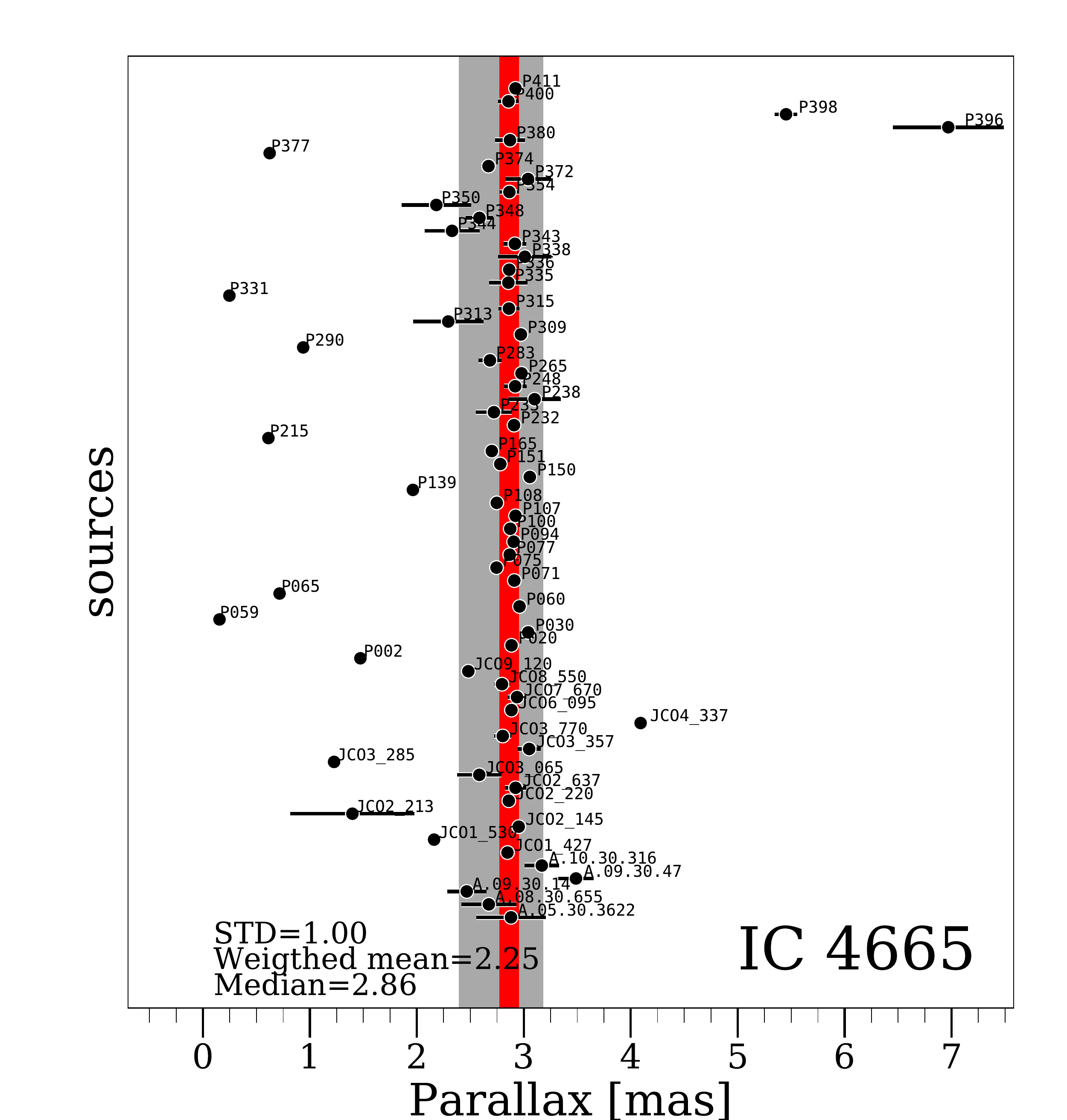}   
      \caption[Parallaxes for IC 4665 members close to the LDB.]
              {Parallaxes for IC 4665 members close to the LDB.
               Details are the same as in Fig. \ref{fig:plxs_alphapersei} but for IC 4665.
              } 
         \label{fig:plxs_ic4665}
\end{figure}

\clearpage
\begin{figure}
     \subfloat[\label{fig:hrd_ic4665_general}]
      {\includegraphics[width=0.5\textwidth,scale=0.50]{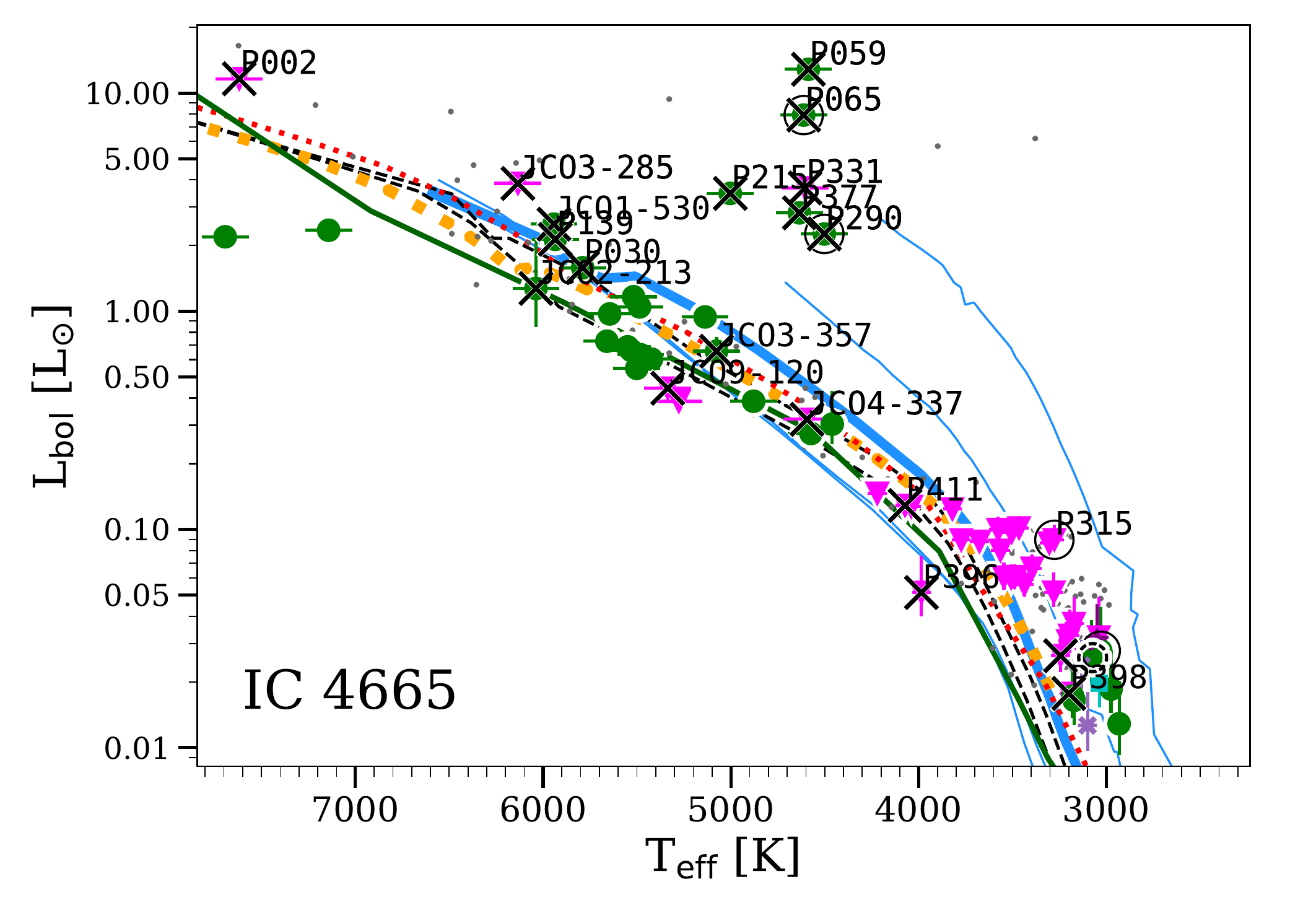}} \quad      
     \subfloat[\label{fig:hrd_ic4665_zoom}]
      {\includegraphics[width=0.5\textwidth,scale=0.50]{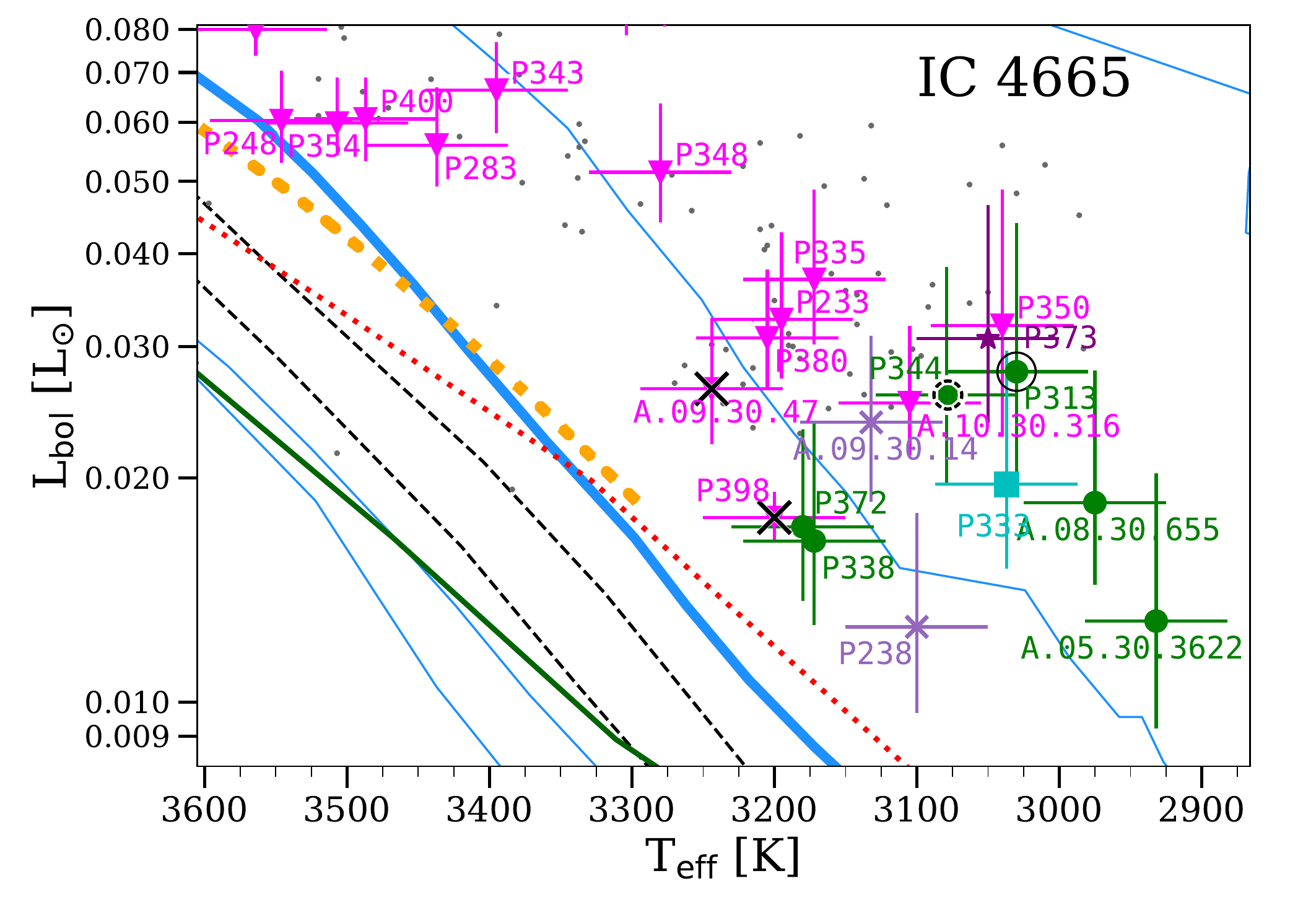}} \\
     \subfloat[\label{fig:hrd_ic4665_zoom_ldb_v1}]     
      {\includegraphics[width=0.5\textwidth,scale=0.50]{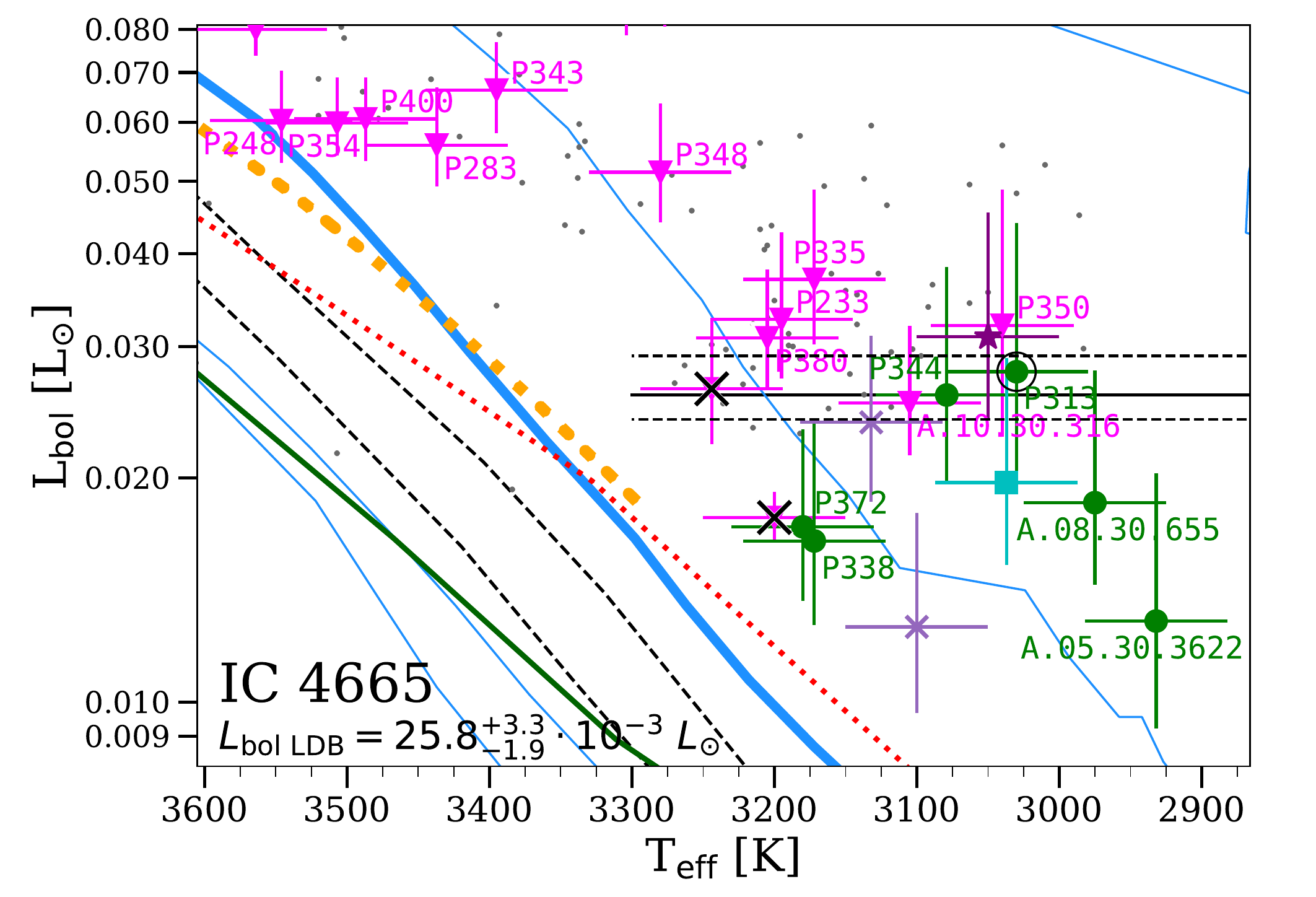}} \quad    
     \subfloat[\label{fig:hrd_ic4665_zoom_ldb_v2}]
      {\includegraphics[width=0.5\textwidth,scale=0.50]{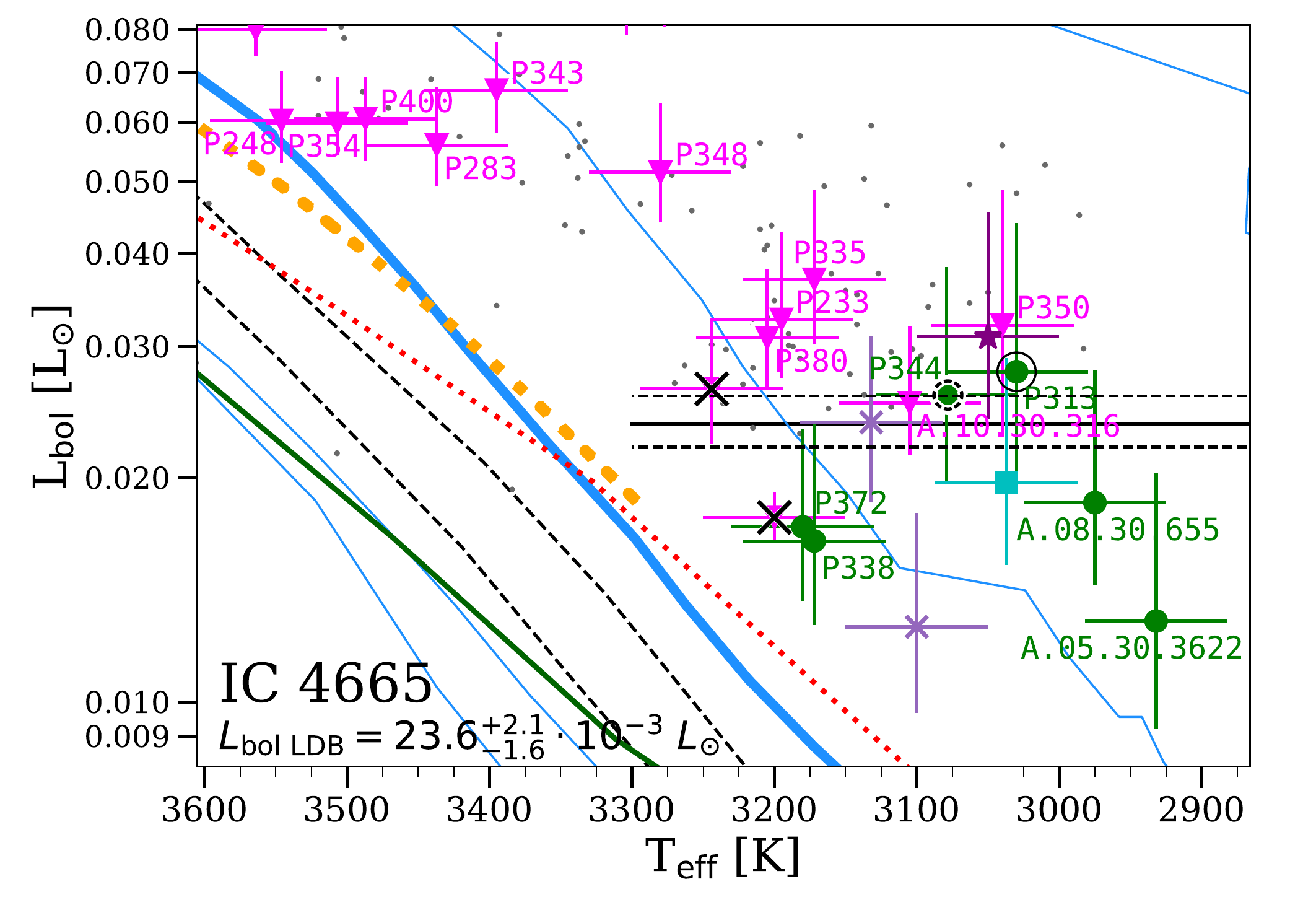}} \\         
    \caption[HRDs and the LDB for IC 4665.]
          {HRDs and the LDB for IC 4665.
           {\bf (a) }the same as in Figs. \ref{fig:hrd_alphapersei_discussion} and \ref{fig:hrds_ngc1960}
            but for IC 4665.
            The violet crosses are sources with undetected or dubious lithium detections due to their low resolution spectrum 
            or other technical disabilities. 
             Thin blue lines correspond to isochrones of 1, 10, 100~Ma, and 1 Ga from the BT-Settl models 
            \citep{f_allard2013}, while the thick blue one to 30~Ma. 
             The figure includes:
            a 20~Ma and a 30~Ma isochrones from \citealt{f_dantona1994} (black dashed lines), 
            a 25~Ma isochrone from \citealt{e_tognelli2011} (orange dashed dot line), 
            a 30~Ma isochrone from \citealt{siess2000} (red dotted line). 
           {\bf (b) }Zoom on the previous plot around the LDB with all the source names.
           {\bf (c) }A black horizontal line marks the $L_{\rm bol\ LDB}$, together with the 16th and 84th percentiles (dashed lines),
            following the first scenario, see Section \ref{sub:ic4665}.             
           {\bf (d) }Same as (b), but the the $L_{\rm bol\ LDB}$ is located following the second scenario, see Section \ref{sub:ic4665}.
            It is $31.6^{+1.3}_{-1.6}$~Ma old using the BT-Settl bolometric luminosity-age relationship \citep{f_allard2012}.
          }  
  \label{fig:hrds_ic4665}
\end{figure}

\clearpage
\begin{figure}
   \centering
   \includegraphics[width=0.45\textwidth,scale=0.50]{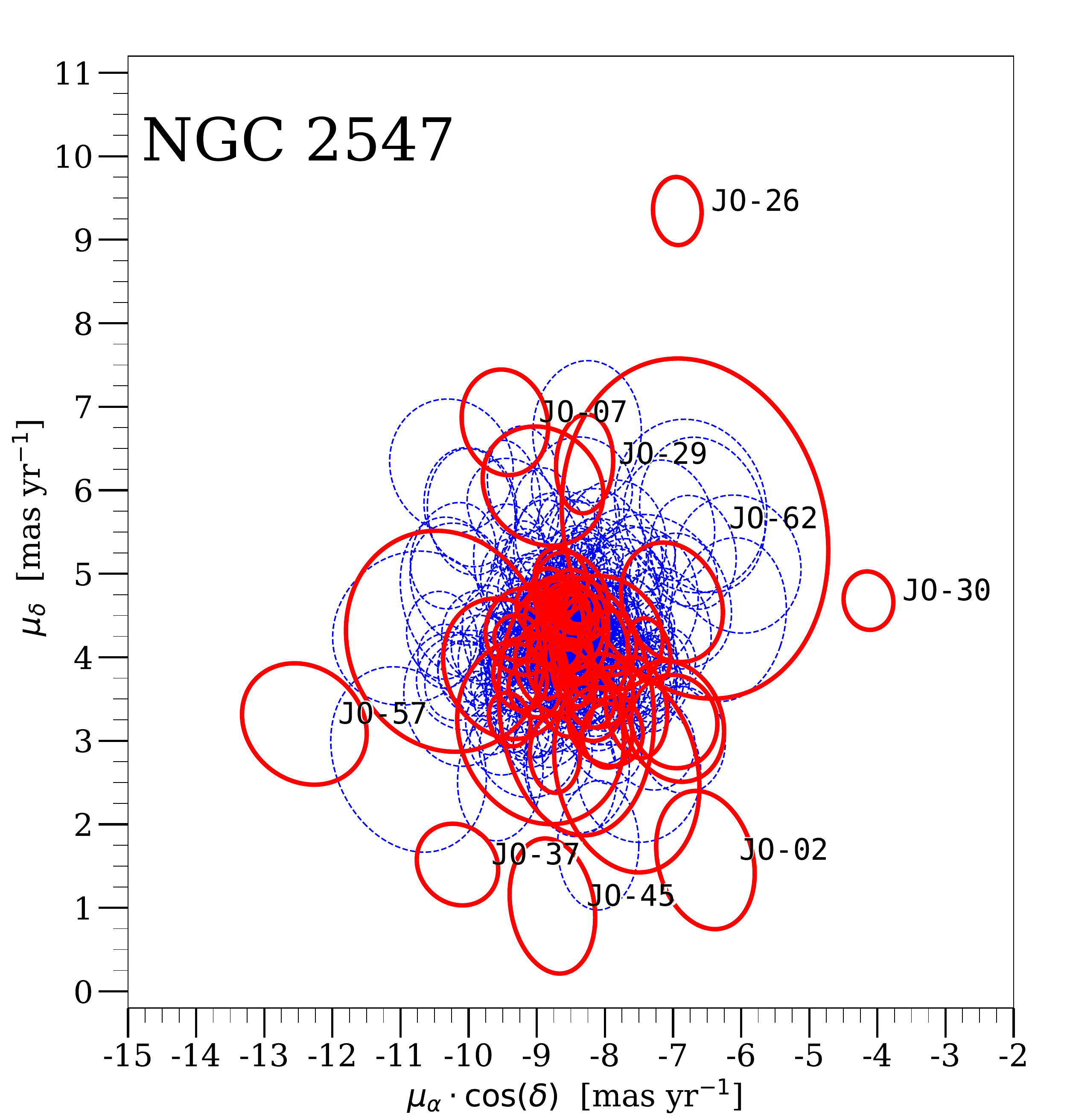}
   \includegraphics[width=0.45\textwidth,scale=0.50]{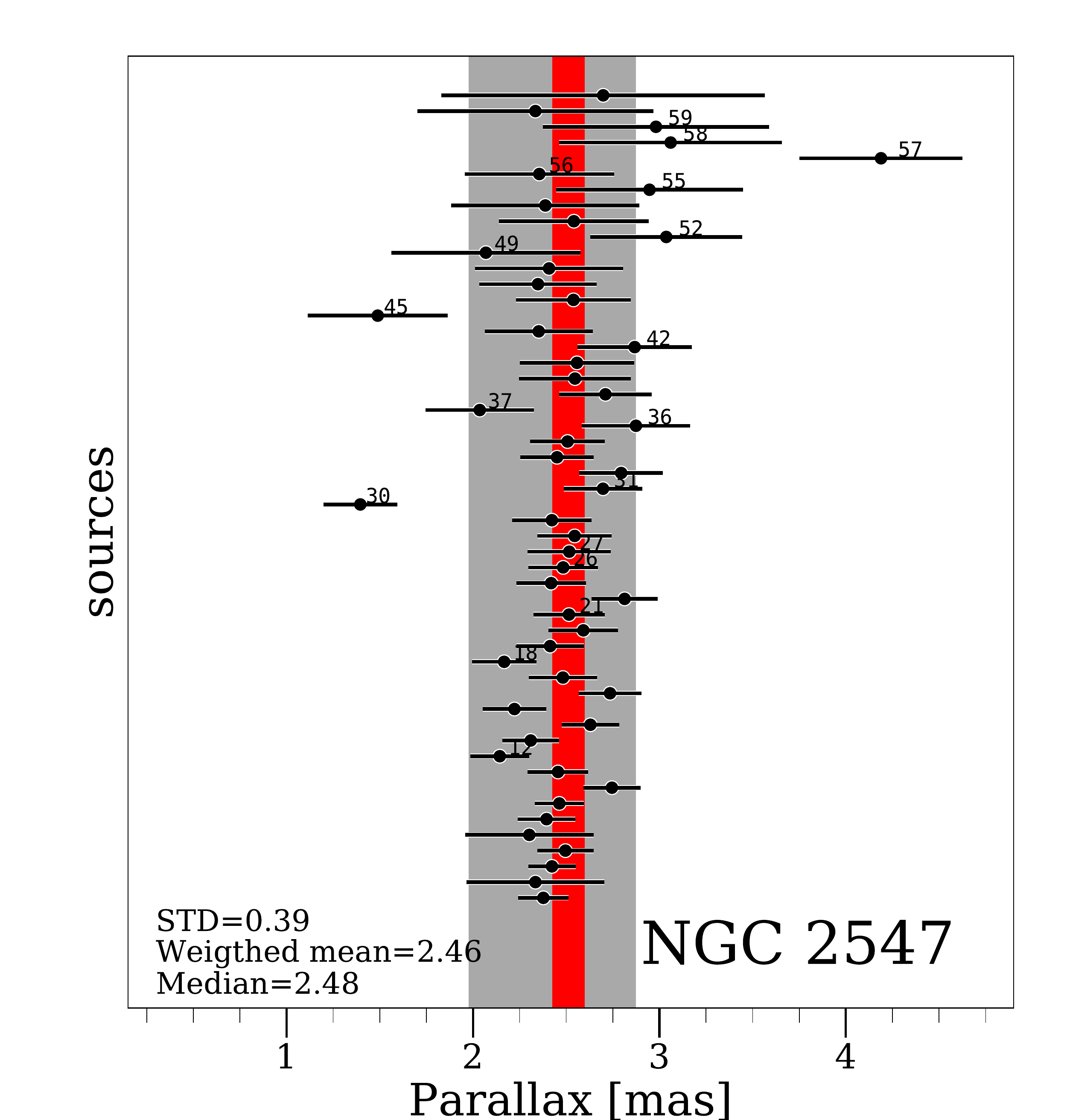}   
      \caption[VPD and parallaxes for NGC 2547.]      
              {VPD and parallaxes for NGC 2547.
               {\bf Left: }Same details as in Fig. \ref{fig:vpd_alphapersei} but for NGC 2547. 
               {\bf Right: }Same as in Fig. \ref{fig:plxs_alphapersei} but for NGC 2547. 
              }
         \label{fig:vpd_plxs_ngc2547}
\end{figure}

\begin{figure}
     \subfloat[\label{fig:hrd_ngc2547_general}]
      {\includegraphics[width=0.5\textwidth,scale=0.50]{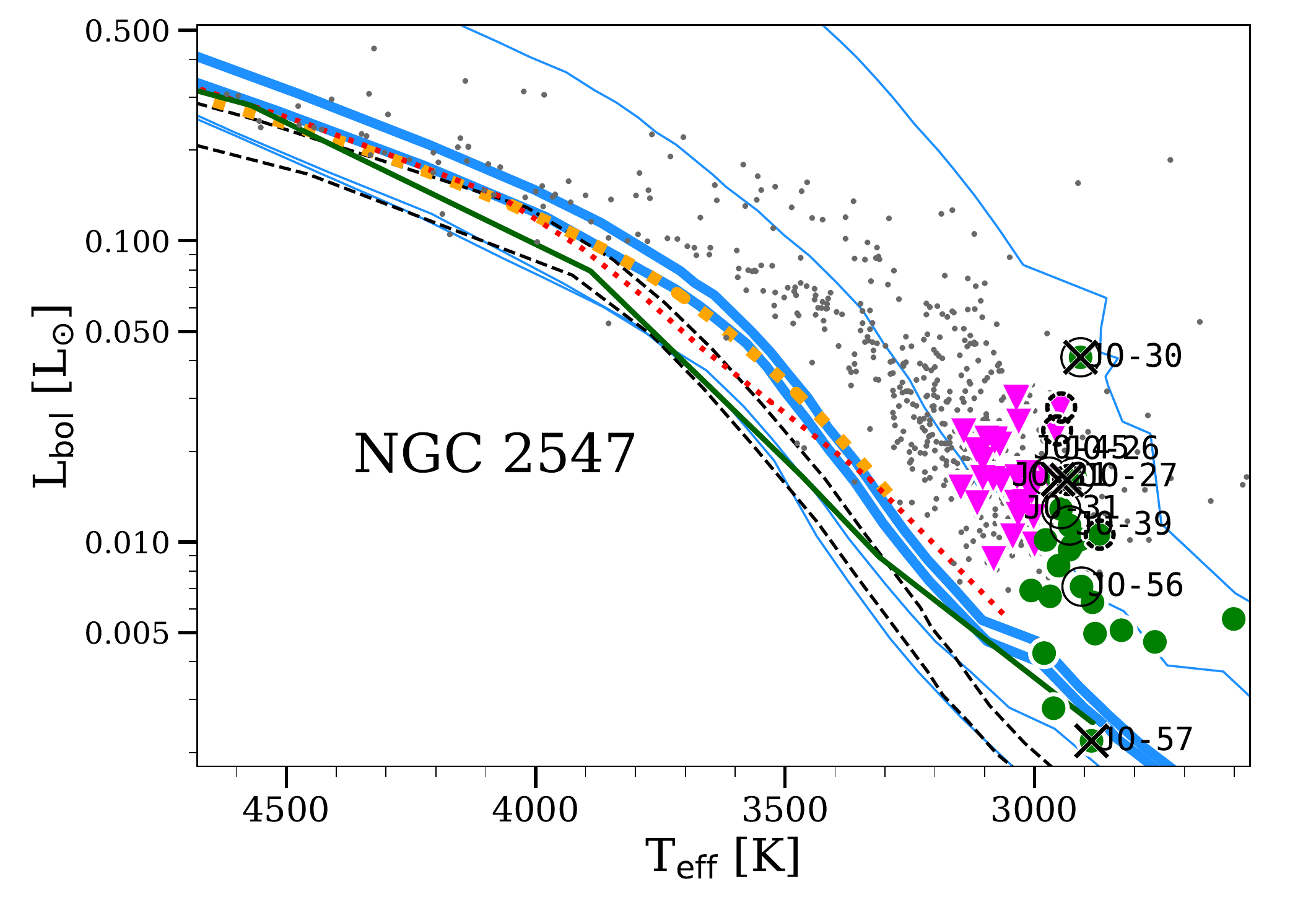}} \quad      
     \subfloat[\label{fig:hrd_ngc2547_zoom}]
      {\includegraphics[width=0.5\textwidth,scale=0.50]{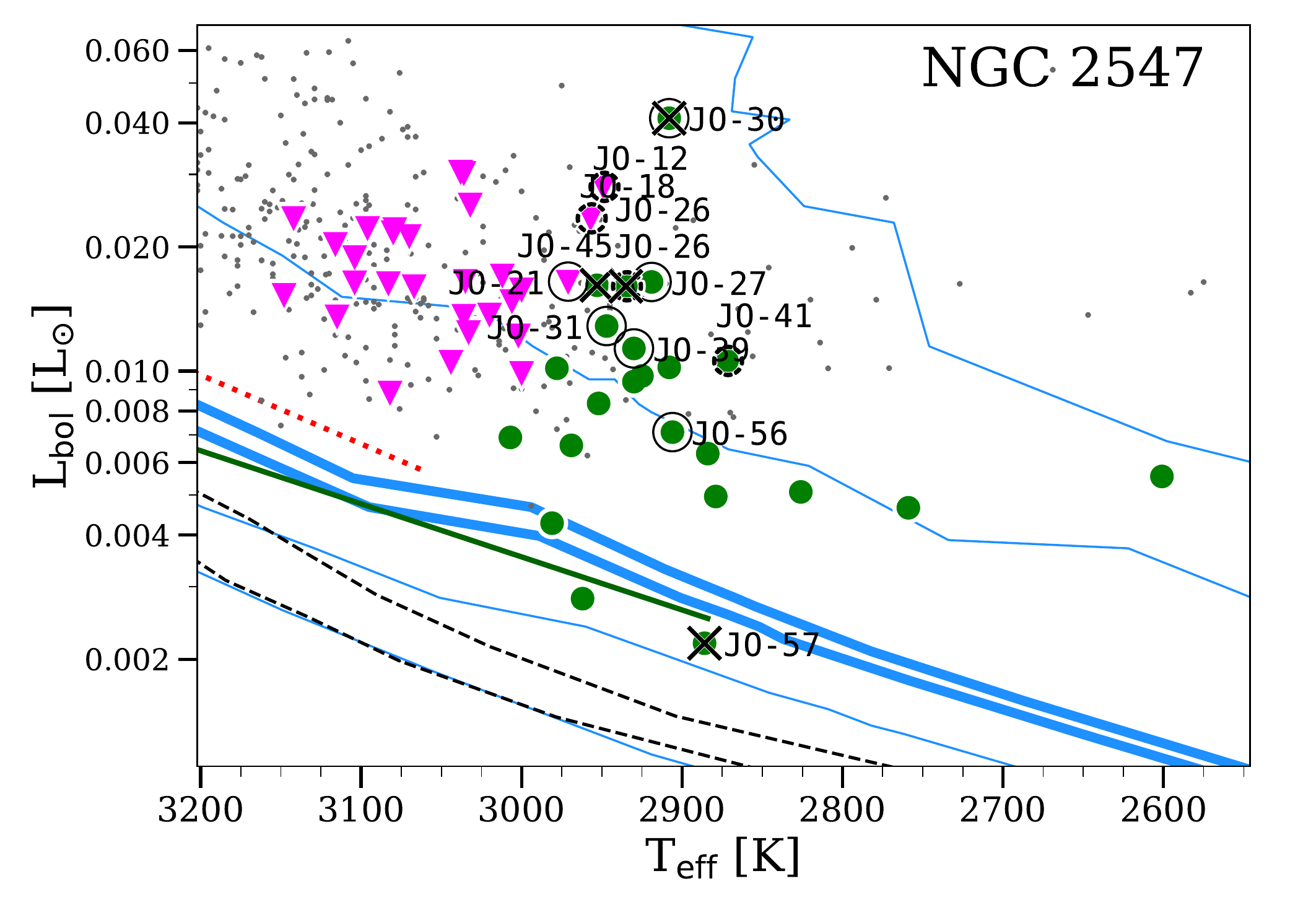}} \\      
     \subfloat[\label{fig:hrd_ngc2547_zoom_ldb}]
      {\includegraphics[width=0.5\textwidth,scale=0.50]{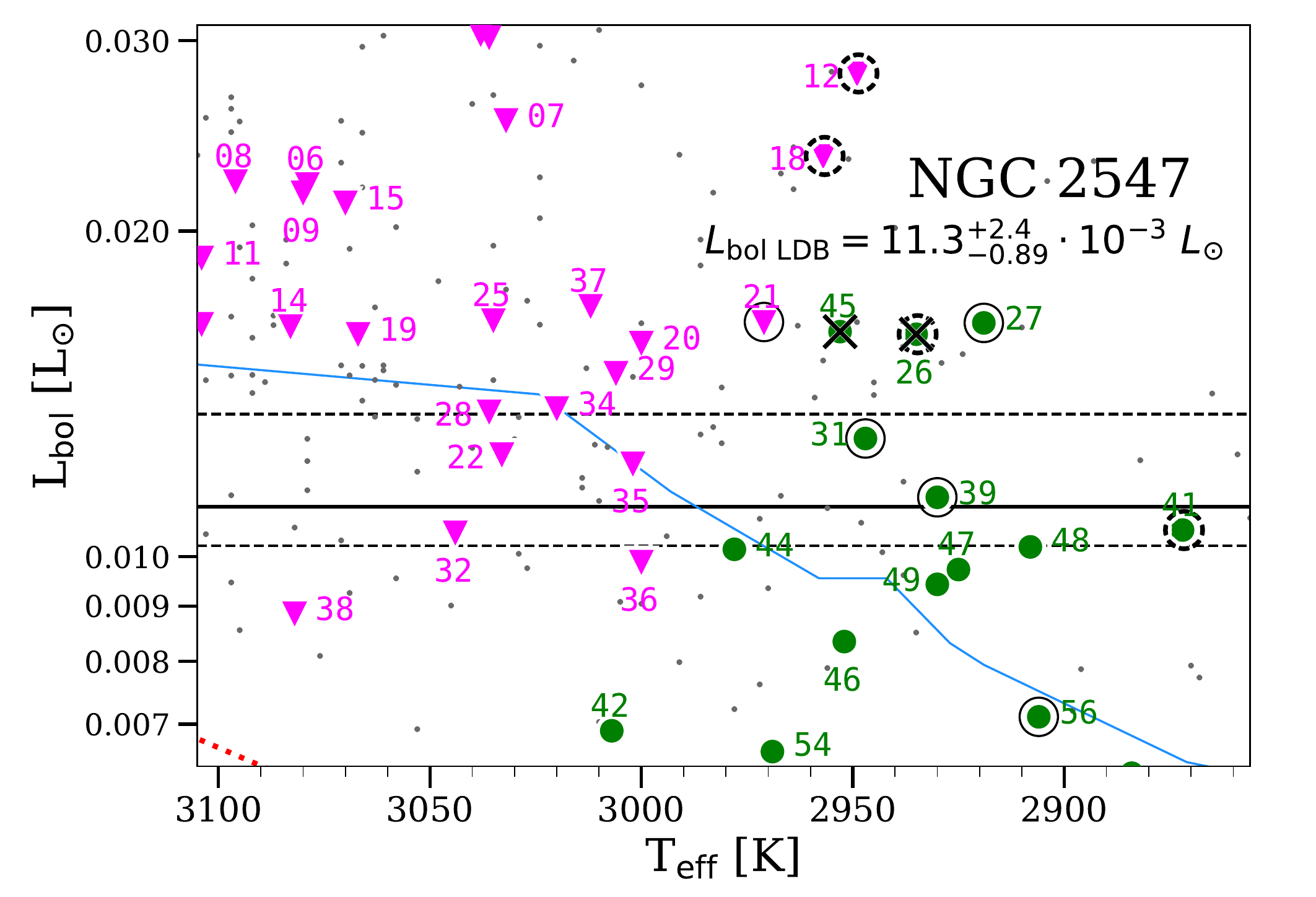}} \quad 
    \caption[HRDs and the LDB for NGC 2547.]
          {HRDs and the LDB for NGC 2547.       
           {\bf (a) }
             the same as in Fig. \ref{fig:hrd_alphapersei_discussion} and \ref{fig:hrds_ngc1960} but for NGC 2547. 
             Thin blue lines correspond to isochrones of 1, 10, 100~Ma, and 1~Ga 
            from the BT-Settl models \citep{f_allard2013}, while the thick blue lines to 40 and 50~Ma. 
              The figure includes:
            a 30~Ma and a 50~Ma isochrones from \citealt{f_dantona1994} (black dashed lines), 
            a 35~Ma isochrone from \citealt{e_tognelli2011} (orange dashed dot line), 
            a 40~Ma isochrone from \citealt{siess2000} (red dotted line). 
           {\bf (b) }Zoom on the (a) plot around the 52 radial velocity members from \cite{jeffries2004} sample.
           {\bf (c) }Zoom around the LDB.  
            NGC 2547 is $43.5^{+1.7}_{-4.0}$~Ma old using the BT-Settl bolometric luminosity-age relationship \citep{f_allard2012}.
            We show the names of all the objects but not the uncertainties (for the purposes of clarity). 
          }
  \label{fig:hrds_ngc2547}
\end{figure}

\clearpage
\begin{figure}
   \centering
   \includegraphics[width=9cm]{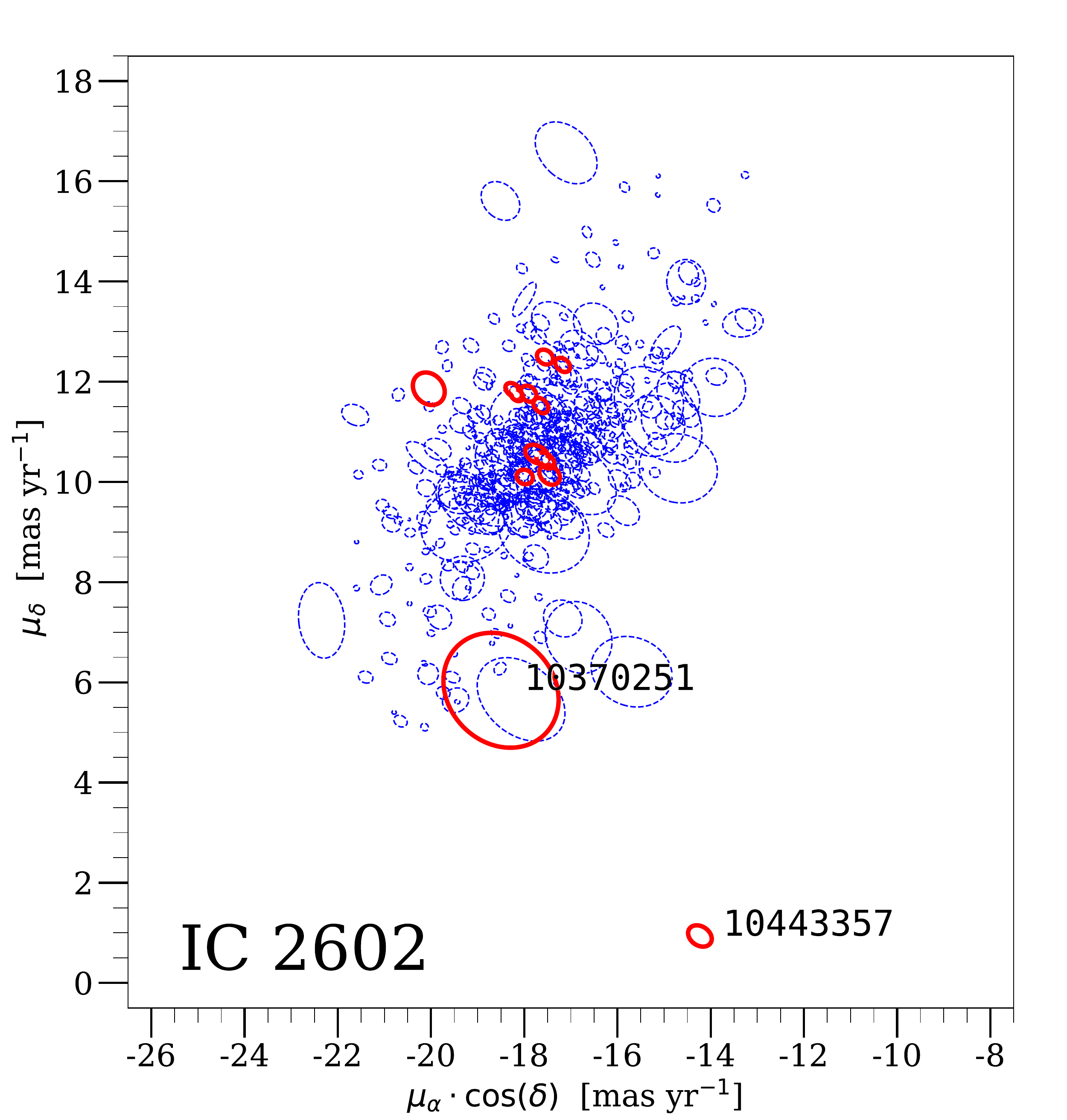}
   \includegraphics[width=9cm]{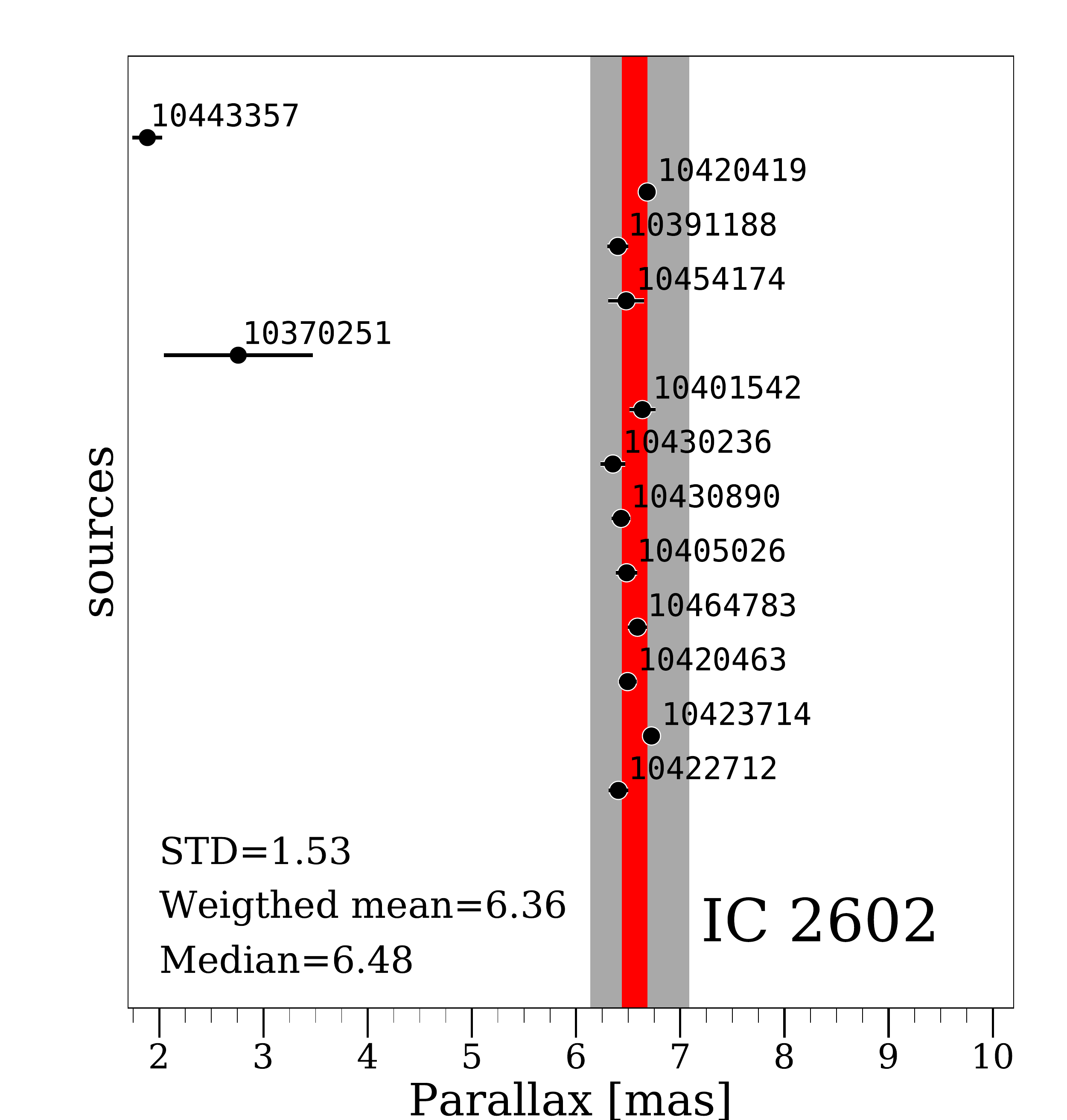}
      \caption[VPD and parallaxes for IC 2602.]
              {VPD and parallaxes for IC 2602.
               {\bf Left: }Same details as in Fig. \ref{fig:vpd_alphapersei} but for IC 2602. 
                Source 10443357 possesses  
                a proper motion far different than the rest of members. 
               {\bf Right: }Same as in Fig. \ref{fig:plxs_alphapersei} but for IC 2602. 
              }
         \label{fig:vpd_plxs_ic2602}
\end{figure}

\begin{figure}
     \subfloat[\label{fig:hrd_ic2602_general}]
     {\includegraphics[width=0.5\textwidth,scale=0.50]{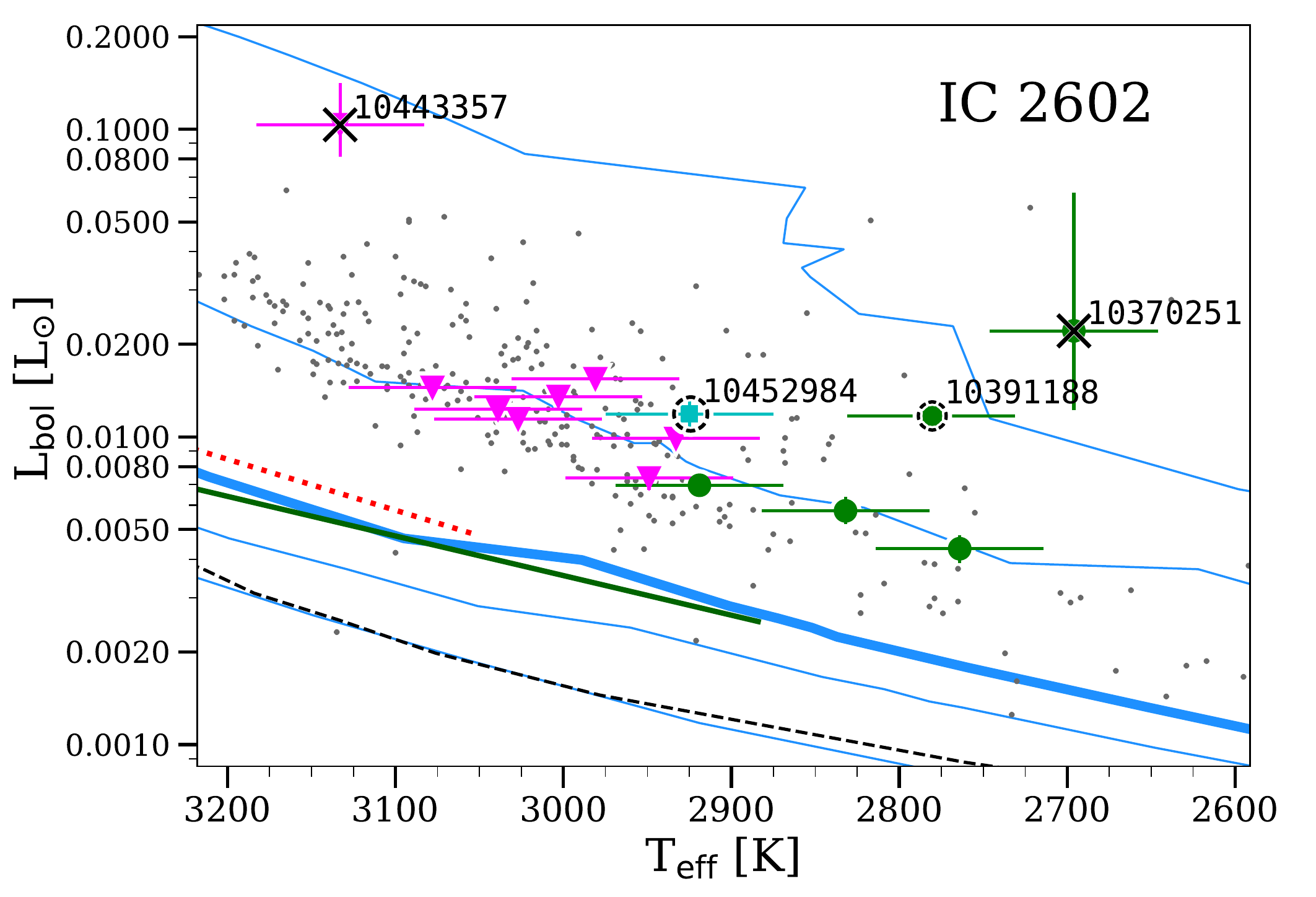}} \quad
     \subfloat[\label{fig:hrd_ic2602_zoom_ldb}]                 
     {\includegraphics[width=0.5\textwidth,scale=0.50]{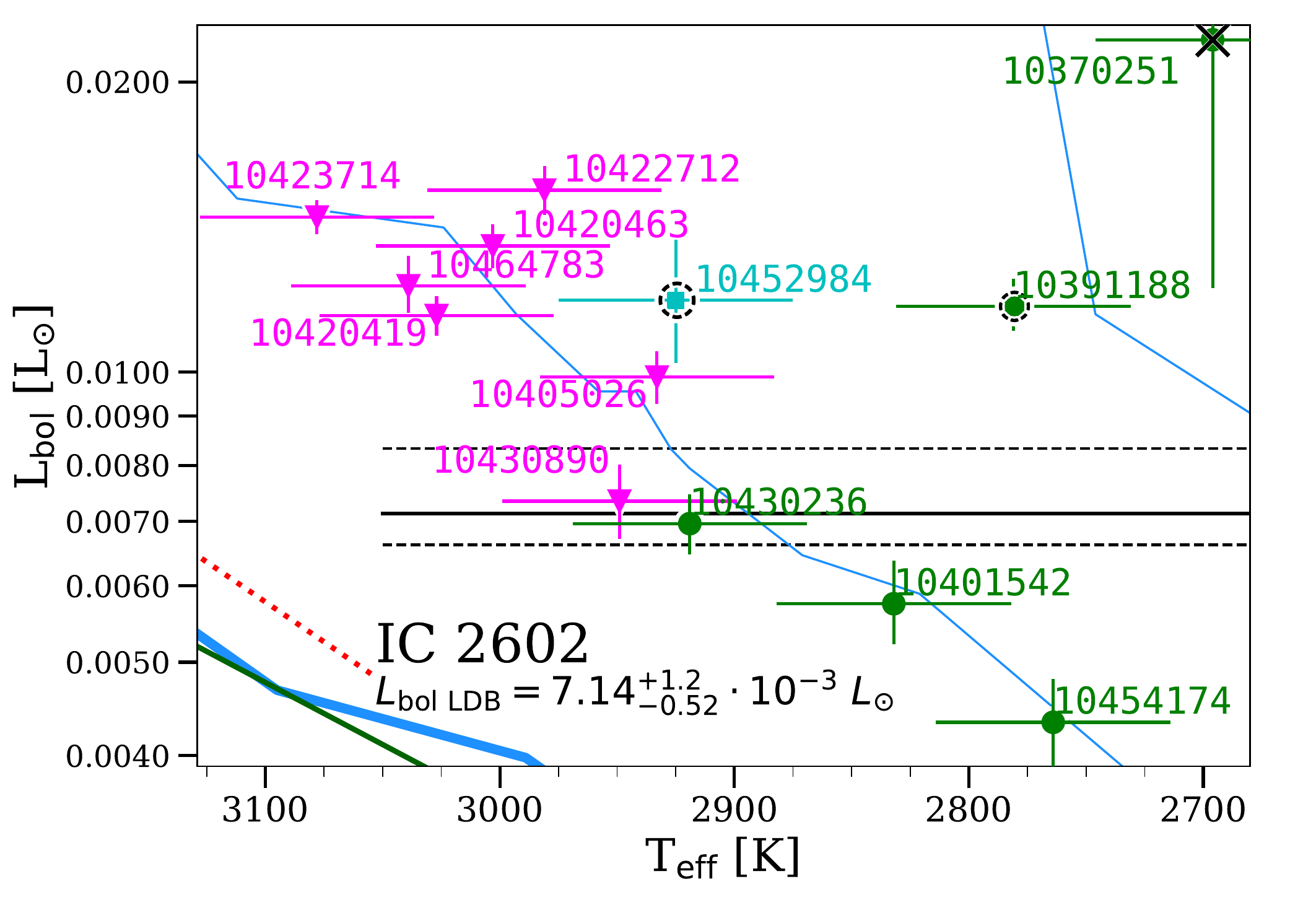}} \\
    \caption[HRDs and the LDB for IC 2602.]
          {HRDs and the LDB for IC 2602.
           {\bf (a): }
             The same as in Fig. \ref{fig:hrd_alphapersei_discussion} but for IC 2602.
            Thin blue lines correspond to isochrones of 1, 10, 100~Ma, and 1~Ga 
            from the BT-Settl models \citep{f_allard2013}, while the thick blue one to 50~Ma. 
             The figure includes \mbox{50~Ma} isochrones from:             
            \citealt{siess2000} (red dotted line), 
            \citealt{f_dantona1994} (black dashed line). 
           {\bf (b): }Zoom on the (a) plot with the LDB. 
            IC 2602 is $52.5^{+2.2}_{-3.7}$~Ma old using the BT-Settl bolometric luminosity-age relationship \citep{f_allard2012}.
          }
  \label{fig:hrds_ic2602}
\end{figure}

\clearpage
\begin{figure}
   \centering
   \includegraphics[width=0.45\textwidth,scale=0.50]{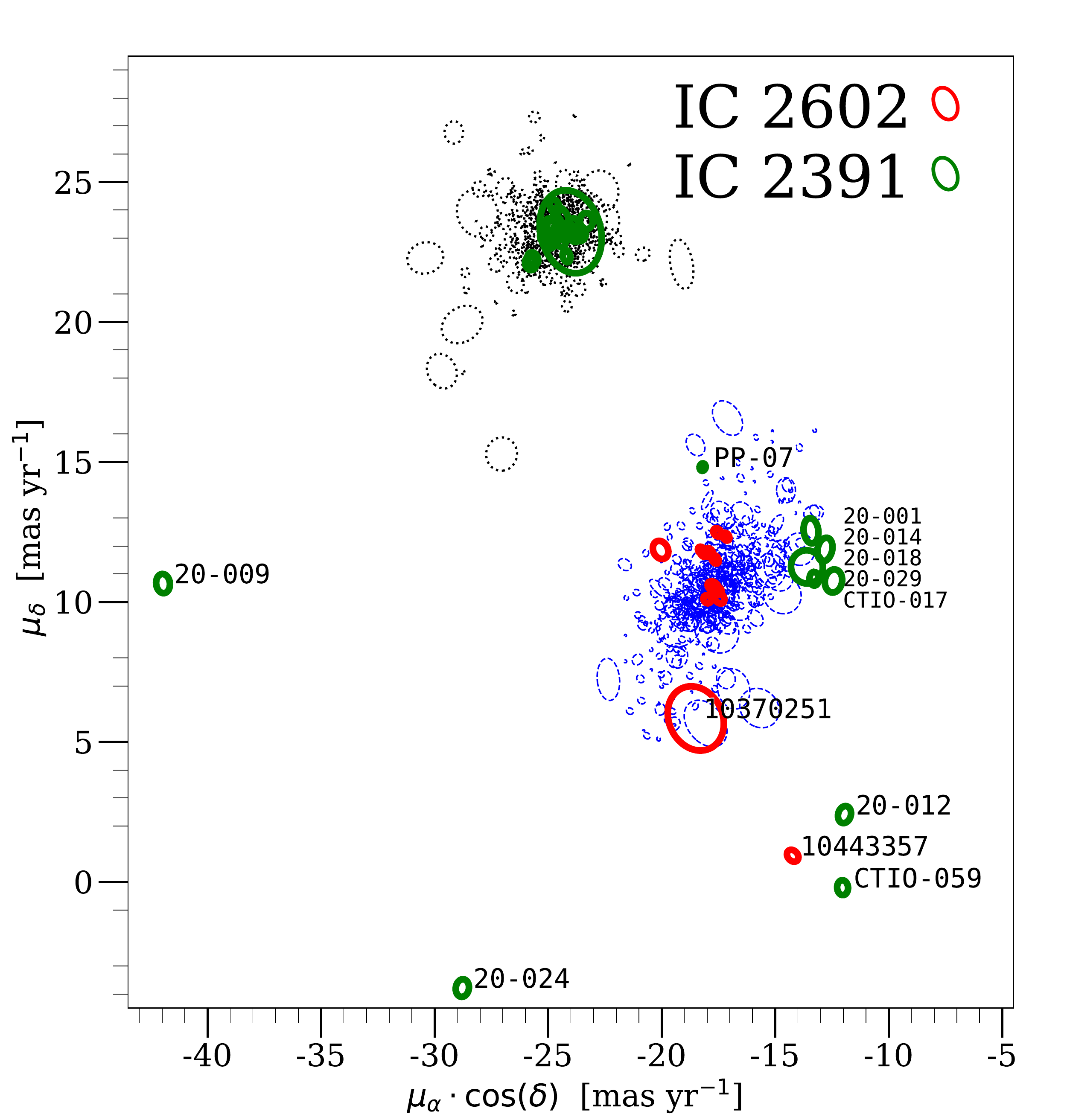}
   \includegraphics[width=0.45\textwidth,scale=0.50]{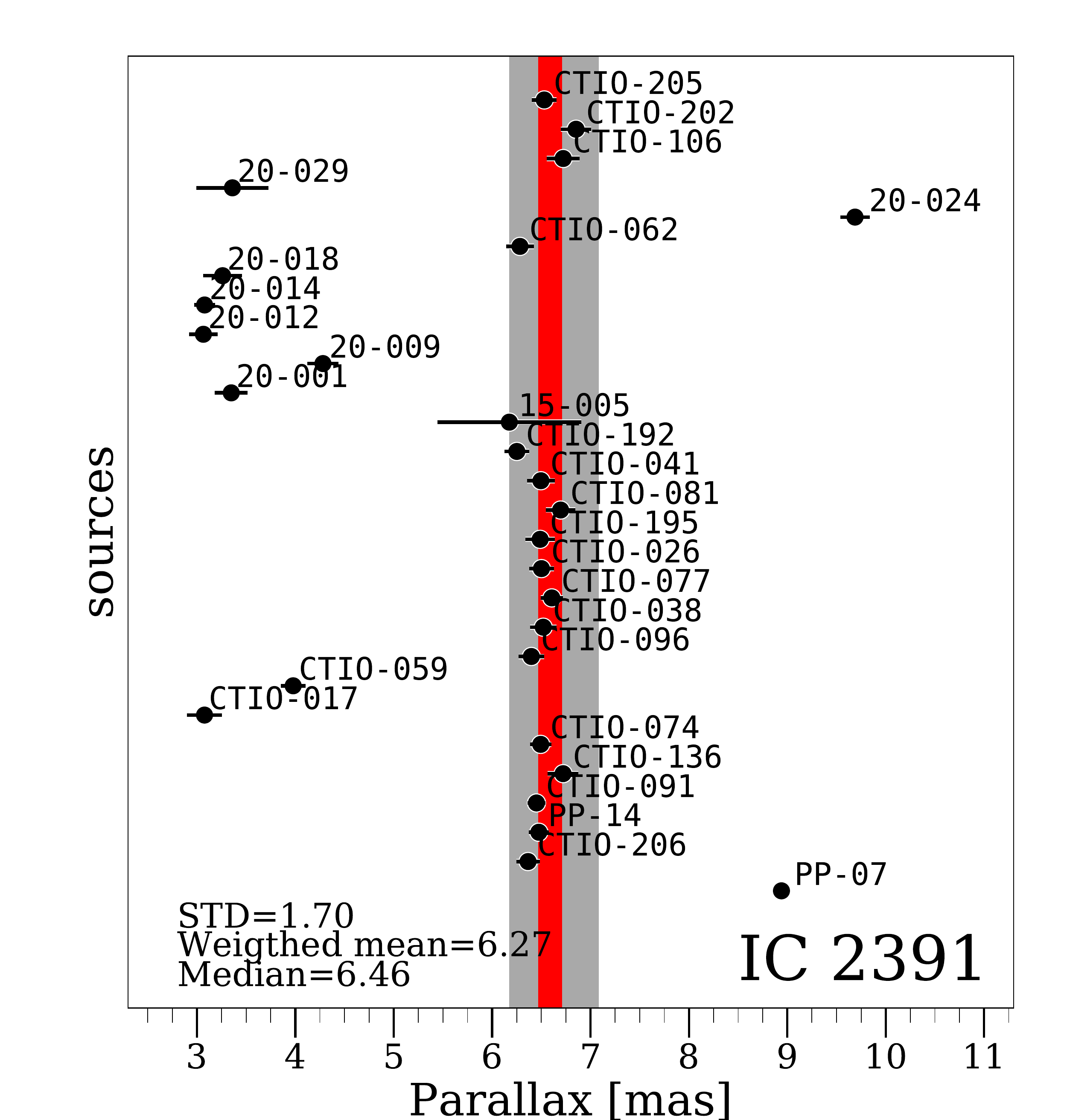}
      \caption[VPD for IC 2391 and IC 2602 members and parallaxes for IC 2391.]
              {VPD for IC 2391 and IC 2602 members and parallaxes for IC 2391.
               {\bf Left: }Same details as in Fig. \ref{fig:vpd_alphapersei} 
                but for IC 2391 and IC 2602.
                Some objects have uncertainties smaller than the size of the markers.               
                Black dotted ellipses are the IC 2391 members and 
                blue dashed ellipses are the IC 2602 members,  
                green solid ellipses are the IC 2391 LDB sample objects,
                and red solid ellipses are the IC 2602 LDB sample objects.  
               {\bf Right: }
                 The same as in Fig. \ref{fig:plxs_alphapersei} but for IC 2391. 
              }   
         \label{fig:vpd_ic2602_ic2391_plxs_ic2391}
   \end{figure}

\begin{figure}
    \subfloat[\label{fig:hrd_ic2391_general}]
      {\includegraphics[width=0.50\textwidth,scale=0.50]{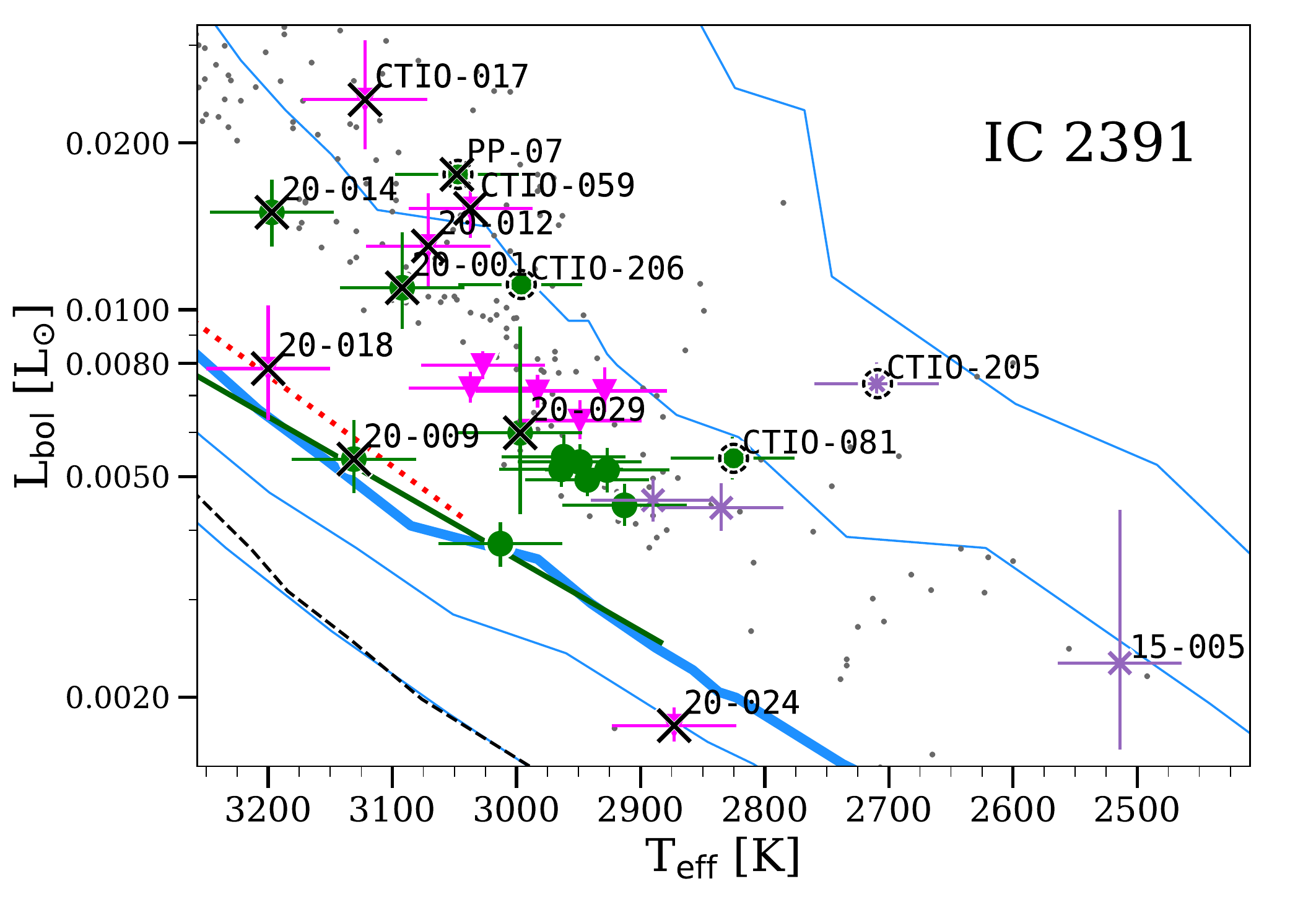}}      \quad
    \subfloat[\label{fig:hrd_ic2391_zoom_ldb}]                   
      {\includegraphics[width=0.50\textwidth,scale=0.50]{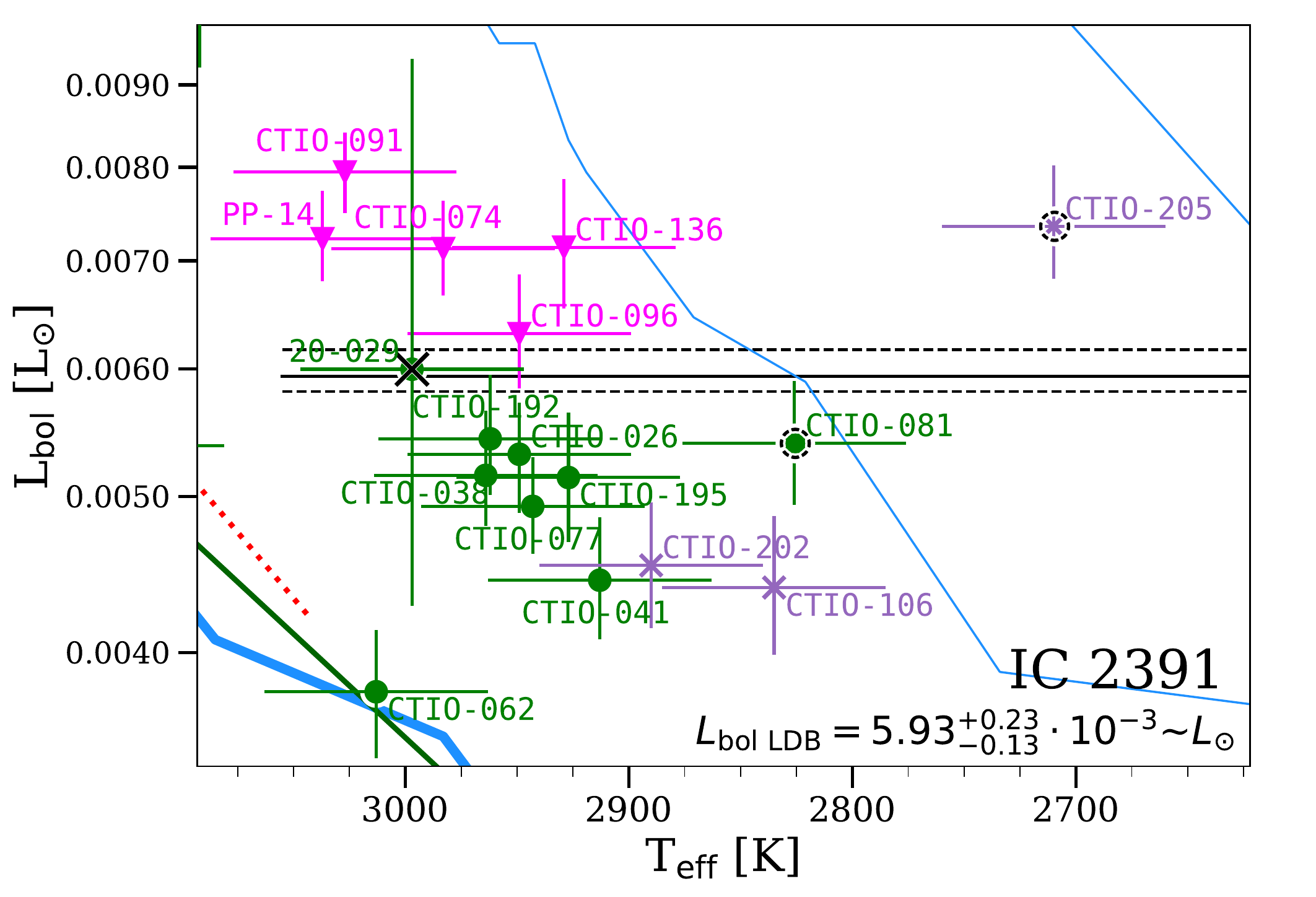}}\\
    \caption[HRDs and the LDB for IC 2391.]
          {HRDs and the the LDB for IC 2391.
           {\bf (a) }
             The same as in Fig. \ref{fig:hrd_alphapersei_discussion} but for IC 2391.
            Thin blue lines correspond to isochrones of 1, 10, 100~Ma, and 1~Ga 
            from the BT-Settl models \citep{f_allard2013}, while the thick blue one to 60~Ma. 
              The figure includes:              
             a 50~Ma isochrones from \citealt{f_dantona1994} (black dashed line), 
             a 60~Ma isochrone from \citealt{siess2000} (red dotted line). 
           {\bf (b) }Zoom on the left plot with the location of the LDB.
            IC 2391 is $57.7^{+0.5}_{-1.0}$~Ma old using the BT-Settl bolometric luminosity-age relationship \citep{f_allard2012}.
           }
  \label{fig:hrds_ic2391}
\end{figure}

\clearpage
\begin{figure}
   \centering
   \includegraphics[width=0.45\textwidth,scale=0.50]{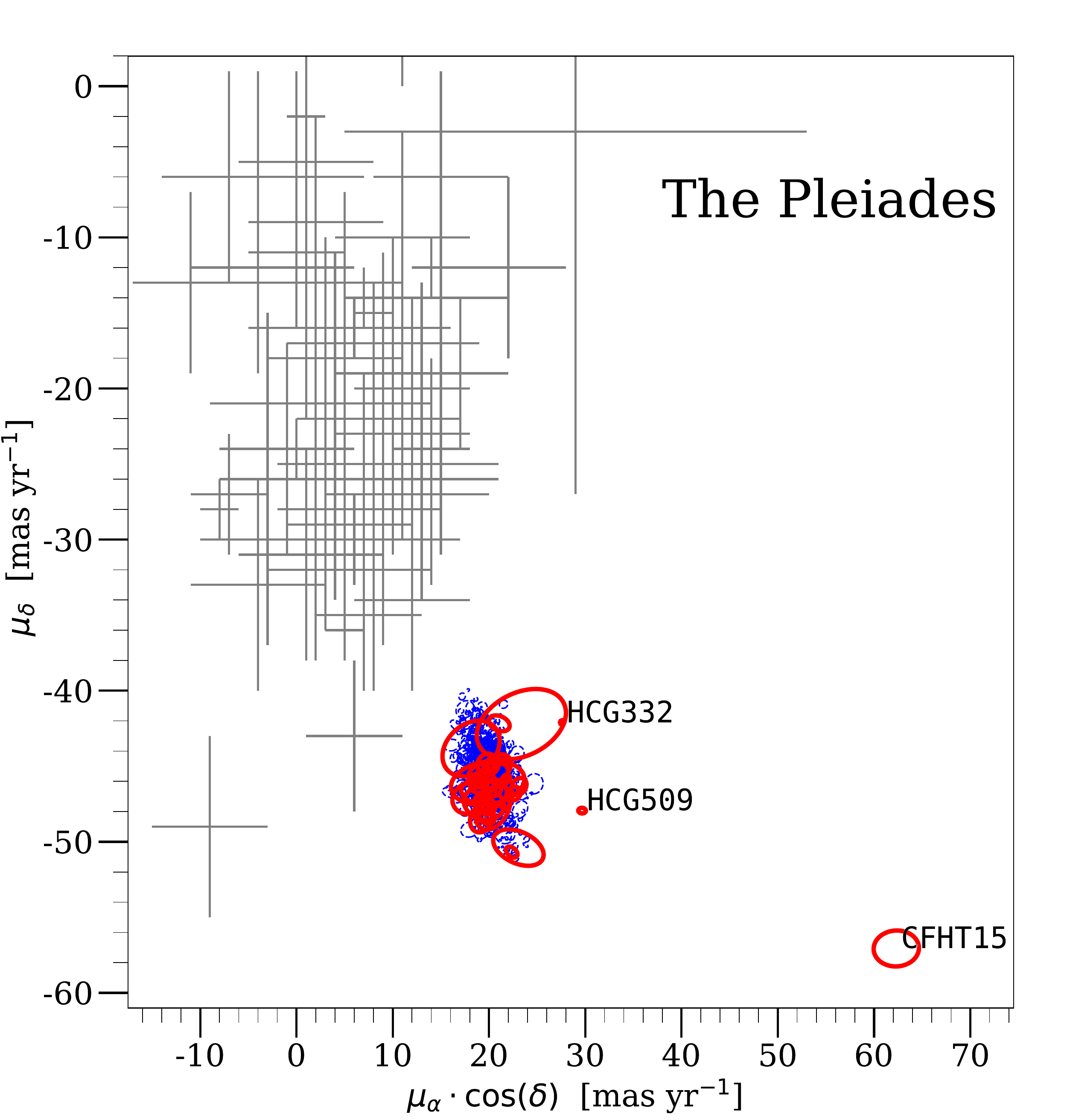}
   \includegraphics[width=0.45\textwidth,scale=0.50]{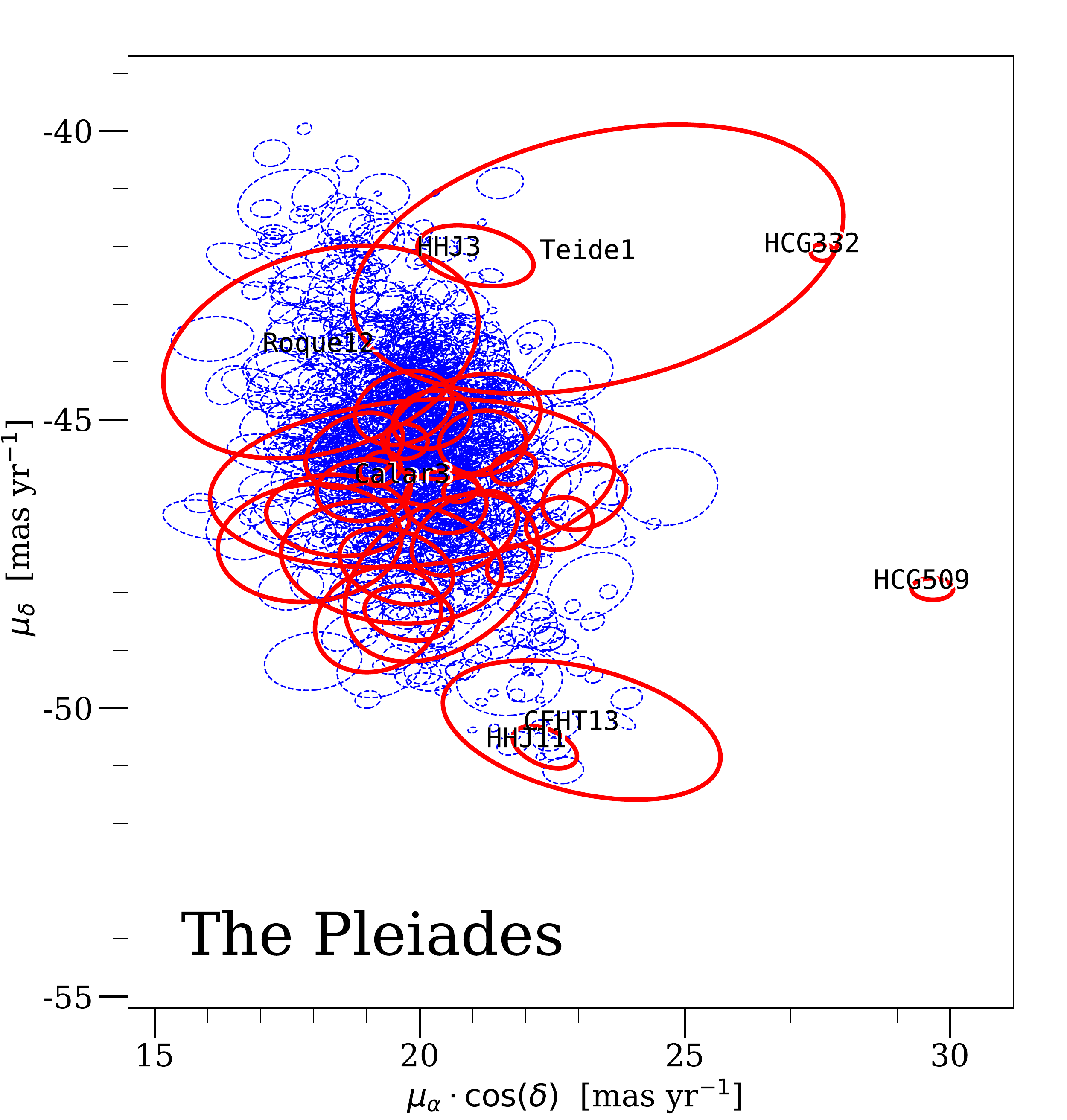}
      \caption[VPD for the Pleiades.]
              {VPD for the Pleiades.
               {\bf Left: }Same details as in Fig. \ref{fig:vpd_alphapersei} but for the Pleiades.
                Light grey plus symbols are the Taurus-Auriga members taken from the 
                \cite{c_ducourant2005} proper motions catalogue.                
               {\bf Right: }Zoom on the left figure. 
              }
         \label{fig:vpds_thepleiades}
\end{figure}

\begin{figure}
   \centering
   \includegraphics[width=0.50\textwidth,scale=0.50]{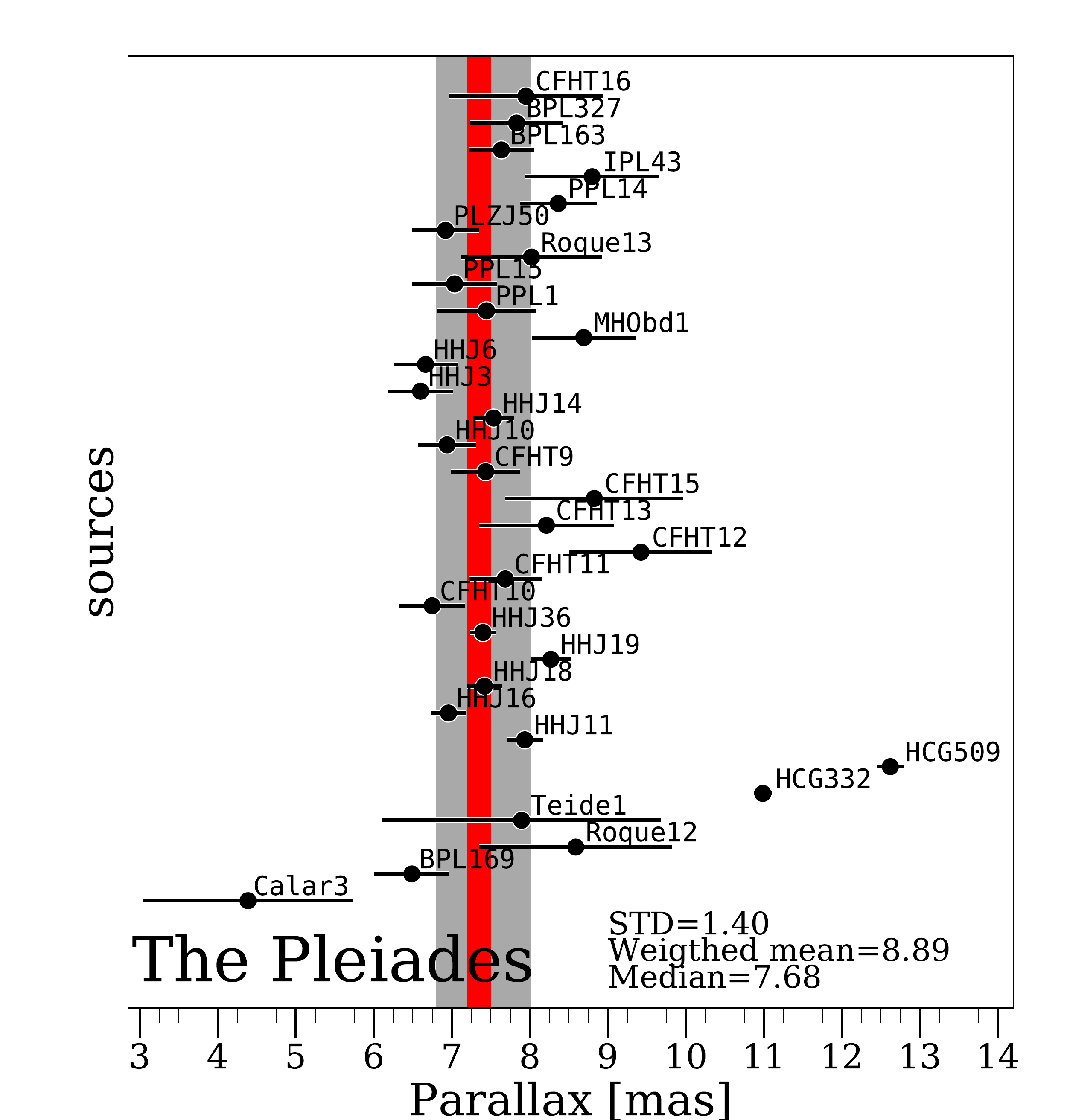}   
      \caption[Parallaxes for the Pleiades members close to the LDB.]
              {Parallaxes for the Pleiades members close to the LDB. 
               Same details as in Fig. \ref{fig:plxs_alphapersei} but for the Pleiades.
              }
         \label{fig:plxs_thepleiades}
\end{figure}

\begin{figure}
    \subfloat[\label{fig:hrd_thepleiades_general}]
      {\includegraphics[width=0.50\textwidth,scale=0.50]{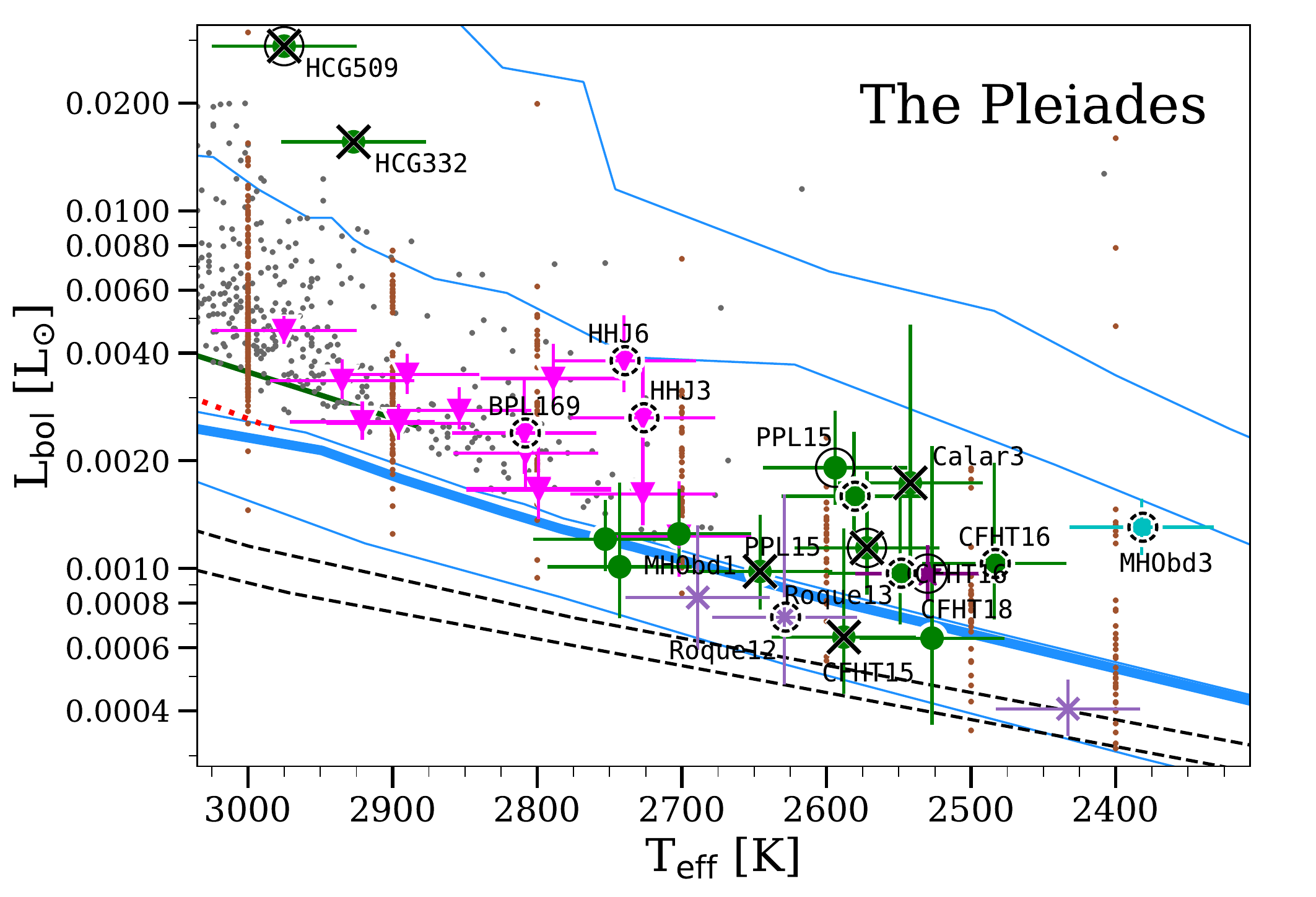}} \quad
    \subfloat[\label{fig:hrd_thepleiades_zoom_ldb}]
      {\includegraphics[width=0.50\textwidth,scale=0.50]{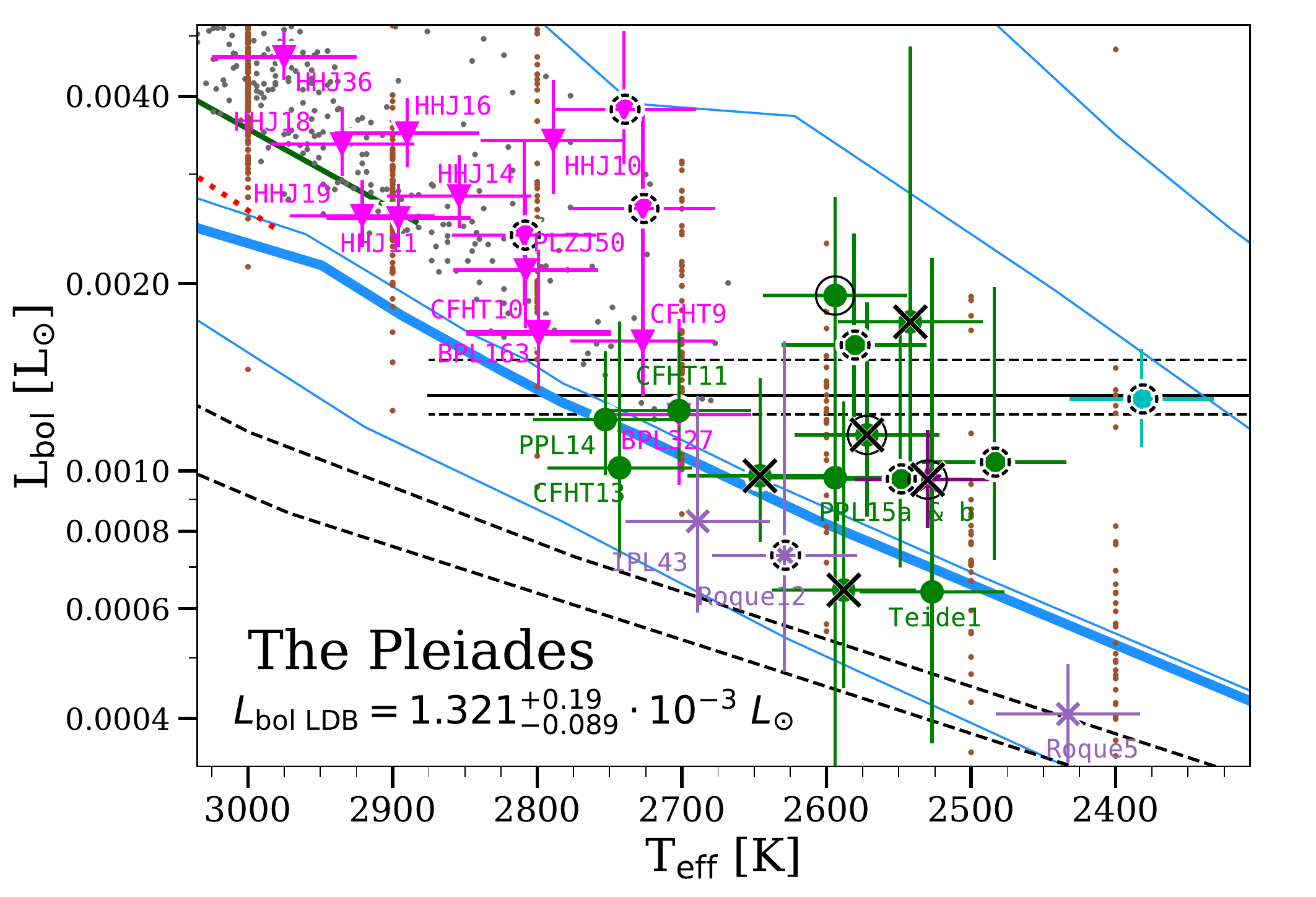}} \\
    \caption[HRDs and the LDB for the Pleiades.]
          {HRDs and the LDB for the Pleiades.
           {\bf (a) }
             The same as in Fig. \ref{fig:hrd_alphapersei_discussion} but for the Pleiades.
            In addition, grey points are known members from \cite{c_babusiaux2018}, 
            and sienna points from \cite{h_bouy2015b}.   
             Thin blue lines correspond to isochrones of 1, 10, 100~Ma, and 1~Ga 
            from the BT-Settl models \citep{f_allard2013}, while the thick blue one to 120~Ma.
             The figure includes:              
            a 100 and 200~Ma isochrones from \citealt{f_dantona1994} (black dashed lines), 
            and a 120~Ma isochrone from \citealt{siess2000} (red dotted line). 
           {\bf (b) }Zoom on the previous plot with the location of the LDB.
            The Pleiades is $127.4^{+6.3}_{-10}$~Ma old using the BT-Settl bolometric luminosity-age relationship \citep{f_allard2012}.
          }
  \label{fig:hrds_thepleiades}
\end{figure}

\clearpage
\begin{figure}
   \subfloat[\label{fig:vpd_blanco1}]
     {\includegraphics[width=0.45\textwidth,scale=0.50]{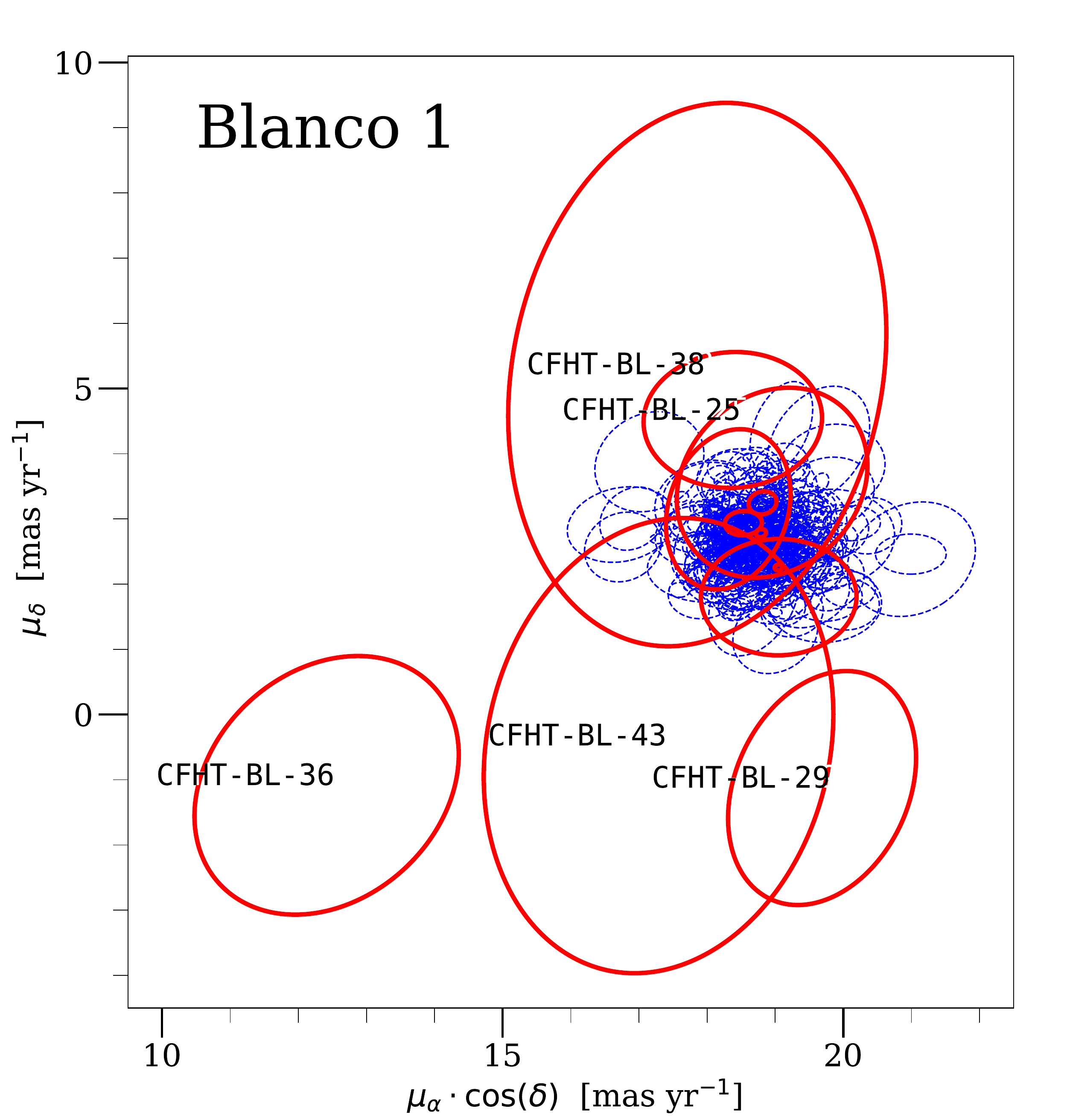}} \quad
   \subfloat[\label{fig:plx_blanco1}]    
     {\includegraphics[width=0.45\textwidth,scale=0.50]{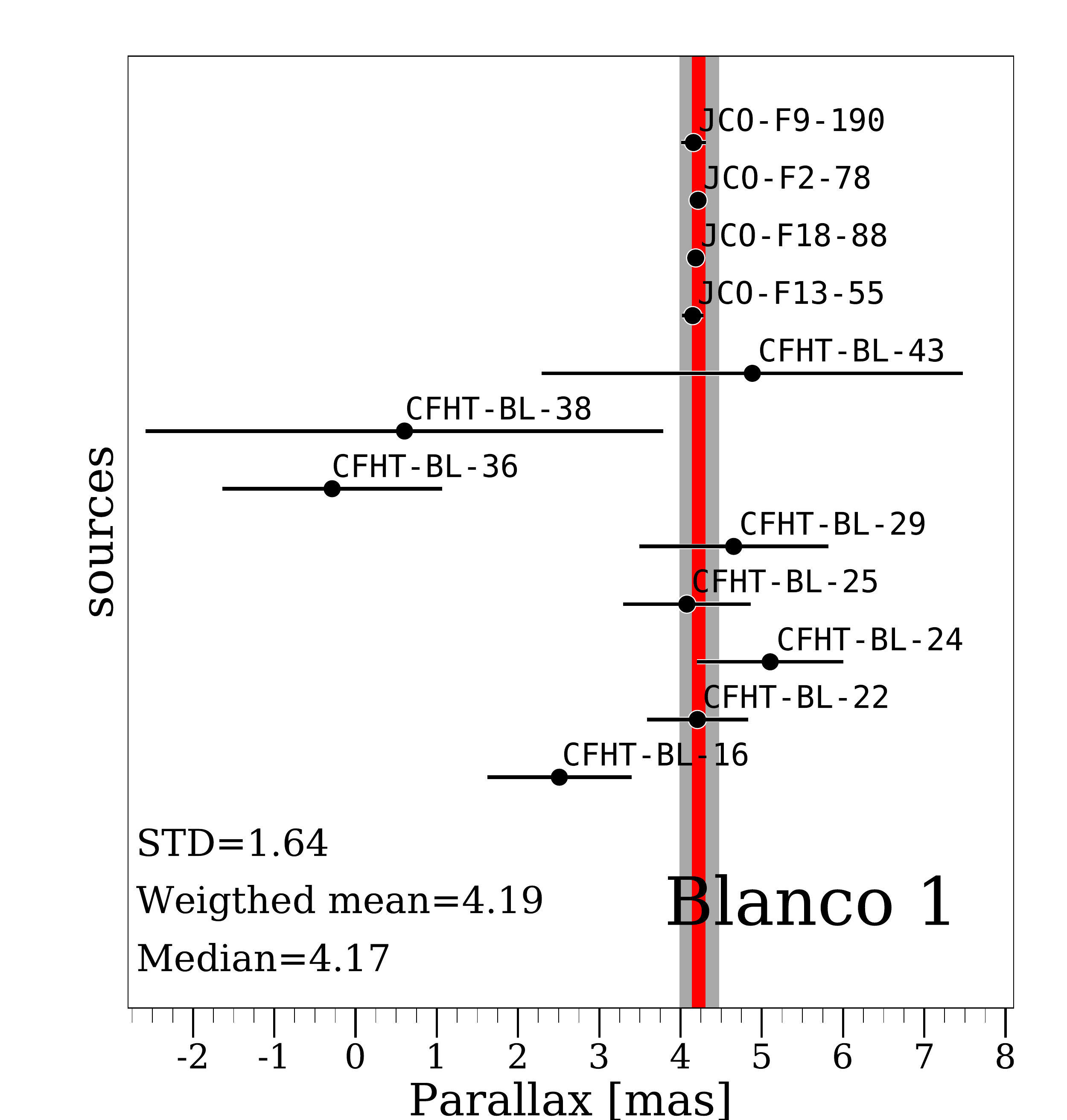}} \\
     \caption[VPD and parallaxes for Blanco 1.]
           {VPD and parallaxes for Blanco 1.
            {\bf Left: }Same details as in Fig. \ref{fig:vpd_alphapersei} but for Blanco 1.
            {\bf Right: }Same as in Fig. \ref{fig:plxs_alphapersei} but for Blanco 1.
                We note that \mbox{CFHT-BL-36} has got negative parallax.
           }
         \label{fig:vpd_plxs_blanco1}
\end{figure}
    
\begin{figure}
     \subfloat[\label{fig:hrd_blanco1_general}]
      {\includegraphics[width=0.5\textwidth,scale=0.50]{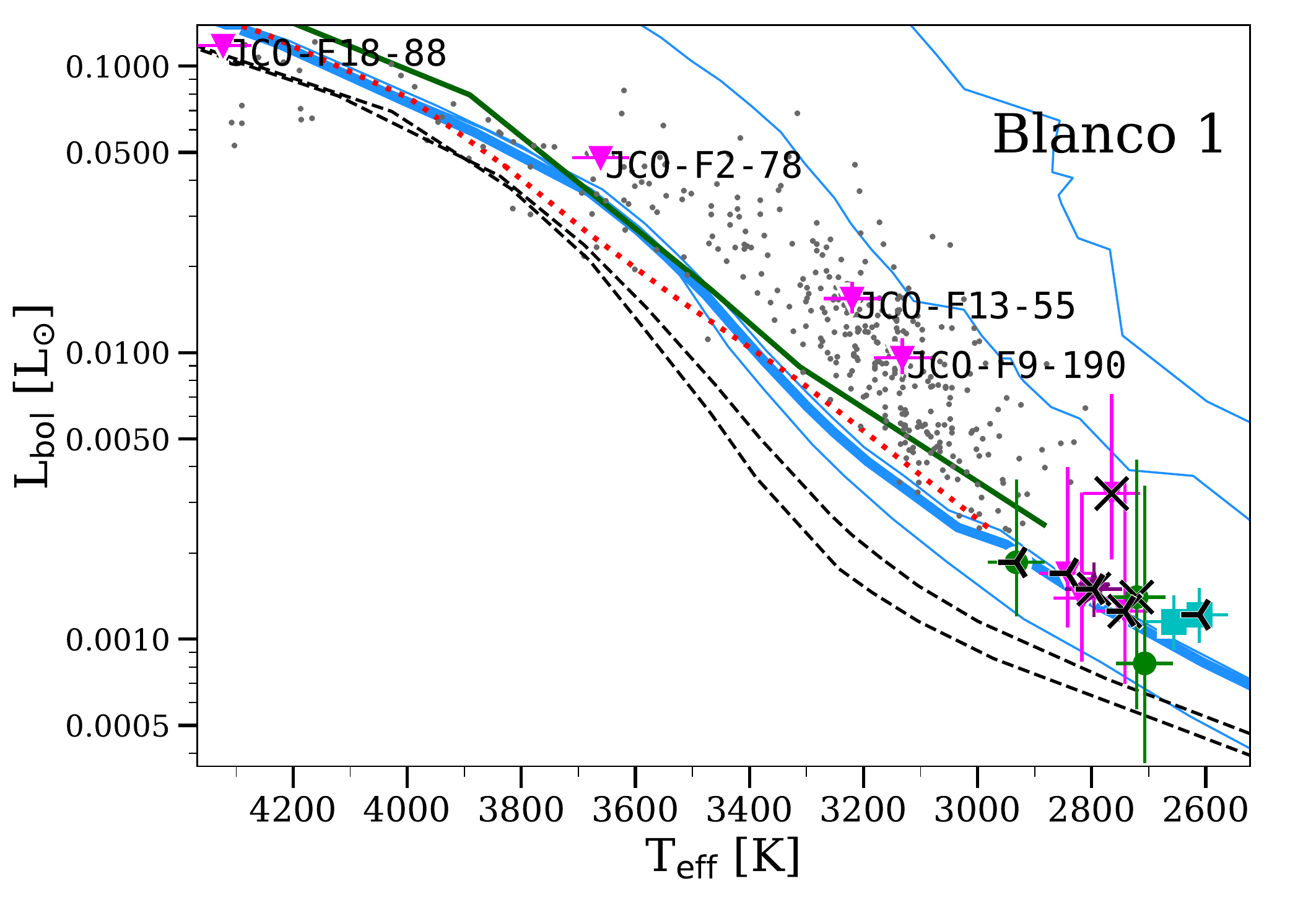}} \quad      
     \subfloat[\label{fig:hrd_blanco1_zoom_ldb_v1}]
      {\includegraphics[width=0.5\textwidth,scale=0.50]{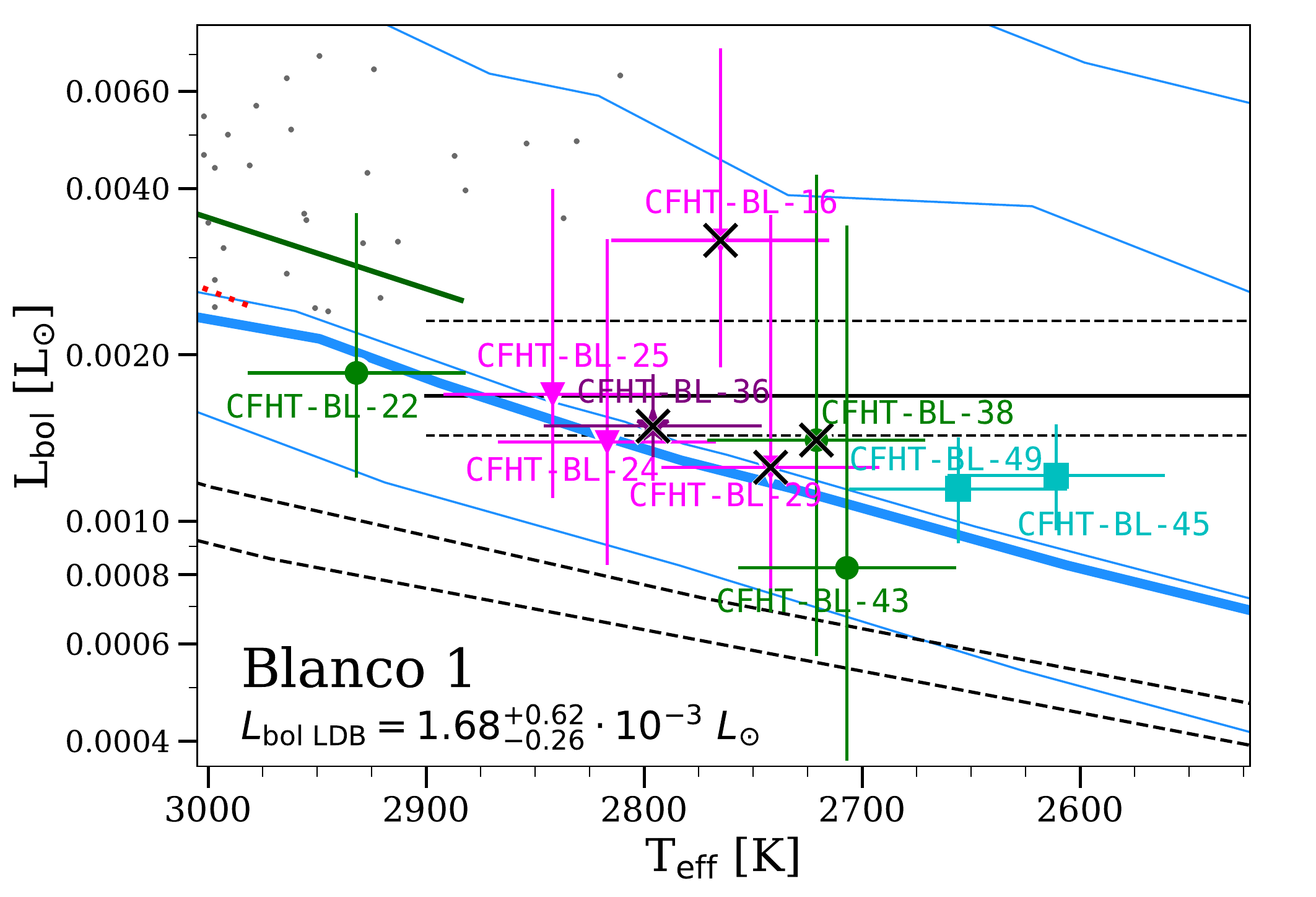}} \\
      \subfloat[\label{fig:hrd_blanco1_zoom_ldb_v2}]
      {\includegraphics[width=0.5\textwidth,scale=0.50]{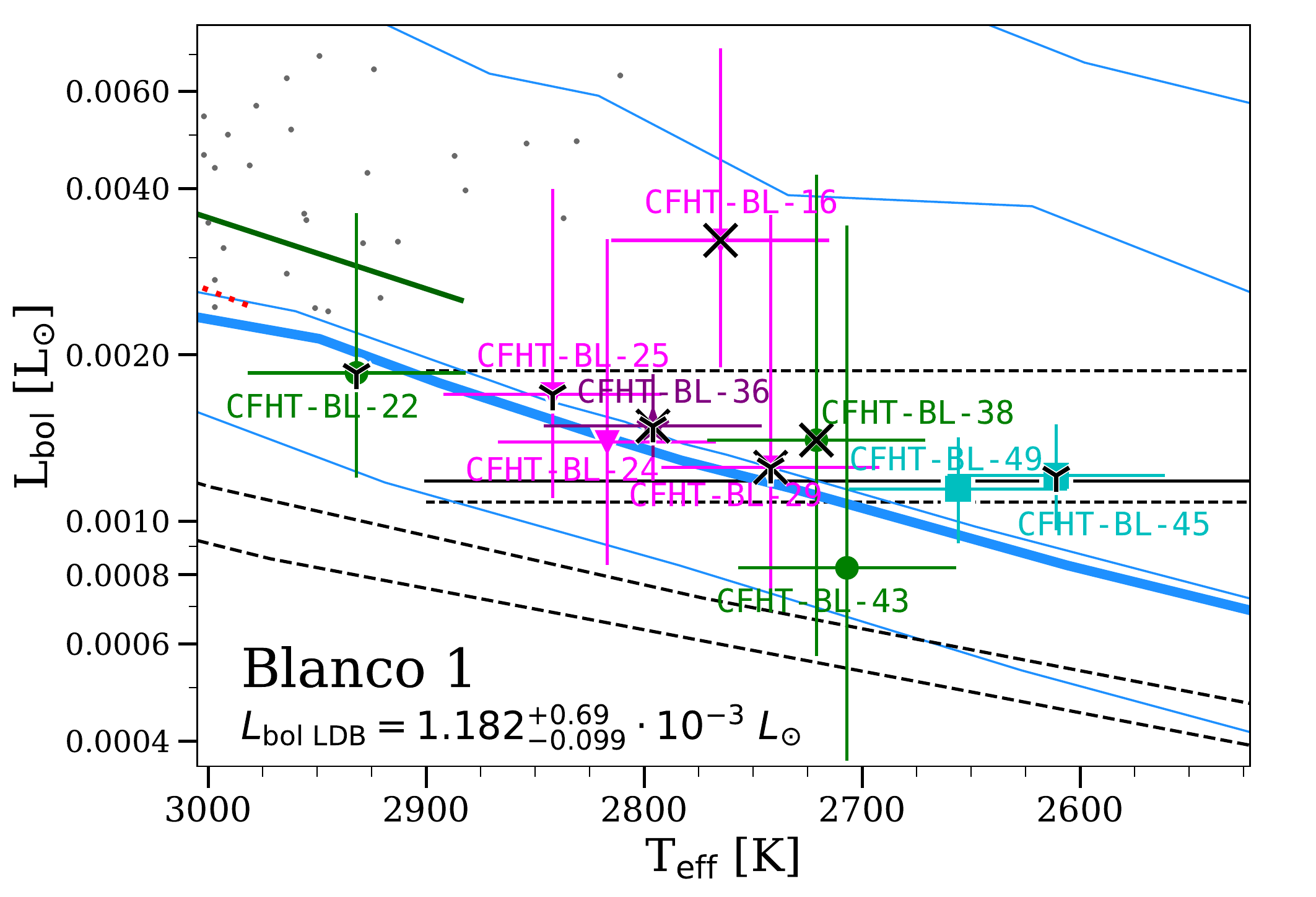}} \quad     
          \caption[HRDs and the LDB for Blanco 1.]
                  {HRDs and the LDB for Blanco 1.                 
                  {\bf (a) }
                    Same as in Fig. \ref{fig:hrd_alphapersei_discussion} but for Blanco 1. 
                   Thin blue lines correspond to isochrones of 1, 10, 100~Ma, and 1~Ga 
                   from the BT-Settl models \citep{f_allard2013}, while the thick blue one to 120~Ma. 
                    The figure includes: 
                   a 120~Ma isochrone from \citealt{siess2000} (red dotted line), 
                   a 100 and a 200~Ma from \citealt{f_dantona1994} (black dashed lines). 
                   The rest of the symbols follow the same convention as previous HRDs.   
                  {\bf (b) }Zoom on  the left plot close to the LDB. 
                   The Blanco 1 age is $110.9^{+9.0}_{-23}$~Ma old using the BT-Settl bolometric luminosity-age relationship \citep{f_allard2012}. 
                  {\bf (c) }Same as (b), but the LDB is located after discarding sources whose 
                   radial velocities are not within the $1\sigma$ radial velocity criterion, see Section \ref{sub:blanco1}.
                   The tri-down symbols are over-imposed in the non-members following this criterion. 
                   The Blanco 1 age is $137.1^{+7.0}_{-33}$~Ma (BT-Settl models from \citealt{f_allard2012}).
          }
  \label{fig:hrds_blanco1}
\end{figure}

\clearpage
   \begin{figure}
  \centering
   \includegraphics[width=0.45\textwidth,scale=0.50]{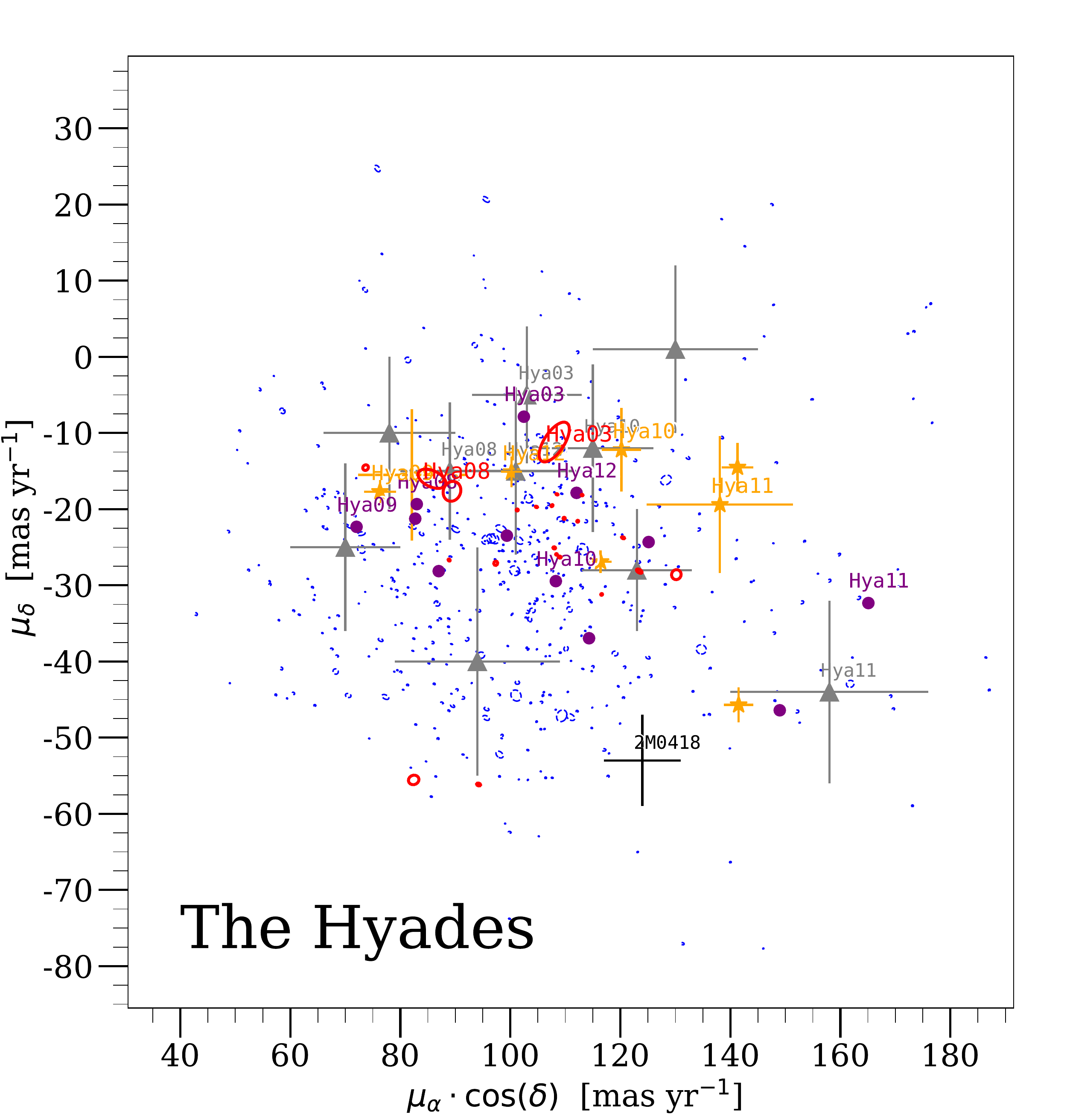}
   \includegraphics[width=0.45\textwidth,scale=0.50]{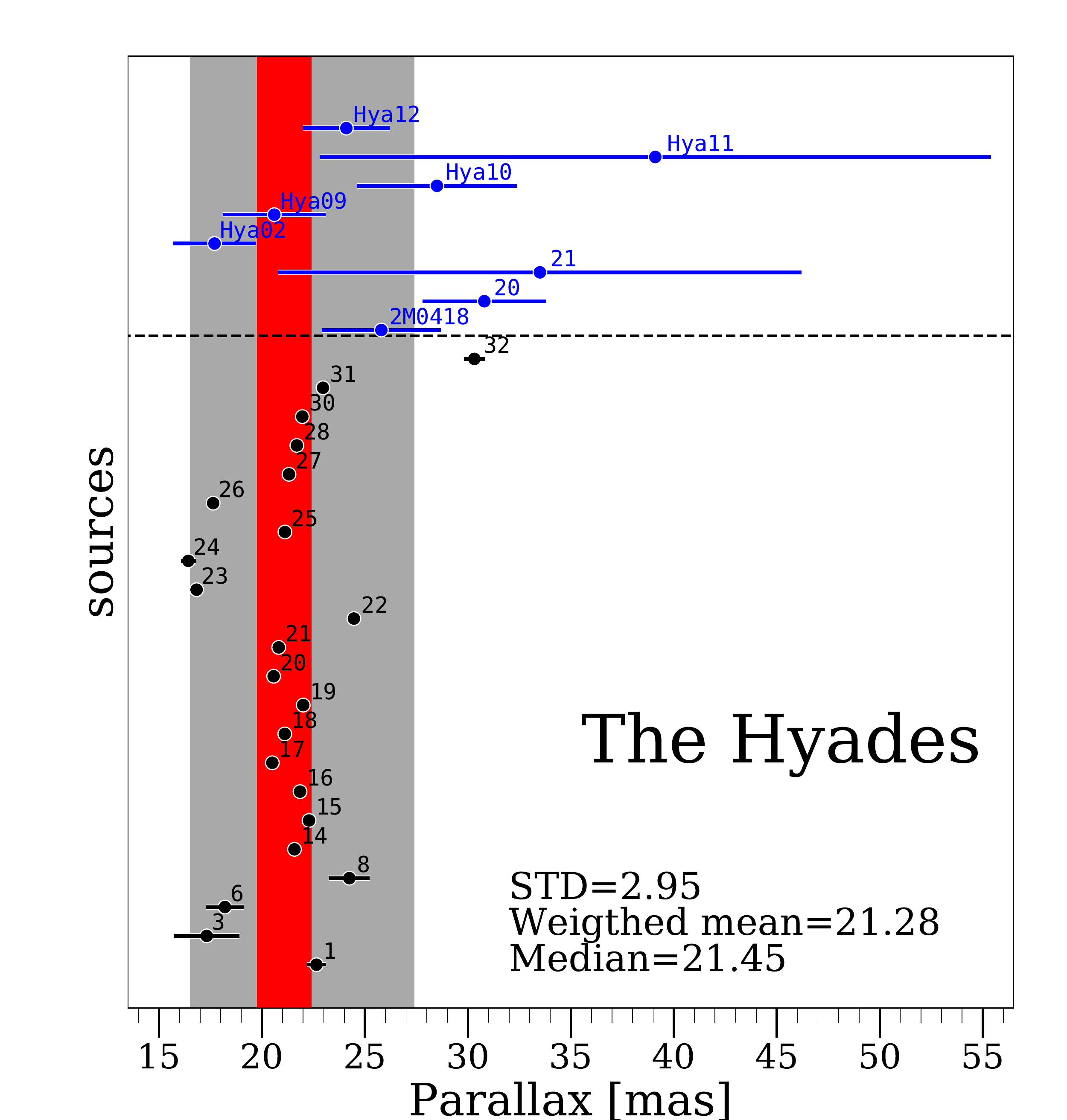}   
      \caption[VPD and parallaxes for the Hyades.]
              {VPD and parallaxes for the Hyades.
               {\bf Left: }Same as in Fig. \ref{fig:vpd_alphapersei} but for the Hyades.
                 We have added proper motions for the LDB sample sources from:
                \citealt{e_hogan2008} (purple filled circles); 
                \citealt{n_lodieu2014c} (grey triangles with their uncertainties); 
                \citealt{a_perezgarrido2017} (the 2M0418, the black plus symbol); 
                \citealt{n_lodieu2019b} (orange stars with their uncertainties).         
               {\bf Right: }Same as in Fig. \ref{fig:plxs_alphapersei} but for the Hyades.
                We have added parallax values from \cite{n_lodieu2019b} in blue, (above the black dashed line). 
               }
         \label{fig:vpd_plxs_thehyades}
   \end{figure}

\begin{figure}
     \subfloat[\label{fig:hrd_thehyades_general}] 
      {\includegraphics[width=0.5\textwidth,scale=0.50]{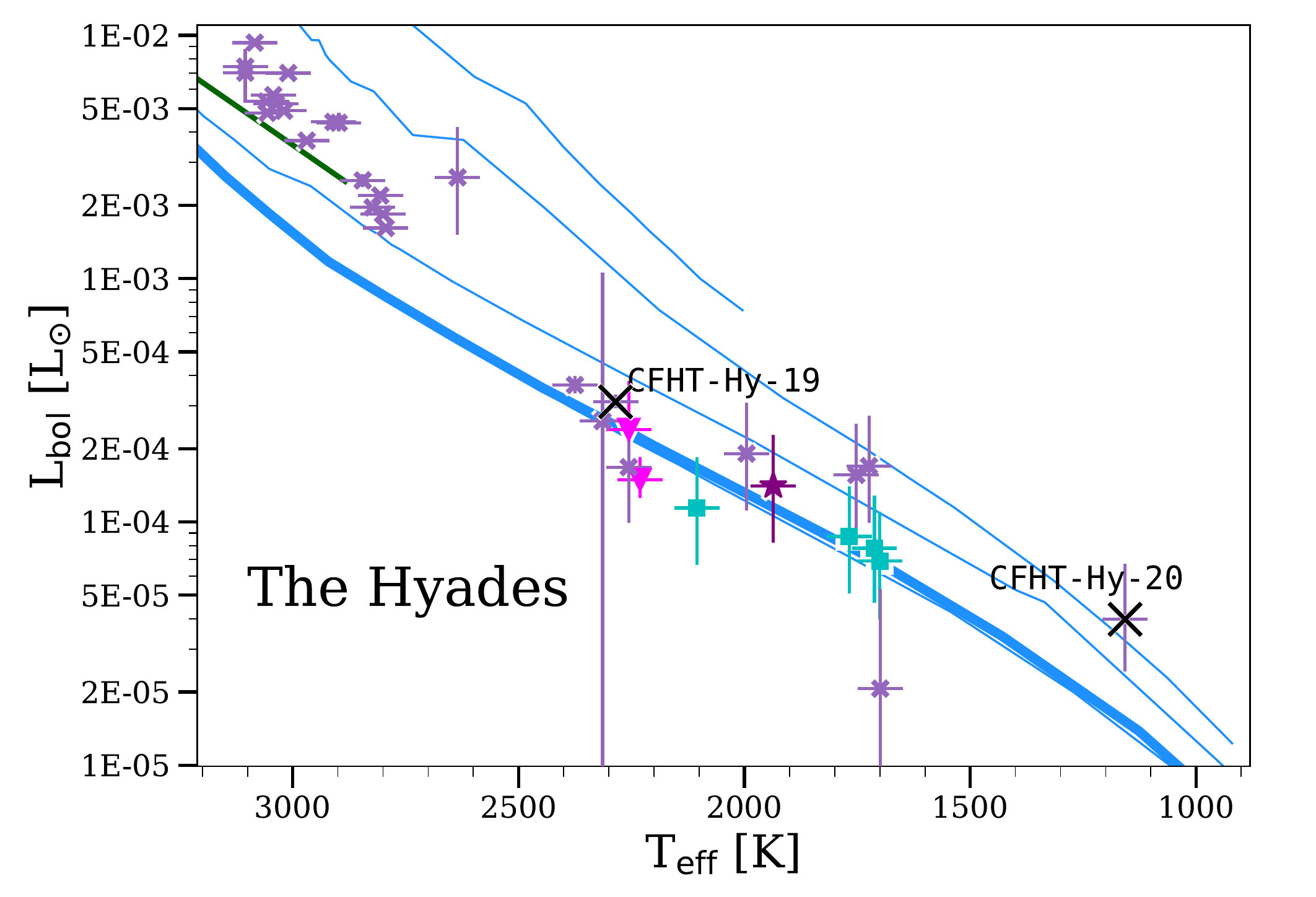}} \quad
     \subfloat[\label{fig:hrd_thehyades_zoom_ldb_v1}] 
      {\includegraphics[width=0.5\textwidth,scale=0.50]{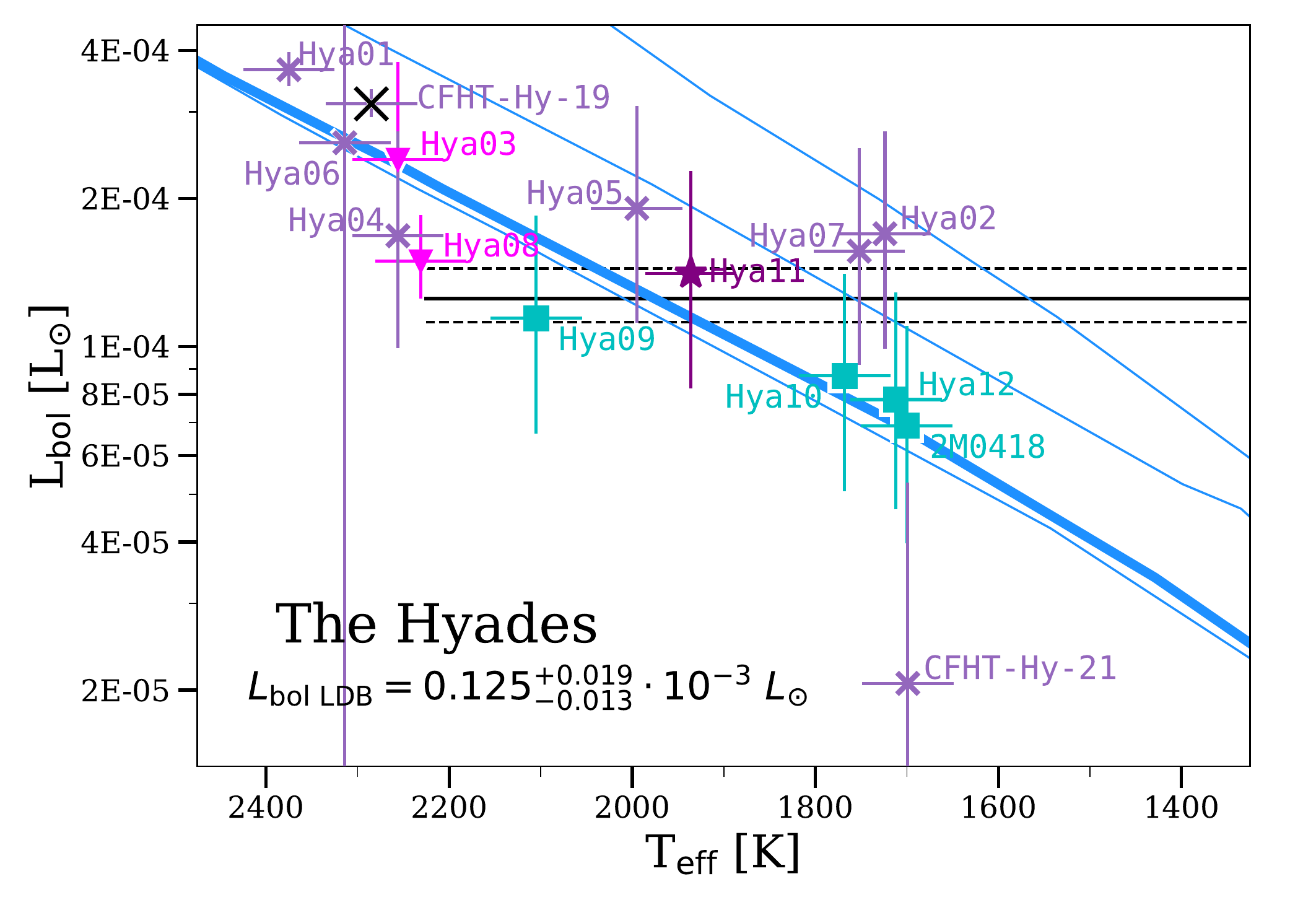}} \quad
     \subfloat[\label{fig:hrd_thehyades_zoom_ldb_v2}]  
      {\includegraphics[width=0.5\textwidth,scale=0.50]{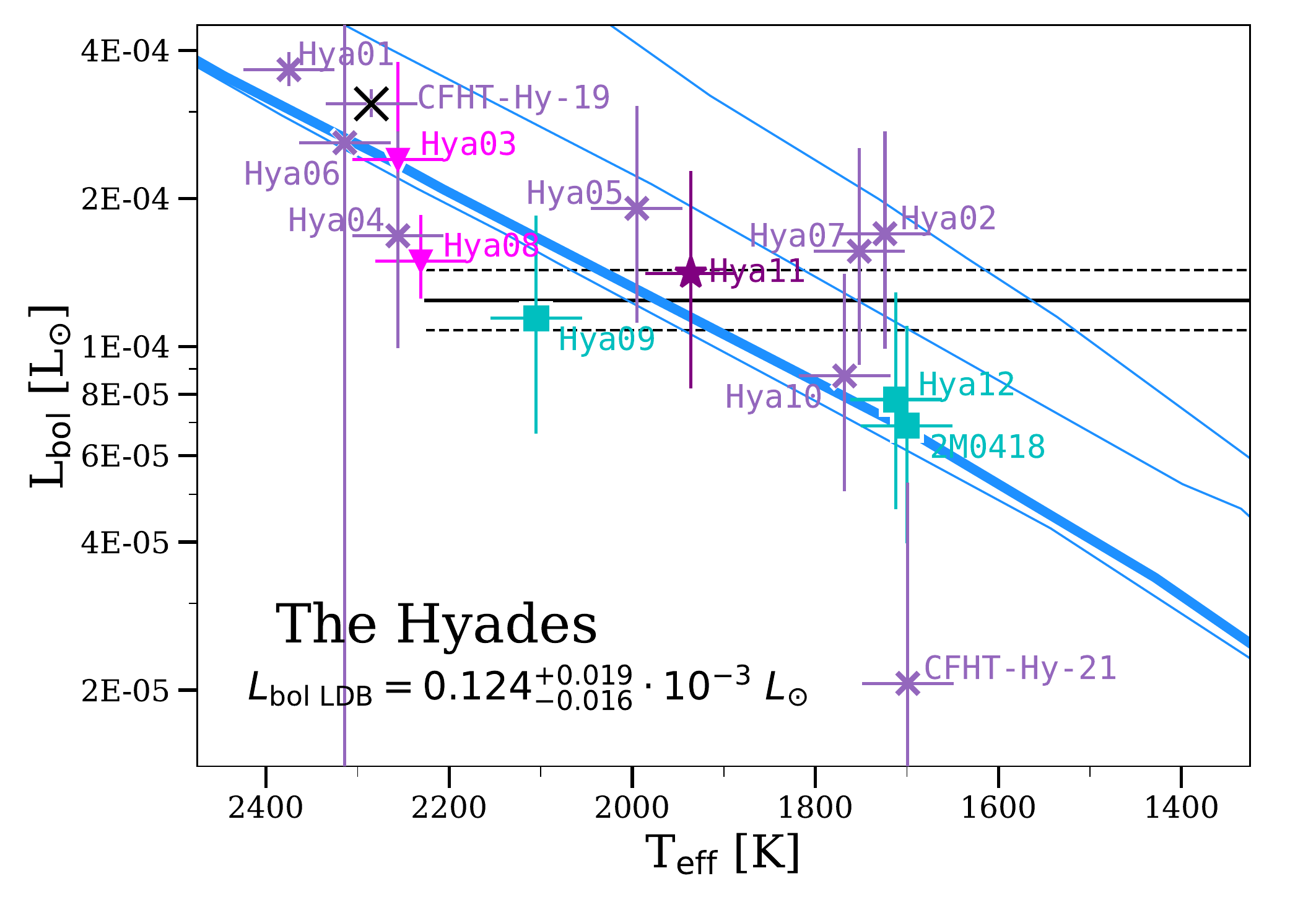}} \\
    \caption[HRDs and the LDB for the Hyades.]
          {HRDs and the LDB for the Hyades.        
           {\bf (a) }
             The same as in Fig. \ref{fig:hrd_alphapersei_discussion} but for the Hyades. 
            Thin blue lines correspond to isochrones of 1, 10, 100~Ma, and 1~Ga 
            from the BT-Settl models \citep{f_allard2013}, while the thick blue one to 700~Ma.  
           {\bf (b) }Zoom on the previous plot close to the LDB. 
           {\bf (c) }Same as the previous plot but, we considered that Hya10 has an unclear lithium detection.
            The Hyades is $695^{+85}_{-67}$~Ma old using the BT-Settl bolometric luminosity-age relationship \citep{f_allard2012}.
          }
  \label{fig:hrds_thehyades}
\end{figure}

\clearpage   
\begin{figure}
     \subfloat[\label{fig:hrd_bpmg_general}]
      {\includegraphics[width=0.5\textwidth,scale=0.50]{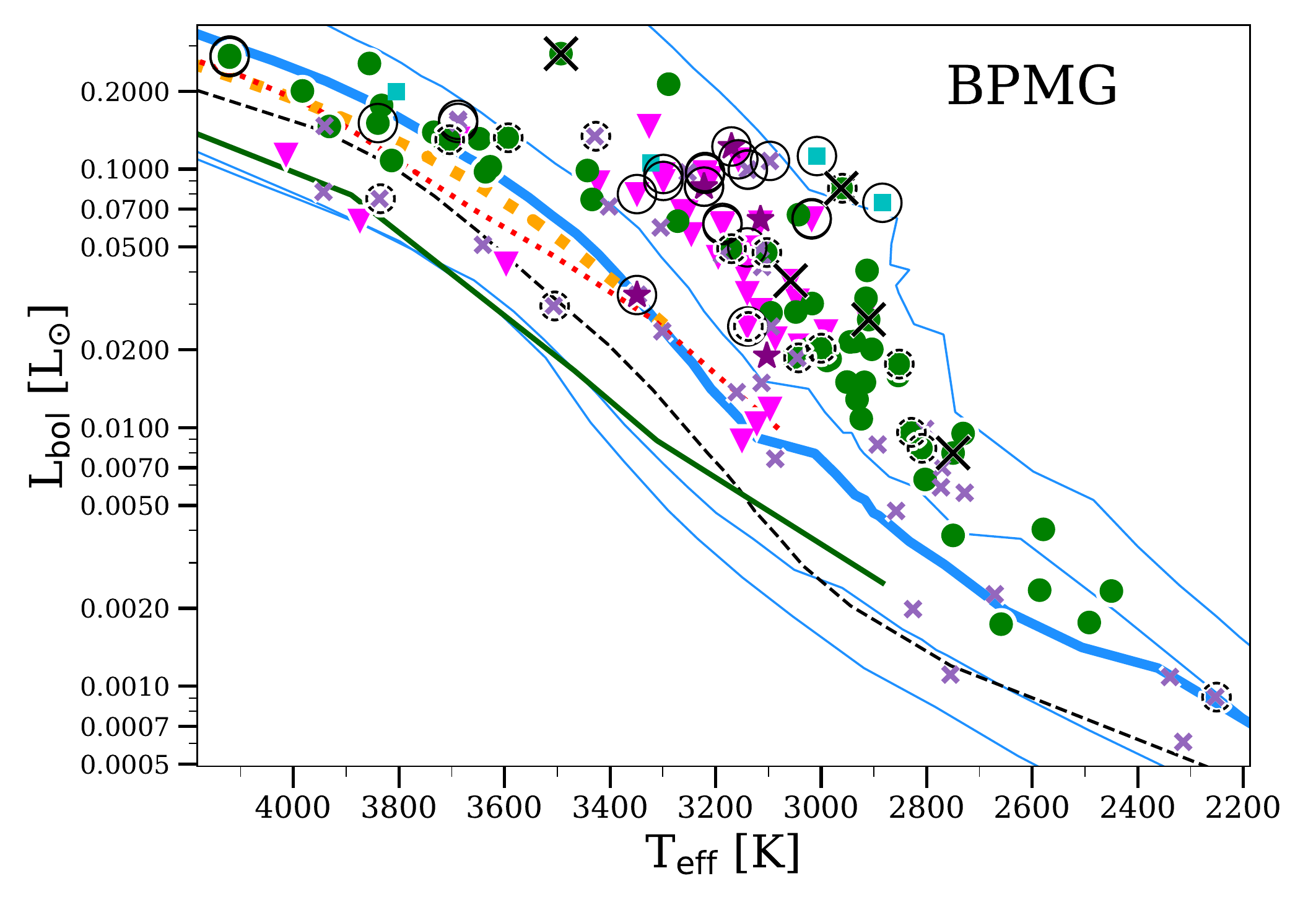}} \quad      
     \subfloat[\label{fig:hrd_bpmg_zoom}]
      {\includegraphics[width=0.5\textwidth,scale=0.50]{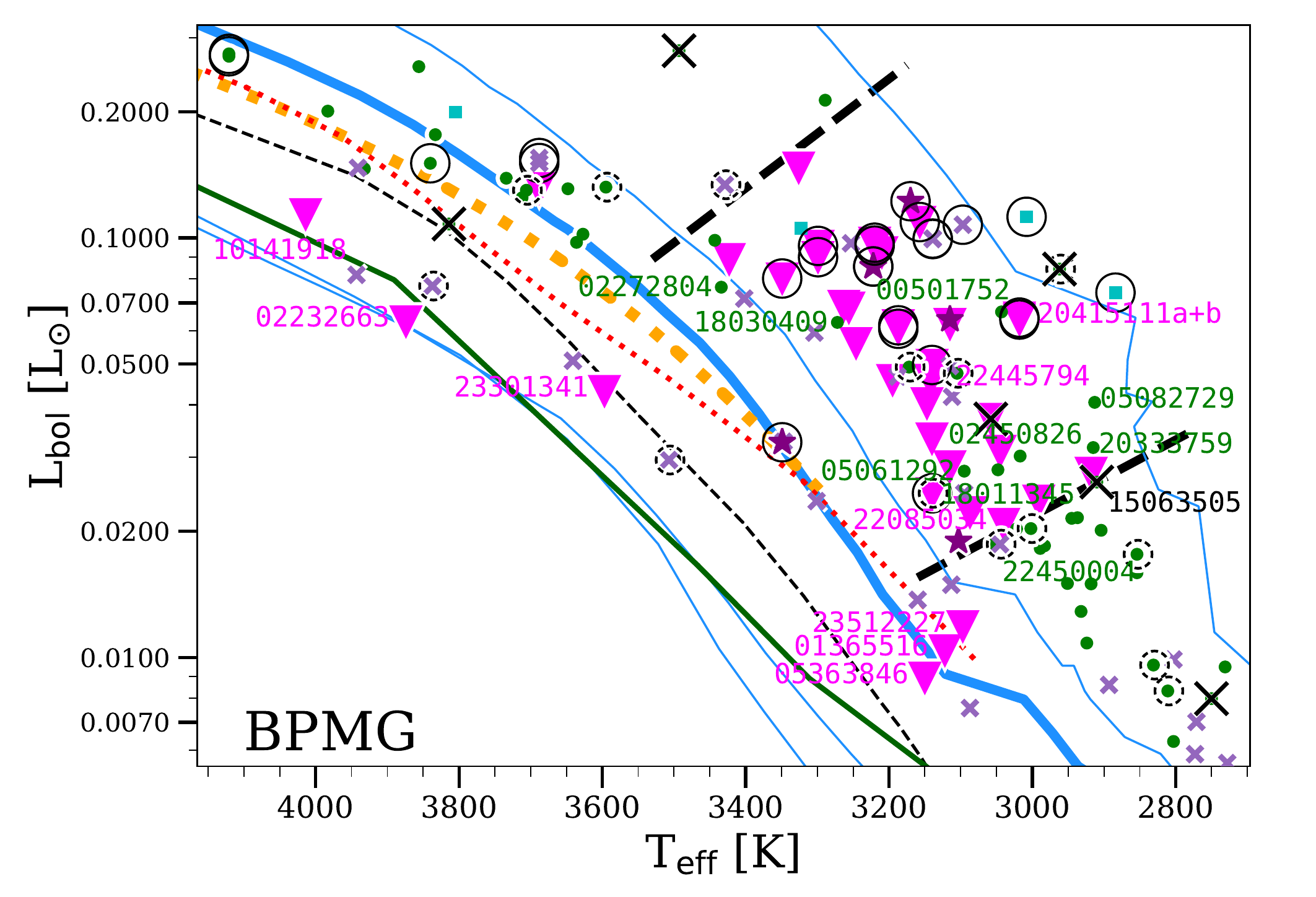}} \\
      \subfloat[\label{fig:hrd_bpmg_zoom_ldb_v5}]
      {\includegraphics[width=0.5\textwidth,scale=0.50]{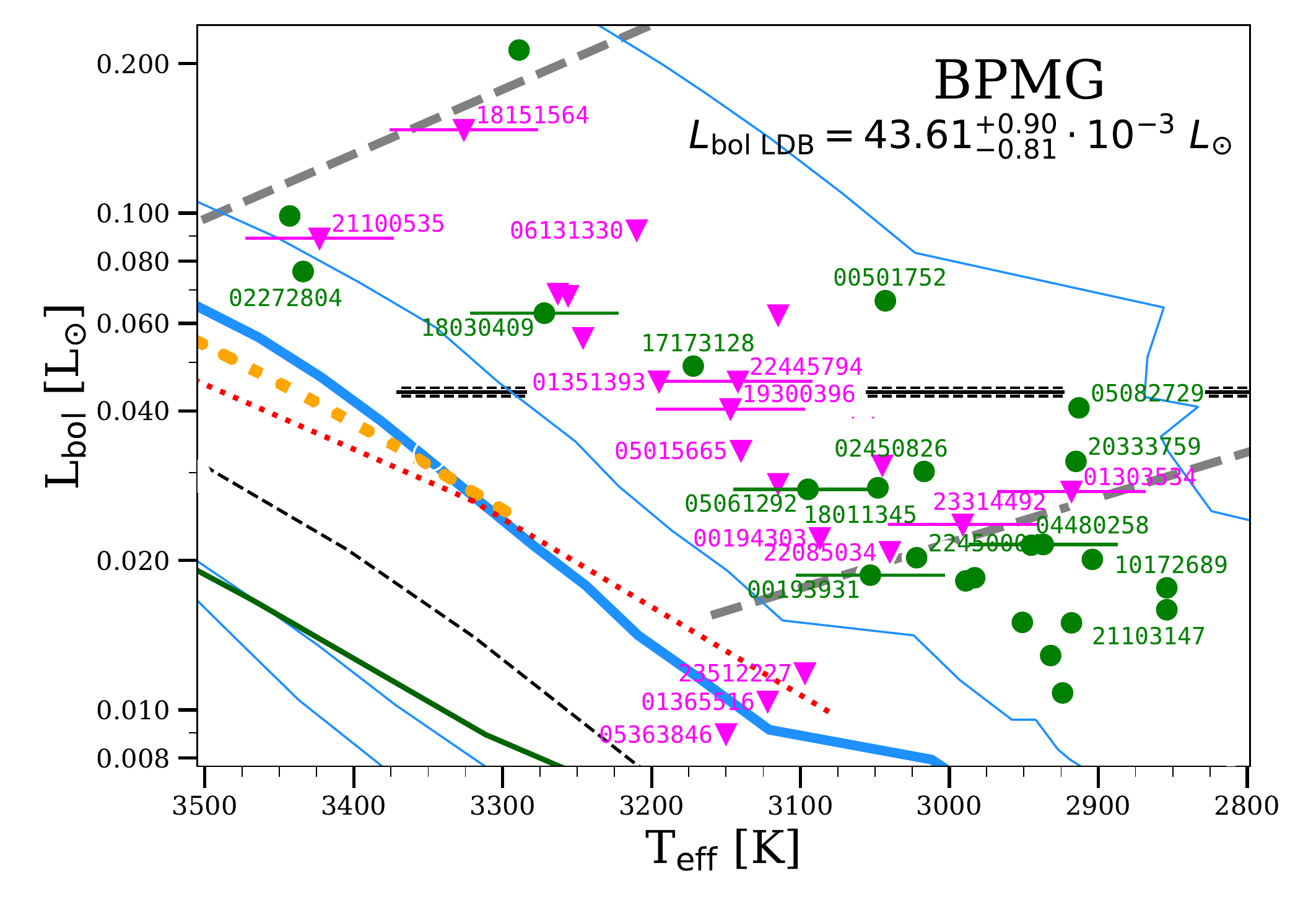}} \quad      
     \subfloat[\label{fig:hrd_bpmg_zoom_ldb_v3}]
      {\includegraphics[width=0.5\textwidth,scale=0.50]{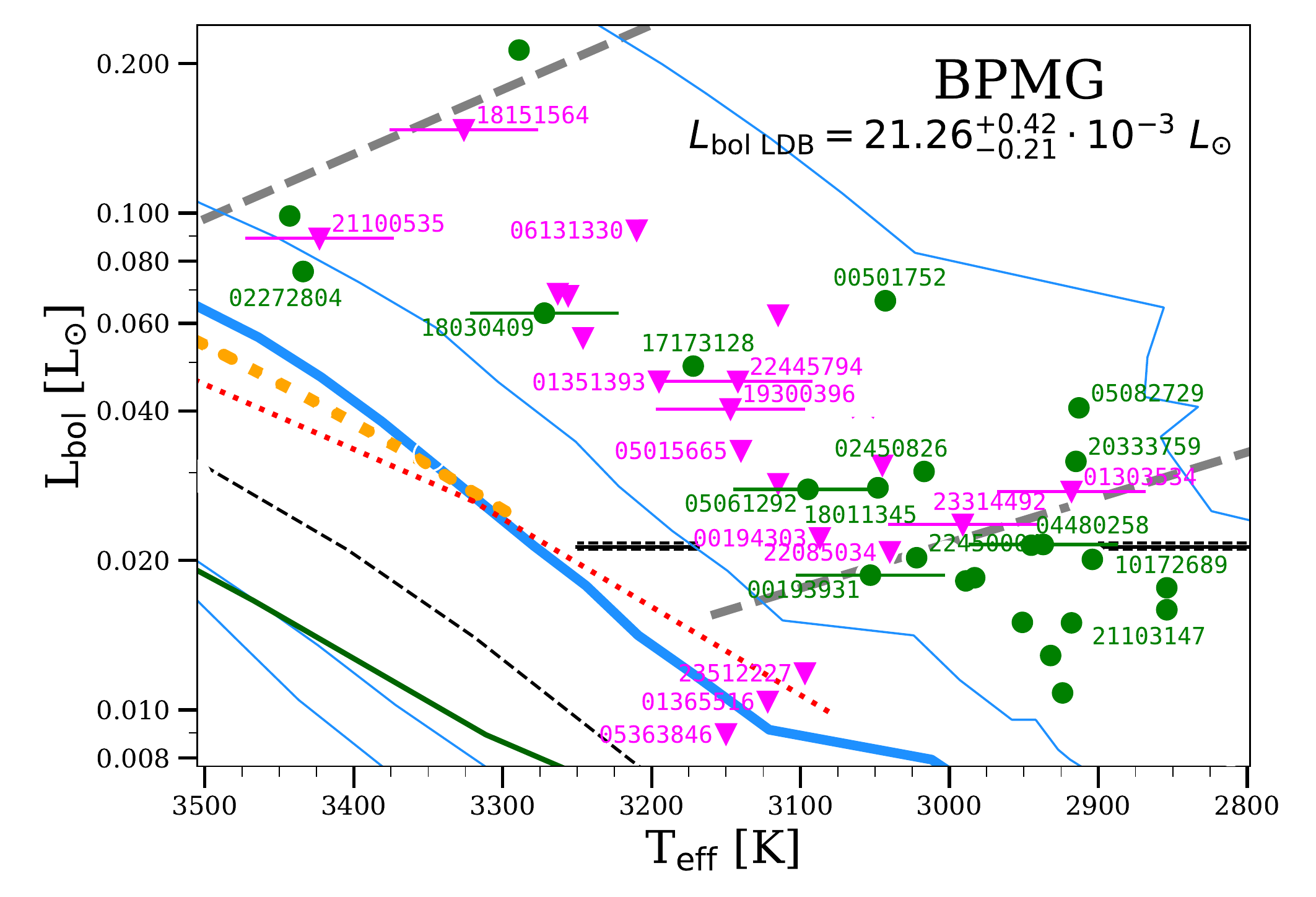}} \\      
          \caption[HRDs and the LDB for the BPMG.]
                  {HRDs and the LDB for the BPMG.                                                      
                  {\bf (a) }
                    The same as in Fig. \ref{fig:hrd_alphapersei_discussion} but for the BPMG. 
                    Empty large black broken line circles are suspected multiple systems or 
                   sources with associated background objects and photometric data blended in some bands.   
                    Thin blue lines correspond to isochrones of 1, 10, 100~Ma, and 1~Ga 
                   from the BT-Settl models \citep{f_allard2013}, while the thick blue one to 20~Ma. 
                    Uncertainties are not shown in order to gain clarity, 
                   although in the bolometric luminosities are usually smaller than the symbols, 
                   with the exception of some sources without parallaxes.
                    The figure includes \mbox{20~Ma} isochrones from:  
                   \citealt{f_dantona1994} (black dashed line),
                   \citealt{siess2000} (red dotted line), and 
                   a 17~Ma from \citealt{e_tognelli2011} (orange dashed dot line).  
                  {\bf (b) }Zoom on the (a) plot around the LDB.  
                   The size of lithium-poor sources has been increased to locate the LDB. 
                   The area delimited by them is marked with gray thick dashed lines.  
                   Some remarkable sources are labelled.                              
                  {\bf (c) }Same as the previous plot with the LDB locus determined following the first scenario. 
                   Since we focus on the sole aim of locating the LDB, we do not show:  
                   sources that are confirmed or suspected multiple systems;
                   sources with two or several associated objects (third configuration) with photometric data blended in some bands; 
                   non-members; 
                   sources with undetected lithium feature; and objects without parallaxes.
                   Effective temperatures uncertainties are shown in some lithium poor objects in order to gain clarity,
                   Bolometric luminosities uncertainties are smaller than the size of the symbols. 
                   The BPMG is $24.3^{+0.3}_{-0.3}$~Ma old using the BT-Settl bolometric luminosity-age relationship \citep{f_allard2012}.
                  {\bf (d) }Same as the previous plot with the LDB locus determined following the second scenario.                                  
          }
  \label{fig:hrds_bpmg}          
\end{figure}

\clearpage   
\begin{figure}
     \subfloat[\label{fig:hrd_thmg_general}]
      {\includegraphics[width=0.5\textwidth,scale=0.50]{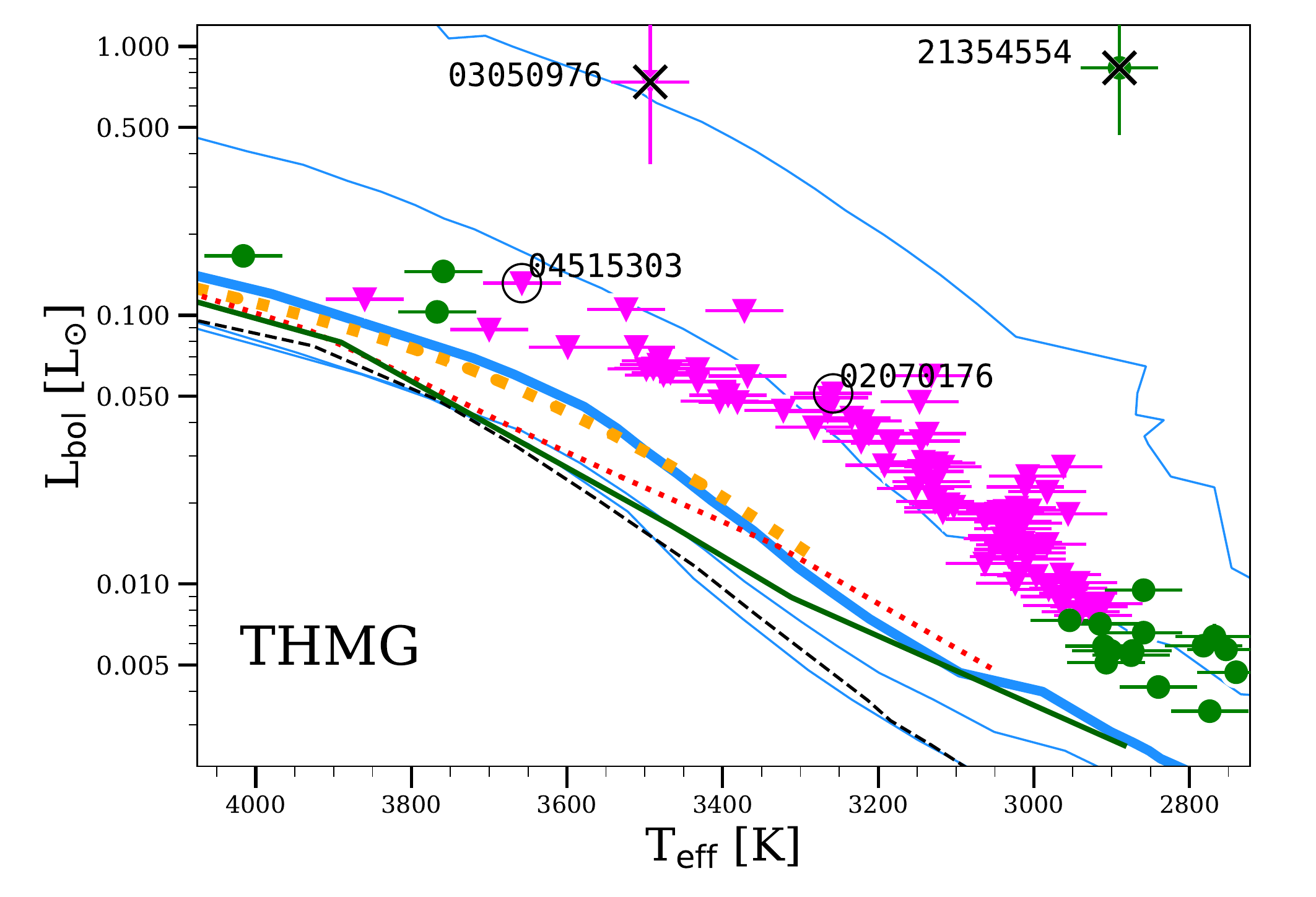}} \quad      
     \subfloat[\label{fig:hrd_thmg_zoom_ldb_v1}]
      {\includegraphics[width=0.5\textwidth,scale=0.50]{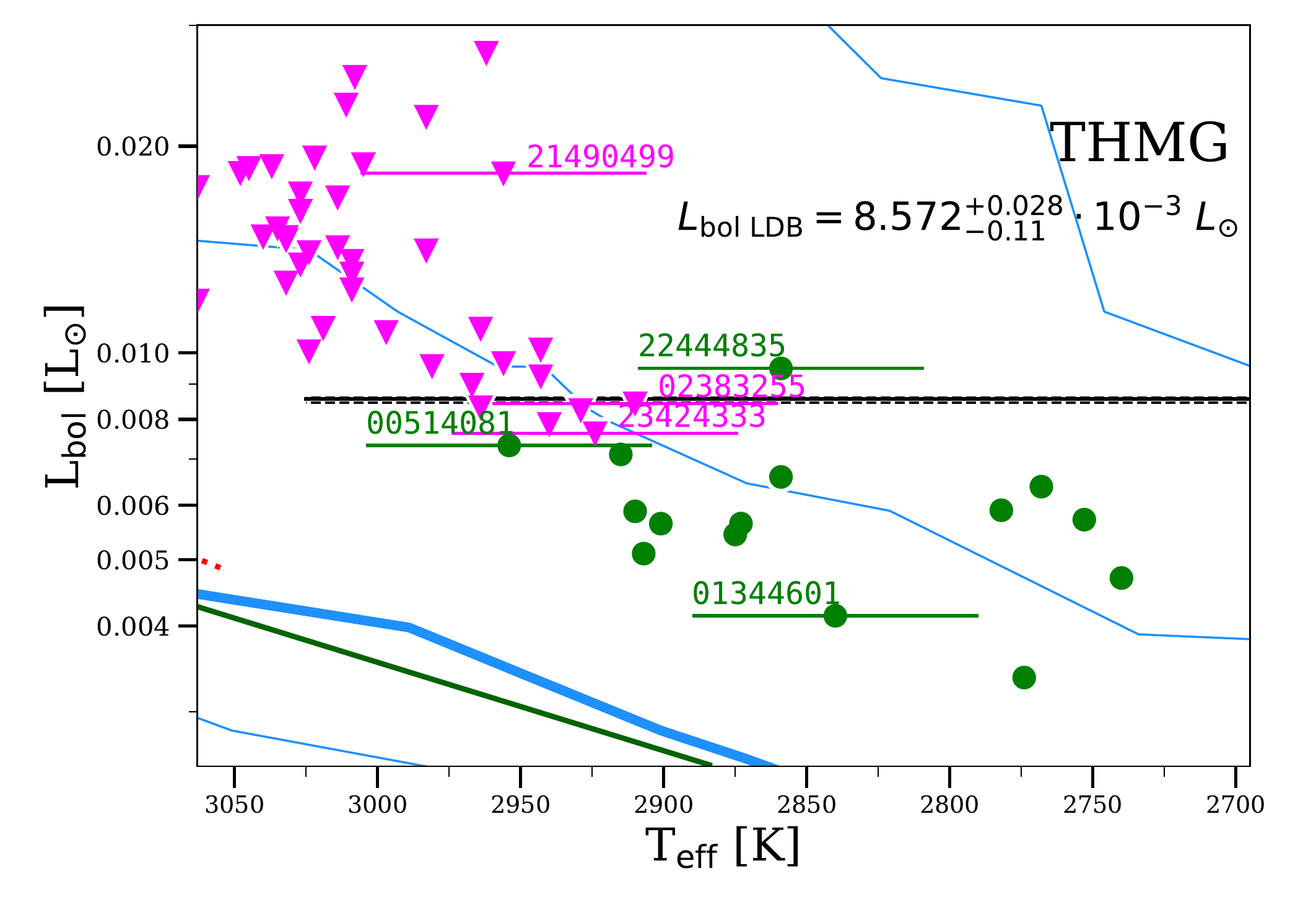}} \\
     \subfloat[\label{fig:hrd_thmg_zoom_ldb_v2}]
      {\includegraphics[width=0.5\textwidth,scale=0.50]{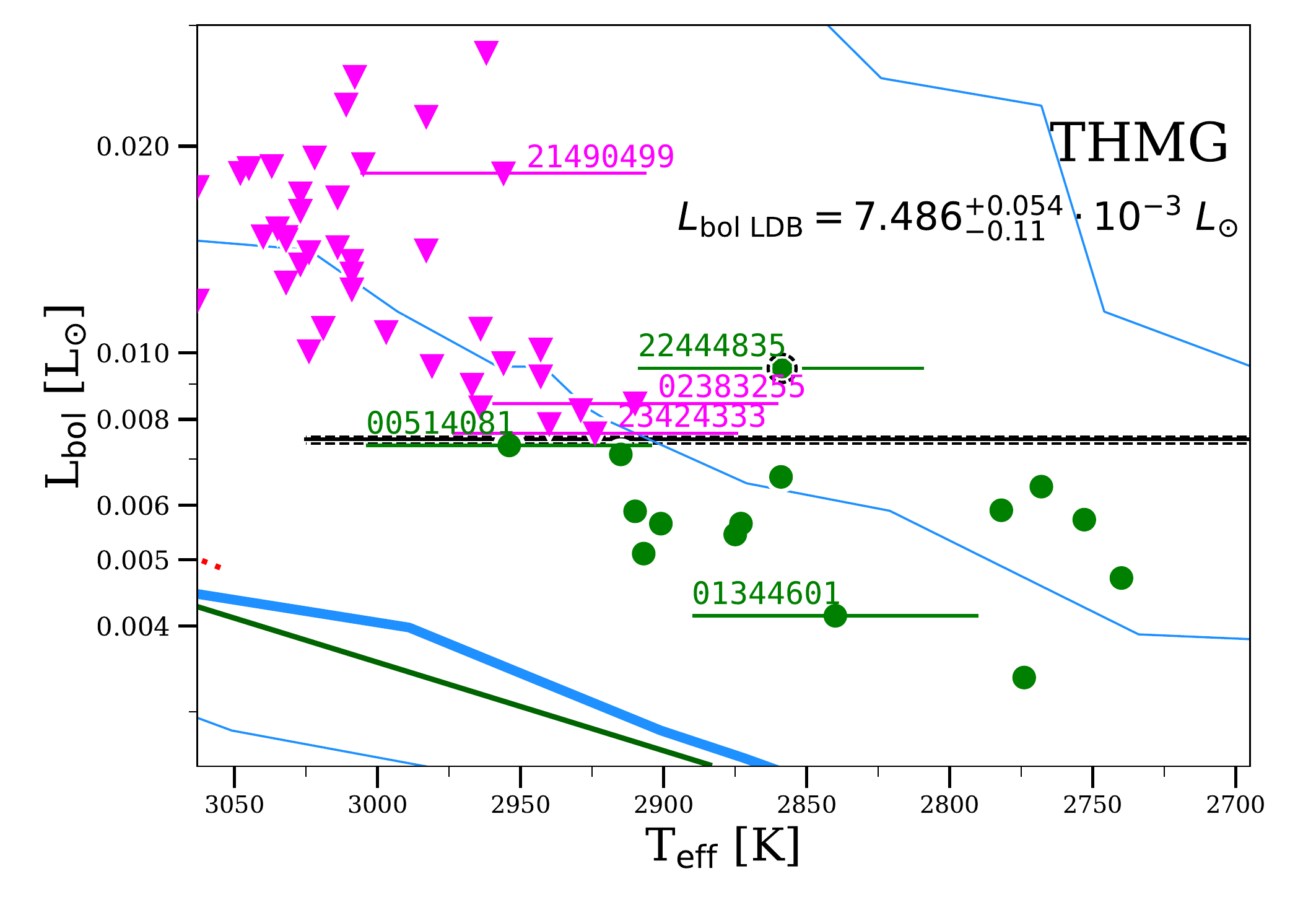}} \quad     
          \caption[HRDs and the LDB for the THMG.]
                  {HRDs and the LDB for the THMG.
                  {\bf (a) }
                    Same as in Fig. \ref{fig:hrd_alphapersei_discussion} but for THMG. 
                   Thin blue lines correspond to isochrones of 1, 10, 100~Ma, and 1~Ga 
                   from the BT-Settl models \citep{f_allard2013}, while the thick blue one to 50~Ma. 
                    The figure includes \mbox{50~Ma} isochrones from:  
                   \citealt{f_dantona1994} (black dashed line), and 
                   \citealt{siess2000} (red dotted line). 
                    It also includes a 40~Ma isochrone from 
                   \citealt{e_tognelli2011} (orange dashed dot line). 
                  {\bf (b) }Zoom on the (a) plot around the LDB following the first scenario.                   
                   To retain clarity,  we have only drawn 
                   the effective temperature uncertainties and the names for the sources that defined the LDB
                   and two additional sources.  
                   The uncertainties in the bolometric luminosities are usually smaller than the size of the symbols. 
                  {\bf (c) }Same as (b), but the LDB is located following the second scenario (Section \ref{sub:thmg}). 
                   The THMG is $51.0^{+0.5}_{-0.2}$~Ma old using the BT-Settl bolometric luminosity-age relationship \citep{f_allard2012}.
          }
  \label{fig:hrds_thmg}          
\end{figure}

\clearpage
\begin{figure}
   \includegraphics[width=0.45\textwidth,scale=0.50]{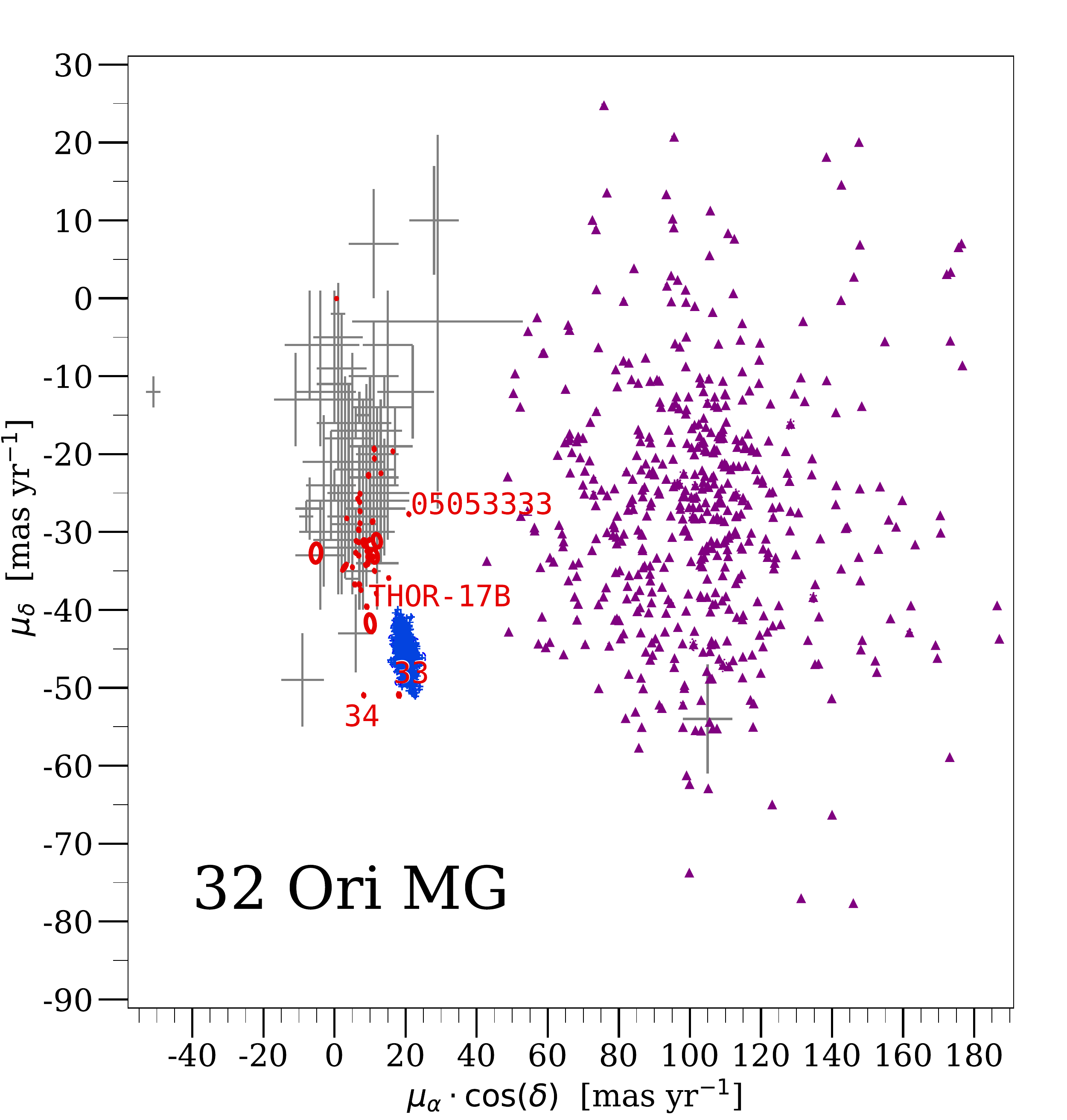}
   \includegraphics[width=0.45\textwidth,scale=0.50]{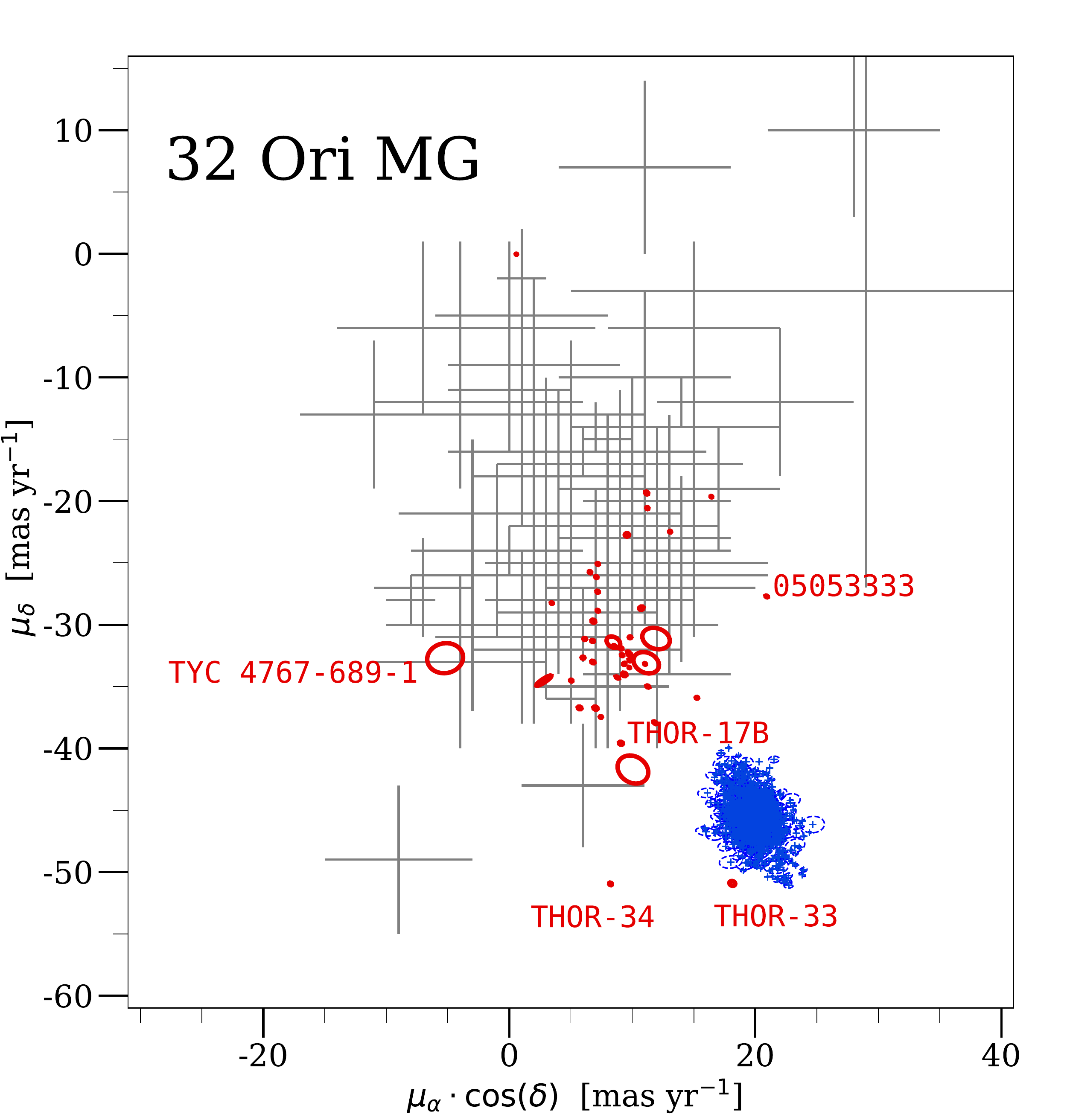}
      \caption[VPDs for the 32 Ori MG.]
              {VPDs for the 32 Ori MG.
               {\bf Left: } 
                 The red ellipses are all 32 Ori MG members \citep{cpmbell2017};  
                the black filled triangles are Hyades members and the blue squares are Pleiades members, 
                (both from \citealt{c_babusiaux2018}); 
                and the light grey plus symbols are the Taurus-Auriga members taken from \cite{c_ducourant2005}.
                The size of the symbols is greater than the size of the uncertainties (shown as ellipses). 
               {\bf Right: }Zoom on the previous plot.               
               }
         \label{fig:vpds_32orimg}
   \end{figure}

\begin{figure}
   \includegraphics[width=0.45\textwidth,scale=0.50]{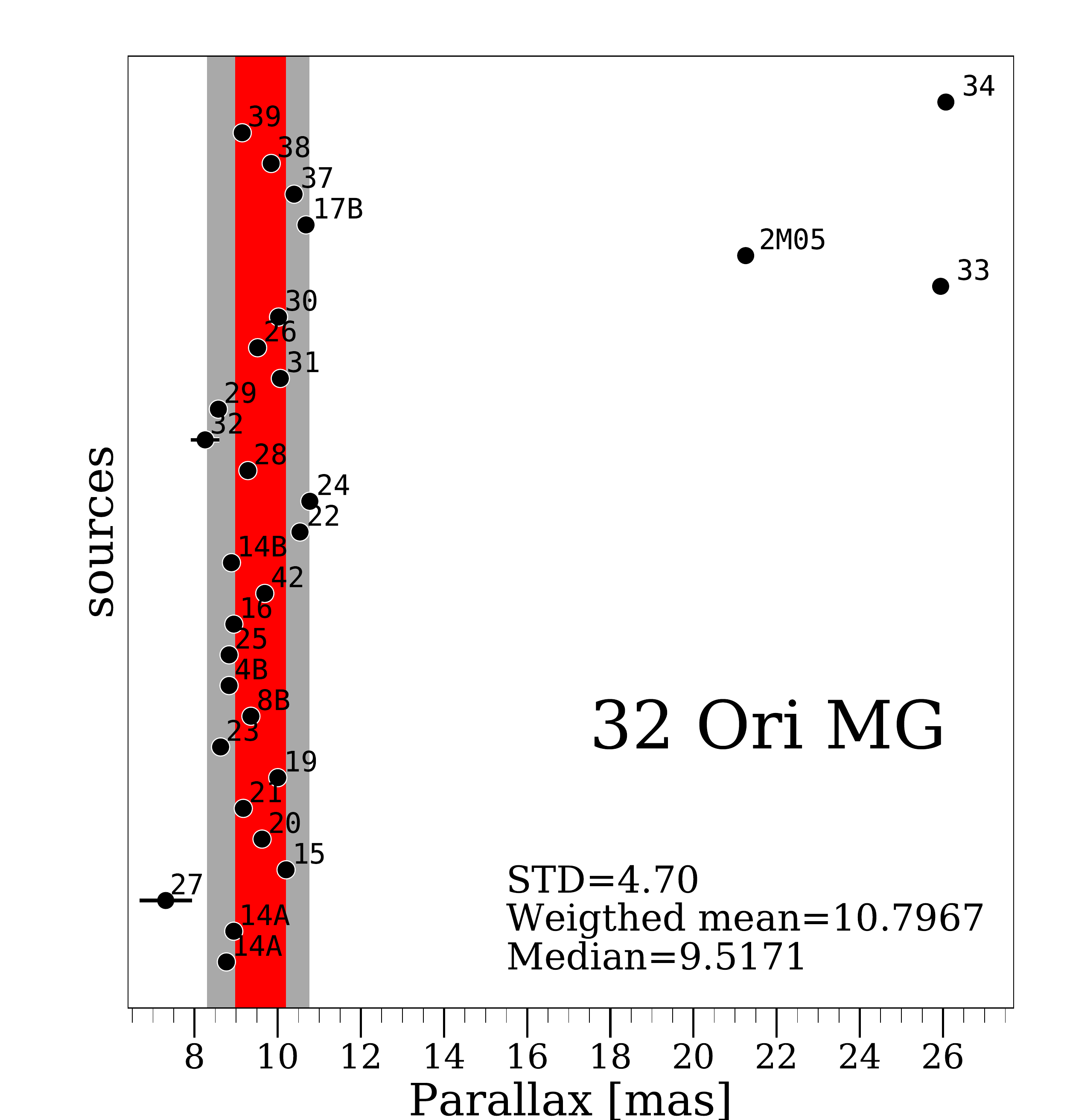}
      \caption[Parallaxes for 32 Ori MG members close to the LDB.]
              {Parallaxes for the 32 Ori MG members close to the LDB.
               Same as in Fig. \ref{fig:plxs_alphapersei} but for the 32 Ori MG. 
               We removed the `THOR-' prefix of each object from the name. 
              }
         \label{fig:plxs_32orimg}
\end{figure}

\begin{figure}
     \subfloat[\label{fig:hrd_32orimg_general}]
      {\includegraphics[width=0.5\textwidth,scale=0.50]{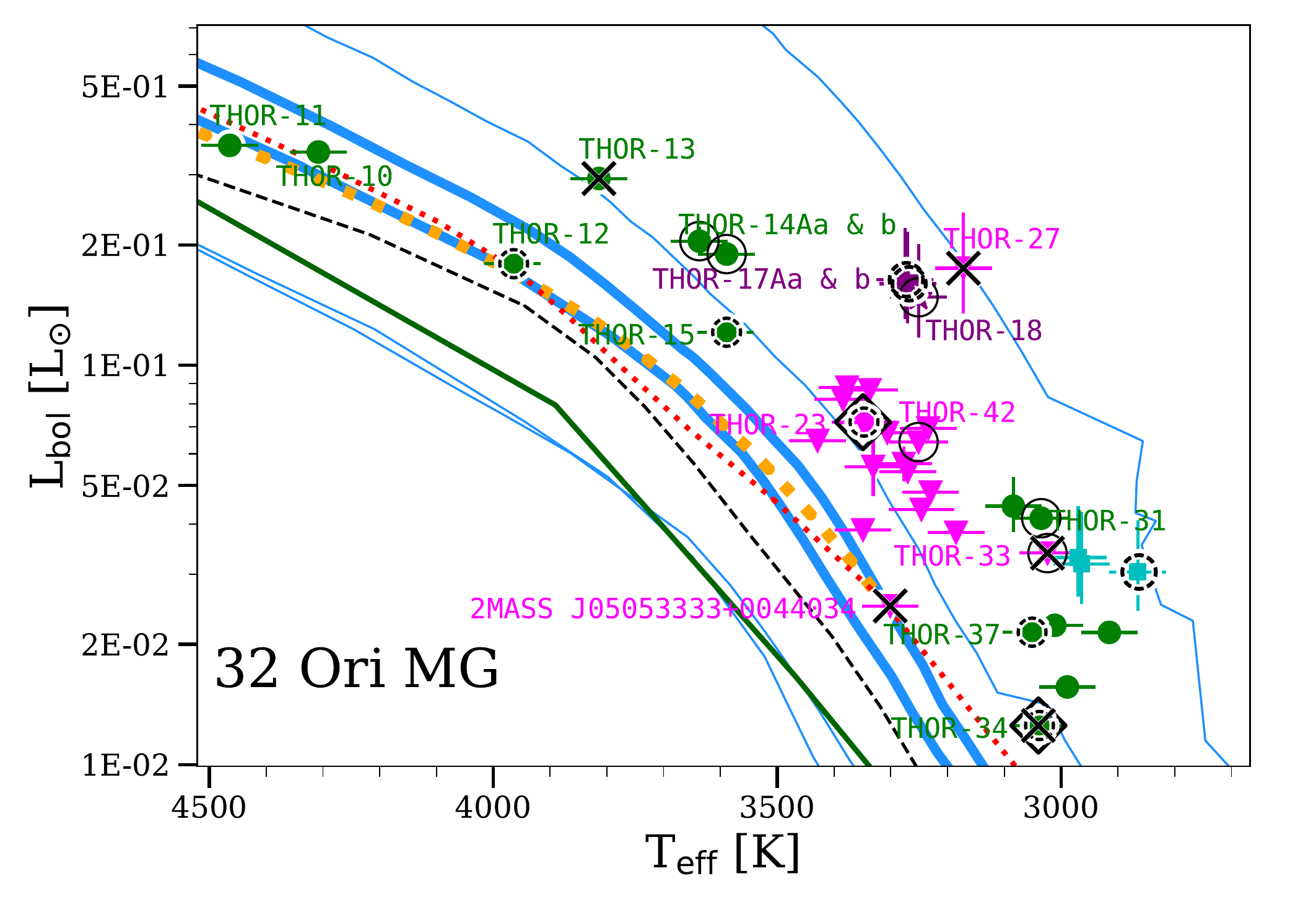}} \quad      
     \subfloat[\label{fig:hrd_32orimg_zoom_ldb_v1}]
      {\includegraphics[width=0.5\textwidth,scale=0.50]{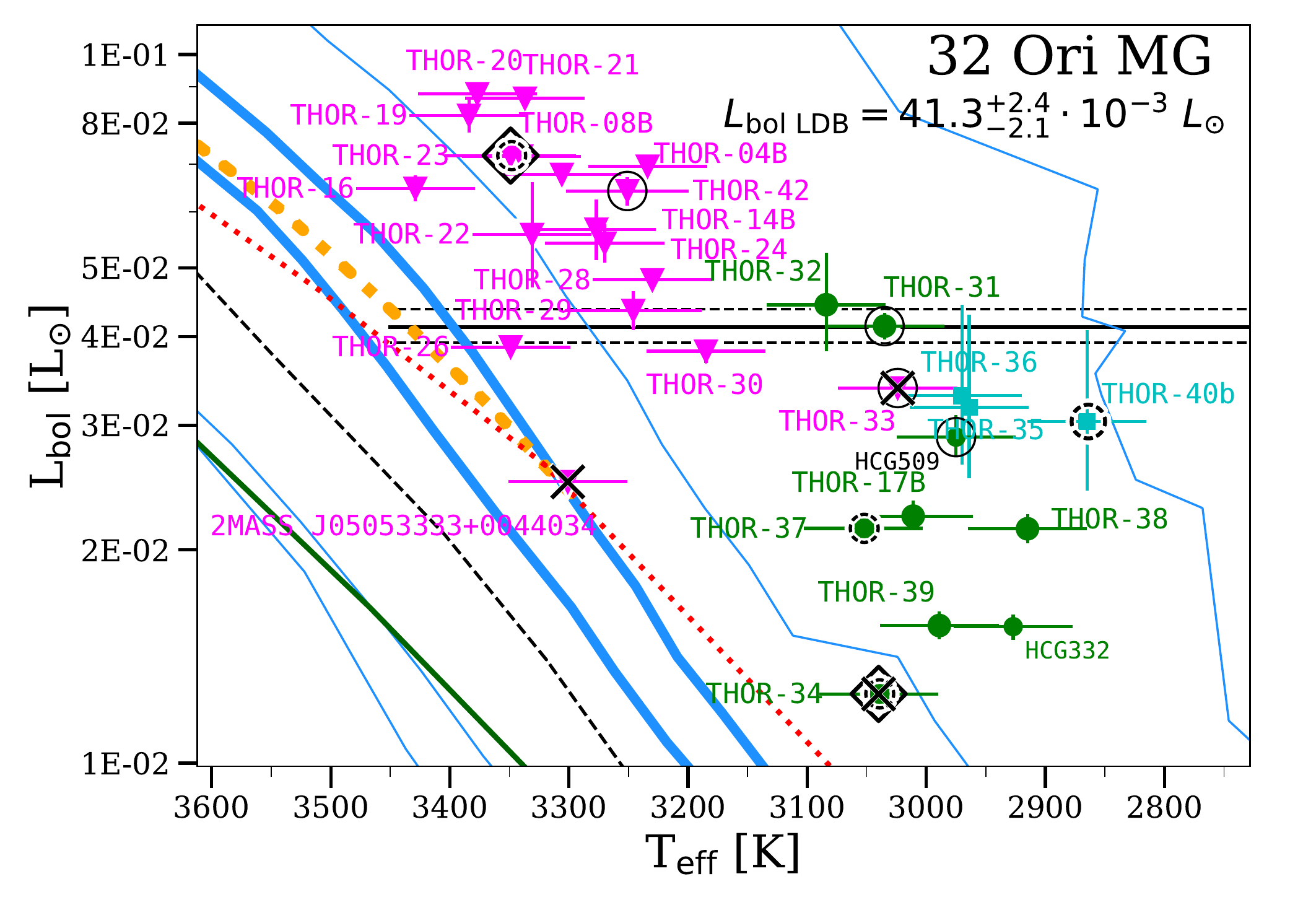}} \\     
     \subfloat[\label{fig:hrd_32orimg_zoom_ldb_pop1}]
      {\includegraphics[width=0.5\textwidth,scale=0.50]{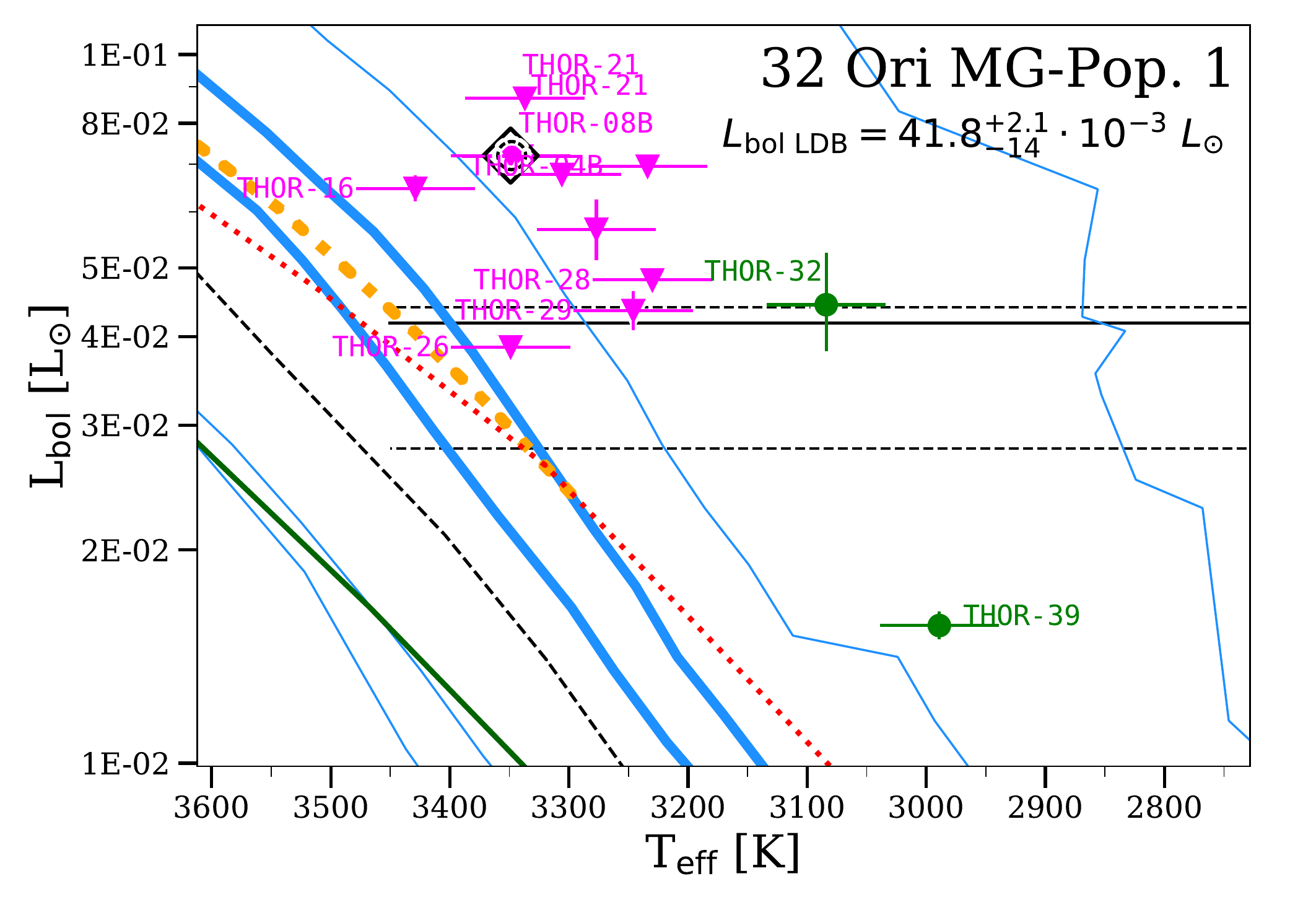}} \quad
     \subfloat[\label{fig:hrd_32orimg_zoom_ldb_pop2}]
      {\includegraphics[width=0.5\textwidth,scale=0.50]{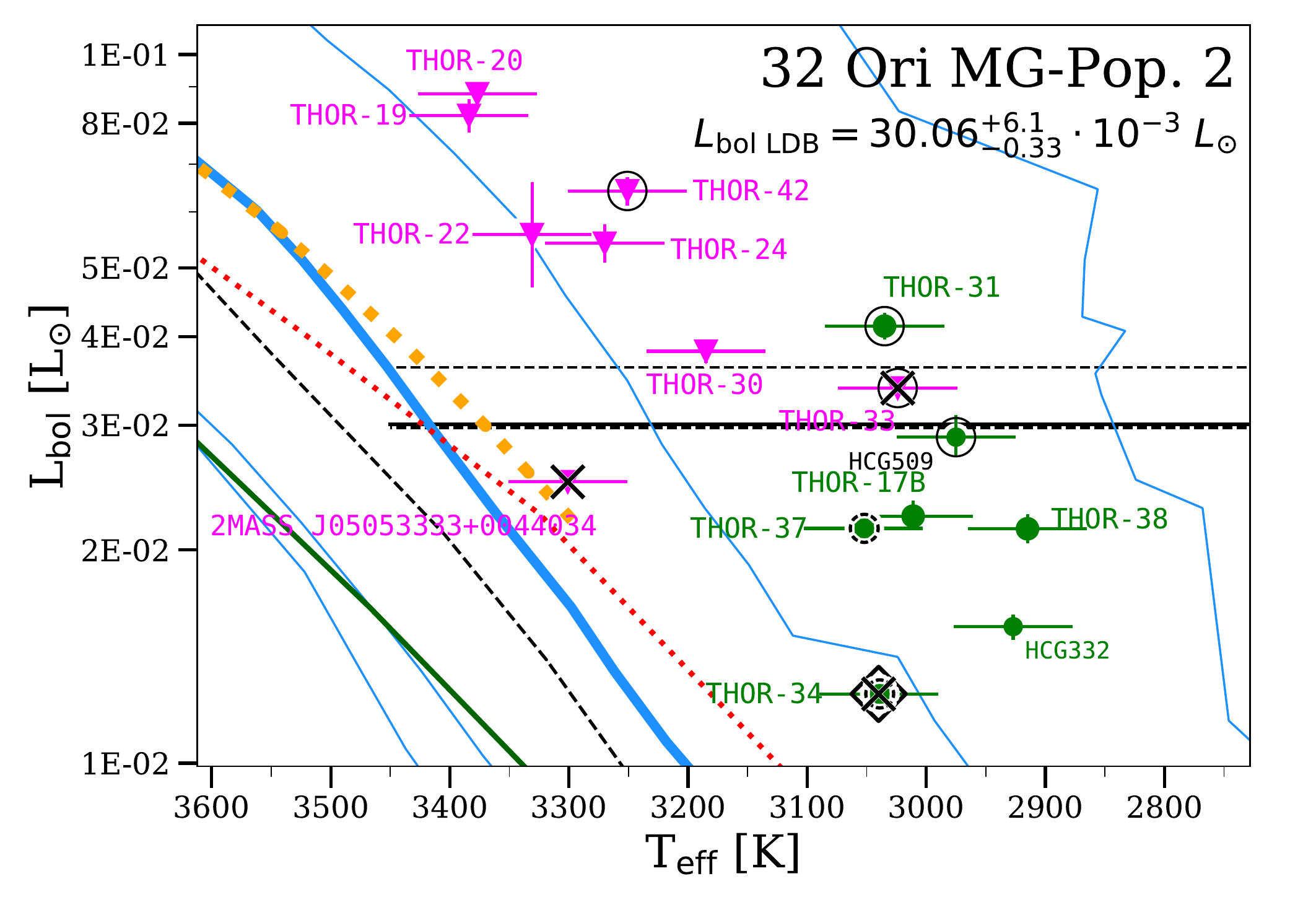}} \\     
          \caption[HRDs and the LDB for the 32 Ori MG.]
                  {HRDs and the LDB for the 32 Ori MG.
                  {\bf (a) }
                    The same as in Fig. \ref{fig:hrd_alphapersei_discussion} but for 32 Ori MG.
                   Empty black diamonds correspond to fast rotators. 
                    Thin blue lines correspond to isochrones of 1, 10, 100~Ma, and 1~Ga 
                   from the BT-Settl models \citep{f_allard2013}, while the thick blue lines to a 20 and 30~Ma.                   
                    The figure includes 20~Ma isochrones from  
                   \citealt{siess2000} (red dotted line), and 
                   \citealt{f_dantona1994} (black dashed line). 
                   It also includes            
                   a \mbox{18~Ma} isochrone from \citealt{e_tognelli2011} (orange dashed dot line).
                  {\bf (b) }Zoom on the previous plot around the LDB. 
                    We added the sources HCG 332 and HCG 509 (see Section \ref{sub:32orimg}). 
                    The 32 Ori MG is $25.0^{+0.7}_{-0.82}$~Ma old using the BT-Settl bolometric luminosity-age relationship \citep{f_allard2012}.                   
                  {\bf (c) }Zoom on the LDB for 32 Ori MG population 1, see Appendix \ref{app:32orimg_one_or_two}.
                    The 32 Ori MG population 1 is $24.9^{+4.5}_{-0.7}$~Ma (BT-Settl models from \citealt{f_allard2012}).
                  {\bf (d) }Zoom on the LDB for 32 Ori MG population 2, see Appendix \ref{app:32orimg_one_or_two}.
                    Thin blue lines correspond to isochrones of 1, 10, 100~Ma, and 1~Ga 
                   from the BT-Settl models \citep{f_allard2013}, while the thick blue one to 30~Ma.
                    The figure includes: 
                   a \mbox{25~Ma} isochrone from \citealt{siess2000} (red dotted line), 
                   a \mbox{20~Ma} isochrone from \citealt{f_dantona1994} (black dashed line), and  
                   a \mbox{20~Ma} isochrone from \citealt{e_tognelli2011} (orange dashed dot line).
                    The 32 Ori MG population 2 is $28.6^{+0.1}_{-2.0}$~Ma (BT-Settl models from \citealt{f_allard2012}).
          }
  \label{fig:hrds_32orimg}
\end{figure}

\clearpage
\twocolumn
\section{Variation of the LDB with the reddening}\label{app:ldb_av}

We adopted two very different reddening values for Alpha Persei: 
$A_{V}=0.279$~mag from \cite{c_babusiaux2018}, and 
$A_{V}=0.055^{+0.018}_{-0.017}$~mag from \cite{d_bossini2019} with $(m-M)_{0}=5.824$~mag, 146~pc.

\begin{figure}
   \includegraphics[width=0.45\textwidth,scale=0.50]{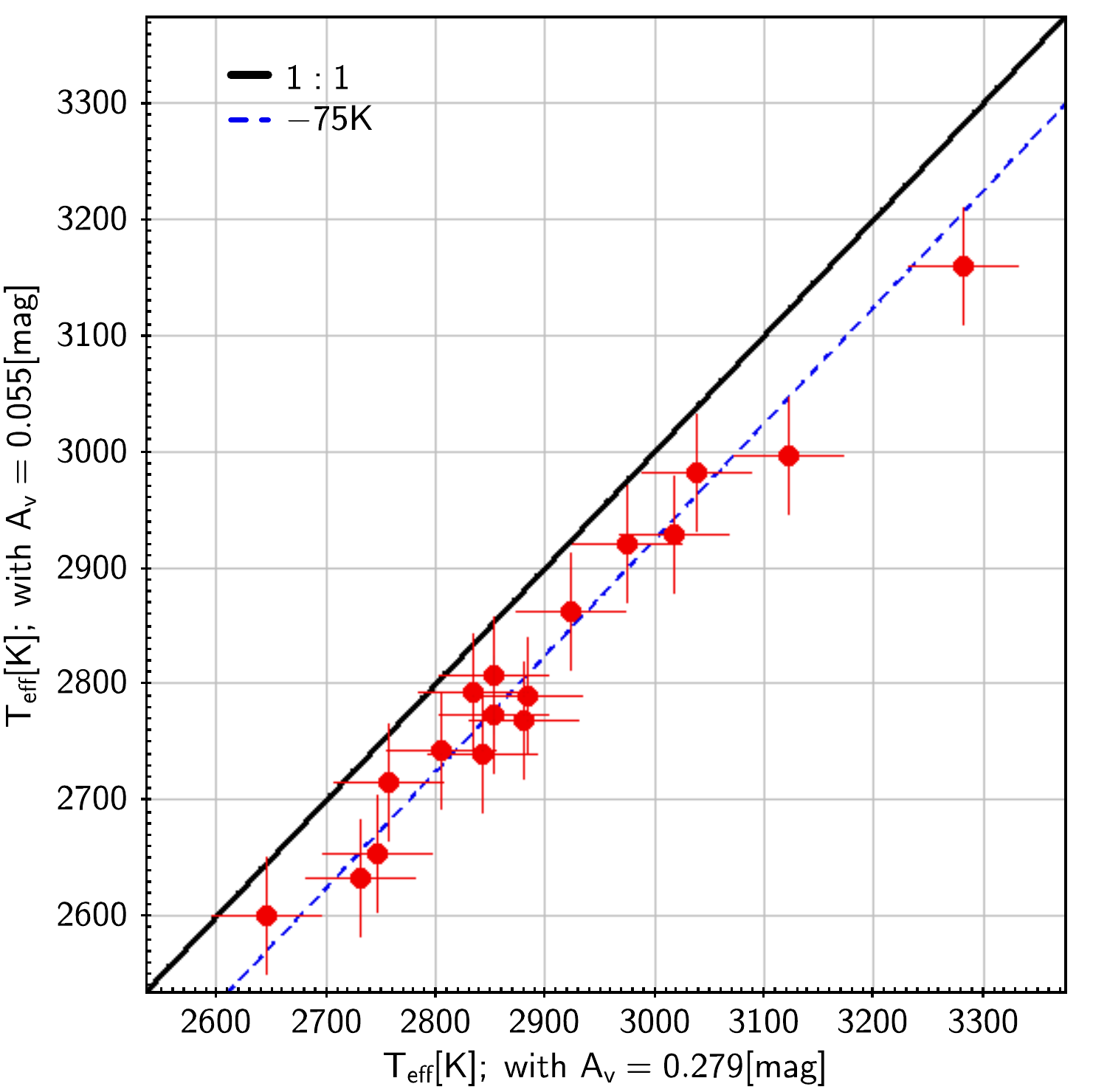}   
   \includegraphics[width=0.45\textwidth,scale=0.50]{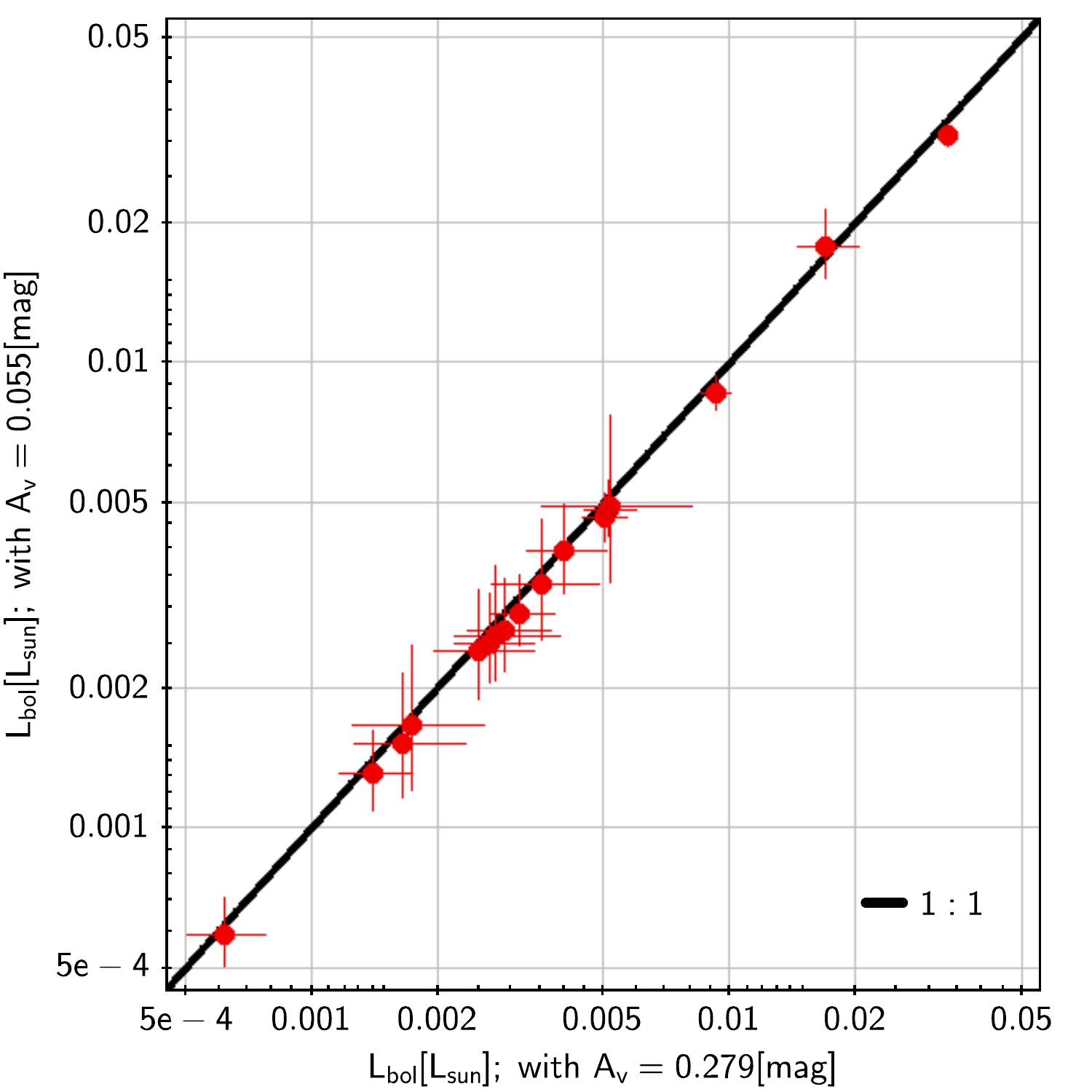}   
      \caption[Comparison between stellar parameters for Alpha Persei LDB sample.]
              {Comparison between stellar parameters for Alpha Persei LDB sample
                using two different reddenings for the objects $A_{V}=0.259$~mag and $A_{V}=0.055$~mag
               {\bf Top: }Comparative analysis between \textbf{effective temperatures }  
                 shown as a 1:1 line and a dashed one shifted 75~K. 
               {\bf Bottom: }Comparison between bolometric luminosities.
               }
         \label{fig:dependence_av_lbol_teff}
   \end{figure}

   We studied the variation of the stellar parameters, 
$L_{\mathrm{bol}}$ and $T_{\mathrm{eff}}$, with both reddening values. 
 We proceeded in the same way as in the whole article:  
we created two input files for VOSA,
one with $A_{V}=0.279$~mag and other with $A_{V}=0.055$~mag. 
We show the results in Figure \ref{fig:dependence_av_lbol_teff} and in 
Table \ref{tab:alphapersei_samples_av}. 

\begin{figure*}
      {\includegraphics[width=0.5\textwidth,scale=0.50]{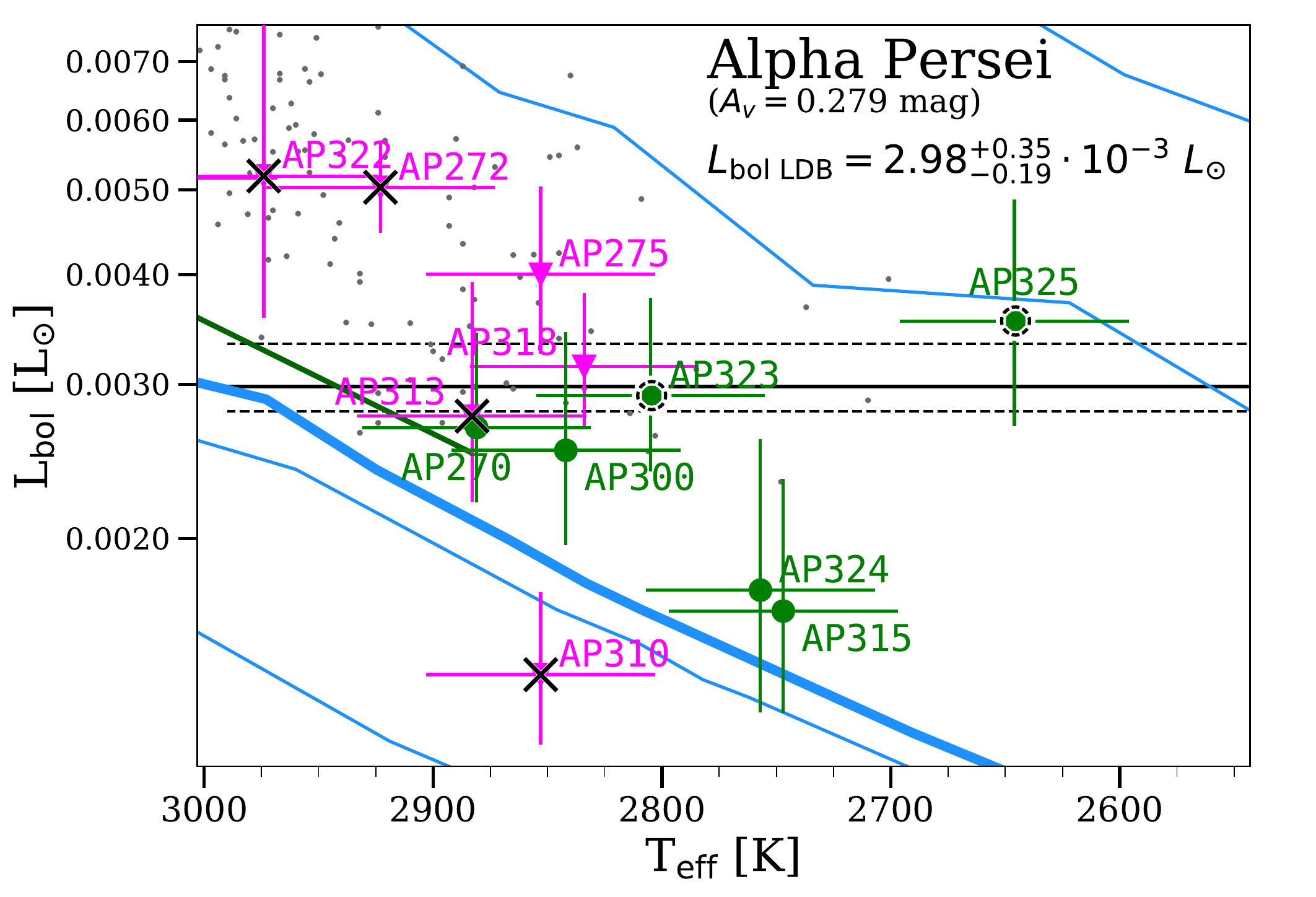}} \quad       
       {\includegraphics[width=0.5\textwidth,scale=0.50]{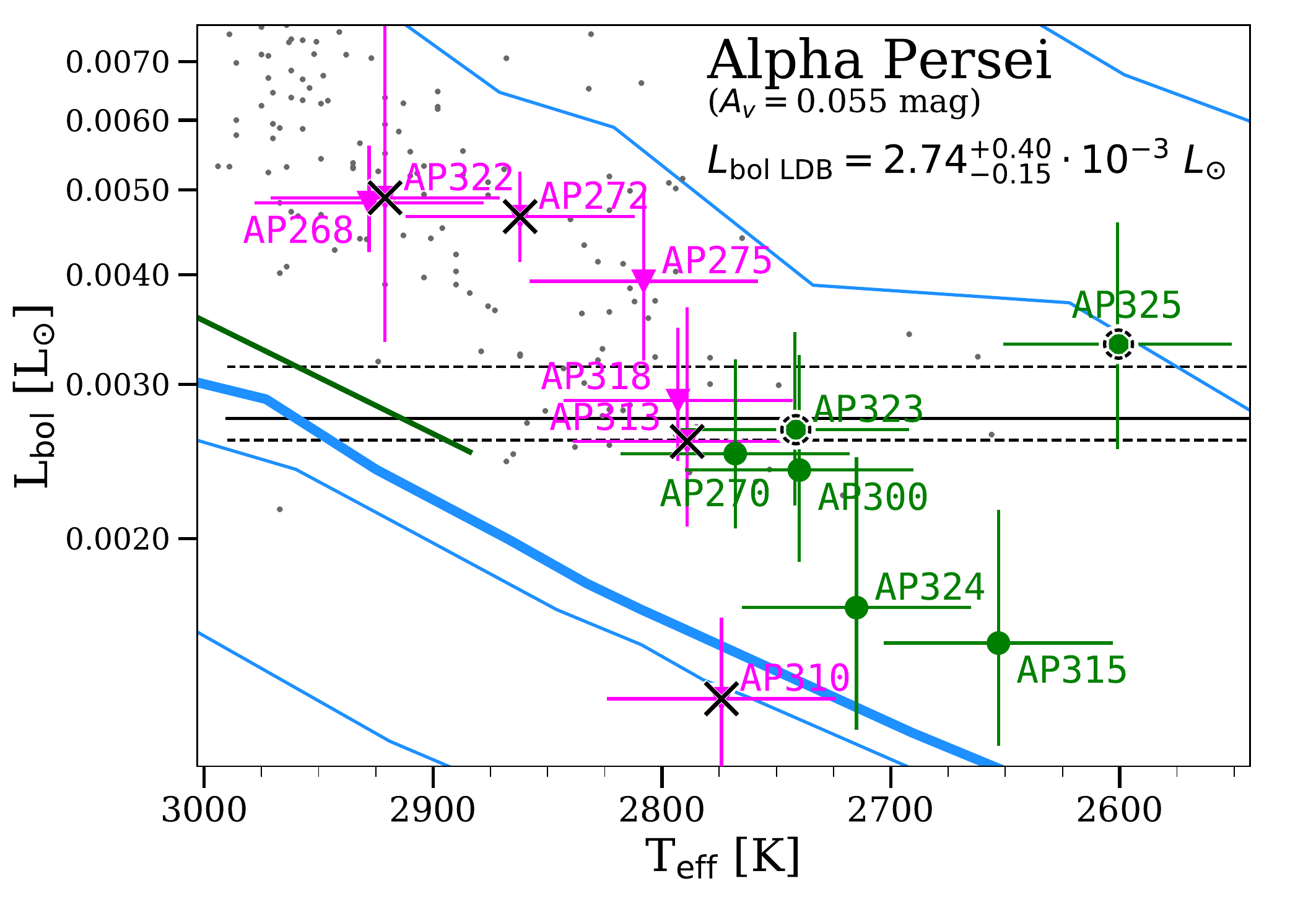}} \\
    \caption[HRDs Alpha Persei different $A_{V}$.]
            {HRDs around the Alpha Persei LDB sample using two different reddening values. 
             the blue isochrones correspond to BT-Settl models \citep{f_allard2013}, 
             with ages of 1, 10, 80, 100~Ma, and 1~Ga, 
             where the thick one corresponds to a 80~Ma. 
             We also show the Zero Age Main Sequence \citep{sw_stahler_f_palla2005} as the green solid line. 
             We have used the same symbols and standards than previous HRDs. 
            {\bf Left: }Reddening used is $A_{V}=0.279$~mag.    
             It is $79.0^{+1.5}_{-2.3}$~Ma old using the BT-Settl bolometric luminosity-age relationship \citep{f_allard2012}. 
            {\bf Right: }Reddening used is $A_{V}=0.055$~mag. 
             It is $81.3^{+2.4}_{-3.4}$~Ma old using the BT-Settl bolometric luminosity-age relationship \citep{f_allard2012}. 
          }
  \label{fig:hrds_alphapersei_av}
\end{figure*}

   A reddening of $A_{V}=$ 0.279~mag produced brighter luminosities 
and hotter effective temperatures than $A_{V}=0.055$~mag: 
a median offset of $0.21\cdot 10^{-3}$ $L_{\sun}$ with standard deviation of $\pm0.53\cdot 10^{-3}$ $L_{\sun}$,  
and a systematic median offset of 79~K with standard deviation of $\pm28$~K. 
In terms of percentages these changes were: $6.3\%$ in bolometric luminosities 
(reaching a maximum of $8.3\%$) and 2.2~\% in $T_{\mathrm{eff}}$.
 This change also affects the location of the LDB:
using $A_{V}=$0.279~mag 
we found
$L_{\rm bol\ LDB}=2.98^{+0.35}_{-0.19}\cdot 10^{-3}$~$L_{\sun}$ 
and $T_{\mathrm{eff\ LDB}}=2\,857\pm74$~K; 
using $A_{V}$=0.055~mag we found 
$L_{\rm bol\ LDB}=2.74^{+0.40}_{-0.15}\cdot 10^{-3}$~$L_{\sun}$
and $T_{\mathrm{eff\ LDB}}=2\,806\pm71$~K (Table \ref{tab:ldb_ages_alphapersei_av}). 
The HRD show the changes (Figure \ref{fig:hrds_alphapersei_av}).

   We conclude that a higher level of reddening produces an LDB at 
brighter bolometric luminosities and hotter effective temperatures and, 
therefore, younger ages. 
Ages obtained using the two reddening values with different evolutionary models 
for Alpha Persei are shown in Table \ref{tab:ldb_ages_alphapersei_av} 
  We estimated that a decrease of $80\%$ in $A_{v}$ 
produces an increase in age between 3 and 6~Ma depending on the model.
These variations may not be the same in stellar associations such as the Hyades (older and closer) or NGC 1960 (younger and further).

\begin{table*}[ht]
\tiny
    \caption[Stellar properties for the Alpha Persei LDB sample derived using two different reddening values.]
            {Stellar properties for the Alpha Persei LDB sample derived using two different reddening values.
            }
    \label{tab:alphapersei_samples_av}        
    \begin{tabular}{l rrrr c rrrr c rrrr}  
\hline \hline    
         & \multicolumn{4}{c}{A$_{V}=0.279$~mag}                                                        & & \multicolumn{4}{c}{A$_{V}=0.055$~mag}                                                                & &          &               &                       &                \\   
\cline{2-5} \cline{7-10}
Object   &$T_{\rm{eff}}\     \chi^{2}$&{$L_{\rm bol}$}     &{$L_{\rm bol}$ 97.5th}&{$L_{\rm bol}$ 2.5th} & &$T_{\rm{eff}}\ \chi^{2}$      &{$L_{\rm bol}$}       & {$L_{\rm bol}$ 97.5th} & {$L_{\rm bol}$ 2.5th} & &diff $T_{\rm eff}$& diff $T_{\rm eff}$ &diff $L_{\rm bol}$&diff $L_{\rm bol}\%$\\
         &       [K]                  &$10^{-3}$ [$L_{\sun}$]&$10^{-3}$ [$L_{\sun}$]&$10^{-3}$ [$L_{\sun}$]& &      [K]                  &{$10^{-3}$ [$L_{\sun}$]}&{$10^{-3}$ [$L_{\sun}$]}&{$10^{-3}$ [$L_{\sun}$]}& &        [K]    &       $\%$         & {$10^{-3}$ [$L_{\sun}$]}&             \\
\hline \hline
  AP165 & 3\,282 & 33.025 & 31.378 & 34.814 & & 3\ 159 & 30.652 & 29.124 & 32.312 & & 123 & 3.74 & 2.372 & 7.18\\  
  AP296 & 3\,122 & 16.978 & 14.557 & 20.346 & & 2\ 997 & 17.648 & 15.127 & 21.156 & & 125 & 4.00 & 0.670 & 3.94\\
  AP284 & 3\,038 &  9.288 &  8.568 & 10.117 & & 2\ 981 &  8.588 &  7.922 &  9.354 & &  57 & 1.87 & 0.700 & 7.54\\
  AP322 & 2\,974 &  5.182 &  3.569 &  8.082 & & 2\ 921 &  4.894 &  3.351 &  7.678 & &  53 & 1.78 & 0.287 & 5.54\\
  AP268 & 3\,018 &  5.155 &  4.525 &  5.991 & & 2\ 928 &  4.831 &  4.241 &  5.614 & &  90 & 2.98 & 0.323 & 6.28\\
  AP272 & 2\,923 &  5.030 &  4.463 &  5.662 & & 2\ 862 &  4.658 &  4.135 &  5.240 & &  61 & 2.08 & 0.371 & 7.38\\  
  AP275 & 2\,853 &  4.004 &  3.252 &  5.044 & & 2\ 808 &  3.930 &  3.192 &  4.951 & &  45 & 1.57 & 0.073 & 1.84\\  
  AP325 & 2\,646 &  3.541 &  2.686 &  4.873 & & 2\ 601 &  3.332 &  2.528 &  4.586 & &  45 & 1.70 & 0.208 & 5.88\\
  AP318 & 2\,834 &  3.143 &  2.683 &  3.809 & & 2\ 793 &  2.873 &  2.452 &  3.481 & &  41 & 1.44 & 0.270 & 8.60\\
  AP323 & 2\,805 &  2.911 &  2.385 &  3.762 & & 2\ 742 &  2.662 &  2.181 &  3.440 & &  63 & 2.24 & 0.249 & 8.56\\
  AP313 & 2\,883 &  2.758 &  2.204 &  3.924 & & 2\ 789 &  2.581 &  2.063 &  3.672 & &  94 & 3.26 & 0.176 & 6.41\\
  AP270 & 2\,881 &  2.677 &  2.200 &  3.432 & & 2\ 768 &  2.500 &  2.055 &  3.204 & & 113 & 3.92 & 0.176 & 6.60\\
  AP300 & 2\,842 &  2.521 &  1.966 &  3.440 & & 2\ 740 &  2.394 &  1.881 &  3.241 & & 102 & 3.58 & 0.127 & 5.03\\
  AP324 & 2\,757 &  1.746 &  1.266 &  2.598 & & 2\ 715 &  1.668 &  1.211 &  2.478 & &  42 & 1.52 & 0.078 & 4.48\\
  AP315 & 2\,747 &  1.653 &  1.264 &  2.339 & & 2\ 653 &  1.520 &  1.160 &  2.155 & &  94 & 3.42 & 0.133 & 8.05\\
  AP310 & 2\,853 &  1.399 &  1.164 &  1.737 & & 2\ 774 &  1.313 &  1.097 &  1.624 & &  79 & 2.76 & 0.085 & 6.13\\
  AP317 & 2\,731 &  0.622 &  0.509 &  0.780 & & 2\ 632 &  0.587 &  0.504 &  0.704 & &  99 & 3.62 & 0.035 & 5.62\\
\hline \hline
$\,$
    \end{tabular}
\end{table*}

\begin{table*} \begin{center}
\caption[The Alpha Persei LDB loci in terms of $L_{\rm bol}$, $T_{\rm {eff}}$ and the ages derived using several evolutionary models.]
        {Alpha Persei LDB loci in terms of $L_{\rm bol}$, $T_{\rm {eff}}$ and the ages derived using several evolutionary models.} 
\label{tab:ldb_ages_alphapersei_av}
\begin{tabular}{lcc cccccc}
\hline \hline
Stellar     &                              &                       &\multicolumn{6}{c}{LDB ages based on $L_{\rm{bol}}$}                                                                                            \\
                                                                    \cline{4-9}                                                                                                                                   
Association &$      L_{\rm{bol\ LDB} }    $&$  T_{\rm{eff\ LDB}}  $& D\&M                  & Burrows               & Siess              & Burke                 &   BT-Settl             &   Pisa                   \\
            &$10^{-3}$ [$L_{\sun}$]        &         [K]           & [Ma]                  & [Ma]                  & [Ma]               & [Ma]                  &   [Ma]                 &   [Ma]                   \\
\hline\hline
Alpha Persei&$ 2.98^{+0.35}_{-0.19}       $&$      2\,857         $&$ 69.8^{+4.5}_{-1.6}  $&$  76.1^{+3.3}_{-5.2} $&$  >70             $&$   77.8^{+2.5}_{-4.9}$&$    79.0^{+1.5}_{-2.3}$&$   70.1^{+2.8}_{-4.2}   $\\
\medskip
$A_{V}=0.279$~mag&                         &                       &   (3\,222)            &   (3\,113)            &   ($<$3\,230)      &    (3\,109)           &     (2\,983)           &    (3\,001)              \\
            
Alpha Persei&$ 2.74^{+0.40}_{-0.15}       $&$      2\,806         $&$ 75.9^{+4.4}_{-6.8}  $&$  80.4^{+3.1}_{-6.8} $&$  >70             $&$   81.2^{+3.8}_{-5.6}$&$    81.3^{+2.4}_{-3.4}$&$   73.8^{+2.6}_{-5.6}   $\\
\medskip
$A_{V}=0.055$~mag&                         &                       &   (3\,210)            &   (3\,096)            &   ($<$3\,230)      &    (3\,089)           &     (2\,961)           &    (2\,985)              \\

\hline
\end{tabular}
\\ 
 \begin{flushleft}
 $\,$
  \textbf{Notes: } 
          The columns and explanations are the same as in Table \ref{tab:ldb_ages_lbol}.
 \end{flushleft}  
\end{center} 
\end{table*} 

\clearpage
\section{Discussion of individual objects from BPMG}\label{app:bpmg_objects}

In this appendix, we provide information and remarks about some objects of the BPMG LDB sample. 
  Some of them are visual binaries or multiple systems only resolved in \textit{Gaia} DR2 and 
their bolometric luminosities are a combination of two or several sources 
in all photometric bands (except for \textit{Gaia}). 
In these cases, we calculated the bolometric luminosities with the photometric bands 
as are given in the catalogues using the parallax of each of the components. 

   \textbf{00164976} has a counterpart in \textit{Gaia} DR2, 386050700157461760, 
without an astrometric solution.
 
   \textbf{00275023} (2MASS J 00275023-3233060, GJ 2006 A) and 
\textbf{00275035} (2MASS J 00275035-3233238, GJ 2006 B) form a wide binary system.

   \textbf{01071194} has a counterpart in \textit{Gaia} DR2, 2353128266975662080,  
without an astrometric solution. 
 
   \textbf{01112542} is a lithium-rich binary system (GJ 3076 AB, LP 467-16)
with a separation between the two main components of $0.41^{\prime\prime}$ \citep{jl_beuzit2004}
or $0.293\pm0.002^{\prime\prime}$ \citep{m_janson2012c}.
The system lacks a counterpart in \textit{Gaia} DR2.
 
   \textbf{01132817} is a binary system that is only resolved in \textit{Gaia} DR2, 
composed of 
4988735051246293504 (01132817a with $\varpi=19.95$~mas) and 
4988735051245195008 (01132817b with $\varpi=19.67$~mas),
both components with similar proper motions and are separated by $1.549^{\prime\prime}$.
 
   \textbf{01535076} is a binary system resolved in \textit{Gaia} DR2: 
5143033331902438656 (01535076a $\varpi=29.54$~mas) and 
5143033331901973888 (01535076b $\varpi=29.53$~mas)
with a separation of $2.792^{\prime\prime}$.  

   \textbf{02335984} is a binary system resolved in \textit{Gaia DR2}: 
5132219016567875200 (02335984a, $G=13.53$~mag and $\varpi=12.41$~mas) and 
5132219012272775808 (02335984b, $G=13.54$~mag without parallax and proper motions).
It was previously identified by \cite{bergfors2010a}.
  
   \textbf{02365171} (GSC 08056-00482 with 25.742~mas) was considered as a possible outlier 
\citep{e_shkolnik2017} but we have discarded this possibility.
   
   \textbf{02485260} is classified as a single line binary (\citealt{l_malo2014a} and \citealt{e_shkolnik2017}).
   
   \textbf{03323578} lacks counterpart in \textit{Gaia} DR2.
   
   \textbf{03363144} has a $\varpi=22.04$~mas and was considered as a 
possible outlier \citep{e_shkolnik2017}, but we have discarded this possibility

   \textbf{03393700} has associated three sources in \textit{Gaia DR2}: 
244734761312058240, with $0.38$~mas, 
244734765606605824 without parallax and 
the right counterpart 244734765608363136 with $\varpi=24.8$~mas, 
a spectroscopic binary \citep{j_skinner2018}.

   \textbf{04593483} (V1005 Ori) is a spectroscopic binary \citep{p_elliott2014}.
   
   \textbf{05015881} is a double system \citep{i_song2003}
or a quadruple system (\mbox{HIP 23418 ABCD}, see \citealt{l_malo2013}).
\cite{bd_mason2018a} calculated a separation between the two main components of $1.234^{\prime\prime}$. 
Two components are resolved in the \textit{Gaia} DR2: 
GJ 3322A (3291643148740384128, LP 476-207 A, 05015881a) and 
GJ 3322B (3291643148741783296, LP 476-207 B, 05015881b). 
\cite{i_song2003} found that the combined system \mbox{HIP 23418}
shows prominent emission lines of Na D $\lambda5890$~\AA\ and $\lambda5896$~\AA, 
\ion{He}{i} $\lambda5876$~\AA and $\lambda6876$~\AA, and H$\alpha$.

   \textbf{05241914} lacks counterpart in \textit{Gaia} DR2, 
\mbox{2MASS J 05241914-1601153}
probably because it is a binary system, M4.5+M5.0, 
with a separation of $0.639\pm0.001^{\prime\prime}$ \citep{bergfors2010a}.
Lithium has been detected in the binary system (unresolved spectra from \citealt{l_malo2013} 
and \citealt{as_binks2014}).

   \textbf{05320450} (2MASS J 05320450-0305291, V1311 Ori) 
has a counterpart in \textit{Gaia} DR2: 3216729573251961856 ($\varpi=28.94$~mas)
and it is identified as a lithium rich member in \cite{l_dasilva2009} just like \cite{l_malo2013}. 
\cite{as_binks2016} also detected lithium but they did not considered it as a member. 
Finally, \cite{m_janson2012c} reported this object as a binary system, M2V+M3.5V, 
with a separation of $0.216\pm0.003^{\prime\prime}$.

   \textbf{10172689} (5355751581627180288 with $\varpi=51.00$~mas)  
could be contaminated by other background sources 
(5355751585943066880 with $\varpi=0.08\pm0.26$~mas and 
5355751585947981568 with $\varpi=1.13\pm1.20$~mas)
and other source detected in 2MASS: 10172697-5354200 
(5355751585943068416 with $\varpi=0.58\pm0.13$~mas).

   \textbf{13545390} (5845972349875278720 $\varpi=43.76$~mas)
has associated the background object \textit{Gaia} DR2 source 5845972349858754688 ($\varpi=0.09$~mas).

   \textbf{15385679} is located in a very crowded region. 
It has associated two \textit{Gaia} DR2 sources: 
the rigth one 5882581895192805632 ($\varpi=25.471$~mas) and 
a background source 5882581929553337216 ($\varpi=0.896$~mas). 

   \textbf{16572029}. There are three sources related that can contaminate the main one 5935776714456619008 ($\varpi=19.75$~mas): 
5935776714456617984 ($\varpi=0.41$~mas) and 
5935776714412775552 ($\varpi=-0.35$~mas).   

   \textbf{17150219} is located in a very crowded region and three sources can contaminate the 
main object 5978985558581221760 ($\varpi=38.59$~mas):
5978985562916750976 ($\varpi=0.05$~mas)  
5978985562873939456 ($\varpi=1.63$~mas). 

   \textbf{17173128} (HD 155555C, 5811866358170877184 with $\varpi=32.95$~mas) 
the photometry might be contaminated with the ghosts from HD 155555, 
by a near background source (5811867835635129728 with $\varpi=-0.14$~mas located at $4.48^{\prime\prime}$) 
or both.
17173128 together with 2MASS J 17172550-6657039 (V824 Ara, HIP 84586) take part in a multiple system. 

   \textbf{17292067} is a source without parallax and proper motions, 
perhaps it is due to two sources detected in 2MASS very near: 
17292067-5014529 and 17292064-5014574.

   \textbf{18142207} the main source, 4045698732855626624 ($\varpi=13.99$~mas) 
could be contaminated by a background source 4045698737270604288 ($\varpi=1.57$~mas). 

   \textbf{18420483} the main source 6649786646225001984 ($\varpi=19.33$~mas) 
could be contaminated by a background source 6649786646219487744 ($\varpi=0.01$~mas). 

   \textbf{18435838} is a lithium-rich object that has associated four counterparts in \textit{Gaia} DR2: 
6730618549223057664 (7.07~mas) the bright and main one, 
6730618549223057920 (0.12~mas), 
6730618549211759104 (-1.03 mas), and 
6730618549211759232 (0.58 mas). 
Its parallax indicates that it is a background object, and 
it is very likely that the bolometric luminosity is a combination of the four sources. 
This object must be treated with caution.  
We  considered it as a non-member and as a multiple system.

   \textbf{18465255}. 18465255 (6631685008336771072, $\varpi=19.78$~mas) 
could be contaminated by a background source, 6631685008335769728 ($\varpi=1.26$~mas).

   \textbf{18471351}. 18471351 (4071532308311834496, $\varpi=16.8$~mas)
could be contaminated by a background source, 4071532312630649344 ($\varpi=-0.29$~mas).

   \textbf{19082195}. 19082195 (4088823159447848064, $\varpi=14.4$~mas) 
could be contaminated by a background source, 4088823163758008192 ($\varpi=1.46$~mas).
 
   \textbf{19102820} is a binary system formed by two sources separated 
by $0.424^{\prime\prime}$ \citep{m_janson2017},
both sources are resolved in \textit{Gaia} DR2 and both without parallaxes:
4080460377082576128 (19102820a, the closest source 19102820a with $G=12.07$~mag)
4080460381377496832 (19102820b with $G=19.42$~mag). 

   \textbf{19260075}. 19260075 (6643851448094862592, $\varpi=20.82$~mas) 
could be contaminated by a background source, 6643851448094200320 ($\varpi=-0.33$~mas).

   \textbf{19560294} (1RXS J195602.8-320720, 6747467431032539008 with $\varpi=18.02$~mas) 
is a spectroscopic binary \citep{p_elliott2014}.

    \textbf{20004841}. 20004841 (6366726276822544768, $\varpi=33.96$~mas) 
could be contaminated by a background source, 6366726276820119040 ($\varpi=0.23$~mas).
     
    \textbf{20085368} is a multiple system with two sources associated, 
a possible binary resolved with \textit{Gaia} DR2 separated $1.426^{\prime\prime}$:
20085368a (6697858840776095616 with $\varpi=22.22$~mas) and 
20085368b (6697858840773435776 with $\varpi=22.43$~mas).

    \textbf{20100002} (SCR J2010-2801) is a source without parallax and proper motions, this is 
probably because it is a double line binary \citep{l_malo2014a}. 
    
    \textbf{20434114} is a visual binary \citep{e_shkolnik2009} as we can confirm using 
\textit{Gaia} DR2 data:
20434114a (6806301370519190912 with $\varpi=22.81$~mas) and 
20434114b (6806301370521783552 with $\varpi=23.51$~mas). 

    \textbf{20415111} (AT Mic, HD 196982) is a binary system formed by two sources \citep{cao_torres2006}:  
20415111a (6792436799475128960, TYC-7460-137-1)
and 
20415111b (6792436799477051904, TYC-7460-137-2) 
both resolved in \textit{Gaia} DR2 and Tycho-2, 
but not resolved in other surveys as 2MASS or WISE. 

    \textbf{20560274} (AZ Cap, HD 358623) is a binary system formed by two sources
separated by $2.181^{\prime\prime}$ \citep{m_janson2017} and also 
it is resolved in \textit{Gaia} DR2:
20560274a (6882838031331951488 with $\varpi=21.77$~mas) and 
20560274b (6882838031330569856 with $\varpi=21.62$~mas). 
Also there is a very close background object: 
it is only detected in \textit{Gaia} DR2: 
6882838031330568960, with $\varpi=0.64$~mas.

    \textbf{21374019} (WDS J21376+0137AB, RX J2137.6+0137)
is a binary system \citep{bd_mason2001} and perhaps due to that 
lacks of parallax and proper motions in \textit{Gaia} DR2. 

    \textbf{23353085} has counterpart in \textit{Gaia} DR2, 2392233119572232064, 
but lacks parallax and proper motions.

\clearpage
\section{32 Ori MG as one  or two associations}\label{app:32orimg_one_or_two}

In this appendix, we report on a significant aspect of 32 Ori MG. 
In contrast to other stellar associations, the moving group 
appears to be composed of two sub-samples.   
  First, we calculated the parallax PDF of the known members 
as it is explained in Section \ref{sub:outliers}.    
We realized that some of them are non-members due to the parallax: 
 2MASS J05572121+0502158 ($\varpi=0.325$~mas) and 
 TYC 4767-689-1 ($\varpi=1.759$~mas)
are background sources; 
 2MASS J05053333+0044034 ($\varpi=21.255$~mas), 
 THOR-33 ($\varpi=25.944$~mas) and 
 THOR-34 ($\varpi=26.069$~mas) 
are foreground sources. 
We removed them and re-did the parallax PDF.

\begin{figure}
   \includegraphics[width=9cm]{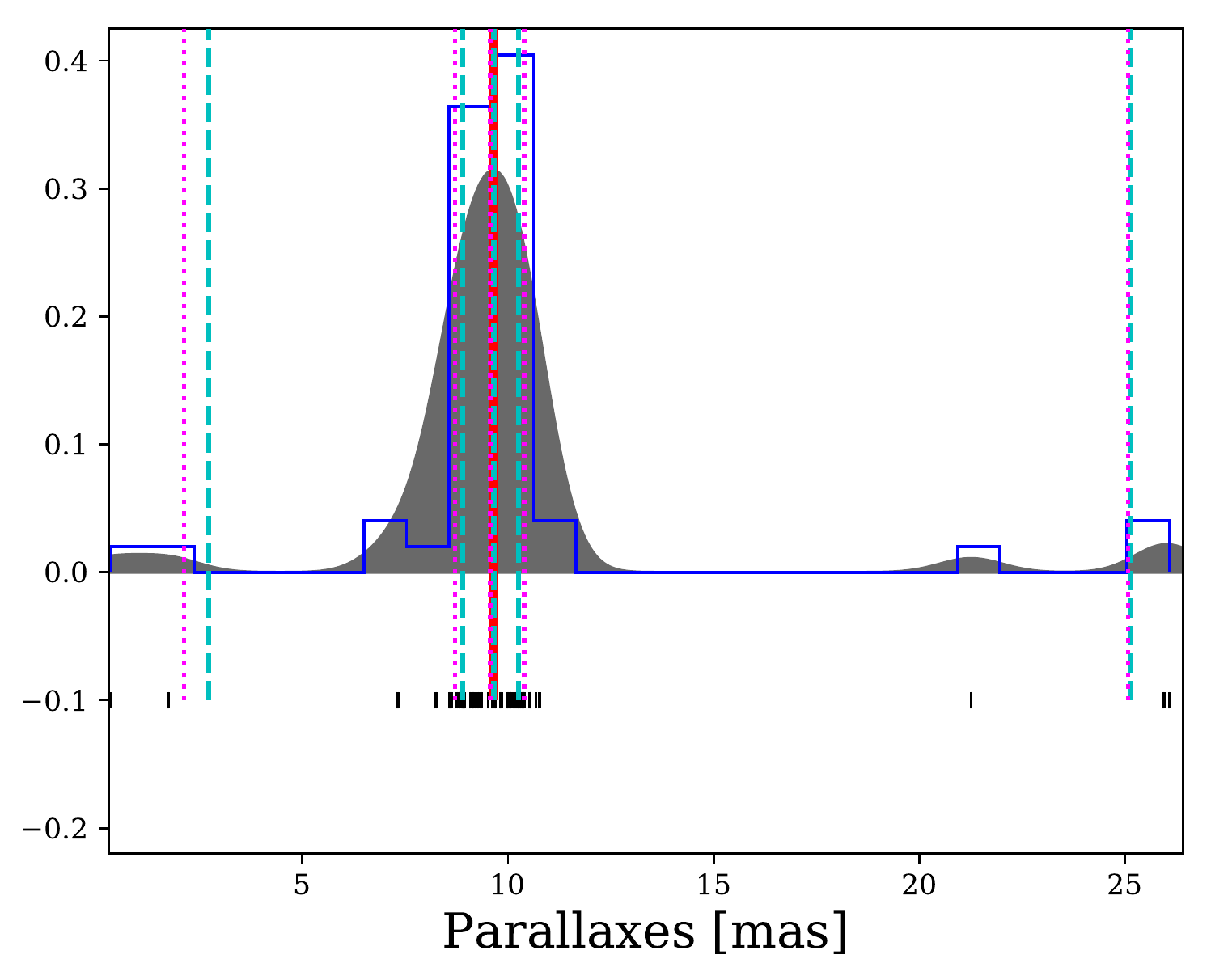}
   \includegraphics[width=9cm]{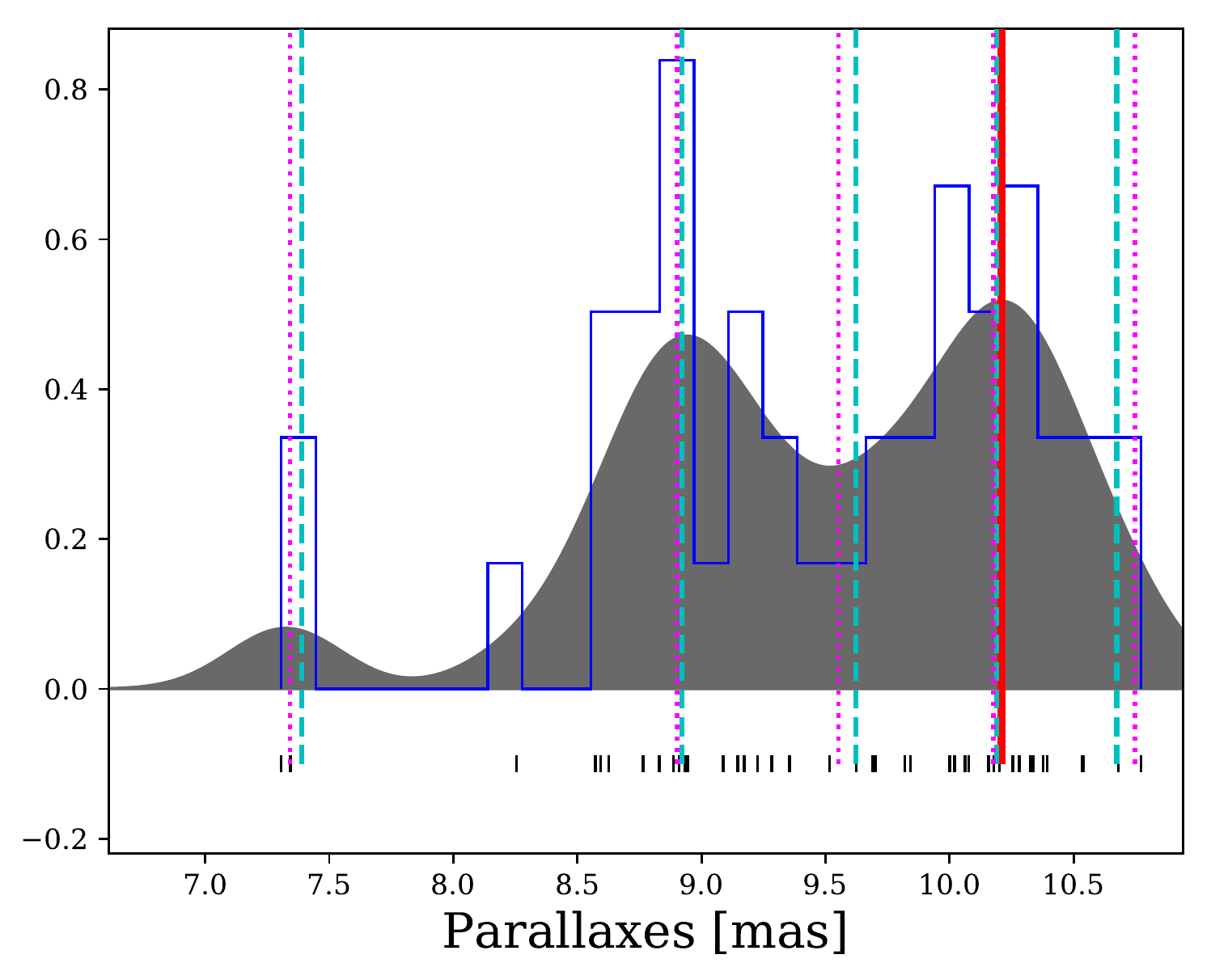}
   \includegraphics[width=9cm]{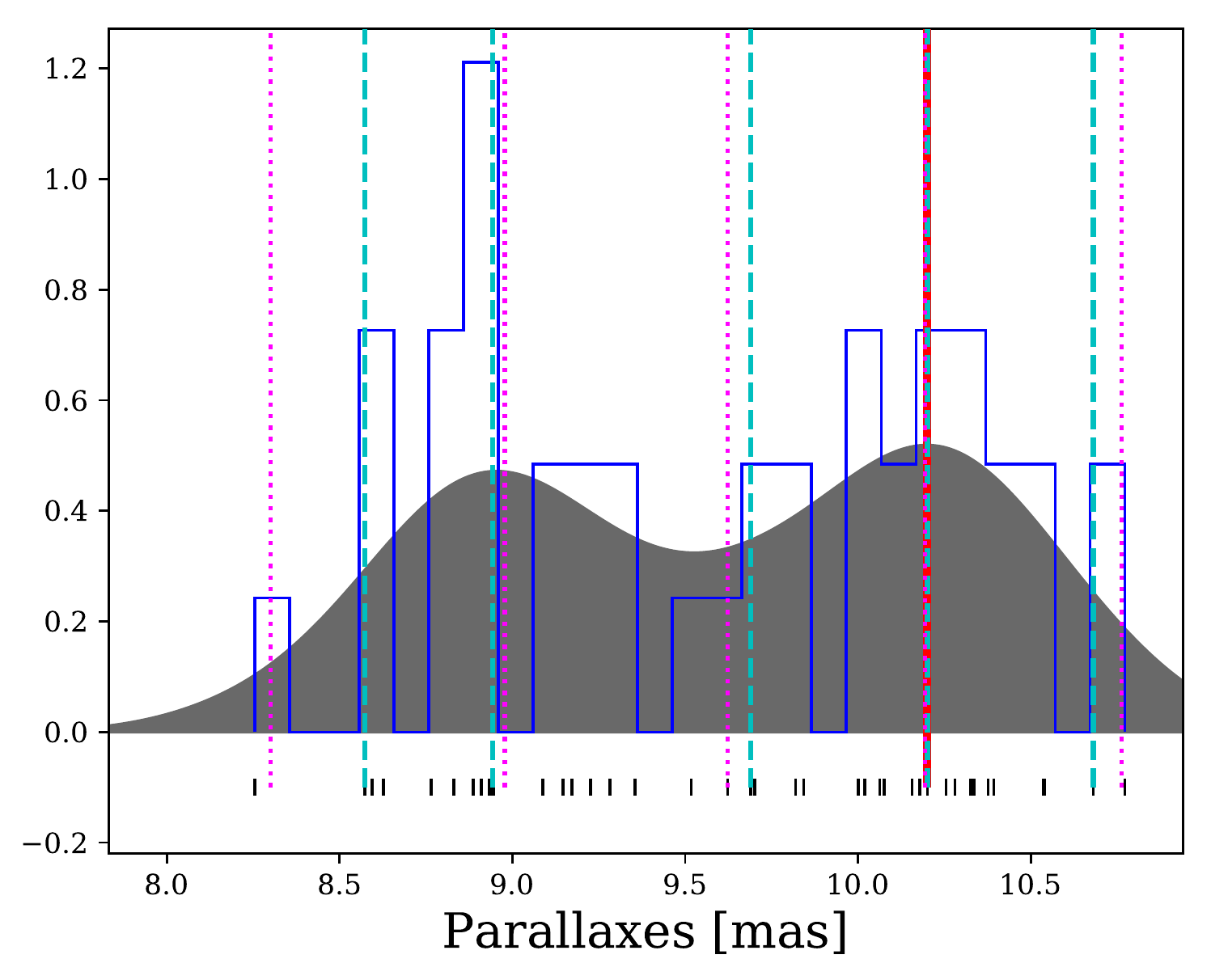}
      \caption[KDEs VPDs and distances The Hyades members.]
              {Parallax PDF for the 32 Ori MG known members 
               (taken from \citealt{cpmbell2017}). 
               The modelled PDF is shown in gray and 
               a superimposed histogram in blue.
               The cyan vertical dashed lines  
                 point the percentiles of the raw data at the 2.5th, 25th, 50th, 75th, and 97.5th.
               The magenta vertical dotted lines  
                 point the same percentiles as before but calculated from the modelled KDE.
               The red vertical line shows the mode of the PDF.                
               The small vertical black lines below the horizontal axis (below $y=0.0$) 
                shows the parallaxes of each source. 
              {\bf Top: } All the previous known members from 
                 \cite{cpmbell2017}. 
              {\bf Middle: }First iteration. 
              {\bf Bottom: }Second iteration.
               }
         \label{fig:kde_plxs_32orimg_populations}
   \end{figure}

    We carried out this process twice and the results are shown Figure \ref{fig:kde_plxs_32orimg_populations}. 
After the first and the second iteration, we can see two maximums in the parallax PDF:
around 8.93 mas and 10.21 mas and 
after the second iteration in 8.95 mas and 10.19 mas.
It appears that the population of 32 Ori MG is composed of two populations. 
This fact is also remarkable with the distances released by \cite{cpmbell2017}. 
We divided the two populations based on parallaxes: 
population 1 with $\varpi<9.52$~mas and 
population 2 with $\varpi>9.52$~mas. 

\begin{figure}
   \includegraphics[width=9cm]{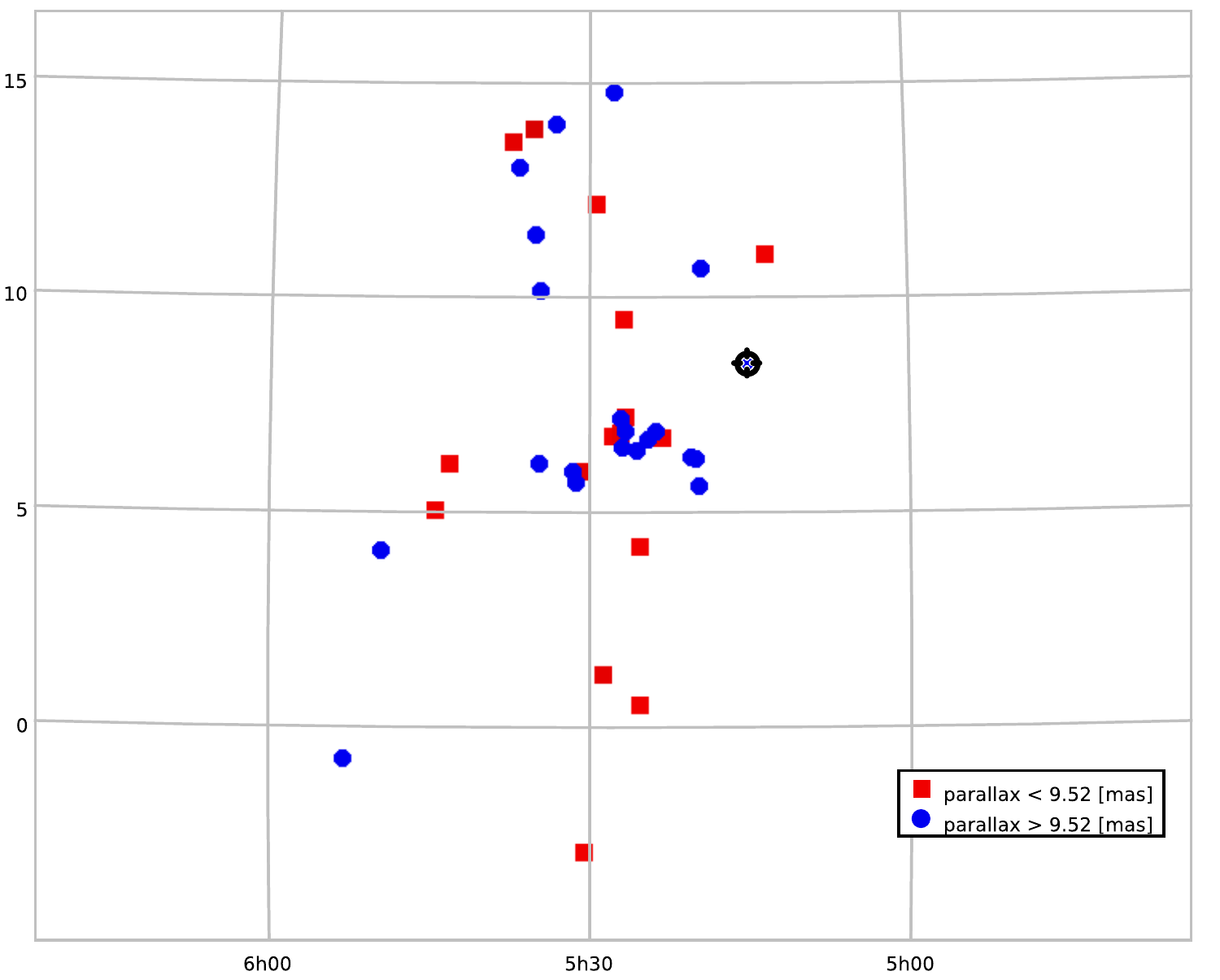}
   \includegraphics[width=9cm]{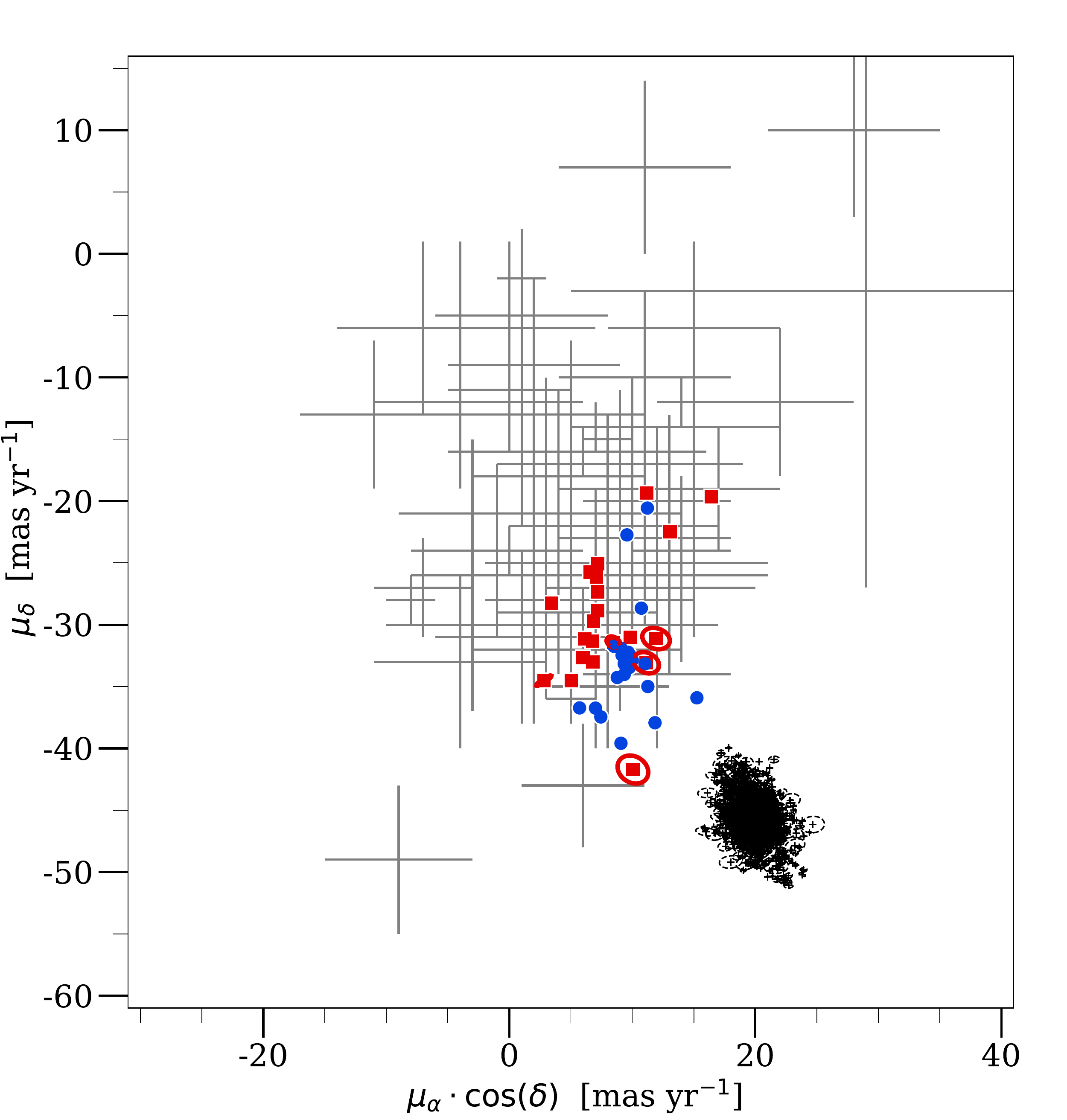}
      \caption[Sky area and VPD of 32 Ori MG.]
              {Spatial distribution and VPD for the 32 Ori MG members. 
               {\bf Top: }Spatial distribution of the known members from \cite{cpmbell2017}, 
                 after removing some outliers. 
                 The two populations based on parallaxes are shown.                
               {\bf Bottom: }VPD of the 32 Ori MG, sub-sample 
                 red squares are population 1 ($\varpi<9.52$~mas)
                 and blue circles population 2 ($\varpi>9.52$~mas).               
                Black dashed ellipses are the Pleiades members taken from \citealt{c_babusiaux2018}
                 and the light grey plus symbols are the Taurus-Auriga members 
                 taken from \cite{c_ducourant2005}.
                The proper motion data have been taken from the \textit{Gaia} DR2 catalogue, 
                 with the exception of the Taurus sample that comes from \cite{c_ducourant2005}.
               }
         \label{fig:spatial_vpd_32orimg}
   \end{figure}

    We do not see any trend across both figures in terms of
the sky distribution and 
the VPD of the two populations in Fig. \ref{fig:spatial_vpd_32orimg}.
  We calculated the parallax PDF using a width validated with cross-validation, 
but this value may produce these two maximums in the parallax PDF and 
we must be cautious with our results.

    In any case, we located the LDB for each sub-sample.
We located the LDB at 
$L_{\rm bol\ LDB}=41.8^{+2.2}_{-14}\cdot 10^{-3}L_{\sun}$ and  
$T_{\mathrm{eff\ LDB}}=3\,157\pm71$~K 
for population 1 (Figure \ref{fig:hrd_32orimg_zoom_ldb_pop1}).
The large uncertainties are due to the scarcity number of sources.
  For population 2, we located the LDB at 
$L_{\rm bol\ LDB}=30.1^{+6.1}_{-0.3}\cdot 10^{-3}$ $L_{\sun}$ and  
$T_{\mathrm{eff\ LDB}}=3\,098\pm71$~K 
(Figure \ref{fig:hrd_32orimg_zoom_ldb_pop2}). 
However, the LDB might be located at fainter bolometric luminosities due to 
the scarcity of sources. 
The $L_{\rm bol\ LDB}$ may be able to reach $23.418\cdot 10^{-3} L_{\sun}$, 
which would result in an age increase of about 2~Ma.

\begin{figure}
     \subfloat[\label{fig:hrd_theia133_zoom_ldb_v1}]
      {\includegraphics[width=0.5\textwidth,scale=0.50]{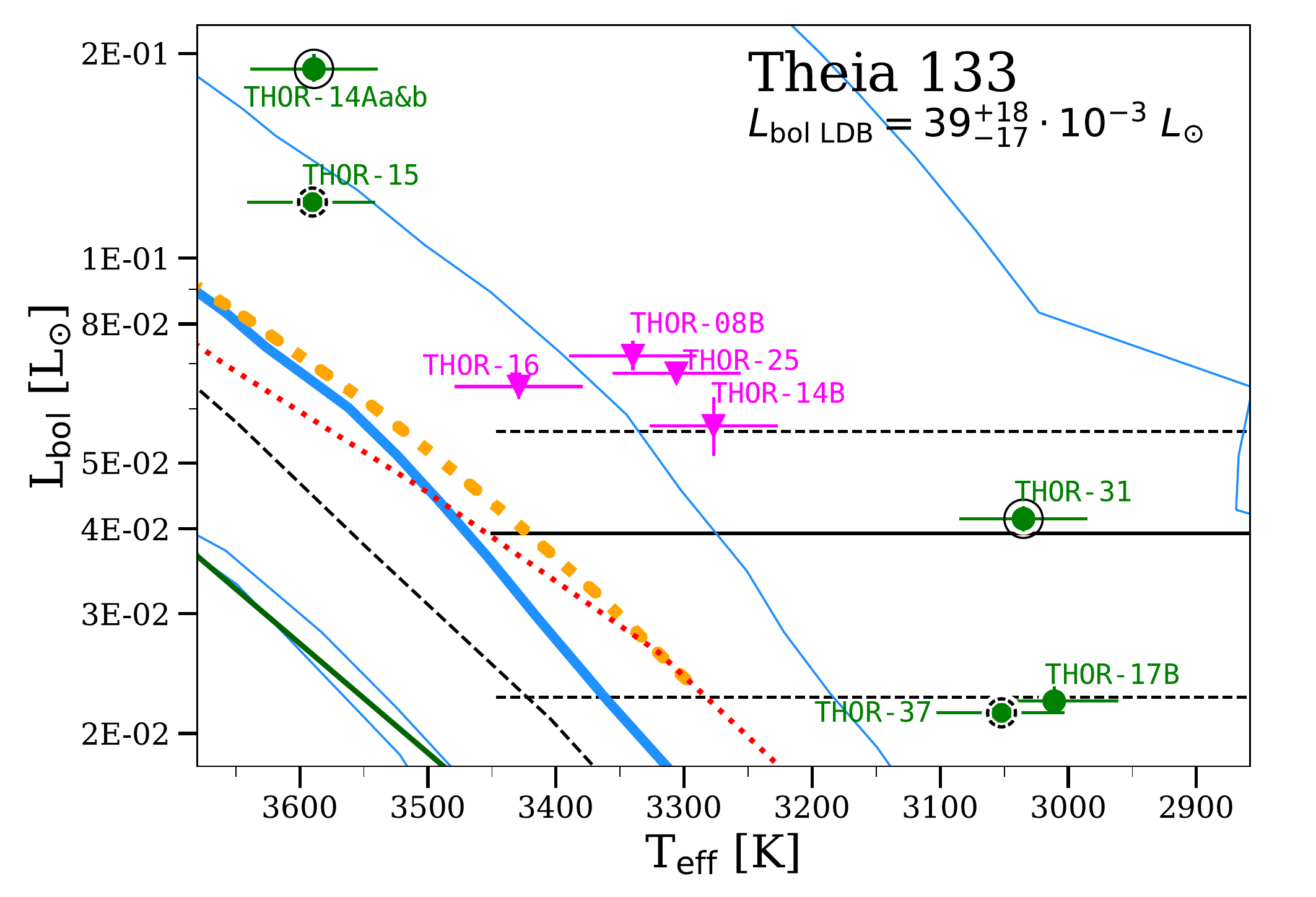}} \quad      
     \subfloat[\label{fig:hrd_theia370_zoom_ldb_v1}]
      {\includegraphics[width=0.5\textwidth,scale=0.50]{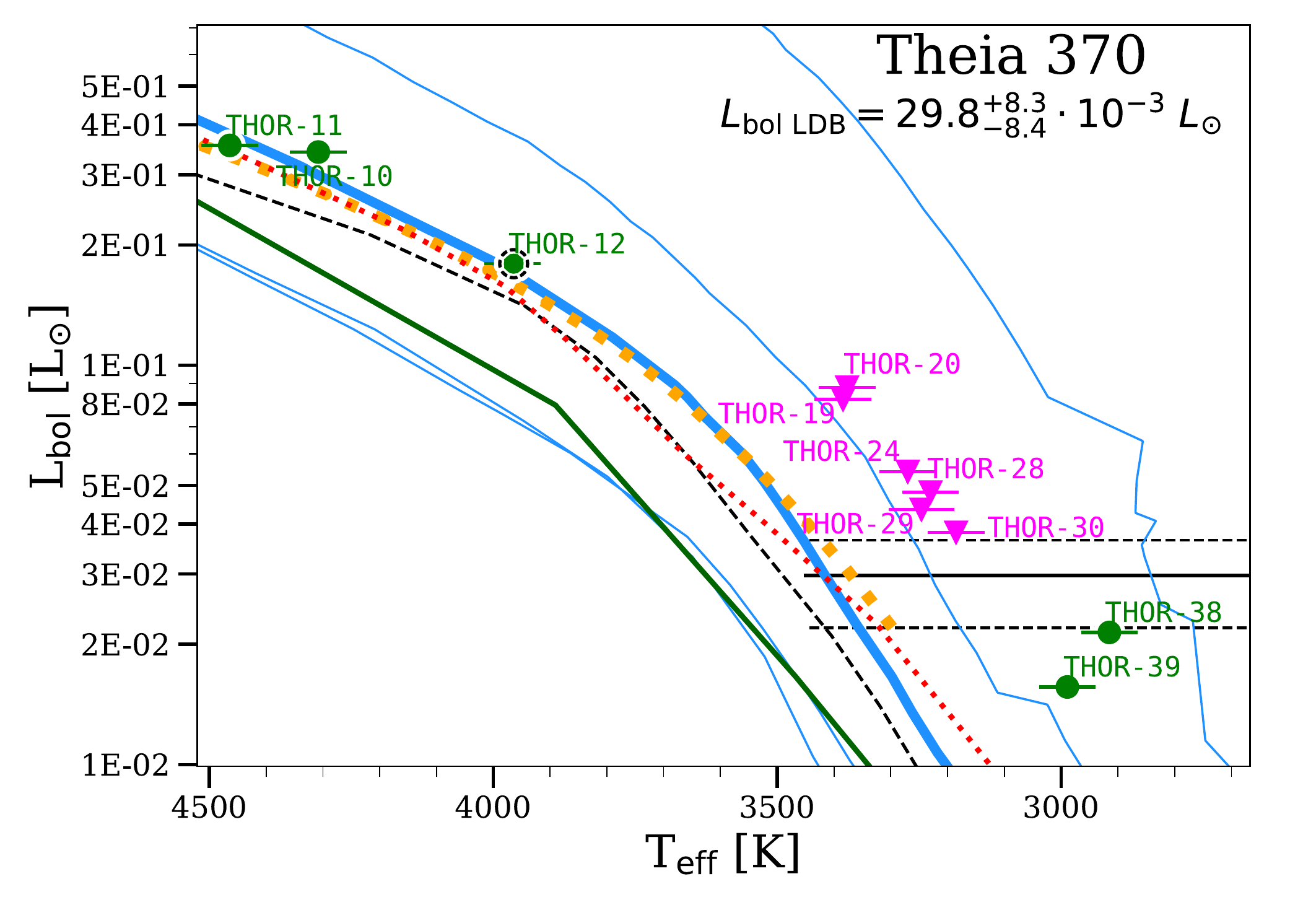}} \\      
          \caption[HRDs and the LDB for Theia 130 and Theia 370.]
                  {HRDs and the LDB for Theia 130 and Theia 370. 
                   All the symbols and lines follow the same convention as Figure \ref{fig:hrds_32orimg}.                    
                   The blue lines correspond to a isochrones of 1, 10, 30, 100~Ma, and 1~Ga 
                   from the BT-Settl models \citep{f_allard2013}, where the thick one corresponds to a 30~Ma.  
                    The figure includes a \mbox{20~Ma} isochrone from \citealt{f_dantona1994} (black dashed line).            
                  {\bf (a): }Theia 133. 
                    The figure includes 
                   a \mbox{20~Ma} isochrone from \citealt{siess2000} (red dotted line) and 
                   a \mbox{18~Ma} isochrone from \citealt{e_tognelli2011} (orange dashed dot line).
                   Theia 133 is $25.6^{+7.0}_{-5.5}$~Ma old using the BT-Settl bolometric luminosity-age relationship \citep{f_allard2012}.                   
                  {\bf (b): }Theia 370. 
                    The figure includes 
                   a \mbox{25~Ma} isochrone from \citealt{siess2000} (red dotted line) and 
                   a \mbox{20~Ma} isochrone from \citealt{e_tognelli2011} (orange dashed dot line). 
                   Theia 370 is $28.7^{+4.7}_{-2.6}$~Ma old using the BT-Settl bolometric luminosity-age relationship \citep{f_allard2012}.                   
          }
  \label{fig:hrds_theia133_theia370}
\end{figure}

    We carried out an extra analysis to analyse the 32 Ori MG population. 
We crossmatched the 32 Ori MG sources with the stellar groups identified in \cite{m_kounkel2019b} work.
The sources from \cite{cpmbell2017} belong to two different stellar groups:
Theia 133, with $\log {\rm age}=7.74419$ (55~Ma) and $\varpi=6.376$~mas, 
and 
Theia 370, with $\log {\rm age}=8.16773$, (147~Ma) and $\varpi=6.552$~mas.
This outcome is contrary to our previous naive population division based on parallaxes. 
Following the \cite{m_kounkel2019b} membership criterion, 
we determined the LDB for these stellar associations. 
  For Theia 133, we located the LDB at 
$L_{\rm bol\ LDB}=39.4^{+3.9}_{-1.4}\cdot 10^{-3}L_{\sun}$ and  
$T_{\mathrm{eff\ LDB}}=3\,139\pm71$~K. 
However, the uncertainties are underestimated due to the source scarcity
between the faintest lithium-poor source, THOR-14B and the lithium-rich one, THOR-17B, 
so, the LDB would be located at
$L_{\rm bol\ LDB}=39^{+18}_{-17}\cdot 10^{-3}L_{\sun}$, 
see Figure \ref{fig:hrd_theia133_zoom_ldb_v1}. 
  In the same way, for Theia 370 we located the LDB at 
$L_{\rm bol\ LDB}=29.8^{+2.4}_{-2.7}\cdot 10^{-3}L_{\sun}$ and 
$T_{\mathrm{eff\ LDB}}=3\,087\pm71$~K, 
(Figure \ref{fig:hrd_theia370_zoom_ldb_v1}).
The $L_{\rm bol\ LDB}$ uncertainties are underestimated, as is also the case with Theia 133, and 
the updated LDB would be $L_{\rm bol\ LDB}=29.7^{+8.3}_{-8.4}\cdot 10^{-3}L_{\sun}$. 

    There are still many unanswered questions about the real existence of this moving group: 
the two peaked parallax distribution together with the evidence of two stellar groups as \cite{m_kounkel2019b} released.     
To develop a full picture of the 32 Ori MG, we need additional membership kinematic studies. 

\clearpage
\section{Additional tables}
\label{app:additional_tables}

\setcounter{table}{0}
\begin{table}[ht] 
  \caption[Priors used in the distance calculation with \textit{Kalkayotl}.]
          {Priors used in the distance calculation with \textit{Kalkayotl}.}  
  \label{tab:kalkayotl_priors}
  \small
  \center
  \begin{tabular}{lccc}
\hline \hline
Stellar     &   &\verb+prior_loc+ &\verb+prior_scale+ \\
Association &   &      [pc]       &        [pc]       \\
\hline
Alpha Persei&   &             175 &  120 \\
NGC 1960    &   &          1\ 197 &  300 \\
IC 4665     &   &             346 &   70 \\
NGC 2547    &   &             394 &  135 \\
IC 2602     &   &             152 &  190 \\
IC 2391     &   &             152 &  115 \\
The Pleiades&   &             136 &   85 \\
Blanco 1    &   &             237 &   85 \\
The Hyades  &   &              48 &   90 \\
BPMG        &   &              41 &  200 \\
THMG        &   &              53 &  200 \\
32 Ori MG   &   &             100 &   25 \\
\hline
    \end{tabular}
\end{table}

Table \ref{tab:associations_stellarproperties_v1}, available at the CDS, 
includes calculated and collected stellar parameters for each object from each LDB sample.
    Parallaxes and proper motions were taken from \textit{Gaia} DR2. Here,
`$\rho_{\mu_{\alpha} \ast \ \mu_{\delta}}$' is the correlation between proper motion in 
right ascension and proper motion in declination; `EW Li' is the equivalent width of \ion{Li}{i} at $6\,707.8$~\AA, 
with its reference `Ref.'. 
Equivalent width of a lithium-poor object is given by an upper limit or without any measure;  
if this is the case, then it is shown as '-'. 
  There are some objects that have undetermined lithium content 
because the data retrieved from the literature have low resolving power, 
low signal-to-noise ratio, or bad quality or they do not cover this part of the spectrum.
These objects are marked with '$\cdots$'.
    The label `Li rich' indicates whether  it is a lithium-rich or lithium-poor object 
based on `EW Li': 
 'Y' indicates a lithium-rich object, 
 'N' indicates a lithium-poor object and 
 '?' indicates an object with undetermined lithium content. 
    '`Other'' means other remarks of the source: 
 'EL' means strong emission line object,    
 'F' is a flaring object,    
 'FR' is a fast rotator,    
 'NM' is no member (based on this work), 
 'SuB' suspected binary or multiple system, 
 'CMS' is a confirmed binary or multiple system. Then,
    `$d_{MAP}$', `$d_{min}$', and `$d_{max}$' are the distances for each object.  
If a source lacks a parallax, proper motions, or both in \textit{Gaia} DR2 or it has a negative parallax, 
we indicated the distances with '-'.

    We complete this appendix with tables that gather photometric bands for each stellar association, 
with the exception of THMG and 32 Ori MG, where all the photometric bands were recovered using VOSA. 
The values in parentheses have not been taken into account in the SED fitting.

\onecolumn

\setcounter{table}{1}
\begin{table*}[ht] 
  \caption[Comparison between stellar parameters derived with VOSA and StarHorse.]
          {Comparison between stellar parameters derived with VOSA and StarHorse.}
  \label{tab:vosa_vs_starhorse}
  \center
  \begin{tabular}{l ccccc r}
\hline \hline
  Stellar   &$\Delta$ [M/H]    &$ \Delta \log g   $&$\Delta A_{V}  $&$\Delta T_{\rm eff}$&$   \Delta\ d    $& \# \\
association &    [dex]         &       [dex]       &     [mag]      &     [K]            &      [pc]        &    \\
\hline \hline 
\multicolumn{7}{l}{ \qquad {\it All sources with }{\tt StarHorse }{\it counterpart} } \\
\hline 
Alpha Persei&$-0.19  \pm0.19  $&$ -0.159 \pm0.040 $&$-0.26 \pm0.28 $&$  -43 \pm 142     $&$   1.7 \pm  1.8 $&  2 \\
NGC 1960    &$ 0.05  \pm0.16  $&$ -0.051 \pm0.064 $&$-1.67 \pm0.48 $&$-1223 \pm 608     $&$-358   \pm855   $&  4 \\
IC 4665     &$ 0.06  \pm0.27  $&$  0.18  \pm0.52  $&$-0.28 \pm0.40 $&$ -143 \pm 481     $&$-141   \pm683   $& 53 \\
NGC 2547    &              --  &             --    &             -- &                 -- &              --  & -- \\
IC 2602     &$-0.13  \pm0.11  $&$ -0.23  \pm0.18  $&$-0.43 \pm0.34 $&$  -97 \pm 140     $&$   0.2 \pm  1.3 $& 10 \\
IC 2391     &$-0.17  \pm0.26  $&$ -0.31  \pm0.17  $&$-0.43 \pm0.33 $&$  -42 \pm 145     $&$   1.5 \pm  2.1 $& 11 \\
The Pleiades&$-0.04  \pm0.44  $&$ -0.050 \pm0.069 $&$-0.055\pm0.059$&$   87 \pm  10     $&$   0.09\pm  0.76$&  2 \\
Blanco 1    &$-0.09  \pm0.15  $&$ -0.268 \pm0.097 $&$-0.02 \pm0.12 $&$   30 \pm  64     $&$   1.5 \pm  1.3 $&  4 \\
The Hyades  &$-0.15  \pm0.17  $&$ -0.449 \pm0.098 $&$-0.15 \pm0.11 $&$   38 \pm  47     $&$   0.07\pm  0.36$& 17 \\
BPMG        &$-0.19  \pm0.23  $&$ -0.03  \pm0.18  $&$-0.53 \pm0.77 $&$ -220 \pm 475     $&$   0.1 \pm  1.3 $&101 \\
THMG        &$-0.10  \pm0.18  $&$ -0.03  \pm0.10  $&$-0.29 \pm0.51 $&$ -105 \pm 358     $&$   0.12\pm  0.22$&108 \\
32 Ori MG   &$-0.22  \pm0.18  $&$ -0.00  \pm0.12  $&$-0.49 \pm0.77 $&$ -232 \pm 467     $&$   0.73\pm  0.75$& 24 \\
All         &$-0.11  \pm0.23  $&$ -0.03  \pm0.28  $&$-0.39 \pm0.61 $&$ -155 \pm 432     $&$ -26   \pm293   $&336 \\
\hline \hline
\multicolumn{7}{l}{ \qquad {\it Subsample with }{\tt SH\_GAIAFLAG=="000"} \& {\tt SH\_OUTFLAG=="00000"} } \\
\hline
Alpha Persei&$-0.0011         $&$ -0.1198         $&$ 0.021        $&$   99             $&$   3.47         $&  1 \\
NGC 1960    &$-0.1020\pm0.0086$&$ -0.050 \pm0.064 $&$-1.44 \pm0.33 $&$ -939 \pm 425     $&$ -30   \pm 72   $&  2 \\
IC 4665     &$ 0.13  \pm0.26  $&$  0.30  \pm0.615 $&$-0.26 \pm0.38 $&$ -151 \pm 476     $&$-220   \pm830   $& 35 \\
NGC 2547    &             --   &               --  &            --  &                --  &              --  &  --\\
IC 2602     &$-0.165 \pm0.096 $&$ -0.23  \pm0.17  $&$-0.38 \pm0.26 $&$  -91 \pm 124     $&$   0.3 \pm1.3   $&  8 \\
IC 2391     &$-0.21  \pm0.18  $&$ -0.23  \pm0.17  $&$-0.35 \pm0.14 $&$  -11 \pm  72     $&$   1.8 \pm2.3   $&  7 \\
The Pleiades&$-0.4915         $&$  0.0186         $&$ 0.003        $&$   96             $&$   0.85         $&  1 \\
Blanco 1    &$-0.08  \pm0.11  $&$ -0.27  \pm0.12  $&$ 0.072\pm0.081$&$   71 \pm  30     $&$   0.9  \pm1.6  $&  2 \\
The Hyades  &$-0.16  \pm0.16  $&$ -0.4061\pm0.0097$&$-0.054\pm0.023$&$   87 \pm  29     $&$  -0.08 \pm0.18 $&  3 \\
BPMG        &$-0.23  \pm0.15  $&$ -0.03  \pm0.18  $&$-0.57 \pm0.60 $&$ -216 \pm 400     $&$   0.16 \pm0.29 $& 50 \\
THMG        &$-0.10  \pm0.14  $&$ -0.047 \pm0.078 $&$-0.25 \pm0.41 $&$  -82 \pm 254     $&$   0.13 \pm0.14 $& 52 \\
32 Ori MG   &$-0.23  \pm0.14  $&$  0.008 \pm0.049 $&$-0.58 \pm0.43 $&$ -249 \pm 306     $&$   1.08 \pm0.60 $& 13 \\
All         &$-0.11  \pm0.22  $&$  0.00  \pm0.34  $&$-0.38 \pm0.49 $&$ -147 \pm 367     $&$ -44   \pm 382  $&174 \\  
\hline \hline
\multicolumn{7}{l}{ \qquad {\it Subsample with sources considered as members after this work} } \\
\hline
Alpha Persei&$-0.19  \pm0.19  $&$ -0.159 \pm 0.040$&$-0.26 \pm0.28 $&$  -42  \pm142     $&$   1.6  \pm  1.7 $&  2 \\    
NGC 1960    &$ 0.05  \pm0.16  $&$ -0.01  \pm 0.11 $&$-1.67 \pm0.49 $&$-1224 \pm 608     $&$-358    \pm855   $&  4 \\    
IC 4665     &$ 0.03  \pm0.262 $&$ -0.01  \pm 0.11 $&$-0.32 \pm0.37 $&$ -115 \pm 499     $&$  10    \pm 15   $& 36 \\   
NGC 2547    &             --   &               --  &            --  &                --  &               --  &  --\\
IC 2602     &$-0.13  \pm0.12  $&$ -0.23  \pm 0.18 $&$-0.43 \pm0.34 $&$  -98  \pm139     $&$   0.2  \pm  1.3 $& 10 \\  
IC 2391     &$-0.26  \pm0.16  $&$ -0.32  \pm 0.15 $&$-0.34 \pm0.13 $&$   -3  \pm 65     $&$   1.3  \pm  1.3 $&  9 \\   
The Pleiades&$ 0.39           $&$ -0.119          $&$-0.11         $&$   77             $&$  -0.67          $&  1 \\   
Blanco 1    &$-0.09  \pm0.15  $&$ -0.268 \pm 0.097$&$-0.02 \pm0.13 $&$   30  \pm 64     $&$   1.4  \pm  1.2 $&  4 \\
The Hyades  &$-0.15  \pm0.17  $&$ -0.449 \pm 0.098$&$-0.15 \pm0.11 $&$   38  \pm 46     $&$   0.07 \pm  0.36$& 17 \\   
BPMG        &$-0.21  \pm0.21  $&$ -0.04  \pm 0.19 $&$-0.45 \pm0.61 $&$ -169  \pm426     $&$   0.1  \pm  1.4 $& 75 \\
THMG        &$-0.10  \pm0.19  $&$ -0.03  \pm 0.10 $&$-0.26 \pm0.42 $&$  -82  \pm279     $&$   0.12 \pm  0.20$&107 \\
32 Ori MG   &$-0.26  \pm0.14  $&$  0.021 \pm 0.068$&$-0.52 \pm0.43 $&$ -201  \pm291     $&$   0.89 \pm  0.76$& 19 \\   
All       &$-0.13  \pm0.22  $&$ -0.074 \pm 0.179$&$-0.35 \pm0.49 $&$ -121  \pm377     $&$  -4    \pm110   $&284 \\  
\hline \hline
\multicolumn{7}{l}{ \qquad {\it Members with }{\tt SH\_GAIAFLAG=="000"} \& {\tt SH\_OUTFLAG=="00000"} } \\
\hline
Alpha Persei&$-0.0011         $&$ -0.119           $&$ 0.021        $&$   99             $&$   3.4           $&  1 \\    
NGC 1960    &$-0.1020\pm0.0086$&$ -0.050 \pm 0.064 $&$-1.44 \pm0.34 $&$ -939  \pm424     $&$ -31    \pm 72   $&  2 \\
IC 4665     &$ 0.11  \pm0.22  $&$  0.00  \pm 0.10  $&$-0.30 \pm0.37 $&$ -147  \pm519     $&$   6.8  \pm  4.3 $& 22 \\   
NGC 2547    &             --   &                --  &            --  &                --  &               --  &  --\\
IC 2602     &$-0.165 \pm0.095 $&$ -0.230 \pm 0.17  $&$-0.38 \pm0.26 $&$  -91  \pm124     $&$   0.3  \pm  1.2 $&  8 \\    
IC 2391     &$-0.22  \pm0.19  $&$ -0.26  \pm 0.16  $&$-0.37 \pm0.14 $&$  -17  \pm 76     $&$   1.0  \pm  1.5 $&  6 \\    
The Pleiades&             --   &                --  &            --  &                --  &               --  &  --\\
Blanco 1    &$-0.08  \pm0.10  $&$ -0.27  \pm 0.12  $&$ 0.072\pm0.081$&$   71  \pm 30     $&$   0.9  \pm  1.6 $&  2 \\    
The Hyades  &$-0.16  \pm0.16  $&$ -0.406 \pm 0.009 $&$-0.054\pm0.023$&$   87  \pm 29     $&$  -0.08 \pm  0.18$&  3 \\ 
BPMG        &$-0.22  \pm0.15  $&$ -0.04  \pm 0.19  $&$-0.53 \pm0.60 $&$  201  \pm 408    $&$   0.19 \pm  0.32$& 40 \\   
THMG        &$-0.10  \pm0.14  $&$ -0.047 \pm 0.078 $&$-0.25 \pm0.41 $&$  -82  \pm 254    $&$   0.13 \pm  0.14$& 52 \\
32 Ori MG   &$-0.23  \pm0.14  $&$  0.0081\pm 0.0490$&$-0.58 \pm0.43 $&$ -249  \pm 306    $&$   1.08 \pm  0.60$& 13 \\
All         &$-0.12  \pm0.19  $&$ -0.06  \pm 0.15  $&$-0.38 \pm0.48 $&$ -141  \pm 365    $&$   0.8  \pm  9.6 $&149 \\
\hline
$\,$
    \end{tabular}
    \\ 
\begin{flushleft}
  {\bf Notes. } `\#' is the number of sources.
\end{flushleft} 
\end{table*}

\setcounter{table}{2}
\begin{table*}[ht] 
  \caption[Comparison between stellar parameters derived with VOSA and from GES \citep{randich2018}.]
          {Comparison between stellar parameters derived with VOSA and from GES \citep{randich2018}.}
  \label{tab:randich2018_vs_vosa}
  \center
\begin{tabular}{l ccc r}
\hline \hline
  Stellar   &$ \Delta \log g  $&$ \Delta E(B-V) $&$\Delta T_{\rm eff}$& \# \\
association &       [dex]      &      [mag]      &     [K]            &    \\
\hline\hline
\multicolumn{5}{l}{ \qquad {\it All \cite{c_babusiaux2018} sources with counterpart in \cite{randich2018}} } \\
\hline 
IC 4665     &$-0.44 \pm 0.60  $&$    0.056      $&$  51 \pm 175      $&  24\\
NGC 2547    &$-0.48 \pm 0.61  $&$    0.040      $&$ 362 \pm 814      $&  87\\
IC 2602     &$-0.22 \pm 0.64  $&$    0.037      $&$ 329 \pm 634      $&  40\\
IC 2391     &$-0.047\pm 0.44  $&$    0.058      $&$-433 \pm4401      $&  23\\
All         &$-0.36 \pm 0.61  $&$0.0477\pm0.0093$&$ 206 \pm1750      $& 174\\
\hline \hline
\multicolumn{5}{l}{ \qquad {\it Sub-sample with sources considered as members after this work.} } \\
\hline 
IC 4665     &$-0.33 \pm 0.44  $&$    0.056      $&$  43 \pm 160      $&  18\\
NGC 2547    &            --    &$    0.040      $&           --       &  --\\
IC 2602     &$ 0.26           $&$    0.037      $&$ 280              $&   1\\
IC 2391     &            --    &$    0.058      $&           --       &  --\\
All         &$-0.303\pm 0.45  $&$0.0477\pm0.0093$&$  55 \pm 164      $&  19\\
\hline
$\,$
    \end{tabular}
    \\ 
\begin{flushleft}
  {\bf Notes. } `\#' is the number of sources. 
                `$\Delta E(B-V)$' is the difference between the excess derived in \cite{randich2018} and 
                   the one used by us. 
\end{flushleft} 
\end{table*}

\setcounter{table}{3}
%
%
\input{tab_associations_stellar_parameters_v2_0.tex}
\label{tab:associations_stellarproperties_v1}
\clearpage

\setcounter{table}{4}
%
%
\input{tab_alphapersei_phot_bands_v0.tex}
\label{tab:alphapersei_phot_bands}
\clearpage

\setcounter{table}{5}
%
%
\input{tab_ngc1960_phot_bands_v0.tex}
\label{tab:ngc1960_phot_bands_v0}
\clearpage

\setcounter{table}{6}
%
%
\input{tab_ic4665_phot_bands_v0.tex}
\label{tab:ic4665_phot_bands_v0}
\clearpage

\setcounter{table}{7}
%
%
\input{tab_ngc2547_phot_bands_v0.tex}
\label{tab:ngc2547_phot_bands_v0}
\clearpage

\setcounter{table}{8}
%
%
\input{tab_ic2602_ic2391_phot_bands_v0.tex}
\label{tab:ic2602_2391_phot_bands_v0}
\clearpage

\setcounter{table}{10}
%
%
\input{tab_thepleiades_phot_bands_part1_v0.tex}
\label{tab:thepleiades_phot_bands_part1_v0}
\clearpage

\setcounter{table}{11}
%
%
\input{tab_thepleiades_phot_bands_part2_v0.tex}
\label{tab:thepleiades_phot_bands_part2_v0}
\clearpage

\setcounter{table}{12}
%
%
\input{tab_thepleiades_ids_v0.tex}
\label{tab:thepleiades_ids_v0}
\clearpage

\setcounter{table}{13}
%
%
\input{tab_blanco1_thehyades_phot_bands_v0.tex}

\label{tab:blanco1_thehyades_phot_bands_v0}
\clearpage

\setcounter{table}{15}
%
%
\input{tab_thehyades_astrometry_ref_BKM08_PLR17_HJC08_LBB14_v0.tex}
\label{tab:thehyades_astrometry_part1_2_3}
\clearpage

\setcounter{table}{18}
%
%
\input{tab_thehyades_astrometry_ref_MPL18_LSP19_v0.tex}
\label{tab:thehyades_astrometry_part4_5}
\clearpage

\setcounter{table}{20}
%
%
\input{tab_32orimg_ids_v0.tex}
\label{tab:32orimg_ids_v0}
\clearpage

\end{appendix}

\end{document}

%% file: tab_associations_stellar_parameters_v2_0.tex
\onecolumn
\begin{landscape}
\tiny
\begin{longtable}{@{\extracolsep{-15pt}}llrr rrrrrrr r rc r rrr r rrr}
\caption{Stellar properties and astrometric data from \textit{Gaia} DR2 for all LDB samples.}\\
\label{tab:associations_stellarproperties_v1}\\
\hline
{Object} & 
{Stellar} & 
{RA J2000} & 
{Dec. J2000} &
  \multicolumn{2}{c}{$\varpi$}&
  \multicolumn{2}{c}{$\mu_{\alpha} \cos(\delta)$} &
  \multicolumn{2}{c}{$\mu_{\delta}$} &
{$\rho_{\mu_{\alpha} \ast \ \mu_{\delta}}$} &
{EW Li}&
{Ref.}&
{Li} &
{Other} & 
{$d_{MAP}$} &
{$d_{min}$} &
{$d_{max}$} &
{$T_{\mathrm{eff}}\ \chi^{2}$} &
{$L_{\rm bol}$} &
{$-\Delta L_{\rm bol}$} &     
{$+\Delta L_{\rm bol}$} \\    
           &
association&
[deg]   &
{[deg]} &
  \multicolumn{2}{c}{[mas]} &
  \multicolumn{2}{c}{[mas yr$^{-1}$]} &
  \multicolumn{2}{c}{[mas yr$^{-1}$]} &
  &
{[\AA]}&
{} & 
{rich} &
{remarks} &
{[pc]} &
{[pc]} &
{[pc]} &
{[K]} &
{$10^{-3}$ [$L_{\sun}$]} &
{$10^{-3}$ [$L_{\sun}$]} &
{$10^{-3}$ [$L_{\sun}$]} \\
\hline
\hline
AP165 &Alpha Persei& 52.0128333& 50.6719722 & 5.680 &$\pm$ 0.074& 21.707 &$\pm$ 0.140 & -25.362 &$\pm$ 0.104 & 0.125 &       $<0.1$&(1)& N &     & 176 & 171 & 181 & 3\ 282 & 33.02 & 31.37 & 34.81 \\
AP296 &Alpha Persei& 53.9184583& 48.4407222 & 5.339 &$\pm$ 0.223& 13.582 &$\pm$ 0.499 & -16.890 &$\pm$ 0.339 & 0.079 &       $<0.1$&(1)& N &  NM & 187 & 173 & 205 & 3\ 122 & 16.97 & 14.55 & 20.34 \\
\hline
\end{longtable}
 \begin{flushleft}
 $\,$
  \textbf{References for ``EW Li'': }  
         (1) \citealt{stauffer1999}; (2) \citealt{BM99}; (3) \citealt{zapateroosorio1996};
         (4) \citealt{jeffries2013};
         (5) \citealt{manzi2008}; (6) \citealt{jeffries2009d};
         (7) \citealt{jeffries2005};
         (8) \citealt{dobbie2010};
         (9) \citealt{barrado2004d}; (10) \citealt{s_boudreault2009};
        (11) \citealt{gw_marcy1994}; (12) \citealt{basri1996}; (13) \citealt{rebolo1996}; (14) \citealt{oppenheimer1997}; (15) \citealt{stauffer1998a}; (16) \citealt{stauffer1998b}; (17) \citealt{el_martin1998c}; (18) \citealt{el_martin2000c}; (19) \citealt{se_dahm2015}; 
        (20) \citealt{pa_cargile2010c}; (21) \citealt{aj_juarez2014};
        (22) \citealt{el_martin2018}; (23) \citealt{n_lodieu2018c}; 
        (24) \citealt{in_reid2002h}; (25) \citealt{i_song2003}; (26) \citealt{cao_torres2006}; (27) \citealt{l_dasilva2009}; (28) \citealt{ll_kiss2011}; (29) \citealt{a_moor2013b}; (30) \citealt{as_binks2014}; (31) \citealt{al_kraus2014c}; (32) \citealt{l_malo2014b}; (33) \citealt{dr_rodriguez2014}; (34) \citealt{as_binks2016}; (35) \citealt{e_shkolnik2017}; (36) \citealt{l_malo2013}; 
        (37) \citealt{cpmbell2017}.
 \\
 $\,$        
  \textbf{Notes: }
         $^{(1)}$ Undetermined lithium detection (see Section \ref{sub:thehyades}).
         $^{(2)}$ Negative $L_{\rm bol}$ 2.5th value, because $\Delta$~\verb+F_tot+ is bigger than \verb+F_tot+.
         $^{(3)}$ 01132817a and 01132817a, 01535076a and 01535076b, and 02335984a and 02335984b are binary systems only resolved in \textit{Gaia} DR2.
         $^{(4)}$ THOR-14Aa and THOR-14Ab are two sources resolved in \textit{Gaia} DR2 but in the rest of photometric surveys considered are blended.
         $^{(5)}$ J05053333 is the short name of 2MASS J05053333+0044034. 
         $^*$ Peculiar objects described in Appendix \ref{app:bpmg_objects}. 
 \\
 $\,$        
  \textbf{``Other''} column means other remarks: 
         EL, strong emission lines in \ion{He}{i} $\lambda 5\ 876$~\AA\ and $\lambda 6\ 876$~\AA; 
         FR, fast rotator;                         
         NM, no member based in this work; 
         CMS, confirmed multiple system; 
         CP, contaminated photometry; 
         SuB, suspected binary or multiple system;
         P1, subsample with $\varpi<9.52$~mas only for the 32 Ori MG; 
         P2, subsample with $\varpi>9.52$~mas only for the 32 Ori MG.         
 \end{flushleft}  
\end{landscape} 

%% file: tab_alphapersei_phot_bands_v0.tex
\clearpage
\onecolumn
\tiny
\begin{landscape}
\begin{longtable}{l lrc lrc lrc rlrlrlrlc lrc lrc}
\caption{Photometry for the Alpha Persei members retrieved from the literature.}\\
\hline
{Object} & 
\multicolumn{2}{c}{$V$}&
{Ref.} & 
\multicolumn{2}{c}{$R_c$}&
{Ref.} & 
\multicolumn{2}{c}{$I_c$}&
{Ref.} & 
\multicolumn{2}{c}{$g$}&
\multicolumn{2}{c}{$r$}&
\multicolumn{2}{c}{$i$}&
\multicolumn{2}{c}{$z$}&
{Ref.} & 
\multicolumn{2}{c}{$J$}&
{Ref.} & 
\multicolumn{2}{c}{$K^{\prime}$}&
{Ref.} 
\\
        &
\multicolumn{2}{c}{[mag]} &
{} & 
\multicolumn{2}{c}{[mag]} &
{} & 
\multicolumn{2}{c}{[mag]} &
{} & 
\multicolumn{2}{c}{[mag]} &
\multicolumn{2}{c}{[mag]} &
\multicolumn{2}{c}{[mag]} &
\multicolumn{2}{c}{[mag]} &
{} & 
\multicolumn{2}{c}{[mag]} &
{} & 
\multicolumn{2}{c}{[mag]} &
   \\ 
\hline
  AP165 & 17.24 & 0.08 &(1)&       &      &   & 14.66 & 0.08 &(1)&  --   & --    &   --  & --      &  --   & --     &    --  & --    &   & --     & --   &   & --    & --   &   \\
        &       &      &   & 16.135& 0.20 &(5)& 14.665& 0.20 &(5)&       &       &       &         &       &        &        &       &   & --     & --   &   & --    & --   &   \\
  AP296 & 18.57 & 0.10 &(4)&   --  & --   &   & 15.35 & 0.10 &(4)& 19.694& 0.014 & 18.042& 0.007   & 16.367& 0.004  & 15.360 & 0.005 &(8)& --     & --   &   & --    & --   &   \\
  AP284 & 19.44 & 0.10 &(4)&       &      &   & 15.60 & 0.10 &(4)&  --   & --    &   --  & --      &  --   &  --    & --     & --    &   & --     & --   &   & --    & --   &   \\
        &       &      &   & 17.810& 0.20 &(5)& 15.717& 0.20 &(5)&       &       &       &         &       &        &        &       &   & --     & --   &   & --    & --   &   \\
  AP322 &       &      &   & 19.74 & 0.10 &(3)& 17.60 & 0.10 &(3)&  --   & --    & 20.00 & 0.09    &17.71  & 0.02   &        &       &(2)& --     & --   &   & 14.57 & 0.055&(3)\\
  AP268 & 20.50 & 0.10 &(4)&   --  & --   &   & 16.89 & 0.10 &(4)&  --   & --    &   --  & --      &  --   &  --    & --     & --    &   & --     & --   &   & --    & --   &   \\
  AP272 & 20.42 & 0.02 &(7)&   --  & --   &   & 16.65 & 0.017&(7)&  --   & --    &   --  & --      &  --   &  --    & --     & --    &   & --     & --   &   & --    & --   &   \\
  AP275 &       &      &   & 19.45 & 0.10 &(3)& 17.25 & 0.10 &(3)&  --   & --    &   --  & --      &  --   &  --    & --     & --    &   & --     & --   &   & --    & --   &   \\
        & 21.10 & 0.10 &(4)&       &      &   & 17.15 & 0.10 &(4)&       &       &       &         &       &        &        &       &   & --     & --   &   & --    & --   &   \\
  AP325 &   --  & --   &   & 19.94 & 0.10 &(3)& 17.65 & 0.10 &(3)&  --   & --    &   --  & --      &  --   &  --    & --     & --    &   & --     & --   &   & 14.14 & 0.05 &(3)\\
  AP318 &   --  & --   &   & 19.61 & 0.10 &(3)& 17.45 & 0.10 &(3)&  --   & --    &   --  & --      &  --   &  --    & --     & --    &   & --     & --   &   & 14.10 & 0.05 &(3)\\
  AP323 &   --  & --   &   & 19.63 & 0.10 &(3)& 17.50 & 0.10 &(3)&  --   & --    &   --  & --      &  --   &  --    & --     & --    &   & --     & --   &   & 14.33 & 0.05 &(3)\\
  AP313 &   --  & --   &   & 19.68 & 0.10 &(3)& 17.55 & 0.10 &(3)&  --   & --    &   --  & --      &  --   &  --    & --     & --    &   & --     & --   &   & --    & --   &   \\
  AP270 & 21.87 & 0.10 &(4)& 19.534& 0.014&(6)& 17.518& 0.007&(6)&  --   & --    &   --  & --      &  --   &  --    & --     & --    &   & --     & --   &   & --    & --   &   \\
        &       &      &   &       &      &   & 17.80 & 0.10 &(4)&       &       &       &         &       &        &        &       &   &        &      &   &       &      &   \\
  AP300 &  --   & --   &   & 20.03 & 0.10 &(3)& 17.85 & 0.10 &(3)&  --   & --    &   --  & --      &  --   &  --    & --     & --    &   & 14.62  & 0.05 &(3)& --    & --   &   \\
  AP324 &  --   & --   &   & 20.46 & 0.10 &(3)& 18.10 & 0.10 &(3)&  --   & --    &   --  & --      &  --   &  --    & --     & --    &   & --     & --   &   & --    & --   &   \\
  AP315 &  --   & --   &   & 20.54 & 0.10 &(3)& 18.20 & 0.10 &(3)&  --   & --    &   --  & --      &  --   &  --    & --     & --    &   & 14.80  & 0.05 &(3)& --    & --   &   \\
        &       &      &   & 20.247& 0.012&(6)& 18.049& 0.009&(6)&       &       &       &         &       &        &        &       &   & 15.899 & 0.022&(7)& 14.982& 0.032&(7)\\
  AP310 &       &      &   & 20.13 & 0.10 &(3)& 17.80 & 0.10 &(3)&  --   & --    &   --  & --      &  --   &  --    & --     & --    &   & 14.55  & 0.05 &(3)& --    & --   &   \\
  AP317 &  --   & --   &   & 20.14 & 0.10 &(3)& 17.85 & 0.10 &(3)&  --   & --    &   --  & --      &  --   &  --    & --     & --    &   & --     & --   &   & 15.00 & 0.05 &(3)\\
\hline
\end{longtable}
 \begin{flushleft}
 $\,$
  \textbf{References: } 
         (1)~\citealt{prosser1992};
         (2) IPHAS DR2 \citep{barentsen2014a};
         (3)~\citealt{stauffer1999}; 
         (4)~\citealt{prosser1994b}; 
         (5)~\citealt{deacon2004}; 
         (6)~\citealt{barrado2002b};
         (7)~\citealt{zapateroosorio1996};
         (8)~SDSS DR12 \citep{s_alam2015a}.         
 \end{flushleft}  
\end{landscape} 

%% file: tab_ngc1960_phot_bands_v0.tex
\clearpage
\onecolumn
\tiny
\begin{landscape}
\begin{longtable}{l rlrl rlrlrl rlrlrl rl}
\caption{Photometry for the NGC 1960 members retrieved from the literature.}\\
\hline
{Object} & 
\multicolumn{2}{c}{$R_c$}&
\multicolumn{2}{c}{$I_c$}&
\multicolumn{2}{c}{$g$}&
\multicolumn{2}{c}{$r$}&
\multicolumn{2}{c}{$i$}&
\multicolumn{2}{c}{$r_{\rm\ IPHAS}$}&
\multicolumn{2}{c}{$i_{\rm\ IPHAS}$}&
\multicolumn{2}{c}{$\rm{H}\alpha_{\rm\ IPHAS}$}&
\multicolumn{2}{c}{$J$}
\\
        &
\multicolumn{2}{c}{[mag]} &
\multicolumn{2}{c}{[mag]} &
\multicolumn{2}{c}{[mag]} &
\multicolumn{2}{c}{[mag]} &
\multicolumn{2}{c}{[mag]} &
\multicolumn{2}{c}{[mag]} &
\multicolumn{2}{c}{[mag]} &
\multicolumn{2}{c}{[mag]} &
\multicolumn{2}{c}{[mag]} 
\\
\hline
   876  & 17.34 & 0.13 & 16.508 & 0.008 & 18.87 & 0.15 &(17.639 & 0.14)&(16.922 & 0.009)& 17.534 & 0.050 & 16.638 & 0.051 & 17.151 & 0.051 & 15.211 & 0.050 \\
   827  & 17.50 & 0.13 & 16.583 & 0.008 & 19.28 & 0.15 &(17.928 & 0.14)&(17.086 & 0.009)& 17.838 & 0.051 & 16.897 & 0.051 & 17.344 & 0.051 & 15.434 & 0.051 \\
   829  & 17.46 & 0.13 & 16.591 & 0.008 & 19.20 & 0.15 &(17.841 & 0.14)&(17.063 & 0.009)& 17.542 & 0.050 & 16.728 & 0.051 & 17.119 & 0.051 & 15.323 & 0.051 \\
   1056 & 18.03 & 0.13 & 16.979 & 0.008 & 19.88 & 0.16 & 18.462 & 0.15 & 17.496 & 0.009 & 18.298 & 0.051 & 17.180 & 0.051 & 17.789 & 0.052 & 15.918 & 0.054 \\
   1269 & 18.43 & 0.13 & 17.284 & 0.009 & 20.42 & 0.17 & 18.962 & 0.15 & 17.848 & 0.010 & 18.832 & 0.052 & 17.592 & 0.053 & 18.282 & 0.055 & 16.031 & 0.052 \\
   1291 & 18.41 & 0.13 & 17.181 & 0.009 & 20.29 & 0.17 & 18.894 & 0.15 & 17.736 & 0.010 & 18.821 & 0.053 & 17.500 & 0.053 & 18.147 & 0.054 & 15.879 & 0.052 \\
   1042 & 17.66 & 0.16 & 16.729 & 0.012 &     - &    - &      - &    - &      - &     - & 17.841 & 0.050 & 16.895 & 0.051 & 17.401 & 0.051 & 15.547 & 0.052 \\
   1833 & 18.57 & 0.13 & 17.357 & 0.009 & 20.54 & 0.17 & 19.107 & 0.15 & 17.919 & 0.010 & 19.075 & 0.054 & 17.809 & 0.054 & 18.542 & 0.060 & 16.043 & 0.057 \\
   1859 & 19.40 & 0.15 & 17.863 & 0.010 & 21.24 & 0.23 & 19.794 & 0.17 & 18.371 & 0.010 & 19.425 & 0.058 & 17.989 & 0.057 & 18.814 & 0.064 & 16.469 & 0.056 \\
   1860 & 19.38 & 0.14 & 17.853 & 0.010 & 21.33 & 0.31 & 19.949 & 0.29 & 18.490 & 0.020 & 19.485 & 0.059 & 17.928 & 0.056 & 18.763 & 0.063 & 16.192 & 0.054 \\
   1878 & 19.24 & 0.14 & 17.712 & 0.010 & 21.33 & 0.20 & 19.803 & 0.16 & 18.306 & 0.010 & 19.722 & 0.064 & 18.042 & 0.057 & 18.958 & 0.067 & 16.171 & 0.052 \\
   1545 & 18.83 & 0.13 & 17.509 & 0.009 & 20.67 & 0.18 &(19.278 & 0.15)&(18.047 & 0.010)& 19.186 & 0.055 & 17.735 & 0.054 & 18.486 & 0.058 & 16.119 & 0.052 \\
   1540 & 19.23 & 0.14 & 17.678 & 0.010 & 21.13 & 0.19 & 19.762 & 0.16 & 18.269 & 0.010 & 19.760 & 0.033 & 17.909 & 0.056 & 18.710 & 0.061 & 16.198 & 0.052 \\
   2171 & 19.81 & 0.16 & 18.191 & 0.011 & 22.50 & 0.56 & 20.538 & 0.37 & 18.999 & 0.024 & 20.351 & 0.079 & 18.540 & 0.065 & 19.529 & 0.087 & 16.815 & 0.057 \\
   1871 & 19.56 & 0.15 & 18.015 & 0.011 & 21.56 & 0.22 & 20.098 & 0.17 & 18.602 & 0.011 & 20.102 & 0.075 & 18.342 & 0.062 & 19.261 & 0.078 & 16.664 & 0.063 \\
   2188 & 19.76 & 0.16 & 18.179 & 0.011 &(21.72 & 0.24)& 20.336 & 0.18 &(18.822 & 0.011)& 19.971 & 0.071 & 18.510 & 0.066 & 19.199 & 0.075 & 16.654 & 0.063 \\
   2249 & 20.01 & 0.17 & 18.270 & 0.012 & 22.18 & 0.28 & 20.605 & 0.18 & 18.928 & 0.011 & 19.966 & 0.087 & 18.450 & 0.078 & 19.936 & 0.169 & 16.753 & 0.062 \\
   2173 & 19.95 & 0.16 & 18.334 & 0.012 & 22.03 & 0.26 & 20.573 & 0.18 & 18.952 & 0.011 & 20.667 & 0.088 & 18.791 & 0.068 & 19.986 & 0.113 & 16.703 & 0.063 \\
   2696 & 20.27 & 0.18 & 18.584 & 0.014 & 22.38 & 0.30 & 20.942 & 0.19 & 19.240 & 0.012 & 21.385 & 0.184 & 18.966 & 0.084 & 20.407 & 0.178 & 17.106 & 0.062 \\
   2214 & 19.80 & 0.16 & 18.213 & 0.012 & 21.87 & 0.24 & 20.424 & 0.17 & 18.856 & 0.011 & 20.460 & 0.093 & 18.514 & 0.066 & 19.493 & 0.089 & 16.855 & 0.058 \\
   2663 & 20.11 & 0.17 & 18.412 & 0.013 & 22.13 & 0.27 & 20.741 & 0.19 & 19.054 & 0.012 & 20.627 & 0.101 & 18.852 & 0.078 & 20.122 & 0.141 & 17.087 & 0.075 \\
   2703 & 20.18 & 0.17 & 18.490 & 0.013 & 22.31 & 0.28 & 20.726 & 0.18 & 19.097 & 0.012 & 20.860 & 0.124 & 19.005 & 0.084 & 20.061 & 0.134 & 17.041 & 0.061 \\
   2672 & 20.39 & 0.18 & 18.658 & 0.014 & 22.40 & 0.40 & 21.063 & 0.35 & 19.305 & 0.022 & 20.863 & 0.120 & 18.985 & 0.085 & 20.106 & 0.139 & 17.174 & 0.065 \\
   3081 & 20.42 & 0.31 & 18.711 & 0.021 & 22.59 & 0.40 & 21.241 & 0.25 & 19.380 & 0.015 & 20.464 & 0.101 & 18.952 & 0.089 & 19.888 & 0.133 & 17.189 & 0.083 \\
   3150 & 20.67 & 0.20 & 18.843 & 0.016 & 22.77 & 0.35 & 21.407 & 0.22 & 19.532 & 0.013 & 21.809 & 0.266 & 19.245 & 0.100 & 20.970 & 0.292 & 17.353 & 0.067 \\
   3080 & 20.69 & 0.20 & 18.795 & 0.016 & 22.97 & 0.39 & 21.333 & 0.22 & 19.442 & 0.013 & 20.488 & 0.106 & 19.215 & 0.111 & 20.149 & 0.174 &     -  &    -  \\
   3590 & 20.97 & 0.22 & 19.107 & 0.019 & 23.56 & 0.49 & 21.751 & 0.26 & 19.779 & 0.015 &     -  &    -  & 19.902 & 0.159 & 21.048 & 0.304 & 17.551 & 0.082 \\
   3073 & 20.58 & 0.20 & 18.785 & 0.016 & 22.61 & 0.32 & 21.289 & 0.22 & 19.460 & 0.013 &(22.952 & 0.674)& 19.022 & 0.082 &(21.819 & 0.591)& 17.884 & 0.139 \\
   3028 & 20.97 & 0.22 & 19.060 & 0.019 & 23.43 & 0.46 & 21.675 & 0.24 & 19.754 & 0.014 & 21.813 & 0.240 & 19.681 & 0.130 & 20.768 & 0.229 & 17.606 & 0.089 \\
   3612 & 21.06 & 0.24 & 19.152 & 0.021 & 23.35 & 0.45 & 21.618 & 0.24 & 19.822 & 0.015 &     -  &    -  & 19.892 & 0.162 &     -  &    -  & 17.745 & 0.085 \\
   3596 & 21.08 & 0.23 & 19.185 & 0.020 & 23.17 & 0.40 & 21.651 & 0.24 & 19.895 & 0.015 & 22.251 & 0.395 & 19.896 & 0.166 & 20.652 & 0.221 & 17.982 & 0.109 \\
   4165 & 21.46 & 0.27 & 19.469 & 0.025 & 23.89 & 0.54 & 22.257 & 0.30 & 20.213 & 0.018 &     -  &    -  & 20.938 & 0.482 &     -  &    -  & 17.978 & 0.102 \\
\hline
\end{longtable}
 \begin{flushleft}
 $\,$
  \textbf{Note: }The photometric values in brackets correspond to cases where we encountered problems in the SED reconstruction procedure.\\
 $\,$
  \textbf{References: } 
           $R_{c}I_{c}$ are taken from \cite{jeffries2013};
           $gri$ from \cite{cpmbell2013}; 
           $ri\rm{H}\alpha_{\rm\ IPHAS}$ from \cite{barentsen2014a}; 
           $J$ from Galindo-Guil et al. (in prep).
 \end{flushleft}  
\end{landscape} 

%% file: tab_ic4665_phot_bands_v0.tex
\onecolumn
\tiny
\begin{longtable}{lc rlrl rlrl}
\caption{Photometry for the IC 4665 members retrieved from the literature.}\\
\hline
{Object} & 
\multicolumn{1}{l}{ID} & 
\multicolumn{2}{c}{$V$}&
\multicolumn{2}{c}{$I_c$}&
\multicolumn{2}{c}{$I_{\rm\ CFHT12}$}&
\multicolumn{2}{c}{$z_{\rm\ CFHT12}$}
\\
        &
\multicolumn{1}{l}{JCO}   &
\multicolumn{2}{c}{[mag]} &
\multicolumn{2}{c}{[mag]} &
\multicolumn{2}{c}{[mag]} &
\multicolumn{2}{c}{[mag]} 
\\
\hline
  P059         &          &     -  &    - &      - &    - & 12.614 & 0.090 & 12.346 & 0.070\\
  P002         & JCO6-240 & 11.658 & 0.30 & 12.249 & 0.30 &      - &     - &      - &   -  \\
  P065         &          &     -  &    - &      - &    - & 12.440 & 0.090 & 12.17 & 0.070 \\
  JCO3-285     &          & 13.336 & 0.30 & 14.156 & 0.30 &      - &     - &      - &   -  \\
  P331         & JCO5-191 & 15.163 & 0.30 & 16.520 & 0.30 & 13.705 & 0.090 & 13.449 & 0.070\\
  P215         &          &     -  &    - &      - &    - & 13.413 & 0.090 & 13.154 & 0.070\\
  P377         & JCO7-066 & 14.976 & 0.30 & 16.236 & 0.30 & 13.680 & 0.090 & 13.403 & 0.070\\
  JCO1-530     &          & 12.518 & 0.30 & 13.305 & 0.30 &      - &     - &      - &   -  \\
  P020         & JCO5-282 & 12.146 & 0.30 & 12.881 & 0.30 &      - &     - &      - &   -  \\
  P290         &          &     -  &    - &      - &    - & 13.443 & 0.090 & 13.134 & 0.070\\
  JCO2-145     &          & 12.272 & 0.30 & 13.040 & 0.30 &      - &     - &      - &   -  \\
  P139         & JCO7-079 & 13.088 & 0.30 & 13.935 & 0.30 &      - &     - &      - &   -  \\
  P030         & JCO2-373 & 12.456 & 0.30 & 13.339 & 0.30 &      - &     - &      - &   -  \\
  JCO2-213     & JCO2-213 & 13.218 & 0.30 & 14.151 & 0.30 &      - &     - &      - &   -  \\
  P107         & JCO8-257 & 12.921 & 0.30 & 13.972 & 0.30 &      - &     - &      - &   -  \\
  JCO1-427     &          & 13.069 & 0.30 & 14.066 & 0.30 &      - &     - &      - &   -  \\
  P150         & JCO7-021 & 13.126 & 0.30 & 14.145 & 0.30 &      - &     - &      - &   -  \\
  P165         & JCO4-053 & 13.528 & 0.30 & 14.692 & 0.30 &      - &     - &      - &   -  \\
  JCO6-095     &          & 13.511 & 0.30 & 14.524 & 0.30 &      - &     - &      - &   -  \\
  P060         &          &     -  &    - &      - &    - & 12.623 & 0.090 & 12.295 & 0.070\\
  P151         & JCO4-226 & 13.781 & 0.30 & 14.781 & 0.30 &      - &     - &      - &   -  \\
  JCO3-357     &          & 13.517 & 0.30 & 14.619 & 0.30 &      - &     - &      - &   -  \\
  P075         &          &     -  &    - &      - &    - & 12.697 & 0.090 & 12.492 & 0.070\\
  P108         & JCO5-296 & 13.940 & 0.30 & 15.021 & 0.30 & 12.718 & 0.090 & 12.542 & 0.070\\
  P071         &          &     -  &    - &      - &    - & 12.696 & 0.090 & 12.493 & 0.070\\
  JCO9-120     &          & 14.405 & 0.30 & 15.506 & 0.30 &      - &     - &      - &   -  \\
  P100         & JCO5-179 & 14.407 & 0.30 & 15.647 & 0.30 & 13.074 & 0.090 & 12.841 & 0.070\\
  P094         & JCO5-280 & 14.243 & 0.30 & 15.337 & 0.30 &      - &     - &      - &   -  \\
  JCO3-065     &          & 15.045 & 0.30 & 16.594 & 0.30 &      - &     - &      - &   -  \\
  JCO4-337     &          & 13.913 & 0.30 & 15.296 & 0.30 &      - &     - &      - &   -  \\
  P077         & JCO9-281 & 14.956 & 0.30 & 16.377 & 0.30 &      - &     - &      - &   -  \\
  P232         & JCO6-111 & 15.993 & 0.30 & 17.773 & 0.30 & 14.147 & 0.090 & 13.804 & 0.070\\
  P411         & JCO4-591 & 16.236 & 0.30 & 18.113 & 0.30 &      - &     - &      - &   -  \\
  JCO2-220     & JCO2-220 & 16.420 & 0.30 & 18.443 & 0.30 &      - &     - &      - &   -  \\
  P374         & JCO4-437 & 16.716 & 0.30 & 18.795 & 0.30 & 14.554 & 0.090 & 14.155 & 0.070\\
  JCO7-670     &          & 17.349 & 0.30 & 19.925 & 0.30 &      - &     - &      - &   -  \\
  JCO2-637     &          & 17.200 & 0.30 & 19.945 & 0.30 &      - &     - &      - &   -  \\
  JCO3-770     &          & 17.561 & 0.30 & 20.260 & 0.30 &      - &     - &      - &   -  \\
  P315         &          &     -  &    - &      - &    - & 14.876 & 0.090 & 14.398 & 0.070\\
  P309         & JCO5-472 & 16.910 & 0.30 & 19.075 & 0.30 &      - &     - &      - &   -  \\
  P265         & JCO5-521 & 17.062 & 0.30 & 19.356 & 0.30 &      - &     - &      - &   -  \\
  JCO8-550     & JCO8-550 & 17.872 & 0.30 & 20.711 & 0.30 &      - &     - &      - &   -  \\
  P336         & JCO8-364 & 17.530 & 0.30 & 20.152 & 0.30 & 14.910 & 0.090 & 14.443 & 0.070\\
  P343         & JCO5-515 & 17.851 & 0.30 & 20.592 & 0.30 & 15.228 & 0.090 & 14.778 & 0.070\\
  P400         & JCO4-459 & 17.893 & 0.30 & 20.470 & 0.30 & 15.269 & 0.090 & 14.815 & 0.070\\
  P248         & JCO8-395 & 17.651 & 0.30 & 20.167 & 0.30 &      - &     - &      - &   -  \\
  P354         &          &     -  &    - &      - &    - & 15.278 & 0.090 & 14.800 & 0.070\\
  P283         &          &     -  &    - &      - &    - & 15.474 & 0.090 & 14.987 & 0.070\\
  P396         & JCO7-088 & 15.411 & 0.30 & 17.293 & 0.30 & 13.844 & 0.090 & 13.457 & 0.070\\
  P348         &          &     -  &    - &      - &    - & 15.741 & 0.090 & 15.177 & 0.070\\
  P335         &          &     -  &    - &      - &    - & 15.902 & 0.090 & 15.318 & 0.070\\
  P233         &          &     -  &    - &      - &    - & 16.074 & 0.090 & 15.501 & 0.070\\
  P350         &          &     -  &    - &      - &    - & 16.475 & 0.090 & 15.854 & 0.070\\
  P380         &          &     -  &    - &      - &    - & 16.116 & 0.090 & 15.544 & 0.070\\
  P373         &          &     -  &    - &      - &    - & 16.169 & 0.090 & 15.537 & 0.070\\
  P313         &          &     -  &    - &      - &    - & 16.736 & 0.090 & 15.995 & 0.070\\
  P344         &          &     -  &    - &      - &    - & 16.650 & 0.090 & 16.047 & 0.070\\
  A.09.30.47   &          &     -  &    - &      - &    - & 15.940 & 0.090 & 15.291 & 0.070\\
  A.10.30.316  &          &     -  &    - &      - &    - & 16.236 & 0.090 & 15.566 & 0.070\\
  A.09.30.14   &          &     -  &    - &      - &    - & 16.840 & 0.090 & 16.118 & 0.070\\
  A.08.30.655  &          &     -  &    - &      - &    - & 16.940 & 0.090 & 16.241 & 0.070\\
  P333         &          &     -  &    - &      - &    - & 16.706 & 0.090 & 16.069 & 0.070\\
  P398         &          &     -  &    - &      - &    - & 15.373 & 0.090 & 14.789 & 0.070\\
  P372         &          &     -  &    - &      - &    - & 16.608 & 0.090 & 16.028 & 0.070\\
  P338         &          &     -  &    - &      - &    - & 16.685 & 0.090 & 16.100 & 0.070\\
  A.05.30.3622 &          &     -  &    - &      - &    - & 17.248 & 0.090 & 16.547 & 0.070\\
  P238         &          &     -  &    - &      - &    - & 16.928 & 0.090 & 16.314 & 0.070\\
\hline
\end{longtable}
 \begin{flushleft}
 $\,$
  \textbf{Notes: }           
           ``ID JCO'' means the identification used in \cite{jeffries2009d}.\\ 
 $\,$
  \textbf{References: } 
           $VI_{c}$ are taken from \cite{jeffries2009d};
           $I_{\rm\ CFHT12}\ z_{\rm\ CFHT12}$ from \cite{wj_dewit2006a}.
 \end{flushleft}  

%% file: tab_ngc2547_phot_bands_v0.tex
\onecolumn
\tiny
\begin{landscape}
\begin{longtable}{l rlrlrl rlrlrl rlrlrlrlrl}
\caption{Photometry for the NGC 2547 members retrieved from the literature.}\\
\hline
{Object} & 
\multicolumn{2}{c}{$B$}&
\multicolumn{2}{c}{$V$}&
\multicolumn{2}{c}{$I_c$}&

\multicolumn{2}{c}{$R_c$}&
\multicolumn{2}{c}{$I_c$}&
\multicolumn{2}{c}{$Z$}&

\multicolumn{2}{c}{[3.6]}&
\multicolumn{2}{c}{[4.5]}&
\multicolumn{2}{c}{[5.8]}&
\multicolumn{2}{c}{[8.0]}&
\multicolumn{2}{c}{[24]}
\\
        &
\multicolumn{2}{c}{[mag]} &
\multicolumn{2}{c}{[mag]} &
\multicolumn{2}{c}{[mag]} &

\multicolumn{2}{c}{[mag]} &
\multicolumn{2}{c}{[mag]} &
\multicolumn{2}{c}{[mag]} &

\multicolumn{2}{c}{[mag]} &
\multicolumn{2}{c}{[mag]} &
\multicolumn{2}{c}{[mag]} &
\multicolumn{2}{c}{[mag]} &
\multicolumn{2}{c}{[mag]} 
\\
\hline
  JO-30 & 22.293 & 0.238 & 20.466 & 0.047 & 17.152 & 0.066 & 19.200 & 0.022 & 17.252 & 0.012 & 16.693 & 0.017 & 14.06 & 0.01 & 13.99 & 0.02 & 14.16 & 0.16 & 14.34 & 0.59 &     - &   - \\
  JO-01 & 21.119 & 0.088 & 19.625 & 0.028 & 16.444 & 0.041 & 18.259 & 0.008 & 16.508 & 0.005 & 16.012 & 0.007 & 13.60 & 0.01 & 13.52 & 0.01 & 13.54 & 0.06 & 13.19 & 0.08 & 10.65 & 0.15\\
  JO-02 & 21.175 & 0.088 & 19.495 & 0.024 & 16.401 & 0.034 & 18.239 & 0.008 & 16.524 & 0.005 & 16.028 & 0.007 & 13.59 & 0.01 & 13.48 & 0.01 & 13.32 & 0.05 & 13.26 & 0.08 &     - &   - \\
  JO-12 & 21.439 & 0.131 & 20.110 & 0.046 & 16.806 & 0.065 & 18.752 & 0.017 & 16.918 & 0.010 & 16.360 & 0.014 & 13.79 & 0.01 & 13.69 & 0.02 & 13.48 & 0.09 & 13.48 & 0.20 &     - &   - \\
\hline
\end{longtable}
 \begin{flushleft}
 $\,$
  \textbf{References: } 
          $BVI_{c}$ are taken from \cite{t_naylor2002};
          $R_{c}I_{c}Z$ from \cite{jeffries2005}; 
          [3.6] [4.5] [5.8] [8.0] [24] from \cite{n_gorlova2007}. 
 \end{flushleft}  
\end{landscape} 

%% file: tab_ic2602_ic2391_phot_bands_v0.tex
\onecolumn
\tiny
\begin{longtable}{lc}
\caption{Photometry for the IC 2602 members retrieved from \cite{dobbie2010}.}\\
\hline
{Object} & 
\multicolumn{1}{l}{$I_{c}/Iwp_{\rm\ ESO\ 845}$} 
\\
        &
\multicolumn{1}{c}{[mag]} 
\\
\hline
  10443357 & (15.36)\\
  10370251 & (16.35)\\
  10422712 & (14.79)\\
  10423714 & (14.85)\\
  10420463 & (14.92)\\
  10464783 & (15.00)\\
  10452984 & (15.11)\\
  10391188 & (15.36)\\
  10420419 & (15.10)\\
  10405026 & (15.33)\\
  10430890 & (15.56)\\
  10430236 & (15.80)\\
  10401542 & (15.98)\\
  10454174 & (16.36)\\
\hline
\end{longtable}
 \begin{flushleft}
 $\,$
  \textbf{Note: }photometric values in brackets correspond to cases 
  where we encountered problems in the SED reconstruction procedure.\\ 
 $\,$
  \textbf{References: } 
           $I_{c}/Iwp_{\rm\ ESO\ 845}$ is taken from \cite{dobbie2010}.\\
 \end{flushleft}   

\bigskip

\setcounter{table}{9}
\tiny
\begin{longtable}{l rlrlrl c rl}
\caption{Photometry for the IC 2391 members retrieved from the literature.}\\
\hline
{Object} & 
\multicolumn{2}{c}{$V$}&
\multicolumn{2}{c}{$R_c$}&
\multicolumn{2}{c}{$I_c$}&
{Ref.} & 
\multicolumn{2}{c}{$Z$}
\\
        &
\multicolumn{2}{c}{[mag]} &
\multicolumn{2}{c}{[mag]} &
\multicolumn{2}{c}{[mag]} &
{ } & 
\multicolumn{2}{c}{[mag]} 
\\
\hline
  CTIO-017 & 19.30 & 0.15 & 17.741 & 0.060 & 15.989 & 0.040 &(1)&    -  &     - \\
  PP-07    & 17.31 & 0.15 & 15.880 & 0.060 & 14.110 & 0.040 &(1)&    -  &     - \\
           & 17.21 & 0.34 & 15.890 & 0.310 & 14.110 & 0.280 &(2)&    -  &     - \\
  CTIO-059 & 19.47 & 0.15 & 17.869 & 0.060 & 16.050 & 0.040 &(1)&15.420 & 0.020 \\
  20-014   &    -  &    - &      - &     - &      - &    -  &   &    -  &     - \\
\hline
\end{longtable}
 \begin{flushleft}
 $\,$
  \textbf{References: }
           $VR_{c}I_{c}$ are taken from (1) \citealt{barrado2001d} or (2) \citealt{patten1999}; 
           $Z$ is taken from  \cite{barrado2001d}.
 \end{flushleft}  

%% file: tab_thepleiades_phot_bands_part1_v0.tex
\onecolumn
\tiny
\begin{landscape}
\begin{longtable}{l rlrlrlrlrl rl rlrlrl}
\caption{Photometry for the Pleaides members retrieved from the literature, part I.}\\
\hline
{Object} & 
\multicolumn{2}{c}{$u$}&
\multicolumn{2}{c}{$g$}&
\multicolumn{2}{c}{$r$}&
\multicolumn{2}{c}{$i$}&
\multicolumn{2}{c}{$z$}&
\multicolumn{2}{c}{$Y$}&
\multicolumn{2}{c}{$J$}&
\multicolumn{2}{c}{$H$}&
\multicolumn{2}{c}{$K$}
\\
        &
\multicolumn{2}{c}{[mag]} &
\multicolumn{2}{c}{[mag]} &
\multicolumn{2}{c}{[mag]} &
\multicolumn{2}{c}{[mag]} &
\multicolumn{2}{c}{[mag]} &
\multicolumn{2}{c}{[mag]} &
\multicolumn{2}{c}{[mag]} &
\multicolumn{2}{c}{[mag]} &
\multicolumn{2}{c}{[mag]} 
\\
\hline
  HCG 509    &(19.143 & 0.006)& 16.846 & 0.002 & 15.575 & 0.001 &      - &     - &      - &     - &      - &     - & 11.160 & 0.022 & 10.587 & 0.021 & 10.289 & 0.018\\
  HCG 332    &(20.278 & 0.005)& 18.093 & 0.001 & 16.793 & 0.001 &      - &     - & 13.797 & 0.050 &      - &     - & 12.149 & 0.040 & 11.632 & 0.027 & 11.282 & 0.060\\
  HHJ 36     &(22.328 & 0.037)& 20.153 & 0.005 & 18.773 & 0.003 & 17.012 & 0.001 & 15.950 & 0.051 & 14.919 & 0.050 & 14.290 & 0.050 & 13.736 & 0.050 & 13.436 & 0.050\\
  HHJ 6      &      - &     - & 21.703 & 0.027 & 20.199 & 0.016 & 17.928 & 0.002 & 16.661 & 0.052 & 15.464 & 0.050 & 14.709 & 0.050 & 14.124 & 0.050 & 13.686 & 0.050\\
  HHJ 16     &(23.628 & 0.114)& 21.029 & 0.012 & 19.641 & 0.008 & 17.632 & 0.001 & 16.523 & 0.051 & 15.408 & 0.050 & 14.750 & 0.050 & 14.157 & 0.050 & 13.818 & 0.050\\
\hline
\end{longtable}
 \begin{flushleft}
 $\,$
  \textbf{Note: }photometric values in brackets correspond to cases 
  where we encountered problems in the SED reconstruction procedure.\\ 
 $\,$
  \textbf{References: } 
           $ugrizYJHK$ are taken from \cite{h_bouy2015b}.
 \end{flushleft}  
\end{landscape} 

%% file: tab_thepleiades_phot_bands_part2_v0.tex
\onecolumn
\tiny
\begin{landscape}
\begin{longtable}{l rlrlc rlrlc rlrlrlc}
\caption{Photometry for the Pleiades members retrieved from the literature, part II.}\\
\hline
{Object} & 
\multicolumn{2}{c}{$V$}&
\multicolumn{2}{c}{$I$}&
\multicolumn{1}{c}{Ref.}&
\multicolumn{2}{c}{$R$}&
\multicolumn{2}{c}{$I$}&
\multicolumn{1}{c}{Ref.}&
\multicolumn{2}{c}{$J$}&
\multicolumn{2}{c}{$H$}&
\multicolumn{2}{c}{$K$}&
\multicolumn{1}{c}{Ref.}
\\
        &
\multicolumn{2}{c}{[mag]} &
\multicolumn{2}{c}{[mag]} &
   &
\multicolumn{2}{c}{[mag]} &
\multicolumn{2}{c}{[mag]} &
   &
\multicolumn{2}{c}{[mag]} &
\multicolumn{2}{c}{[mag]} &
\multicolumn{2}{c}{[mag]} &
\\
\hline
  HCG 509    &(16.20 & 0.10)&(15.30 & 0.10)&(1)& 14.89 & 0.21 & 13.20 & 0.12 &(2)& 11.10 & 0.02 & 10.59 & 0.02 & 10.26 & 0.04 &(2)\\
  HCG 332    &(17.40 & 0.10)&(16.30 & 0.10)&(1)& 15.86 & 0.21 & 14.18 & 0.12 &(2)& 12.21 & 0.02 & 11.51 & 0.02 & 11.30 & 0.04 &(2)\\
  HHJ 36     &     - &    - &     - &    - &   & 18.14 & 0.21 & 16.30 & 0.12 &(2)& 13.32 & 0.04 & 12.76 & 0.04 & 12.49 & 0.04 &(2)\\
  HHJ 6      &     - &    - & 16.99 & 0.20 &(4)& 19.34 & 0.21 & 17.15 & 0.12 &(2)& 14.67 & 0.04 & 14.06 & 0.04 & 13.71 & 0.04 &(2)\\
  HHJ 16     &     - &    - &     - &    - &   &     - &    - &     - &    - &   &    -  &    - &     - &    - &     - &    - &   \\
\hline
\end{longtable}
 \begin{flushleft} 
 $\,$
  \textbf{References: }
           (1) \cite{nc_hambly1993a};        
           (2) \cite{ia_steele1993};
           (3) \cite{stauffer1994e}; 
           (4) \cite{stauffer1998b}; 
           (5) \cite{j_bouvier1998}.   
 \end{flushleft}  
\end{landscape} 

%% file: tab_thepleiades_ids_v0.tex
\onecolumn
\tiny
\begin{landscape}
\begin{longtable}{lrc rrr}
\caption{Other identifiers for the Pleiades LDB sample.}\\
\hline
{Object} & 
\multicolumn{1}{r}{Other id.} & 
\multicolumn{1}{c}{DANCe id.} & 
\multicolumn{1}{r}{$P_{B}$} & 
\multicolumn{1}{r}{$P_{O}$} & 
\multicolumn{1}{r}{$P_{\rm emb}$} \\ 
\hline  
  HCG 509    & HHJ430, V V692 Tau,                  & J035630.43+241718.5 & 0.1576 & 0.173278 & 0.855910\\
  HCG 332    & SK362, HHJ339, V348 Tau,             & J034826.59+231129.7 & 0.8864 & 0.658344 & 0.188240\\
  HHJ 36     &                                      & J034915.62+232249.3 & 0.9995 & 0.999211 & 0.007580\\
  HHJ 6      &                                      & J034142.41+235457.2 & 0.9994 & 0.998524 & 0.956460\\
  HHJ 16     &                                      & J034729.93+233315.1 & 0.9987 & 0.998292 & 0.001710\\
  HHJ 10     &                                      & J034835.19+225342.3 & 0.9926 & 0.999504 & 0.129050\\
\hline
\end{longtable}
 \begin{flushleft}
 $\,$
  \textbf{Notes: } 
           ``Other id.'' compiles other identifiers from several previous works.
           ``DANCe id.'' are the DANCe \citep{h_bouy2015b} identifier.  
           $P_{B}$ is the membership probability calculated in \cite{h_bouy2015b}.
           $P_{O}$ is the membership probability calculated in \cite{j_olivares2018b}.
           $P_{\rm eqm}$ is the equal-mass binaries membership probability calculated in \cite{j_olivares2018b}.
 \end{flushleft}  
\end{landscape} 

%% file: tab_blanco1_thehyades_phot_bands_v0.tex
\onecolumn
\tiny
\begin{longtable}{l rl rl rl rl rl c}
\caption{Photometry for the Blanco 1 members retrieved from the literature.}\\
\hline
{Object} & 
\multicolumn{2}{c}{$B$}&
\multicolumn{2}{c}{$V$}&
\multicolumn{2}{c}{$I_c$}&
\multicolumn{2}{c}{$z_{\rm\ CFHT12}$}&
\multicolumn{2}{c}{$K_{s}$}&
\multicolumn{1}{c}{Ref.} \\
        &
\multicolumn{2}{c}{[mag]} &
\multicolumn{2}{c}{[mag]} &
\multicolumn{2}{c}{[mag]} &
\multicolumn{2}{c}{[mag]} &
\multicolumn{2}{c}{[mag]} &
        \\
\hline
  JCO-F18-88 & 16.085 & 0.070 & 14.829 & 0.022 & 13.345& 0.003&    -  &   -  & 11.622& 0.029&(2)\\
  JCO-F2-78  &     -  &    -  &      - &     - & 14.339& 0.018&    -  &   -  & 12.373& 0.038&(2)\\
  JCO-F13-55 &     -  &    -  &      - &     - & 15.757& 0.033&    -  &   -  & 13.594& 0.069&(2)\\
  JCO-F9-190 &     -  &    -  &      - &     - & 16.230& 0.089&    -  &   -  & 14.114& 0.140&(2)\\
  CFHT-BL-16 &     -  &    -  &      - &     - & 18.330& 0.050& 17.60 & 0.05 & 15.460& 0.030&(1)\\
  CFHT-BL-22 &     -  &    -  &      - &     - & 18.500& 0.050& 17.72 & 0.05 & 15.580& 0.030&(1)\\
  CFHT-BL-25 &     -  &    -  &      - &     - & 18.650& 0.050& 17.86 & 0.05 & 15.570& 0.030&(1)\\
  CFHT-BL-38 &     -  &    -  &      - &     - & 19.010& 0.050& 18.19 & 0.05 & 15.890& 0.030&(1)\\
  CFHT-BL-36 &     -  &    -  &      - &     - & 18.990& 0.050& 18.10 & 0.05 & 15.860& 0.030&(1)\\
  CFHT-BL-24 &     -  &    -  &      - &     - & 18.540& 0.050& 17.80 & 0.05 & 15.570& 0.030&(1)\\
  CFHT-BL-29 &     -  &    -  &      - &     - & 18.800& 0.050& 17.99 & 0.05 & 15.750& 0.030&(1)\\
  CFHT-BL-45 &     -  &    -  &      - &     - & 19.330& 0.050& 18.44 & 0.05 & 16.040& 0.030&(1)\\
  CFHT-BL-43 &     -  &    -  &      - &     - & 19.110& 0.050& 18.28 & 0.05 & 15.960& 0.030&(1)\\
  CFHT-BL-49 &     -  &    -  &      - &     - & 19.490& 0.050& 18.57 & 0.05 & 15.910& 0.030&(1)\\
\hline
\end{longtable}
 \begin{flushleft}
 $\,$
  \textbf{References: }
           $BVI_{c}K$ are taken from (1) \cite{pa_cargile2010c} and \cite{pa_cargile2009}, 
           $I_{c}\ z_{\rm\ CFHT12}K_{s}$ from (2) \cite{e_moraux2007}.
 \end{flushleft}  

\bigskip

\setcounter{table}{14}
\tiny
\begin{longtable}{l rlrl llc rlrl}
\caption{Photometry for the Hyades members retrieved from the literature.}\\
\hline
{Object} & 
\multicolumn{2}{c}{$I_{\rm\ CFHT12}$}&
\multicolumn{2}{c}{$z_{\rm\ CFHT12}$}&
\multicolumn{2}{c}{$J$}&
{Ref.} & 
\multicolumn{2}{c}{$H$}&
\multicolumn{2}{c}{$K$}
\\
        &
\multicolumn{2}{c}{[mag]} &
\multicolumn{2}{c}{[mag]} &
\multicolumn{2}{c}{[mag]} &
        &
\multicolumn{2}{c}{[mag]} &
\multicolumn{2}{c}{[mag]} 
\\
\hline
  Hya03      &   -   &   -  &   -   &   -  & 15.647& 0.076&(1)&    -  &   -  &  -    &  -   \\
  Hya08      &   -   &   -  &   -   &   -  & 15.463& 0.044&(1)&    -  &   -  &  -    &  -   \\
  Hya11      &   -   &   -  &   -   &   -  &   -   &  -   &   &    -  &   -  &  -    &  -   \\
  Hya09      &   -   &   -  &   -   &   -  &   -   &  -   &   &    -  &   -  &  -    &  -   \\
  Hya10      &   -   &   -  &   -   &   -  & 16.506& 0.059&(1)&    -  &   -  &  -    &  -   \\
  Hya12      &   -   &   -  &   -   &   -  & 16.778& 0.069&(1)&    -  &   -  &  -    &  -   \\
  2M0418     &   -   &   -  &   -   &   -  &   -   &  -   &   &    -  &   -  &  -    &  -   \\
\hline
  CFHT-Hy-1  &(12.87 & 0.06)& 12.33 & 0.06 &   -   &  -   &   &    -  &   -  & 10.37 & 0.05 \\
  CFHT-Hy-2  &(12.91 & 0.06)& 12.45 & 0.06 &   -   &  -   &   &    -  &   -  & 10.50 & 0.05 \\
  CFHT-Hy-4  &(13.35 & 0.06)& 12.78 & 0.06 &   -   &  -   &   &    -  &   -  & 10.74 & 0.05 \\
  CFHT-Hy-3  &(13.29 & 0.06)& 12.70 & 0.06 &   -   &  -   &   &    -  &   -  & 10.60 & 0.05 \\
  CFHT-Hy-10 &(14.17 & 0.06)& 13.57 & 0.06 &   -   &  -   &   &    -  &   -  & 11.49 & 0.05 \\
  CFHT-Hy-5  &(13.40 & 0.06)& 12.85 & 0.06 &   -   &  -   &   &    -  &   -  & 10.95 & 0.05 \\
  CFHT-Hy-7  &(13.53 & 0.06)& 13.01 & 0.06 &   -   &  -   &   &    -  &   -  & 11.03 & 0.05 \\
  CFHT-Hy-6  &(13.48 & 0.06)& 12.96 & 0.06 &   -   &  -   &   &    -  &   -  & 11.07 & 0.05 \\
  CFHT-Hy-8  &(13.59 & 0.06)& 13.07 & 0.06 &   -   &  -   &   &    -  &   -  & 11.16 & 0.05 \\
  CFHT-Hy-11 &(14.39 & 0.06)& 13.82 & 0.06 &   -   &  -   &   &    -  &   -  & 11.74 & 0.05 \\
  CFHT-Hy-13 &(14.52 & 0.06)& 13.81 & 0.06 &   -   &  -   &   &    -  &   -  & 11.51 & 0.05 \\
  CFHT-Hy-9  &(13.59 & 0.06)& 12.97 & 0.06 &   -   &  -   &   &    -  &   -  & 11.00 & 0.05 \\
  CFHT-Hy-12 &(14.44 & 0.06)& 13.80 & 0.06 &   -   &  -   &   &    -  &   -  & 11.74 & 0.05 \\
  CFHT-Hy-16 &(14.81 & 0.06)& 13.99 & 0.06 &   -   &  -   &   &    -  &   -  & 11.62 & 0.05 \\
  CFHT-Hy-15 &(14.78 & 0.06)& 14.07 & 0.06 &   -   &  -   &   &    -  &   -  & 12.02 & 0.05 \\
  CFHT-Hy-14 &(14.71 & 0.06)& 14.02 & 0.06 &   -   &  -   &   &    -  &   -  & 11.95 & 0.05 \\
  CFHT-Hy-17 &(14.83 & 0.06)& 14.12 & 0.06 &   -   &  -   &   &    -  &   -  & 12.07 & 0.05 \\
  CFHT-Hy-18 &(15.20 & 0.08)& 14.45 & 0.06 &   -   &  -   &   &    -  &   -  & 12.03 & 0.05 \\
  Hya01      &   -   &   -  &   -   &      &   -   &  -   &   &    -  &   -  &  -    &  -   \\
  CFHT-Hy-19 &(17.49 & 0.10)& 16.31 & 0.08 &   -   &  -   &   &    -  &   -  & 12.90 & 0.05 \\
  Hya06      &   -   &   -  &   -   &   -  & 15.462& 0.081&(1)&    -  &   -  &  -    &  -   \\
  Hya05      &   -   &   -  &   -   &   -  & 15.729& 0.071&(1)&    -  &   -  &  -    &  -   \\
  Hya02      &   -   &   -  &   -   &   -  &   -   &  -   &   &    -  &   -  &  -    &  -   \\
  Hya04      &   -   &   -  &   -   &   -  & 15.595& 0.044&(1)&    -  &   -  &  -    &  -   \\
  Hya07      &   -   &   -  &   -   &   -  & 15.956& 0.073&(1)&    -  &   -  &  -    &  -   \\
  CFHT-Hy-20 &(21.58 & 0.08)& 19.79 & 0.06 & 17.02 & 0.05 &(2)& 16.51 & 0.05 & 16.08 & 0.05 \\
  CFHT-Hy-21 &(22.16 & 0.10)& 20.80 & 0.08 & 18.48 & 0.05 &(2)& 17.36 & 0.05 & 16.59 & 0.05 \\
\hline
\end{longtable}
 \begin{flushleft}
 $\,$
  \textbf{Note: }photometric values in brackets correspond to cases 
  where we encountered problems in the SED reconstruction procedure.\\
 $\,$
  \textbf{References: } 
           $I_{\rm\ CFHT12}\ z_{\rm\ CFHT12}HK^{\prime}$ are taken from \citealt{j_bouvier2008a};
           $J$ is taken from two different works: (1) \citealt{n_lodieu2014c} and (2) \citealt{j_bouvier2008a}.
 \end{flushleft}  

%% file: tab_thehyades_astrometry_ref_BKM08_PLR17_HJC08_LBB14_v0.tex
\onecolumn
\tiny
\begin{longtable}{l rrrrc ccc c}
\caption{Quantities obtained from \cite{j_bouvier2008a} and \cite{a_perezgarrido2017}.}\\
\hline
{Object} & 
\multicolumn{2}{c}{$\mu_{\alpha} \cos(\delta)$} &
\multicolumn{2}{c}{$\mu_{\delta}$} &
{$\rho_{\mu_{\alpha} \ast \ \mu_{\delta}}$} &
{$d_{MAP}$} &
{$d_{min}$} &
{$d_{max}$} &
{Ref.} 
\\
{} & 
\multicolumn{2}{c}{[mas yr$^{-1}$]} &
\multicolumn{2}{c}{[mas yr$^{-1}$]} &
{} &
{[pc]} & 
{[pc]} &
{[pc]} &
{} 
\\
\hline
  2M0418     & 124 &  7 & -53 &  6 & - & 48.8 & 44.8 & 52.8 &(2)\\
\hline
  CFHT-Hy-1  & 111 & 10 &  -9 & 10 & - &   -  &   -  &   -  &(1)\\
  CFHT-Hy-2  & 123 & 10 & -14 & 10 & - &   -  &   -  &   -  &(1)\\
  CFHT-Hy-4  &  84 & 10 &   4 & 10 & - &   -  &   -  &   -  &(1)\\
  CFHT-Hy-3  & 114 & 10 & -65 & 10 & - &   -  &   -  &   -  &(1)\\
  CFHT-Hy-10 & 102 & 10 &  -7 & 10 & - &   -  &   -  &   -  &(1)\\
  CFHT-Hy-5  &  87 & 10 & -11 & 10 & - &   -  &   -  &   -  &(1)\\
  CFHT-Hy-7  & 106 & 10 & -17 & 10 & - &   -  &   -  &   -  &(1)\\
  CFHT-Hy-6  &  97 & 10 &   0 & 10 & - &   -  &   -  &   -  &(1)\\
  CFHT-Hy-8  & 103 & 10 &  -1 & 10 & - &   -  &   -  &   -  &(1)\\
  CFHT-Hy-11 &  82 & 10 &   0 & 10 & - &   -  &   -  &   -  &(1)\\
  CFHT-Hy-13 &  85 & 10 &  -1 & 10 & - &   -  &   -  &   -  &(1)\\
  CFHT-Hy-9  & 113 & 10 &  -7 & 10 & - &   -  &   -  &   -  &(1)\\
  CFHT-Hy-12 &  94 & 10 &  -7 & 10 & - &   -  &   -  &   -  &(1)\\
  CFHT-Hy-16 & 104 & 10 &  -6 & 10 & - &   -  &   -  &   -  &(1)\\
  CFHT-Hy-15 &  93 & 10 & -16 & 10 & - &   -  &   -  &   -  &(1)\\
  CFHT-Hy-14 &  99 & 10 & -57 & 10 & - &   -  &   -  &   -  &(1)\\
  CFHT-Hy-17 &  96 & 10 & -15 & 10 & - &   -  &   -  &   -  &(1)\\
  CFHT-Hy-18 & 123 & 10 & -38 & 10 & - &   -  &   -  &   -  &(1)\\
  CFHT-Hy-19 &  99 & 10 & -28 & 10 & - &   -  &   -  &   -  &(1)\\
  CFHT-Hy-20 & 135 & 10 & -9  & 10 & - &   -  &   -  &   -  &(1)\\
  CFHT-Hy-21 &  79 & 10 & -18 & 10 & - &   -  &   -  &   -  &(1)\\
\hline
\end{longtable}
 \begin{flushleft}
 $\,$
  \textbf{References: }
           (1) \citealt{j_bouvier2008a};
           (2) \citealt{a_perezgarrido2017}.\\
 \end{flushleft}   

\medskip

\setcounter{table}{16}
\tiny
\begin{longtable}{l ccrlrlc rrr r rrr}
\caption{Stellar properties for the Hyades LDB sample. Quantities obtained from \cite{e_hogan2008}.}\\
\hline
{Object} & 
\multicolumn{2}{c}{$\varpi$}&
\multicolumn{2}{c}{$\mu_{\alpha} \cos(\delta)$} &
\multicolumn{2}{c}{$\mu_{\delta}$} &
{$\rho_{\mu_{\alpha} \ast \ \mu_{\delta}}$} &
{$d_{MAP}$} &
{$d_{min}$} &
{$d_{max}$} &
{$T_{\mathrm{eff}}\ \chi^{2}$} &
{L$_{\rm bol}$} &
{$-\Delta L_{\rm bol}$} &     
{$+\Delta L_{\rm bol}$} \\    
      &
\multicolumn{2}{c}{[mas]} &
\multicolumn{2}{c}{[mas yr$^{-1}$]} &
\multicolumn{2}{c}{[mas yr$^{-1}$]} &
       &  
{[pc]} &
{[pc]} &
{[pc]} &
{[K]} &
{[$10^{-3} L_{\sun}$]} &
{[$10^{-3} L_{\sun}$]} &
{[$10^{-3} L_{\sun}$]} \\
\hline
  Hya03 & - & - & 102.46 & - &  -7.86 & - & - & 54.60 & 50.88 & 58.32 & 2\ 298 & 0.217 & 0.160 & 0.285 \\
  Hya08 & - & - &  82.99 & - & -19.34 & - & - & 44.80 & 41.12 & 48.48 & 1\ 993 & 0.171 & 0.106 & 0.254 \\
  Hya11 & - & - & 165.09 & - & -32.33 & - & - & 37.60 & 36.04 & 39.16 & 1\ 512 & 0.083 & 0.071 & 0.095 \\
  Hya09 & - & - &  72.04 & - & -22.32 & - & - & 57.80 & 52.44 & 63.16 & 2\ 000 & 0.174 & 0.142 & 0.209 \\
  Hya10 & - & - & 108.27 & - & -29.45 & - & - & 47.80 & 44.82 & 50.78 & 1\ 771 & 0.084 & 0.069 & 0.101 \\
  Hya12 & - & - & 112.04 & - & -17.86 & - & - & 41.40 & 38.84 & 43.96 & 1\ 811 & 0.061 & 0.045 & 0.080 \\
\hline  
  Hya01 & - & - & 148.98 & - & -46.41 & - & - & 36.20 & 34.57 & 37.83 & 2\ 384 & 0.247 & 0.220 & 0.276 \\
  Hya06 & - & - &  99.37 & - & -23.48 & - & - & 50.90 & 47.41 & 54.39 & 2\ 324 & 0.206 & 0.170 & 0.247 \\
  Hya05 & - & - & 125.14 & - & -24.32 & - & - & 46.60 & 44.04 & 49.16 & 2\ 000 & 0.189 &-0.058 & 0.496 \\
  Hya02 & - & - & 114.31 & - & -36.95 & - & - & 52.80 & 49.72 & 55.88 & 1\ 970 & 0.216 & 0.181 & 0.255 \\
  Hya04 & - & - &  82.71 & - & -21.25 & - & - & 54.40 & 49.94 & 58.86 & 1\ 986 & 0.219 & 0.180 & 0.263 \\
  Hya07 & - & - &  86.98 & - & -28.14 & - & - & 53.00 & 48.94 & 57.06 & 1\ 535 & 0.196 & 0.165 & 0.229 \\
\hline
\end{longtable}
 \begin{flushleft}
 $\,$
  \textbf{Note: } 
           The distances and proper motions were taken from \cite{e_hogan2008}. 
           Bolometric luminosities were calculated using these distances.\\ 
 \end{flushleft}  

\medskip

\setcounter{table}{17}
\tiny
\begin{longtable}{l ccrrrrc rrr r rrr}
\caption{Stellar properties for the Hyades LDB sample. Quantities obtained from \cite{n_lodieu2014c}.}\\
\hline
{Object} & 
\multicolumn{2}{c}{$\varpi$}&
\multicolumn{2}{c}{$\mu_{\alpha} \cos(\delta)$} &
\multicolumn{2}{c}{$\mu_{\delta}$} &
{$\rho_{\mu_{\alpha} \ast \ \mu_{\delta}}$} &
{$d_{MAP}$} &
{$d_{min}$} &
{$d_{max}$} &
{$T_{\mathrm{eff}}\ \chi^{2}$} &
{L$_{\rm bol}$} &
{$-\Delta L_{\rm bol}$} &     
{$+\Delta L_{\rm bol}$} \\    
{} & 
\multicolumn{2}{c}{[mas]} &
\multicolumn{2}{c}{[mas yr$^{-1}$]} &
\multicolumn{2}{c}{[mas yr$^{-1}$]} &
{} &  
{[pc]} &
{[pc]} &
{[pc]} &
{[K]}  &
{[$10^{-3} L_{\sun}$]} &
{[$10^{-3} L_{\sun}$]} &
{[$10^{-3} L_{\sun}$]} 
\\
\hline
  Hya03      & - & - & 103 & 10 & -5 &  9 & - & 62.1 & 57.2 & 66.8 & 2\ 298 & 0.281 & 0.202 & 0.374\\
  Hya08      & - & - &  89 &  9 &-15 &  9 & - & 57.9 & 53.3 & 62.3 & 1\ 993 & 0.286 & 0.179 & 0.419\\
  Hya11      & - & - & 158 & 18 &-44 & 12 & - & 62.4 & 57.0 & 68.4 & 1\ 512 & 0.228 & 0.178 & 0.292\\
  Hya09      & - & - &  -  &  - &  - &  - & - & 73.2 & 65.3 & 78.7 & 2\ 000 & 0.279 & 0.220 & 0.325\\
  Hya10      & - & - & 115 & 11 &-12 & 11 & - & 82.6 & 76.0 & 89.4 & 1\ 771 & 0.252 & 0.200 & 0.314\\
  Hya12      & - & - & 101 & 11 &-15 & 11 & - & 57.3 & 51.5 & 62.6 & 1\ 811 & 0.117 & 0.079 & 0.162\\
\hline    
  Hya01      & - & - &  -  &  - &  - &  - & - & 48.9 & 45.5 & 52.7 & 2\ 384 & 0.451 & 0.381 & 0.536\\
  Hya06      & - & - &  78 & 12 &-10 & 10 & - & 64.3 & 59.6 & 68.8 & 2\ 324 & 0.330 & 0.270 & 0.395\\
  Hya05      & - & - & 130 & 15 &  1 & 11 & - &173.1 &120.1 &177.1 & 2\ 000 & 2.620 &-0.437 & 6.437\\
  Hya02      & - & - & 123 & 10 &-28 &  8 & - & 57.7 & 52.7 & 62.6 & 1\ 970 & 0.258 & 0.203 & 0.320\\
  Hya04      & - & - &  70 & 10 &-25 & 11 & - & 67.1 & 62.2 & 71.8 & 1\ 986 & 0.334 & 0.279 & 0.392\\
  Hya07      & - & - &  94 & 15 &-40 & 15 & - &443.8 &328.8 &495.1 & 1\ 535 &13.761 & 7.441 &17.309\\
  CFHT-Hy-20 & - & - &  -  &  - &  - &  - & - & 28.8 & 28.5 & 29.1 & 1\ 070 & 0.016 &-0.006 & 0.039\\
  CFHT-Hy-21 & - & - &  -  &  - &  - &  - & - & 55.5 & 55.4 & 55.9 & 1\ 811 & 0.031 & 0.030 & 0.032\\
\hline
\end{longtable}
 \begin{flushleft}
 $\,$
  \textbf{Notes: }
           The proper motion and distances were taken from \cite{n_lodieu2014c}. 
           Bolometric luminosities were calculated using these distances.\\  
 \end{flushleft}  

%% file: tab_thehyades_astrometry_ref_MPL18_LSP19_v0.tex
\onecolumn
\begin{longtable}{l ccccrrc ccc r rrr}
\caption{Stellar properties for the Hyades LDB sample. Quantities obtained from \cite{el_martin2018}.}\\
\hline
{Object} & 
\multicolumn{2}{c}{$\varpi$}&
\multicolumn{2}{c}{$\mu_{\alpha} \cos(\delta)$} &
\multicolumn{2}{c}{$\mu_{\delta}$} &
{$\rho_{\mu_{\alpha} \ast \ \mu_{\delta}}$} &
{$d_{MAP}$} &
{$d_{min}$} &
{$d_{max}$} &
{$T_{\mathrm{eff}}\ \chi^{2}$} &
{$L_{\rm bol}$} &
{$-\Delta L_{\rm bol}$} &     
{$+\Delta L_{\rm bol}$} \\    
      &
\multicolumn{2}{c}{[mas]} &
\multicolumn{2}{c}{[mas yr$^{-1}$]} &
\multicolumn{2}{c}{[mas yr$^{-1}$]} &
       &  
{[pc]} &
{[pc]} &
{[pc]} &
{[K]}  &
{[$10^{-3} L_{\sun}$]} &
{[$10^{-3} L_{\sun}$]} &
{[$10^{-3} L_{\sun}$]} \\
\hline
  Hya03 & - & - & - & - & - & - & - & 48 & 39 & 57 & 2\,298 & 0.168 & 0.094 & 0.272 \\
  Hya08 & - & - & - & - & - & - & - & 45 & 36 & 54 & 1\,993 & 0.173 & 0.081 & 0.315 \\
  Hya11 & - & - & - & - & - & - & - & 50 & 43 & 57 & 1\,512 & 0.146 & 0.101 & 0.203 \\
  Hya09 & - & - & - & - & - & - & - & 57 & 49 & 65 & 2\,000 & 0.169 & 0.124 & 0.221 \\
  Hya10 & - & - & - & - & - & - & - & 58 & 50 & 66 & 1\,771 & 0.124 & 0.086 & 0.171 \\
  Hya12 & - & - & - & - & - & - & - & 56 & 48 & 64 & 1\,811 & 0.112 & 0.069 & 0.170 \\
\hline
\end{longtable}
 \begin{flushleft}
 $\,$
  \textbf{Notes: } 
           The distances were taken from \cite{el_martin2018}. 
           Bolometric luminosities were caluclated using these distances.\\
 \end{flushleft}  

\bigskip

\tiny
\begin{longtable}{l rrrrrrc ccr r rrr}
\caption{Stellar properties for the Hyades LDB sample. Quantities obtained from \cite{n_lodieu2019b}.}\\
\hline
{Object} & 
\multicolumn{2}{c}{$\varpi$}&
\multicolumn{2}{c}{$\mu_{\alpha} \cos(\delta)$} &
\multicolumn{2}{c}{$\mu_{\delta}$} &
{$\rho_{\mu_{\alpha} \ast \ \mu_{\delta}}$} &
{$d_{MAP}$} &
{$d_{min}$} &
{$d_{max}$} &
{$T_{\mathrm{eff}}\ \chi^{2}$} &
{$L_{\rm bol}$} &
{$-\Delta L_{\rm bol}$} &     
{$+\Delta L_{\rm bol}$} \\    
      &
\multicolumn{2}{c}{[mas]} &
\multicolumn{2}{c}{[mas yr$^{-1}$]} &
\multicolumn{2}{c}{[mas yr$^{-1}$]} &
       &  
{[pc]} &
{[pc]} &
{[pc]} &
{[K]}  &
{[$10^{-3} L_{\sun}$]} &
{[$10^{-3} L_{\sun}$]} &
{[$10^{-3} L_{\sun}$]} \\
\hline
  Hya11      & 39.1 &16.3 &138.1 &13.3 &-19.4 & 9.0 & - & 29 & 17 &192 & 1\ 512 & 0.052 & 0.016 & 2.308 \\
  Hya09      & 20.6 & 2.5 & 76.3 & 2.9 &-17.7 & 1.5 & - & 48 & 40 & 67 & 2\ 000 & 0.123 & 0.083 & 0.237 \\
  Hya10      & 28.5 & 3.9 &120.2 & 3.6 &-12.2 & 5.5 & - & 34 & 28 & 50 & 1\ 771 & 0.043 & 0.028 & 0.102 \\
  Hya12      & 24.1 & 2.1 &100.2 & 1.9 &-15.1 & 2.0 & - & 41 & 35 & 50 & 1\ 811 & 0.061 & 0.038 & 0.108 \\
  2M0418     & 25.8 & 2.9 &141.5 & 2.7 &-45.7 & 2.3 & - & 38 & 32 & 51 & 1\ 814 & 0.043 & 0.029 & 0.077 \\
\hline  
  Hya02      & 17.7 & 2.0 &116.4 & 2.0 &-26.9 & 1.5 & - & 56 & 47 & 75 & 1\ 970 & 0.250 & 0.164 & 0.461 \\
  CFHT-Hy-20 & 30.8 & 3.0 &141.3 & 2.9 &-14.5 & 3.2 & - & 32 & 27 & 41 & 1\ 070 & 0.020 &-0.006 & 0.079 \\
  CFHT-Hy-21 & 33.5 &12.7 & 82.1 & 9.8 &-15.5 & 8.6 & - & 32 & 20 &180 & 1\ 811 & 0.010 & 0.004 & 0.341 \\
\hline
\end{longtable}
 \begin{flushleft}
 $\,$
  \textbf{Notes: } 
           The parallaxes and proper motions were taken from \cite{n_lodieu2019b}. 
           Bolometric luminosities were calculated using these distances derived from these parallaxes.\\
 \end{flushleft}   

%% file: tab_32orimg_ids_v0.tex
\onecolumn
\begin{landscape}
\begin{longtable}{ll cc l cc}
\caption{Other identifiers for the 32 Ori MG LDB sample.}\\
\hline
{Object} & 
\multicolumn{1}{l}{Other id.} & 
\multicolumn{1}{l}{\textit{Gaia} DR2} & 
\multicolumn{1}{l}{Pan-STARRS DR1} & 
\multicolumn{1}{l}{DENIS} & 
\multicolumn{1}{l}{2MASS} & 
\multicolumn{1}{l}{WISE} \\ 
\hline
  THOR-14Ab & -                     & 3341625126275164032 & 124680838233826256 &                - & J05351761+1354180 & J053517.61+135417.5\\
  THOR-14Aa & -                     & 3341625121978020736 & 124680838233826256 &                - & J05351761+1354180 & J053517.61+135417.5\\
  THOR-27   & 1RXS J052748.7+064544 & 3238015740407127936 & 116110819523395693 &                - & J05274855+0645459 & J052748.55+064545.6\\
  THOR-17Ab & 1RXS J052743.4+144609 & 3389908598858685696 & 125720819297114362 &                - & J05274313+1446121 & J052743.14+144611.7\\
  THOR-17Aa & 1RXS J052743.4+144609 & 3389908598860532096 & 125720819297114362 &                - & J05274313+1446121 & J052743.14+144611.7\\
  THOR-18   & SCR0522-0606          & 3209175443035096320 & 100670806696252366 & J052240.7-060623 & J05224069-0606238 & J052240.71-060624.1\\
  THOR-15   & 1RXS J053306.7+140011 & 3389588293083264128 & 124810832739272523 &                - & J05330574+1400365 & J053305.75+140036.1\\
  THOR-20   & 1RXS J051930.4+103812 & 3386885525999518976 & 120760798726343064 &                - & J05192941+1038081 & J051929.42+103807.8\\
  THOR-21   & 1RXS J052515.0+003027 & 3221837251438378624 & 108600813132168138 & J052515.1+003023 & J05251517+0030232 & J052515.17+003022.9\\
  THOR-19   & 1RXS J052532.3+062534 & 3237973752806799360 & 115710813856941688 &               -  & J05253253+0625336 & J052532.54+062533.4\\
  THOR-23   & 1RXS J054424.7+050153 & 3320524914022156288 & 114040861019574221 &               -  & J05442447+0502114 & J054424.48+050211.4\\
  THOR-08B  & -                     & 3341556922194703232 & 124310843359170753 &               -  & J05372061+1335310 & J053720.61+133530.6\\
  THOR-04B  & -                     & 3221986647580887296 & 109470821687871498 &               -  & J05284050+0113333 & J052840.51+011333.1\\
  THOR-25   & -                     & 3387055125667571968 & 121150783596485041 &               -  & J05132631+1057439 & J051326.31+105743.7\\
  THOR-16   & 1RXS J054304.3+060646 & 3333228804511967488 & 115330857648411792 &               -  & J05430354+0606340 & J054303.55+060633.9\\
  THOR-42   & CRTS J055255.7-004426 & 3218460376351485056 & 107110882322151434 &               -  & J05525572-0044266 & J055255.72-004426.9\\
  THOR-14B  & 1RXS J053516.6+135404 & 3341625087618092672 & 124680838177560174 &               -  & J05351625+1353594 & J053516.26+135359.0\\
  THOR-22   & 1RXS J054926.3+040541 & 3319628434087776000 & 112910873597313009 &               -  & J05492632+0405379 & J054926.32+040537.6\\
  THOR-24   & -                     & 3237508762467678336 & 114700799332671108 &               -  & J05194398+0535021 & J051943.99+053501.8\\
  THOR-28   & -                     & 3337724677494561792 & 119360817036292205 &               -  & J05264886+0928055 & J052648.86+092805.2\\
  THOR-32   & -                     & 3334165249117507968 & 116640816697198918 &               -  & J05264073+0712255 & J052640.73+071225.3\\
  THOR-29   & 1RXS J052315.0+064412 & 3241048880671509888 & 116070808099508068 &               -  & J05231438+0643531 & J052314.38+064352.9\\
  THOR-31   & -                     & 3341238918520502912 & 123610841538702692 &               -  & J05363692+1300369 & J053636.92+130036.6\\
  THOR-26   & -                     & 3216793104408043520 & 104470826061711843 &               -  & J05302546-0256255 & J053025.47-025625.6\\
  THOR-30   & 1RXS J053132.6+055639 & 3237712515716135936 & 115140828871060328 &               -  & J05313290+0556597 & J053132.90+055659.4\\
  THOR-33   & -                     & 3216729878197029120 & 104320829911177914 & J053157.8-030336 & J05315786-0303367 & J053157.96-030338.3\\
  THOR-36   & 1RXS J052355.2+110110 & 3339253170454118272 & 121220809819151364 &               -  & J05235565+1101027 & J052355.65+110102.5\\
  THOR-35   & -                     & 3338707537810168448 & 119960794567877802 &               -  & J05174962+0958221 & J051749.63+095821.8\\
  THOR-40b  & -                     & 3341538436653898240 & 124190843750381886 &               -  & J05373000+1329344 & J053730.00+132934.1\\
  THOR-40a  & -                     & 3341538436655619840 & 124190843750381886 &               -  & J05373000+1329344 & J053730.00+132934.1\\
  J05053333 & -                     & 3228582201223411200 & 108880763889481502 & J050533.3+004403 & J05053333+0044034 & J050533.35+004403.2\\
  THOR-17B  & -                     & 3389908598860532224 & 125710819335229833 &               -  & J05274404+1445584 & J052744.05+144558.0\\
  THOR-37   & -                     & 3339935348702249472 & 121710837538854498 &               -  & J05350092+1125423 & J053500.93+112542.0\\
  THOR-38   & -                     & 3238048652740273408 & 116010811254372030 &               -  & J05243009+0640349 & J052430.10+064034.6\\
  THOR-39   & -                     & 3238063569162793472 & 116210817764702963 &               -  & J05270634+0650377 & J052706.34+065037.4\\
  THOR-34   & -                     & 3216753556349327232 & 104370830250595136 & J053205.9-030115 & J05320596-0301159 & J053205.95-030116.4\\
\hline
\end{longtable}
\end{landscape} 